# nEXO Pre-Conceptual Design Report

## Abstract

The projected performance and detector configuration of nEXO are described in this pre-Conceptual Design Report (pCDR). nEXO is a tonne-scale neutrinoless double beta ($0\nu\beta\beta$) decay search in $^{136}$Xe, based on the ultra-low background liquid xenon technology validated by EXO-200. With $\simeq 5000$ kg of xenon enriched to 90% in the isotope 136, nEXO has a projected half-life sensitivity of approximately $10^{28}$ years. This represents an improvement in sensitivity of about two orders of magnitude with respect to current results. Based on the experience gained from EXO-200 and the effectiveness of xenon purification techniques, we expect the background to be dominated by external sources of radiation. The sensitivity increase is, therefore, entirely derived from the increase of active mass in a monolithic and homogeneous detector, along with some technical advances perfected in the course of a dedicated R&D program. Hence the risk which is inherent to the construction of a large, ultra-low background detector is reduced, as the intrinsic radioactive contamination requirements are generally not beyond those demonstrated with the present generation $0\nu\beta\beta$ decay experiments. Indeed, most of the required materials have been already assayed or reasonable estimates of their properties are at hand. The details described herein represent the base design of the detector configuration as of early 2018. Where potential design improvements are possible, alternatives are discussed.

This design for nEXO presents a compelling path towards a next generation search for $0\nu\beta\beta$, with a substantial possibility to discover physics beyond the Standard Model.

# The nEXO Collaboration

S. Al Kharusi,[1] A. Alamre,[2] J.B. Albert,[3] M. Alfaris,[4] G. Anton,[5] I.J. Arnquist,[6] I. Badhrees,[2,a] P.S. Barbeau,[7] D. Beck,[8] V. Belov,[9] T. Bhatta,[10] F. Bourque,[11] J.P. Brodsky,[12] E. Brown,[13] T. Brunner,[1,14] A. Burenkov,[9] G.F. Cao,[15,b] L. Cao,[16] W.R. Cen,[15] C. Chambers,[17] S.A. Charlebois,[11] M. Chiu,[18] B. Cleveland,[19,c] R. Conley,[20] M. Coon,[8] M. Côté,[11] A. Craycraft,[17] W. Cree ,[2,d] J. Dalmasson,[21] T. Daniels,[22] D. Danovitch,[11] L. Darroch,[1] S.J. Daugherty,[3] J. Daughhetee,[10] R. DeVoe,[21] S. Delaquis,[20,e] A. Der Mesrobian-Kabakian,[19] M.L. Di Vacri,[6] J. Dilling,[14,23] Y.Y. Ding,[15] M.J. Dolinski,[24] A. Dragone,[20] J. Echevers,[8] L. Fabris,[25] D. Fairbank,[17] W. Fairbank,[17] J. Farine,[19] S. Ferrara,[6] S. Feyzbakhsh,[4] P. Fierlinger,[26] R. Fontaine,[11] D. Fudenberg,[21] G. Gallina,[23,14] G. Giacomini,[18] R. Gornea,[2,14] G. Gratta,[21] G. Haller,[20] E.V. Hansen,[24] D. Harris,[17] J. Hasi,[20] M. Heffner,[12] E.W. Hoppe,[6] J. Hößl,[5] A. House,[12] P. Hufschmidt,[5] M. Hughes,[27] Y. Ito,[1,f] A. Iverson,[17] A. Jamil,[28] C. Jessiman,[2] M.J. Jewell,[21] X.S. Jiang,[15] A. Karelin,[9] L.J. Kaufman,[20,g] C. Kenney,[20] R. Killick,[2] D. Kodroff,[4] T. Koffas,[2] S. Kravitz,[21,h] R. Krücken,[23,14] A. Kuchenkov,[9] K.S. Kumar,[29] Y. Lan,[23,14] A. Larson,[10] B.G. Lenardo,[21] D.S. Leonard,[30] C.M. Lewis,[4,i] G. Li,[21] S. Li,[8] Z. Li,[28] C. Licciardi,[19] Y.H. Lin,[24] P. Lv,[15] R. MacLellan,[10] K. McFarlane,[2] T. Michel,[5] B. Mong,[20] D.C. Moore,[28] K. Murray,[1] R.J. Newby,[25] T. Nguyen,[1] Z. Ning,[15] O. Njoya,[29] F. Nolet,[11] O. Nusair,[27] K. Odgers,[13] A. Odian,[20] M. Oriunno,[20] J.L. Orrell,[6] G.S. Ortega,[6] I. Ostrovskiy,[27] C.T. Overman,[6] S. Parent,[11] M. Patel,[21] A. Peña-Perez,[20] A. Piepke,[27] A. Pocar,[4] J.-F. Pratte,[11] D. Qiu,[16] V. Radeka,[18] E. Raguzin,[18] T. Rao,[18] S. Rescia,[18] F. Retière,[14] A. Robinson,[19] T. Rossignol,[11] P.C. Rowson,[20] N. Roy,[11] J. Runge,[7] R. Saldanha,[6] S. Sangiorgio,[12] S. Schmidt,[5] J. Schneider,[5] A. Schubert,[21,j] J. Segal,[20] K. Skarpaas VIII,[20] A.K. Soma,[27] K. Spitaels,[4,k] G. St-Hilaire,[11] V. Stekhanov,[9] T. Stiegler,[12] X.L. Sun,[15] M. Tarka,[4] J. Todd,[17] T. Tolba,[15] T.I. Totev,[1] R. Tsang,[6] T. Tsang,[18] F. Vachon,[11] B. Veenstra,[2] V. Veeraraghavan,[27] G. Visser,[3] P. Vogel,[31] J.-L. Vuilleumier,[32] M. Wagenpfeil,[5] Q. Wang,[16] M. Ward,[14] J. Watkins,[2] M. Weber,[21] W. Wei,[15] L.J. Wen,[15] U. Wichoski,[19] G. Wrede,[5] S.X. Wu,[21] W.H. Wu,[15] Q. Xia,[28] L. Yang,[8] Y.-R. Yen,[24] O. Zeldovich,[9] X. Zhang,[15,l] J. Zhao,[15] Y. Zhou,[16] and T. Ziegler[5]

[1] *Physics Department, McGill University, Montréal, Québec H3A 2T8, Canada*
[2] *Department of Physics, Carleton University, Ottawa, Ontario K1S 5B6, Canada*
[3] *Department of Physics and CEEM, Indiana University, Bloomington, IN 47405, USA*
[4] *Amherst Center for Fundamental Interactions and Physics Department, University of Massachusetts, Amherst, MA 01003, USA*

[a] Also at King Abdulaziz City for Science and Technology, Riyadh, Saudi Arabia
[b] Also at University of Chinese Academy of Sciences, Beijing, China
[c] Also at SNOLAB, Ontario, Canada
[d] Now at Canadian Department of National Defense
[e] Deceased
[f] Now at JAEA, Ibaraki, Japan
[g] Also at Indiana University, Bloomington, IN, USA
[h] Now at Lawrence Berkeley National Lab, Berkeley, CA, USA
[i] Now at University of Chicago, Chicago, IL, USA
[j] Now at OneBridge Solutions, Boise, ID, USA
[k] Now at Raytheon
[l] Now at Tsinghua University, Beijing, China

[5] *Erlangen Centre for Astroparticle Physics (ECAP), Friedrich-Alexander University Erlangen-Nürnberg, Erlangen 91058, Germany*
[6] *Pacific Northwest National Laboratory, Richland, WA 99352, USA*
[7] *Department of Physics, Duke University, and Triangle Universities Nuclear Laboratory (TUNL), Durham, NC 27708, USA*
[8] *Physics Department, University of Illinois, Urbana-Champaign, IL 61801, USA*
[9] *Institute for Theoretical and Experimental Physics named by A. I. Alikhanov of National Research Center "Kurchatov Institute", Moscow 117218, Russia*
[10] *Department of Physics, University of South Dakota, Vermillion, SD 57069, USA*
[11] *Université de Sherbrooke, Sherbrooke, Québec J1K 2R1, Canada*
[12] *Lawrence Livermore National Laboratory, Livermore, CA 94550, USA*
[13] *Department of Physics, Applied Physics and Astronomy, Rensselaer Polytechnic Institute, Troy, NY 12180, USA*
[14] *TRIUMF, Vancouver, British Columbia V6T 2A3, Canada*
[15] *Institute of High Energy Physics, Chinese Academy of Sciences, Beijing 100049, China*
[16] *Institute of Microelectronics, Chinese Academy of Sciences, Beijing 100029, China*
[17] *Physics Department, Colorado State University, Fort Collins, CO 80523, USA*
[18] *Brookhaven National Laboratory, Upton, NY 11973, USA*
[19] *Department of Physics, Laurentian University, Sudbury, Ontario P3E 2C6 Canada*
[20] *SLAC National Accelerator Laboratory, Menlo Park, CA 94025, USA*
[21] *Physics Department, Stanford University, Stanford, CA 94305, USA*
[22] *Department of Physics and Physical Oceanography, University of North Carolina at Wilmington, Wilmington, NC 28403, USA*
[23] *Department of Physics and Astronomy, University of British Columbia, Vancouver, British Columbia V6T 1Z1, Canada*
[24] *Department of Physics, Drexel University, Philadelphia, PA 19104, USA*
[25] *Oak Ridge National Laboratory, Oak Ridge, TN 37831, USA*
[26] *Technische Universität München, Physikdepartment and Excellence Cluster Universe, Garching 80805, Germany*
[27] *Department of Physics and Astronomy, University of Alabama, Tuscaloosa, AL 35487, USA*
[28] *Department of Physics, Yale University, New Haven, CT 06511, USA*
[29] *Department of Physics and Astronomy, Stony Brook University, SUNY, Stony Brook, NY 11794, USA*
[30] *IBS Center for Underground Physics, Daejeon 34126, Korea*
[31] *Kellogg Lab, Caltech, Pasadena, CA 91125, USA*
[32] *LHEP, Albert Einstein Center, University of Bern, Bern CH-3012, Switzerland*

# Acknowledgments

This work has been supported by the Offices of Nuclear and High Energy Physics within DOE's Office of Science, and NSF in the United States; by NSERC, CFI, FRQNT, NRC, and the McDonald Institute (CFREF) in Canada; by SNF in Switzerland; by IBS in Korea; by RFBR in Russia; and by CAS and ISTCP in China. This work was supported in part by Laboratory Directed Research and Development (LDRD) programs at Brookhaven National Laboratory (BNL), Lawrence Livermore National Laboratory (LLNL), Oak Ridge National Laboratory (ORNL), and Pacific Northwest National Laboratory (PNNL).

# Contents

# 1 Introduction

Neutrinoless double-beta ($0\nu\beta\beta$) decay has been recognized to be one of the most sensitive avenues in the search for physics beyond the Standard Model [1–8]. Indeed the observation of $0\nu\beta\beta$ decay would, in one stroke, discover lepton number violation and elementary Majorana fermions. Since the decay also requires neutrino masses to be non-zero, the recent discovery of neutrino oscillations [9–15] has increased the interest in the search for this process worldwide. Most recently, in the US the search for $0\nu\beta\beta$ decay was assigned the highest priority for new initiatives by the 2015 Long Range Planning [1] for the nuclear physics community.

Over the last four years, the nEXO collaboration has developed a conceptual design of the nEXO detector, a tonne-scale $0\nu\beta\beta$ decay experiment conceived to push the search to the next frontier. This process was enhanced by some initial R&D, aiming to retire much of the risk inherent to the scaling of the EXO-200 design to the tonne-scale. At this time the collaboration is comfortable that the factor of $\sim 25$ increase in xenon mass represents a natural progression, achieving the sensitivity standard set by the community [1], while, at the same time, providing a conservative design.

This pre-Conceptual Design Report (pCDR) aims to describe the result of this process, which involved substantial work of performance optimization, instrumentation R&D, materials characterization and simulation. The pCDR is organized as follows: Section 2 provides the physics motivation for nEXO; Section 3 discusses the overarching choices made in designing the nEXO detector, along with the connections with the very successful precursor, EXO-200, and the background model resulting in the projected sensitivity; Section 4 describes the detector design in substantial detail; Section 5 discusses general aspects of the conventional facilities supporting the detector. Section 6 describes the tools and the process that are necessary to screen materials for radioactive contaminations and ensure that the detector will comply with the very strict background requirements. While much of this measurement infrastructure has been already exercised to produce the design described here, projections of the sensitivity and throughput required to execute the project are also provided. Section 7 frames the issues related to the procurement of the isotopically enriched xenon for the experiment, and Section 8 describes the preliminary plans for the handling and preservation of the data collected by nEXO. Finally, a brief description of the remaining R&D required before finalizing the baseline design is offered in Section 9.

While engineering of the various hardware components will be done in the future, and it is expected that details will change, sufficient work has been done to validate the plausibility of all components described. The resulting background model, described in Section 3.3 and, in more detail in a separate paper [16], is derived from an inventory of materials (see Section 6.7) reflecting contaminations actually measured for most entries. Material quantities and locations in the detector are informed by the engineering. This results in a model which, while not final, is complete and conservative by construction.

## 2 Searching for New Physics with $0\nu\beta\beta$ decay

The discovery of neutrino oscillations has provided the first hint of new physics since the establishment of the Standard Model (SM) of strong, weak and electromagnetic interactions. The resulting unambiguous conclusion that at least two neutrinos have non-zero masses has profound ramifications. As the only electrically neutral elementary fermions, neutrinos may have two distinct terms in the Lagrangian formalism describing their mass. These two mass terms include the conventional Dirac mass term as well as a Majorana mass term. The latter is associated with 2-component, spin $1/2$ fermions for which the antiparticle state coincides with the particle state. Such a possibility requires violation of lepton number conservation, which has not yet been observed in experimental measurements to date. Experimental efforts are now positioned to determine the contribution of the Dirac and Majorana mass terms in the theoretical description of neutrinos. Regardless of the outcome, answering this question is of great interest to a number of subfields of physics research. In one case, the presence of a Majorana mass term would imply additional physics beyond the SM—a major discovery. Alternatively, concluding the conventional Dirac mass term is the only contributing mass term would be extraordinary—essentially enforcing lepton number conservation for "accidental reasons," unrelated to any underlying physical symmetry. Nature has provided a nuclear decay process, $0\nu\beta\beta$ decay, that provides one of the most sensitive probes of these profound questions in nuclear and particle physics research.

At the same time, it is stunning that the same nuclear decay process, for which the current limit on the half-life is $\sim 10^{16}$ times longer than the estimated age of the universe, has the potential to provide a window into the nature of the universe in the earliest moments after the Big Bang. From the equal parts matter and antimatter produced according to the Big Bang hypothesis, a matter dominated universe has resulted from some process creating an imbalance of matter to antimatter on the scale of approximately 1 part in 10 billion. A potential explanation of these circumstances was outlined by Andrei Sakharov in 1967 [1], providing three conditions required to produce a matter dominated universe. At the time, neutrinos—and their fundamental nature—were not considered part of Sakharov's ideas. However, as the understanding of the SM of particle physics has improved dramatically in the last five decades, the role of neutrinos in cosmic evolution has become more compelling. Notional theories developed to explain the smallness of neutrino mass have solidified into the basis for considering a fundamentally new structure of mass generation beyond the Yukawa couplings of fermions to the Higgs field. While the identification of a new mass generation mechanism for neutrinos is itself scientifically revolutionary, it may also be inextricably linked to the explanation of the matter dominated universe. Seeking experimental evidence in support of these concepts and connections is a key driver for the nuclear physics community searching for $0\nu\beta\beta$ decay. The tools and techniques of nuclear physics research have the potential to open a new window into the understanding of the neutrino as a fundamental particle and may provide a basis for an improved understanding of why the universe is matter dominated. While many caveats exist to this chain of hypotheses, researchers are pursuing neutrino science on all fronts as the last two decades have revealed that neutrinos are far more intriguing than implied by the simple description of massless neutrinos in the SM.

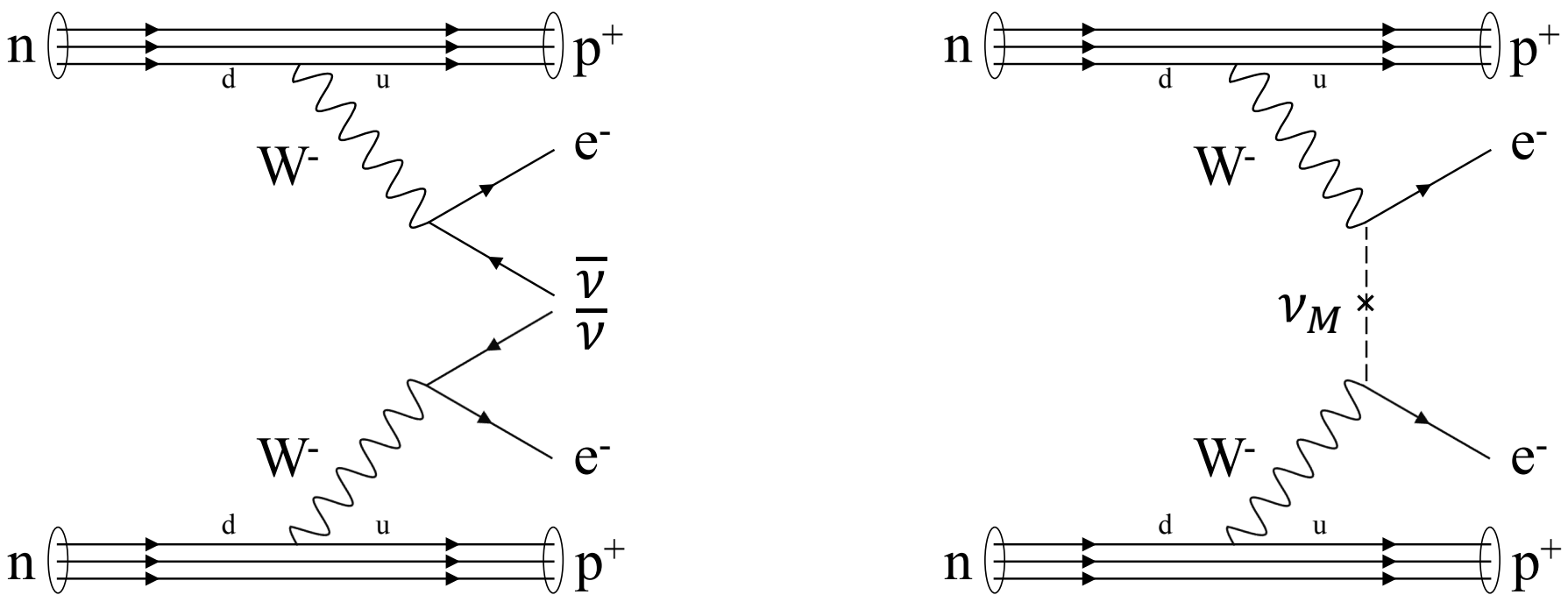

**Figure 2.1:** Adapted from [3], these Feynman diagrams represent two-neutrino double beta ($2\nu\beta\beta$) decay (left) and neutrinoless double beta ($0\nu\beta\beta$) decay (right). The $2\nu\beta\beta$ decay process conserves both baryon and lepton number separately, adhering to the precepts of the SM, and has been observed in about a dozen isotopes. The $0\nu\beta\beta$ decay process, however, is explicitly a leptogenic process, creating two out-going leptons, and is only possible if neutrinos are Majorana-type particles. The $0\nu\beta\beta$ decay process has not been observed, with current searches setting half-life limits generally greater than $10^{25}$ years.

The search for $0\nu\beta\beta$ decay is now at an advanced stage of development and potentially able to shed light on a number of these broad scientific questions. As evident from Figure 2.1, $0\nu\beta\beta$ decay would immediately demonstrate the explicit existence of lepton number violation. In addition, its observation would confirm the existence of Majorana neutrinos, regardless of the decay mechanism [2]. Such a discovery would be a fundamental departure from the picture provided by the SM, as all experimental measurements to date have shown lepton number is a conserved quantity. Next generation $0\nu\beta\beta$ decay experiments, grounded in the methods of nuclear physics research, may be on the verge of a discovery that would reshape the understanding of neutrinos, their role in the SM, and their impact on the history of the evolution of the universe. The following sub-sections will expand on these fascinating connections between different areas of physics and on the prospects for future $0\nu\beta\beta$ decay experiments in nuclear physics research.

## 2.1 Theory Underpinnings

The experimental observation of neutrino oscillations [4–6], which require that neutrinos have mass [7], indicates that the SM's description of neutrinos must be extended. As the SM only includes neutrinos as massless [8], the "simple" extension to minimally accommodate neutrino masses adds a Dirac mass term for the neutrino in the SM Lagrangian, $\mathcal{L}^D$, as:

$$-\mathcal{L}^D = m_D\, \overline{\nu_{\ell,R}}\, \nu_{\ell,L} + \text{h.c.} \quad \text{with} \quad \ell = e, \mu, \tau \tag{2.1}$$

where the neutrino and antineutrino 4-spinor fields $\nu_{\ell,L}$ and $\overline{\nu_{\ell,R}}$ are left- and right-handed chiral projections, respectively. Written as such, massive Dirac-type neutrinos require—through the Hermitian conjugate (h.c.) term—the introduction of experimentally unobserved and non-interacting right-handed neutrinos $\nu_{\ell,R}$[1]. These Dirac-type neutrinos gain their mass, $m_D$, through the same Higgs coupling[2] as the other fermions in the SM, but leave the relative smallness of the neutrino masses at best unexplained, and at worst at odds with expectations of their Higgs coupling. In fact, current experimental constraints require that the neutrino masses are at least $10^6$ times smaller than the lowest mass charged lepton, i.e. the electron [7]. Furthermore, taken at face value, an order unity Higgs coupling suggests neutrino masses approximately $10^{13}$ larger than allowed by experimental evidence [10]. These features (non-interacting states and unexplained small relative mass) make the introduction of massive Dirac-type neutrinos theoretically unappealing from perspectives of parsimony and naturalness.

However, the introduction of Majorana-type particle fermions provides a theoretically compelling "solution" to the need to incorporate massive neutrinos into the SM. Considering only a single generation (and suppressing the $\ell$ index for simplicity), an additional Majorana mass term, with coefficient $m_R$, can be added to the Lagrangian:

$$-2\mathcal{L}^M = m_R\,\overline{(\nu_R)^c}\,\nu_R + \text{h.c.} \tag{2.2}$$

where $(\nu_R)^c$ is the charge-symmetry conjugate of the right-handed neutrino field. Note that a similar term for $\nu_L$ does not appear because the corresponding Majorana mass term would not satisfy gauge invariance [11, 12]. Explicitly including also the charge-symmetry conjugate fields for the Dirac mass term, and grouping the terms appearing in the Hermitian conjugate for simplicity, gives the most general set of neutrino mass terms in the Lagrangian:

$$-2\mathcal{L} = m_D\,\overline{\nu_L}\,\nu_R + m_D\,\overline{(\nu_R)^c}\,(\nu_L)^c + m_R\,\overline{(\nu_R)^c}\,\nu_R + \text{h.c.} \tag{2.3}$$

which can be rewritten in matrix form:

$$-2\mathcal{L} = \left(\overline{\nu_L}, \overline{(\nu_R)^c}\right)\begin{pmatrix} 0 & m_D \\ m_D & m_R \end{pmatrix}\begin{pmatrix} (\nu_L)^c \\ \nu_R \end{pmatrix} + \text{h.c.} \tag{2.4}$$

The physical masses are the eigenvalues of the $2 \times 2$ mass matrix of the preceding equation, as given by:

$$m_\pm = \frac{1}{2}\left(m_R \pm \sqrt{m_R^2 + 4m_D^2}\right) \tag{2.5}$$

Further making the assumption $m_R \gg m_D$, results in the following values[3]:

$$|m_-| = \frac{m_D^2}{m_R} \quad \text{and} \quad m_+ = m_R\left(1 + \frac{m_D^2}{m_R^2}\right) \approx m_R \tag{2.6}$$

This would explain both the absence of observational evidence for a heavy, right-handed neutrino and dramatically suppress the $m_-$ mass. Choosing the mass scale for $m_R$ to be $10^{14}$–$10^{16}$ GeV, close to the grand unification (GUT) scale, and $m_D \sim 100$ GeV results in $m_-$ residing roughly in the observed neutrino mass splitting range of $\sim$ 1–100 meV. This effect of a very massive, right-handed neutrino state driving down the mass of the lower mass, left-handed neutrino state is colloquially referred to as the "see-saw" mechanism [13–15].

[1] And, likewise, experimentally unobserved and non-interacting left-handed antineutrinos $\overline{\nu}_{\ell,L}$.

[2] Dirac neutrino mass acquired through neutrino Yukawa coupling, $Y_\nu$, to the Higgs field: $m_D = Y_\nu\langle H\rangle$ and $\langle H\rangle =$ 174 GeV [9].

[3] Although the eigenvalue $m_-$ is negative, the unphysical negative mass can be avoided by considering the states $\gamma_5\psi$ instead of $\psi$. They obey the same Dirac equation but with $m_-$ replaced by $-m_-$.

Although this picture of adding massive Majorana-type neutrinos to the SM is presented as a "simple" theoretical extension, the resulting theory contains several fundamentally new concepts. First, the neutrino mass generation mechanism is no longer directly tied to their Higgs coupling, but also involves an entirely new mass scale, $m_R$. Second, Majorana-type fermions are an entirely new class of fundamental particle type, since all other fundamental fermions have distinct particle and antiparticle species.

Observational evidence for these new features of particle physics theory is directly accessible if the process of $0\nu\beta\beta$ decay is observed, since this nuclear decay process is only possible when neutrinos are Majorana-type particles. To reiterate the caption of Figure 2.1, discovery of $0\nu\beta\beta$ decay is a leptogenic phenomenon, violating lepton number conservation through the explicit creation of a net increase in lepton number. Looking beyond the immediate consequences of the discovery of $0\nu\beta\beta$ decay described above, the see-saw mechanism for neutrino mass generation has the potential of connecting to an explanation of the matter asymmetry of the universe, as will be discussed in the next section.

The PMNS neutrino mixing matrix [16], $U$, describing the relationship between the neutrino flavor eigenstates and the neutrino mass eigenstates generically contains a CP-violating term, $\delta_{\mathrm{CP}}$:

$$\begin{pmatrix}\nu_e\\ \nu_\mu\\ \nu_\tau\end{pmatrix} = U\begin{pmatrix}\nu_1\\ \nu_2\\ \nu_3\end{pmatrix} = \begin{pmatrix} c_{12}c_{13} & s_{12}c_{13} & s_{13}e^{-i\delta_{\mathrm{CP}}} \\ -s_{12}c_{23} - c_{12}s_{23}s_{13}e^{i\delta_{\mathrm{CP}}} & c_{12}c_{23} - s_{12}s_{23}s_{13}e^{i\delta_{\mathrm{CP}}} & s_{23}c_{13} \\ s_{12}s_{23} - c_{12}c_{23}s_{13}e^{i\delta_{\mathrm{CP}}} & -c_{12}s_{23} - s_{12}c_{23}s_{13}e^{i\delta_{\mathrm{CP}}} & c_{23}c_{13}\end{pmatrix} P \begin{pmatrix}\nu_1\\ \nu_2\\ \nu_3\end{pmatrix} \tag{2.7}$$

Here $c_{ij} = \cos\theta_{ij}$, $s_{ij} = \sin\theta_{ij}$, and $\theta_{ij}$ are the mixing angles between the neutrino mass states, of which there are three unique and independent values ($\theta_{12}$, $\theta_{23}$, $\theta_{13}$). The matrix, $P$, depends on the Dirac or Majorana nature of the neutrino, using the unit matrix for Dirac neutrinos or for Majorana neutrinos a $3 \times 3$ matrix with the three diagonal elements 1, $e^{i\alpha}$, $e^{i\beta}$, and off-diagonal elements equal to zero. It is unknown whether the CP-violating term in the neutrino mixing matrix [17] or the additional Majorana phases are either relevant [18, 19] or "large enough" to produce the required conditions identified by Sakharov to explain the matter-antimatter asymmetry in the universe and discussed in the next section. However, current best fit neutrino data suggests the neutrino sector may have 1000 times greater CP-asymmetry than is present in the quark sector, as studied by the B-factories[4].

The foregoing description is a so-called Type-I see-saw mechanism relying on the introduction of heavy right-handed neutrinos, and for which $0\nu\beta\beta$ decay is mediated by "light" neutrinos. Assuming this simple (Type-I) see-saw mechanism of Majorana neutrinos described above and the formalism of neutrino oscillations, an "effective Majorana mass," $\langle m_{\beta\beta}\rangle$, is defined as:

$$\langle m_{\beta\beta}\rangle = \left| \, m_1 c_{12}^2 c_{13}^2 + m_2 s_{12}^2 c_{13}^2 e^{i\alpha} + m_3 s_{13}^2 e^{i\beta} \, \right| \tag{2.8}$$

following the same notation as Equation 2.7, including the three neutrino mass eigenvalues given as $m_1$, $m_2$, and $m_3$ [23], and where the unknown phase $\beta$ is defined to also incorporate the effect of $\delta_{CP}$. The effective Majorana mass is a key parameter directly related to the underlying neutrino theory and is connected to the nuclear physics observable relevant to $0\nu\beta\beta$ decay, as described in Section 2.3.

[4] The Jarlskog invariant [20], which quantifies the level of CP-violation present in the quark sector, is now determined at the $\sim 7\%$-level. Its value of $J = 3.04 \times 10^{-5}$ [9] can be compared to the theoretical maximum of $1/(6\sqrt{3}) \sim 0.1$. The equivalent neutrino sector Jarlskog invariant measure of CP-violation [21] is near maximal [17], with magnitude of order $3 \times 10^{-2}$ [22].

**Other theoretical scenarios** The Type-I see-saw mechanism for neutrino mass generation described above is generally considered the most economical possibility. It relies on a fermionic singlet (heavy, right-handed neutrino) to describe the mechanism in concrete particle physics terms. In the limit where $m_R \gg m_D$, the only effect of the new right handed states is to generate a local dimension-5 operator [24] suppressed by one power of $m_R$, which after electroweak symmetry breaking reduces to a Majorana mass term for the neutrinos with $m_\nu \sim (m_D)^2/m_R$.

However, the exchange of a virtual light Majorana neutrino between the two nucleons undergoing the transition, described by Equation 2.9, is only one of many possibilities to generate such a gauge-invariant lepton-number-violating dimension-5 operator. Other theoretical prescriptions include Type-II (scalar (Higgs) triplet) and Type-III (fermionic triplet) [25] see-saw mechanisms. These models have the same essential result of relying on Majorana mass terms and Majorana-type particles to augment the SM neutrino sector, resulting in a relatively small value for the lowest mass neutrino state, as observed by experiments.

An additional set of contributions to $0\nu\beta\beta$ could arise by relaxing the assumption that the scale where lepton number is violated (i.e., the lepton number violation [LNV] scale) is well above the electroweak scale. Embedding see-saw scenarios into complete high-energy models (e.g., supersymmetric theories with R-parity violation or left–right symmetric theories) typically results in many new contributions to $0\nu\beta\beta$ [26]. If the LNV scale is in the range 1–100 TeV these effects can compete with light Majorana neutrino exchange, and in certain cases can substantially enhance the expected rate of $0\nu\beta\beta$. Finally, one should also consider the possibility that lepton number might be violated at a scale much lower than the weak scale—in the extreme case corresponding to light sterile Majorana neutrinos. In this case, the light sterile exchange cannot be neglected and potentially affects the phenomenology in a dramatic way [27].

Though the properties of the particle spectrum are unique to each model, the connection to cosmology is the same through the see-saw model's ability to provide a means for an excess of lepton number, which is then converted to an excess of baryon number. In addition, the "black-box" Schechter-Valle theorem [2, 28] proves any observation of $0\nu\beta\beta$ decay also implies the existence of Majorana neutrino mass terms in the formalism and hence new physics.

## 2.2 Connections with Cosmology

The matter-dominated nature of the universe is a puzzling outcome following a naïve evaluation of two considerations. The simple explanation follows the logic of (1) the Big Bang produced equal parts matter and antimatter and (2) as the universe expanded and cooled, matter and antimatter would annihilate into electromagnetic radiation. Following these initial considerations to their logical conclusion, one expects a universe devoid of matter (or antimatter), counter to objective reality. Andrei Sakharov proposed three conditions required to allow for the evolution of a matter dominated universe:

- Violation of baryon number conservation
- Violation of both charge (C) and charge-parity (CP) symmetries
- Interactions occurring out of thermal equilibrium

Modern physics research has shown that all three conditions can be present in the frameworks of the SM and the description of the universe's cosmological evolution. While direct violation of baryon number conservation has never been observed in the laboratory[5], the details of the electroweak formalism describing the unification of electromagnetism and weak-nuclear phenomenon provide an accepted basis for believing that baryon number (and lepton number) changing interactions are nevertheless possible at high energies. This is described as a sphaleron [29] process allowing (at minimum) the conversion of three baryons into three antileptons (or three antibaryons into three leptons), as the quantum number composed of the number of baryons *minus* the number of leptons, remains conserved even at these high energies [19, 30].

Early experimental work studying parity (P) symmetry violation in weak decays in 1957 [31–33] implied the presence of C-symmetry violation through a theorem from Lee and Yang [34]. Later in 1964 studies of the neutral K-mesons showed the existence of CP-symmetry violation [35]. Together these existing observations demonstrated the second of Sakharov's three conditions is also present in Nature. It is worth noting these observations in mesons (two-quark composite particles) are tied to fundamental symmetries governing strong-force hadrons, implying baryons (three-quark composite particles) share the same underlying symmetry properties.

Regarding Sakharov's third condition, a simple way of understanding the "out-of-thermal-equilibrium" requirement is to consider an initial high-energy density state of fixed energy, expanding adiabatically. Initially a full range of energetic particle (and antiparticle) states are kept in equilibrium through annihilation and creation reactions, thereby maintaining equal populations of particle and antiparticle states[6]. However, as the universe expands, the energy density is reduced through expansion. If particle (or antiparticle) decay rates are *slower* than the expansion, then the particle (or antiparticle) decay properties will become the determining factor in the resulting populations of particles and antiparticles. While the details of this model are uncertain, the general concept of the rapid expansion of the early universe allowing for out-of-thermal-equilibrium conditions to exist is supported by a number of plausible cosmology scenarios developed in concert with high-energy particle theory.

By the arrival of the 1980's it appeared that the SM and the cosmology of the Big Bang hypothesis had all the ingredients *necessary* for producing a matter dominated universe. However, experimental research on K-mesons showed the magnitude of the CP-violation present in K-meson systems was not *sufficient* to explain the matter-antimatter asymmetry of the universe [38–40]. Three plausible additional sources of CP-violation were proposed as extensions to the CP-violation of the K-mesons, and a thorough ("over-constrained") search of the Cabibbo-Kobayashi-Maskawa (CKM) matrix[7] was proposed to explore these other sources of CP-violation in the SM [41]. This was the launching point for a major program in B-meson physics [42], resulting in construction in the 1990's of the BaBar and Belle detectors. These several hundred million dollar experimental investments [43] were in part designed to definitively test whether sufficient CP-violation was present in the quark sector to generate a matter dominated universe [44, 45]. However, the culmination of two decades of research at B-meson factories indicate the observed CP-violation is not sufficient to account for the observed baryon asymmetry [46]. Thus the search continues for a process to fulfill the requisite magnitude of CP-violation, potentially in some other area of the SM, such as the lepton sector where neutrinos remain less than fully understood. In fact, the last two decades of developments in neutrino physics research have revealed a number of intriguing results. The simple, massless neutrinos of the SM have actually been found to (1) be massive particles, (2) violate lepton-*flavor* conservation, and (3) potentially contain CP violation hidden within the complete formal description of flavor-oscillating, massive neutrinos. Quite possibly the neutrino will also provide clues to the matter-antimatter asymmetry in the universe.

[5]There is no experimental evidence for direct violation of baryon number conservation from proton decay experiments or from the Large Hadron Collider (LHC) below about 13 TeV.

[6]Assuming zero initial matter-antimatter asymmetry is not required [36, 37] due to an effect known as "washout".

[7]The CKM matrix describes the flavor-changing nature of weak interactions of quarks in the SM [9].

**Caveats** Theoretical considerations have shown that CP-violation in the laboratory (*i.e.*, at "low energy") may not be related to the CP-violation needed at early times in the universe to satisfy Sakharov's second condition. Specifically, low energy ($<$ 13 TeV) searches for CP-violation in the neutrino sector may have no relation to CP-violation in the GUT-scale neutrino theory describing heavy, right-handed neutrinos [19]. Detailed theoretical evaluations of leptogenesis suggest there is likely no connection between the lower energy $\delta_{\mathrm{CP}}$ phase in the PMNS neutrino mixing matrix and the CP-violation required in the early moments of the universe, since that would be relevant to the heavy, right-handed neutrinos described earlier. However, this is an entirely generic issue, beyond consideration of neutrino physics: Lacking direct access to the energies of the GUT-scale in the laboratory it is impossible to confirm if "low energy" CP-violation is related to Sakharov's second condition. Thus, the best one can do is search for plausible CP-violation mechanisms and their potential for appearing or affecting lower energy physics processes, in much the same fashion as the research explorations performed by the B-meson factories.

## 2.3 Phenomenology of $0\nu\beta\beta$ decay

Connecting the physics of Majorana neutrinos to the experimentally observable $0\nu\beta\beta$ decay rate requires nuclear physics. Nuclear physics is a broad and diverse subject concerned with questions framed as "central to the field as a whole, that reach out to other areas of science, and that together animate nuclear physics today." This was the perspective of the authors of the 2013 report [47] by the National Research Council Committee on the Assessment of and Outlook for Nuclear Physics. The first overarching scientific question posed in the 2013 report is, "How did visible matter come into being and how does it evolve?" As described in previous sections, $0\nu\beta\beta$ decay research is likely central to answering this question. Indeed, recent technical progress in experiments at the $\sim$100 kg-scale is setting the stage for the exciting possibility of a discovery with a further two orders of magnitude improvement in half-life sensitivity. This prospect for a scientific breakthrough led the nuclear physics community to recommend a tonne-scale double-beta decay experiment as the top priority for new initiatives, as articulated in the 2015 Long Range Plan for Nuclear Science [48].

The experimentally observable phenomenon is the "$0\nu\beta\beta$ decay rate," which is proportional to the inverse of the half-life of the nuclear decay process. Assuming $0\nu\beta\beta$ decay is mainly driven by the exchange of light neutrinos, the decay rate is proportional to the squared effective Majorana mass $\langle m_{\beta\beta}\rangle$ of neutrinos,

$$\frac{1}{T_{1/2}} = G^{0\nu} \cdot \left|\mathrm{g}_{\mathrm{A}}^2 \cdot M_{\mathrm{GT}}^{0\nu} - \mathrm{g}_{\mathrm{V}}^2 \cdot M_{\mathrm{F}}^{0\nu} + \mathrm{g}_{\mathrm{A}}^2 \cdot M_{\mathrm{T}}^{0\nu}\right|^2 \cdot \frac{\langle m_{\beta\beta}\rangle^2}{m_e^2}, \tag{2.9}$$

where $G^{0\nu}$ denotes a leptonic phase space factor [49, 50], $\mathrm{g}_{\mathrm{A}}$ and $\mathrm{g}_{\mathrm{V}}$ are the weak interaction axial-vector and vector coupling constants, and $M_{\mathrm{GT}}^{0\nu}$, $M_{\mathrm{F}}^{0\nu}$, and $M_{\mathrm{T}}^{0\nu}$ are the Gamow-Teller, Fermi, and tensor nuclear matrix elements. The tensor contribution is neglected in several models and is typically rather small when included. The effective Majorana mass, $\langle m_{\beta\beta}\rangle$, defined from theoretical considerations above in Equation 2.8, is directly related to the nuclear decay half-life through the coefficient terms in Equation 2.9. As noted above in the theoretical considerations, $\langle m_{\beta\beta}\rangle$ is a key parameter of scientific interest for neutrino physics research. Any theoretical inference from the measurement of a $0\nu\beta\beta$ decay half-life must take into account uncertainties in the coefficient terms. Conversely, to target a potential $0\nu\beta\beta$ decay half-life, $T_{1/2}$, for a given $\langle m_{\beta\beta}\rangle$-scale, one must rely on calculations of the phase factor and nuclear matrix elements terms. As described later in this section, this mathematical relationship is particularly germane to planning next generation $0\nu\beta\beta$ decay experiments, specifically in assessing the scale of the experiment in terms of the number of candidate $0\nu\beta\beta$-decay nuclei under observation. Thus, a thorough understanding of the relationship between $T_{1/2}$ and $\langle m_{\beta\beta}\rangle$ through the phase space factor, coupling constants, and nuclear matrix elements is of paramount importance.

| Isotope | $G^{0\nu}$ ($10^{-15}$ year$^{-1}$) | Ratio to $G^{0\nu}$ value for $^{136}$Xe |
|---|---|---|
| $^{76}$Ge | 2.363 | 0.162 |
| $^{130}$Te | 14.22 | 0.97 |
| $^{136}$Xe | 14.58 | 1 |

**Table 2.1:** Selected values from Table III of Ref. [49] demonstrating the relative impact of the phase space factor on the $0\nu\beta\beta$ decay rate.

While the previous discussion assumes the standard decay mechanism for concreteness, if the $0\nu\beta\beta$ decay is caused by other mechanisms (such as some of those described at the end of Section 2.1), there would be a different relationship between $T_{1/2}$ and $\langle m_{\beta\beta}\rangle$. In certain models, the corresponding half-life could be the same as for $\langle m_{\beta\beta}\rangle \sim 100$ meV, which is within reach of next generation detectors.

**Phase space factor** The translation of the physical observable, $T_{1/2}$, into a quantity of interest, $\langle m_{\beta\beta}\rangle$, as described by equation 2.9, is often limited to and focused on the matrix elements alone. However, the comparison of one decaying nuclide to another entails the calculation of decay phase space factors, $G^{0\nu}$, for the nuclei under study. These decay-energy and nuclear-charge dependent factors show a considerable nuclide–to–nuclide variation. Table 2.1 reproduces selected $G^{0\nu}$ values for various $0\nu\beta\beta$ decay candidate isotopes and presents their ratio relative to the value for $^{136}$Xe, thus showing the significant impact across different isotopes.

**Nuclear matrix elements** The calculation of the nuclear matrix elements, $M^{0\nu}_{GT}$, $M^{0\nu}_{F}$, and $M^{0\nu}_{T}$, appearing in Equation 2.9 is complex. References [51, 52] provide reviews of the status of approaches to determining these values. To summarize, nuclear structure calculations employ various approximations to solve the nuclear many body problem including: the nuclear shell model (NSM) [53, 54], the quasiparticle random phase approximation (QRPA) [55–61], the interacting boson model (IBM) [62–65], and the energy density functional theory (EDF) [66–68]. The angular momentum projected Hartree-Fock-Bogoliubov (PHFB) method has also been utilized, though the authors are not aware of a recent calculation for $^{136}$Xe employing this method.

While a direct test of the correctness of calculated nuclear matrix elements for $0\nu\beta\beta$ decay is not possible, comparisons among the various models allow an assessment of their relative consistency.

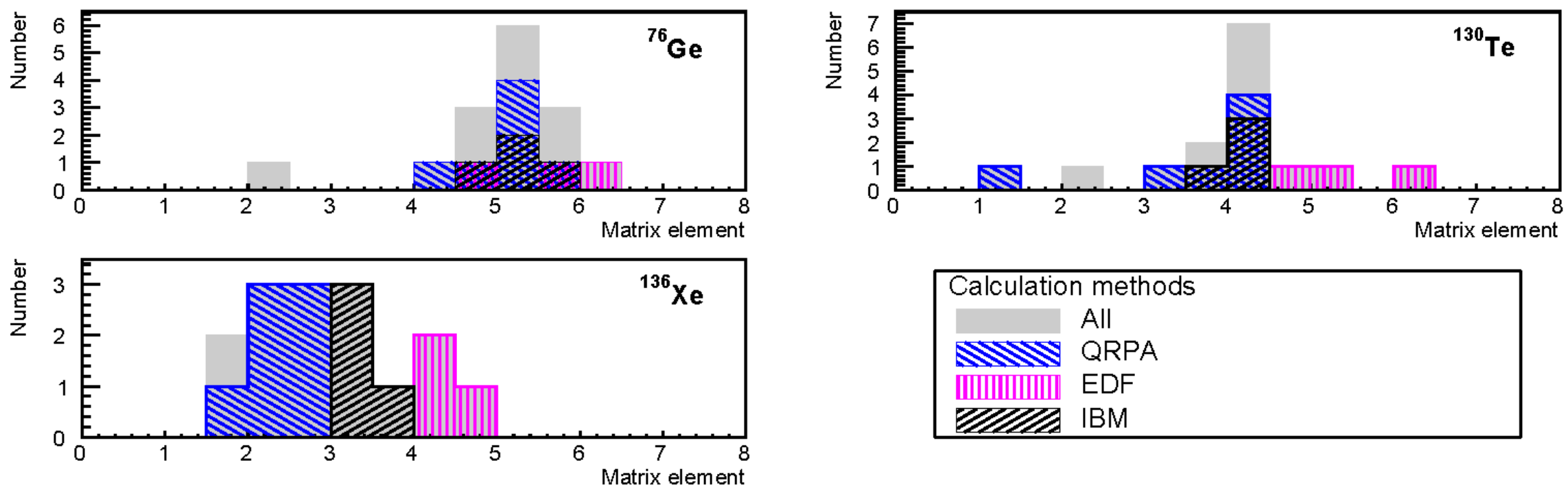

**Figure 2.2:** Frequency distribution of various nuclear matrix elements, taken from publications cited in the text body. The histograms show the distribution of all models (gray) and of the QRPA (blue), EDF (magenta), and IBM (black) calculations separately.

Figure 2.2 presents the distribution of matrix elements found in previously cited papers. These data show grouped distributions with several outlying values. While a review by the Particle Data Group accurately states that there is a "factor ∼2 to 3 spread in the calculated nuclear matrix elements" (See Aug. 2017 revision of Ref. [9], Section 69, *Neutrinoless Double-$\beta$ decay*), often outliers drive such spread.

The above description assumes the standard decay mechanism, although as discussed previously, other mechanisms for Majorana masses are also possible (see Section 2.1). For other mechanisms, the matrix elements are quite different, and so is their uncertainty. Since, in these cases, the mechanisms have nothing to do with weak interactions, the problem of beta quenching, explained next, does not arise.

**Beta quenching** The coupling constants $g_A$ and $g_V$ in Equation 2.9 quantify the translation of quark currents to nucleon currents. As seen in Equation 2.9, both the Gamow-Teller and tensor matrix elements enter into the calculation of $0\nu\beta\beta$-decay rates. In contrast to the Fermi matrix element, scaled by the vector coupling $g_V = 1.00$, the contribution to the $0\nu\beta\beta$ decay rate from the Gamow-Teller and tensor matrix elements depend on the fourth power of the effective axial-vector coupling strength. Accordingly, if this coupling is not equal to the free nucleon value, the effect on the corresponding half-life can be significant.

Within nuclei, Fermi-transitions (vector currents) are unaffected by the presence of the strong interaction. This is assured by the conserved vector current theorem resulting in the vector coupling constant equal to unity: $g_V = 1.00$. However, Gamow-Teller transitions (axial-vector currents) are subject to modification due to the inner structure of the hadrons (composed of quarks), which can alter the effective strength of the weak interaction coupling inside nuclei, resulting in an axial-vector coupling value different from 1. Experimentally, the ratio of the axial-vector coupling strength to vector coupling strength is precisely determined for the single nucleon system of the neutron. The ratio is obtained from the neutron-spin electron-momentum correlation coefficient, also called the beta asymmetry correlation coefficient in neutron decay studies. These measurements from the study of neutron decay determine the axial-vector coupling strength for the free nucleon system of the neutron, $g_A^{\text{free}} = -1.2723 \pm 0.0023$ [9, 69].

No matter which method is used for the theoretical evaluation of the Gamow-Teller matrix elements, a limited Hilbert space of states is necessarily used. Therefore, strictly speaking, the Gamow-Teller operator $\sigma\tau$ should be replaced by an "effective operator" that accounts for such a restriction. It is common to use, instead, an effective value $\mathrm{g_A^{eff}}$ such that $\mathrm{g_A} \to \mathrm{g_A^{eff}} = q \cdot \mathrm{g_A^{free}}$ where $q$ is defined as a "beta-strength" quenching factor.

The relevance of this quenching factor for $0\nu\beta\beta$ remains unclear. As described in a recent, detailed review of nuclear matrix element calculations for $0\nu\beta\beta$ decay [52], the nuclear shell model typically fails to properly describe single Gamow-Teller $\beta$-decay rates, where calculated rates are generally larger than determined from experimental data [70]. This problem has long been remedied by empirically scaling the value of $\mathrm{g_A^{free}}$ to an effective value dependent on the nuclei under study[8]. While this problem would also naively imply a similar quenching for $0\nu\beta\beta$ decay, it has been further observed that simple rules for the summed Gamow-Teller strengths, which can be measured in charge-exchange experiments, are not fulfilled when summing up to the excitation energy where they can be reliably observed[9] [52]. Thus, a similar over-prediction of strength occurs in charge-exchange reactions where the weak interaction plays *no role*. This coincidence suggests that the quenching required to obtain agreement between calculated and measured single $\beta$ decay rates may have other underlying reasons than an altered strength of the weak force in the nuclear system, as often expressed through a modified value of $\mathrm{g}_A$.

Finally, even if the observed quenching does arise from the altered strength of the weak force within nuclei, $\beta$- and $2\nu\beta\beta$-decays are Gamow-Teller transitions involving momentum transfers of the order of several MeV (constrained by the $Q$-value), in contrast to $0\nu\beta\beta$ decay transitions involving momentum transfers of the order of 100 MeV. Thus, much higher nuclear energy states are involved, and a naïve fourth-power scaling (suggested by Equation 2.9) may not apply to determining the $0\nu\beta\beta$ decay rates. Case in point: Calculations of higher momentum transfer nuclear phenomena, such as nuclear muon capture, do not require quenching to match experimental results [75]. This may indicate transitions involving higher multipoles, like $0\nu\beta\beta$-decay, are not affected by this apparent suppression.

While the above discussion indicates that the relevance of $\beta$-quenching for $0\nu\beta\beta$ remains uncertain, various hypotheses for its effects are present in the literature:

1. The coupling constant could experience modification in complex nuclei due to the presence of more nucleons than in the neutron system. In this case the rate would have to be scaled in the fourth power of $\mathrm{g_A}$. The Gamow-Teller matrix elements would have to be scaled by a factor of approximately $(1.27/0.94)^2 = 1.8$ and the decay rate by a factor $\sim 3.3$, depending on the size of the Fermi contribution.
2. Two-particle, two-hole excitations, missing in the nuclear models, may shift Gamow-Teller strength to high excitation energies and therefore to states not included in the models. Short-range, high-energy nucleon–nucleon correlations are not always included in the models. Studies of this effect on calculated $0\nu\beta\beta$ decay rates indicated a 20% to 30% enhancement (not quenching) in calculated $0\nu\beta\beta$ decay rates, while also producing reduced $\beta$- and $2\nu\beta\beta$-matrix elements [76].
3. Missing correlations, in form of many-nucleon currents, can explain some of the observations. Non-nucleonic degrees of freedom (e.g., the $\Delta$-isobars and pions, which are not covered by the Ikeda sum rule), are known to reduce calculated $\beta$-decay rates, although not to the full extent of observations. Calculations have shown $\beta$- and $2\nu\beta\beta$-decay rates are re-duced when these many-nucleon currents are included, but the resulting reduction of the $0\nu\beta\beta$ decay matrix element is of order only 30%, considerably milder than suggested by the fourth power scaling of $g_A$ [77].

[8] For nuclides with A (number of nucleons) less than 16, $\mathrm{g_A^{eff}}$=1.04 is used [71]. For A between 16 and 40 $\mathrm{g_A^{eff}}$=0.98 brings calculations in line with observations [72]. For A between 40 and 50 $\mathrm{g_A^{eff}}$=0.94 seems to work [73].

[9] For example, in the case of the Ikeda sum rule, derived only from simple commutation relations and applicable to nucleonic degrees of freedom (neutrons and protons) [74].

Finally, the effect of $\beta$ quenching depends on the specific model used to evaluate nuclear matrix elements. Within QRPA the strength of the isoscalar particle-particle interaction, $g_{pp}$, is usually adjusted so that the experimental half-life of the $2\nu\beta\beta$-decay is correctly reproduced. As a consequence, the effect of quenching within QRPA is substantially reduced compared to the $g_A^4$ scaling [60, 78]. Models not as well suited to calculate $2\nu\beta\beta$ decay, such as IBM, cannot take advantage of this tuning.

It is clear that $\beta$ quenching is important to translate a possible discovery into a value of the effective Majorana mass, $\langle m_{\beta\beta}\rangle$. This issue is being addressed through renewed efforts in nuclear theory. However, the observation that high momentum transfer nuclear phenomena like nuclear muon capture do not require quenching; the apparent relevance of the two-body weak nucleon currents; and the calculated anti-correlation of $g_A$ with QRPA-derived $0\nu\beta\beta$ decay matrix elements together suggest $\beta$-strength quenching is perhaps on the order of 1.

## 2.4 Planning Next Generation $0\nu\beta\beta$ decay Experiments

From the fundamental equations describing radioactive nuclear decay the number of detectable $0\nu\beta\beta$ decays, $N_{0\nu\beta\beta}$, occurring in a mass, $m$, of an element containing a candidate $0\nu\beta\beta$ decay isotope is related to the isotope's $0\nu\beta\beta$ decay half-life, $T_{1/2}$, as follows,

$$N_{0\nu\beta\beta} = \ln 2 \cdot \frac{a \cdot m}{M_{\beta\beta}} N_A \cdot \epsilon_{\text{det}} \cdot \frac{\epsilon_{\text{live}} \cdot t}{T_{1/2}} \quad , \tag{2.10}$$

where $M_{\beta\beta}$ is the molar mass of the candidate $0\nu\beta\beta$ decay isotope, $a$ is the isotopic abundance, $N_A$ is Avogadro's constant, $\epsilon_{\text{det}}$ is an instrument-specific detection efficiency to observe $0\nu\beta\beta$ decay events, and $t$ is the elapsed observation time with a live-time fraction of $\epsilon_{\text{live}}$. Through the relationship between the $0\nu\beta\beta$ decay half-life, $T_{1/2}$, and the effective Majorana mass, $\langle m_{\beta\beta}\rangle$, given in Equation 2.9, or equivalent relationships for the other Majorana mass models, it is possible to estimate the amount of a candidate $0\nu\beta\beta$ decay isotope required for sensitivity to the effective Majorana mass. Such estimates rely on experiment-specific details including the choice of isotope, the experimental design, as well as backgrounds that may potentially obscure the $0\nu\beta\beta$ decay signature in an experimental apparatus. However, in general next-generation experiments will require several tonnes of detector mass with minimal backgrounds, as described in Section 3.3.9.

To understand the prospects for future $0\nu\beta\beta$ decay experiments, the community often presents experimental reach in terms of the effective Majorana mass, $\langle m_{\beta\beta}\rangle$, assuming a Type-I see-saw mechanism. The effective Majorana mass is compared against the lowest mass neutrino, as shown in Figure 2.3. Knowledge of the neutrino oscillation mixing angles, $\theta_{ij}$, uncertainty on the measured $\theta_{ij}$ values, and spread allowed by the undetermined Majorana phases $\alpha$ and $\beta$ produce allowed bands under the assumption that neutrinos are Majorana particles, for two different neutrino mass hierarchy scenarios (normal or inverted). As shown in Figure 2.3, current generation $0\nu\beta\beta$ decay experiments are probing down to $\langle m_{\beta\beta}\rangle \sim 100$ meV. The vertical widths of the bands denoting experimental limits are a result of theoretical uncertainties in the nuclear decay matrix elements. Figure 2.3 also demarcates the target level of sensitivity for next generation $0\nu\beta\beta$ decay experiments at the roughly $\langle m_{\beta\beta}\rangle \sim 15$ meV level. This is a natural next step in $0\nu\beta\beta$ decay searches as it corresponds to probing another order of magnitude lower in effective Majorana neutrino mass while simultaneously covering the allowed parameter space of the inverted hierarchy of neutrino mass. Recent overviews [79, 80] present projected discovery potential for a number of current and future $0\nu\beta\beta$ decay experiments, showing the nEXO experiment (described in this document) has excellent reach into the $\langle m_{\beta\beta}\rangle \sim 15$ meV band, as shown in Figure 2.4.

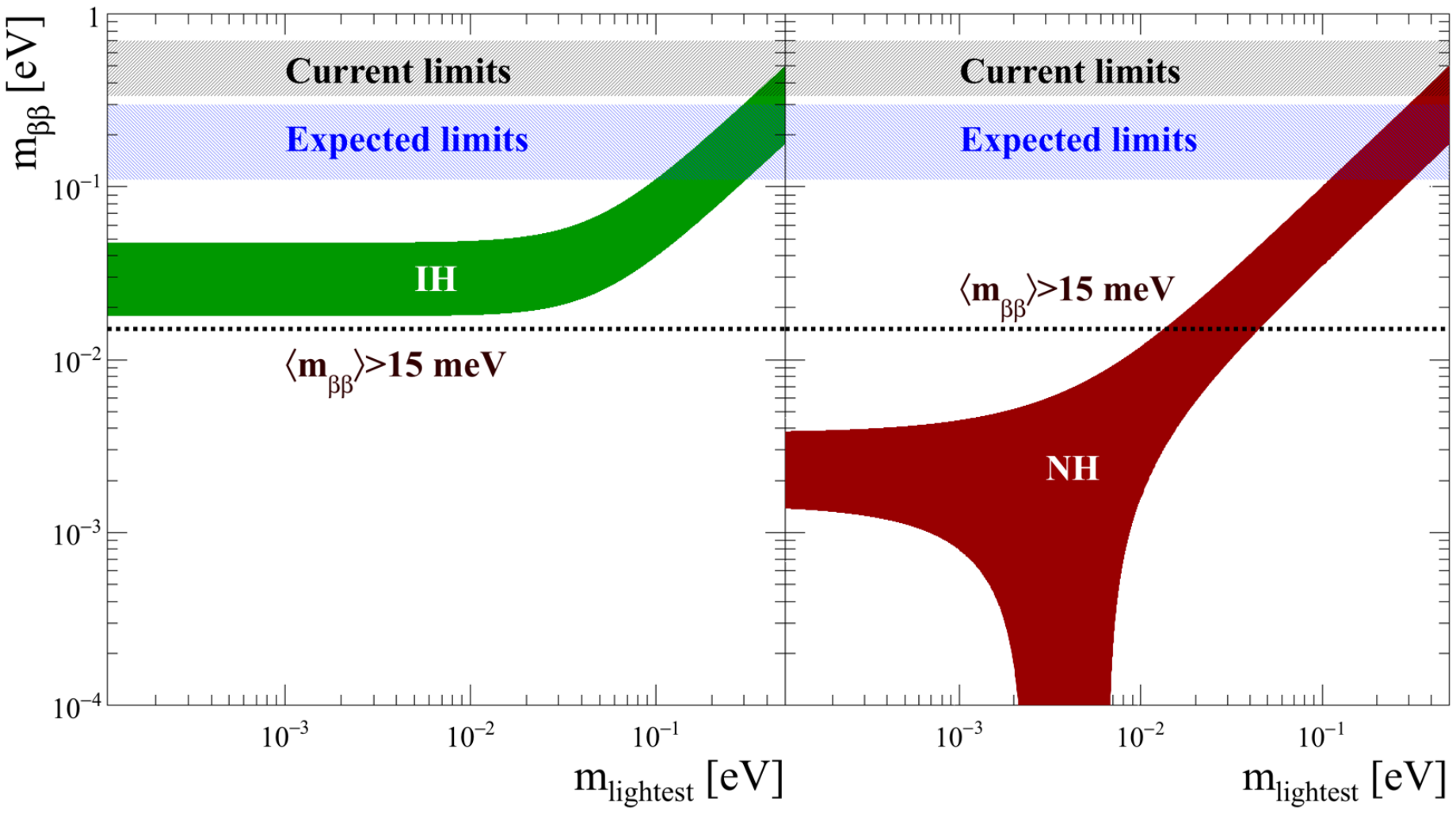

**Figure 2.3:** Effective average neutrino mass from neutrinoless double beta decay *vs.* the mass of the lightest neutrino. Current limits and expected limits from ongoing experiments are shown as gray and blue horizontal bands. The green (for inverted hierarchy) and red (for normal hierarchy) bands show the expected ranges within the light Majorana neutrino exchange mechanism. Next-generation ton-scale experiments aim to probe effective Majorana neutrino masses down to 15 meV, shown as the horizontal dashed line. (*Figure and caption taken from the 2015 Long Range Plan for Nuclear Science [48].*)

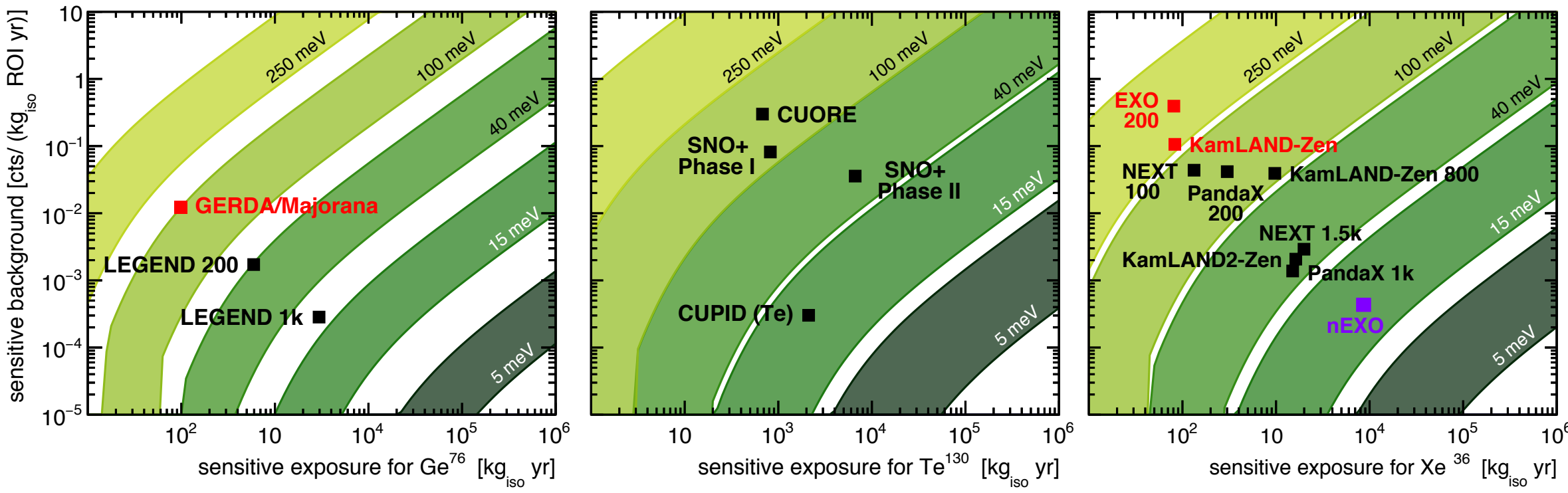

**Figure 2.4:** Adapted from [79]. Projected sensitivity reach of a number of proposed $0\nu\beta\beta$ experiments, plotted over green-shaded bands representing effective Majorana masses, $\langle m_{\beta\beta}\rangle$. Each band's width represents the spread due to nuclear matrix element uncertainties.

While relying on the Type-I see-saw formalism, these observations are useful to provide a concrete goal and to compare experimental sensitivities. In addition, it is appropriate to emphasize, as described in Section 2.1, that other mass generation mechanisms could produce $0\nu\beta\beta$ decay from the existence of Majorana neutrinos. Hence, a broader point of view simply states that increasing the half-life sensitivity to this process directly translates into a greater discovery potential, invoking the "black box" theorem already mentioned in Section 2.1. This is analogous to searching for new physics with a new accelerator, providing access to a new range of energies and interactions.

## 2.5 Neutrino Physics Research in Context

Searches for $0\nu\beta\beta$ decay take place within the larger context of neutrino physics research. The experimental discovery of neutrino oscillations in the late 1990's and early 2000's [81] set the stage for the current questions neutrino research is pursuing today [82]. These questions include:

1. Are neutrinos Dirac or Majorana particles?
2. What is the absolute mass scale of neutrinos?
3. What is the hierarchy of neutrino masses?
4. Is there CP-violation in the neutrino sector?

The answer to the first question is the focus of searches for $0\nu\beta\beta$ decay. Direct measurements of neutrino mass, such as performed by tritium beta decay experiments [83, 84], are seeking to answer the second question. Ongoing precision refinement of the PMNS matrix of neutrino flavor mixing and long-baseline neutrino oscillation studies are seeking to answer the third and fourth questions. And although not included in the above list of questions, astrophysical observations of large scale structure and evolution of the visible universe are pursuing indirect measurements of the total sum of the neutrino masses, $\Sigma m_\nu$. In the coming years, experiments are planned to study all of these open questions. The combination of these research activities is likely to revolutionize our fundamental understanding of neutrinos.

The additional information from other experimental approaches in measuring the properties of neutrinos is complementary to $0\nu\beta\beta$ decay. For instance, tritium endpoint kinematic neutrino mass experiments are pursuing measurement techniques with sensitivity reach down to 40 meV [84]. These measurements of the neutrino mass provide constraints that push from right to left along the $m_{\text{lightest}}$ axis of Figure 2.3, complementing the $0\nu\beta\beta$ decay search parameter space with independent measurements. In addition, long baseline neutrino experiments aiming to determine the neutrino mass hierarchy [85], may contribute to deciding between the inverted (red) or normal (green) hierarchy bands shown in Figure 2.3, even though $0\nu\beta\beta$ decay experiments explore the allowed parameter space for *both* hierarchies simultaneously. Long baseline neutrino experiments also seek to reveal the magnitude of CP-violation in the neutrino sector, opening the door for answering the matter-antimatter asymmetry of the universe question, *if*—and likely only if—neutrinos are Majorana type particles. Finally, fits from astrophysical observations during the CMB Stage-4 era [86] are expected to have sensitivity to the full range of $\Sigma m_\nu$, given the knowledge of neutrino mass splittings from neutrino oscillation experiments. Furthermore, there is a quantifiable relationship between $\Sigma m_\nu$ and $m_{\text{lightest}}$, again through knowledge of neutrino mass splittings from neutrino oscillation experiments, allowing for astrophysical observations to provide input on where in $\langle m_{\beta\beta}\rangle$-parameter space to search [87]. Despite the breadth of complementarity between the differing neutrino experiments described above, only $0\nu\beta\beta$ decay is able to address the key question of whether neutrinos are Dirac or Majorana type particles. Thus, it is nearly assured that as results from other neutrino experiments accumulate, the urgency of answering this fundamental question of the particle nature of the neutrino will only increase.

## 2.6 Scientific Impact of a $0\nu\beta\beta$ decay Discovery

In summary, the search for $0\nu\beta\beta$ decay is *discovery-focused* scientific research. If $0\nu\beta\beta$ decay is observed, there are both immediate and potential far-reaching impacts on the understanding of the nature of matter and the evolution of the universe. To make this more clear, consider three possible outcomes of this quest:

- Irrefutable evidence for:
  - Direct and explicit violation of lepton number conservation.
  - A new type of fundamental fermion, Majorana neutrinos.
- Plausible connections to:
  - A new mass generation mechanism beyond couplings to the Higgs.
  - A new class of GUT-scale heavy, right-handed neutrino fermions.
- Compelling new pictures of physics:
  - A potential window into the early time periods of the universe's expansion.
  - A possible explanation for the matter-antimatter asymmetry of the universe.

The immediacy and potential far-reaching impact of the observation of $0\nu\beta\beta$ decay undergirds the second recommendation of the 2015 Long Range Plan for Nuclear Science [48]:

> *The excess of matter over antimatter in the universe is one of the most compelling mysteries in all of science. The observation of neutrinoless double beta decay in nuclei would immediately demonstrate that neutrinos are their own antiparticles and would have profound implications for our understanding of the matter-antimatter mystery.*
>
> **We recommend the timely development and deployment of a U.S.-led ton-scale neutrinoless double beta decay experiment.**
>
> *A ton-scale instrument designed to search for this as-yet unseen nuclear decay will provide the most powerful test of the particle-antiparticle nature of neutrinos ever performed. With recent experimental breakthroughs pioneered by U.S. physicists and the availability of deep underground laboratories, we are poised to make a major discovery.*

The nEXO experiment described in the remainder of this document is a mature approach to searching for $0\nu\beta\beta$ decay and, when built, is expected to have excellent scientific reach, fully addressing the implicit goals set by the second recommendation of the 2015 Long Range Plan for Nuclear Science.

# 3 nEXO Overview

## 3.1 Instrument Overview and Design Drivers

The nEXO concept is based on a Time Projection Chamber (TPC) filled with five tonnes of liquid xenon (LXe) enriched to 90% in the isotope with atomic mass number $A = 136$ ($^{\mathrm{enr}}\mathrm{Xe}$). This choice is directly derived from the success of EXO-200 and is motivated by the ability of large homogeneous detectors to identify and measure background and signal simultaneously. While the range of few MeV electrons in LXe is too short to be directly measured in a large detector, the most insidious background to $0\nu\beta\beta$ decay derives from $\gamma$ rays and, at the energies of interest, such backgrounds can be identified and separated from electrons of similar energies by the multiple Compton scattering they are likely to undergo. A fully efficient and monolithic detector is ideal for this purpose. This technique acquires more power as the linear size of the detector becomes large compared to the $\gamma$-ray attenuation length that, at 2.4 MeV energy in LXe, is $\lambda_{att} \simeq 8.7$ cm. For detectors with linear size substantially larger than $\lambda_{att}$, backgrounds due to external $\gamma$s are more attenuated the deeper the location is in the detector. Therefore, as quantitatively shown in Section 3.3, in a detector such as nEXO with linear dimensions exceeding 1 m, the external background can be identified over a wide range of energies, fit simultaneously with the signal, and rejected. According to the nEXO background model $\gamma$ radiation emitted by sources external to the xenon constitutes the dominant background component. Backgrounds originating inside the LXe have been found to be entirely negligible at the 100 kg scale [1]. Because of the efficiency with which a noble element such as Xe can be purified, this is expected to also be the case at the tonne scale. Backgrounds from $\alpha$ emission either from Rn dissolved in the LXe or from contaminations in the materials in direct contact with the LXe (such as the various electrodes in the TPC and the SiPMs), are identified and rejected using their characteristically large scintillation-to-ionization ratio, as well as their spatial location for those on detector surfaces.

This intrinsic capability to simultaneously determine signal and background, coupled with careful material selection, is the primary strength of nEXO. Its design is conservative, because few assumptions of material radio-purity beyond what has already been experimentally demonstrated are required.

Energy resolution around the $Q$-value of the decay ($Q_{\beta\beta}^{^{136}\mathrm{Xe}} = 2458.07 \pm 0.31$ keV [1]) is important, but not as crucial as with other types of detectors. Yet, the energy measurement remains the only handle capable of separating the $0\nu\beta\beta$ mode from the SM $2\nu\beta\beta$ decay, as discussed in detail in Section 3.3.

nEXO is designed to optimize the unique features mentioned above, along with providing the best possible energy resolution for a LXe TPC. The total amount of $^{\mathrm{enr}}\mathrm{Xe}$ in the system, 5109 kg, is chosen in such a way as to achieve a sensitivity consistent with the Nuclear Physics community's 2015 Long Range Plan [2], as discussed in Section 3.3. A conceptual sketch of nEXO is shown in Figure 3.1 and the principal parameters of the experiment are presented in Table 3.1.

[1]Except for $^{137}$Xe which is cosmogenically produced and hence depends on the depth of the experimental site. This background is expected to be subdominant at the depths considered for nEXO.

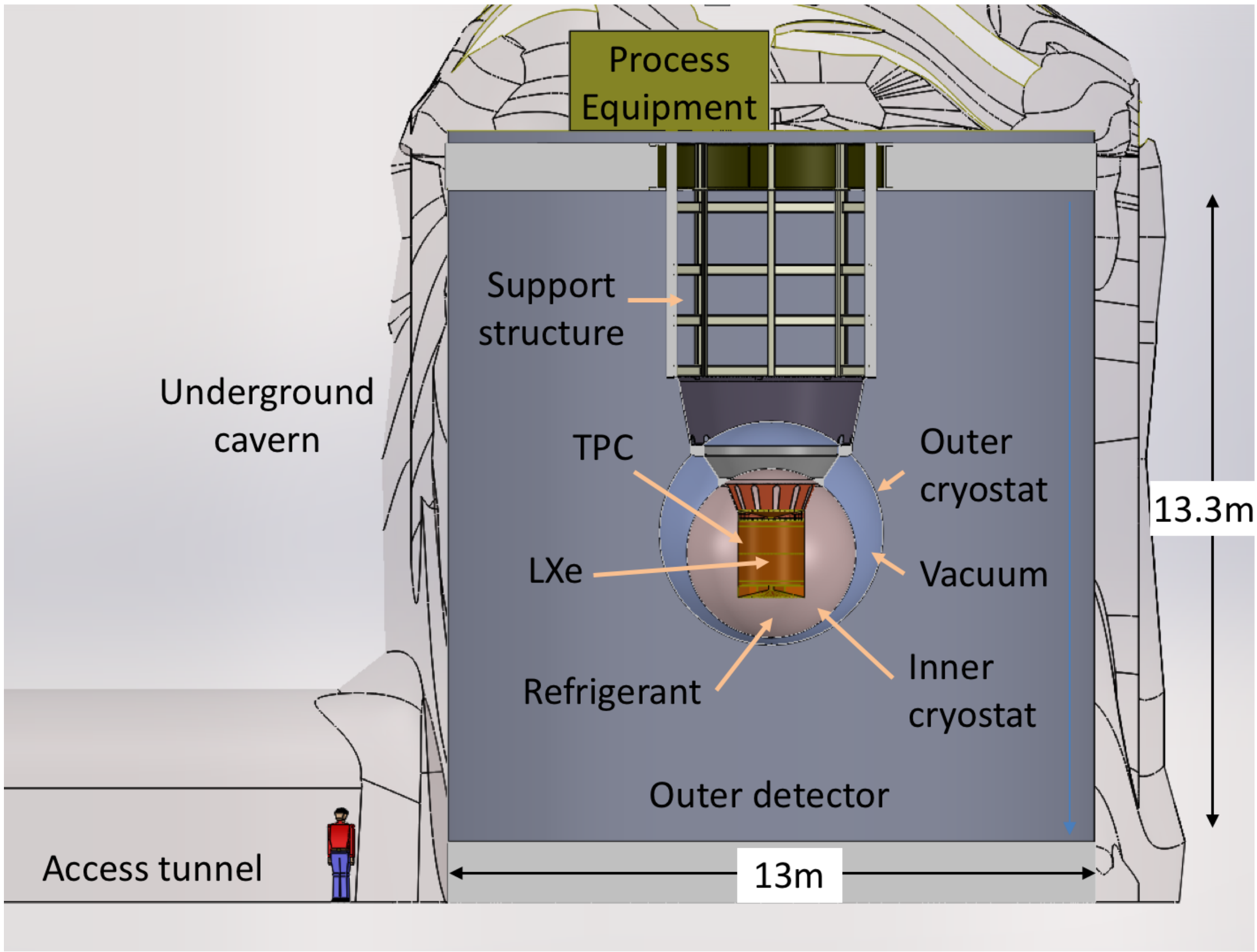

**Figure 3.1:** Sketch of the nEXO detector concept, showing the LXe TPC located inside a vacuum insulated cryostat filled with HFE-7000 refrigerant fluid, doubling as the innermost $\gamma$-ray shield. An outer detector is composed of a large water tank, providing a substantially thicker $\gamma$-ray shield, and read out as a Cherenkov detector, tagging cosmic-ray events. While the host lab for nEXO has not been chosen, for concreteness, the sketch assumes that the detector will be located in the Cryopit at SNOLAB.

As in EXO-200, the nEXO TPC is a LXe single-phase apparatus, a choice dictated by simplicity, resulting in fewer components and lower background. Since optimizing the energy resolution near the $Q$-value is the primary goal (rather than achieving the lowest possible threshold) additional amplification of the charge signal in a two-phase detector is not required. The LXe volume is a cylinder with equal height and diameter, to optimize the surface-to-volume ratio of the detector. The TPC axis is oriented vertically (unlike EXO-200, and under the assumption that the underground space allows for assembly in this configuration) and with a single charge drift volume. This last choice derives from physics as well as technical considerations. From the point of view of the science, a single volume maximizes the bulk of LXe, where the dominant external background is lowest. Technically, a single volume allows the placement of the cathode, sitting at high voltage but otherwise passive, at the bottom of the TPC. Experience with EXO-200 has shown that charged Rn daughters can drift to, and accumulate on, the electrodes, so placing both the anode and cathode at the edges of the volume minimizes the impact of backgrounds from these sources. The drawback of this choice is the doubling of the high voltage required to maintain a certain operating drift field, a more stringent requirement on the electron lifetime ($\tau_e$), and a larger effect of diffusion for events occurring near the cathode. A vigorous R&D program on high voltage (HV), still in progress, provides encouragement that a field of 400 V/cm, similar to the field used in EXO-200, can be comfortably reached and that, quite possibly, a substantially lower field may have negligible effects on detector performance. While it is generally difficult to predict values for $\tau_e$, the substantial reduction of the use of plastics in the TPC compared to EXO-200, along with the three-fold increase of the volume-to-surface ratio, suggest that a $\tau_e$ increase from the 3–5 ms commonly observed in EXO-200 to 10 ms should be comfortably achieved. At the same time, the cathode at one end of the detector simplifies, to some extent, the measurement of $\tau_e$ using external $\gamma$-ray sources since such sources can illuminate the full area of the TPC in the vicinity of the cathode.

A sketch of the nEXO TPC is shown in Figure 3.2. The nEXO TPC is designed to read out, with good efficiency, both ionization and scintillation light, in order to exploit the anti-correlation between these two channels and obtain the best possible energy resolution [4]. Charge collection is achieved at the top of the TPC, with the primary design calling for silica "tiles" patterned with crossed metallic strips. This option eliminates the risk, inherent to a grid, of long (up to 130 cm) wires subject to substantial temperature variations. It also makes the system modular, with front end and digitization electronics built on each tile that can then operate (and be tested) independently from the others. Scintillation readout is obtained with VUV-sensitive photodetectors installed behind the field-shaping rings. This location allows for larger coverage compared to the case of EXO-200, where photodetectors were installed behind the anode grids (an option that is harder to achieve with the opaque anode plane on one side and the cathode on the other). Silicon PhotoMultipliers (SiPMs) sensitive to the 175 nm wavelength of Xe scintillation are being perfected as part of the nEXO R&D program. Custom-built ASIC electronics are being developed to be installed directly behind the charge collection tiles and near subsets of the SiPM arrays. This solution minimizes the input capacitance and reduces the risk of noise deriving from electromagnetic pickup. Great care will be required to reduce the mass and number of components and certify them from the stand-point of radioactivity and LXe purity (not to degrade $\tau_e$). It is expected that all functions except some capacitive bypassing will be achieved by a single chip (likely different for the charge and scintillation channels) that will send only digital data outside of the detector. This solution also minimizes the amount of cabling required inside the TPC, and reduces the required number of electrical feedthroughs. In the primary design of the TPC most HV components are made out of copper, with well understood radiopurity and electrical properties. However, given the substantial energy stored in the system, high resistivity (and ultra-pure) materials such as intrinsic silicon are being investigated for the field cage and the cathode electrodes. Such materials would mitigate the risk of accidental breakdowns producing large and potentially destructive currents.

All materials composing the TPC require the most rigorous screening to ensure sufficiently low radioactive contamination and good $\tau_e$. The material testing and quality assurance program to be developed for this purpose is a direct extension of the one successfully completed for EXO-200. While the engineering of the TPC will be done at a later time, examples of all materials have already been procured and characterized, providing the most crucial input for the background model. It is expected that most of the conductors, including the vessel containing the LXe, will be made out of copper, closely following the EXO-200 experience. Low background quartz and sapphire will represent the bulk of the dielectric structural elements, with the only plastics used being Kapton, as cabling substrate, and very small amounts of selected epoxy adhesive. Silicon will be used in the ASIC chips as well as in the special bypass capacitors.

| Parameter | Primary Value |
|---|---|
| $^{enr}$Xe inventory | 5109 kg |
| Maximum fiducial $^{enr}$Xe | 4038 kg |
| $^{136}$Xe isotope abundance | 90% |
| $^{enr}$Xe operating temperature | 165 K |
| $^{enr}$Xe liquid density$^{a}$ | 3.057 g/cm$^3$ |
| Electric drift field | 400 V/cm |
| Maximum electron drift distance | 125 cm |
| Electron lifetime in LXe | $>$ 10 ms |
| Xe recirculation rate | 350 slpm |
| Diameter of drift volume | 116 cm |
| Charge read-out strip pitch | 3 mm |
| LXe scintillation light wavelength | 175 nm |
| Photodetector area | 4.5 m$^2$ |
| Photodetector dark noise rate | 50 Hz/mm$^2$ @ 0.1 p.e. threshold |
| Overall light detection efficiency | >3% |
| RMS electronics noise (charge channel) | 200-250 e$^-$ |
| RMS electronics noise (scintillation channel) | 0.1 p.e. |
| Inner Cryostat Vessel diameter | 338 cm |
| Outer Cryostat Vessel diameter | 446 cm |
| HFE-7000 mass | 32,000 kg |
| Cool-down/warm-up time | 30 days |
| Minimum HFE-7000 shielding thickness | 0.76 m |
| Water Tank height | 13.3 m |
| Water Tank diameter | 13 m |
| Minimum water shielding thickness | 4.25 m |
| $\beta\beta$ decay Q-value | $2458.07 \pm 0.31$ keV |
| Energy resolution $\sigma/Q_{\beta\beta}$ | $\leq 1.0\%$ |
| MS rejection at $Q_{\beta\beta}$ | $> 10:1$ |
| SS background rate inner 3000 kg | $8.6 \times 10^{-4}$ events/(FWHM·kg·yr) |
| SS background rate inner 2000 kg | $3.6 \times 10^{-4}$ events/(FWHM·kg·yr) |
| SS background rate inner 1000 kg | $1.4 \times 10^{-4}$ events/(FWHM·kg·yr) |
| 90% CL sensitivity $T_{1/2}$ | $9.2 \cdot 10^{27}$ yr (10 yr data) |
| 90% CL sensitivity $\langle m_{\beta\beta} \rangle$ | 5.7-17.7 meV (10 yr data) |
| $3 \cdot \sigma$ discovery potential $T_{1/2}$ | $5.7 \cdot 10^{27}$ yr (10 yr data) |
| $3 \cdot \sigma$ discovery potential $\langle m_{\beta\beta} \rangle$ | 7.3-22.3 meV (10 yr data) |

$^{a}$ The $^{enr}$Xe density is determined by scaling the EXO-200 value [3] to account for the 90% isotope abundance in nEXO.

**Table 3.1:** Compilation of primary design values of some important nEXO parameters. The explanation of symbols and the description of functions can be found in various sections of this document.

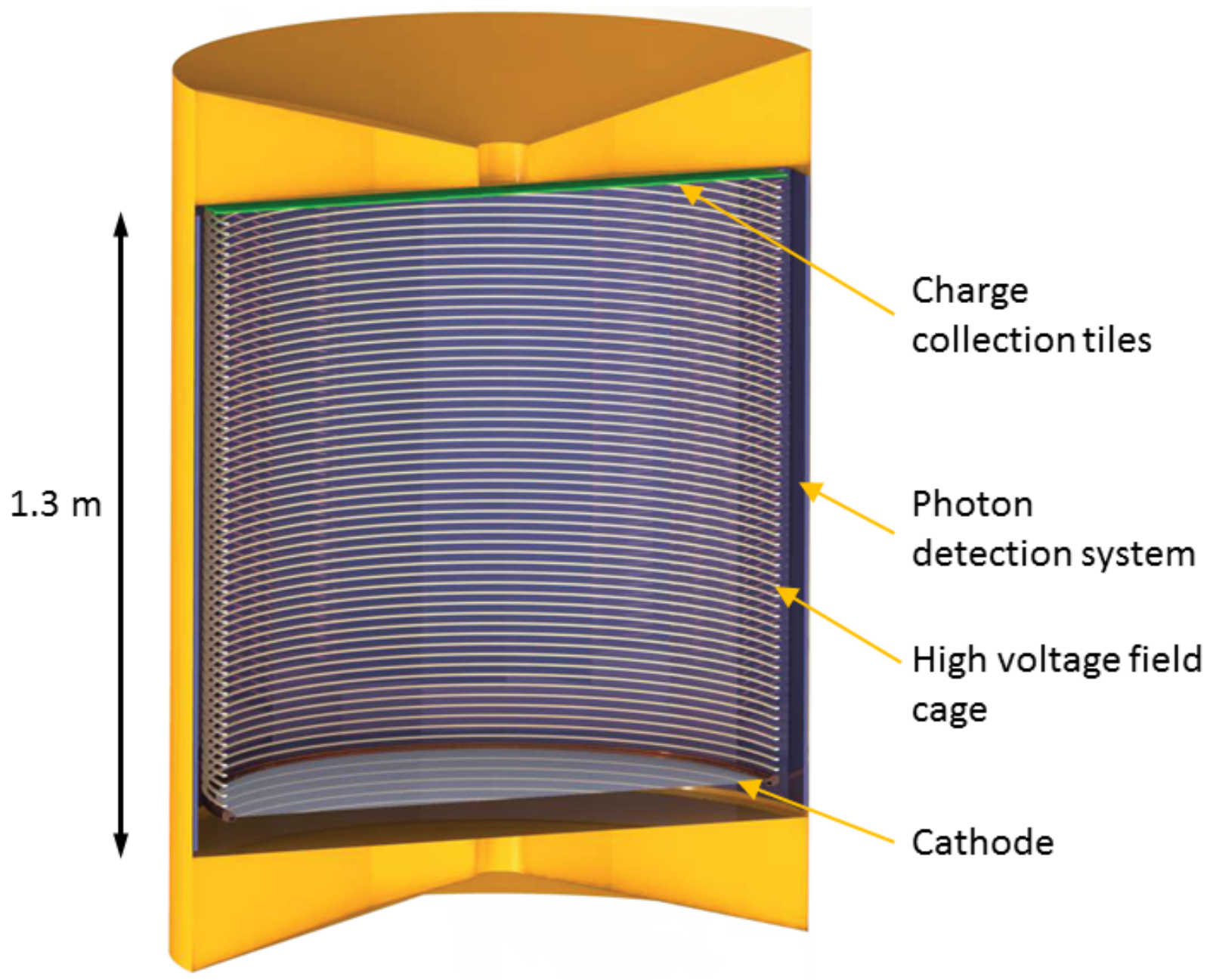

**Figure 3.2:** Sketch of the nEXO TPC. The copper vessel, cathode, charge collection anode and photodetectors, behind the field-shaping rings are schematically shown. For simplicity, this sketch does not show the high voltage and other electrical and Xe recirculation feedthroughs.

The LXe TPC is at the center of different active and passive shielding layers, each containing components made of materials which are progressively lower in radioactive contamination deeper in the detector. As successfully done in EXO-200, the innermost shielding layer will consist of a bath of HFE-7000 [2], at least 76 cm thick in all directions. The HFE-7000 results in very small temperature gradients across the TPC vessel, which is too thin to achieve this goal conductively. With a density of $\sim$ 1.8 g/cm$^3$ at 170 K, this fluid is an efficient $\gamma$-ray shield and is one of the most radiopure materials identified by the EXO-200 and nEXO screening campaigns. The large cold mass, mainly composed of HFE-7000 further provides substantial thermal inertia, making the cryogenic system intrinsically stable. The cryostat is composed of two nested vessels, separated by vacuum insulation. In the primary design these vessels are made of carbon composite, which simplifies the underground construction and does exist with sufficient radiopurity. Beyond the cryostat, a large shield, using purified water, attenuates the bulk of the $\gamma$ radiation emitted by the rock (here assumed to have U and Th concentrations found at SNOLAB [5]), as well as fast neutrons. The water is instrumented with photomultipliers (PMTs) to double as a cosmic-ray veto detector. Modeling of the transmission of $\gamma$-radiation originating in the rock indicates that a tank diameter and height of 10 m and 9 m, respectively, would suffice for shielding purposes.

[2] 3M Novec 7000 Engineered Fluid, 1-methoxyheptafluoropropane,
https://multimedia.3m.com/mws/media/1213720/3m-novec-7000-engineered-fluid-tds.pdf

Mechanical engineering studies for the tank exist with dimensions of 13 m and 13.3 m, respectively (as shown in Figure 3.1), which would entirely fill the SNOLAB Cryopit and provide a generous safety margin. A decision on the final tank dimensions, taking into account the detector location and the placement of xenon and water process equipment, will be made at a later time.

Substantial infrastructure is required to cool down and maintain cryogenic conditions in the LXe and HFE-7000, recirculate the Xe in gas and liquid phases for purification and to handle the HFE-7000 fluid. Care must be applied to provide intrinsic reliability and equipment redundancy in order to ensure operational safety and safeguard the detector and the $^{\mathrm{enr}}\mathrm{Xe}$ inventory. The system envisaged here closely follows the one built for EXO-200 (that successfully recovered from a sudden and protracted loss of access to the WIPP underground laboratory), with important differences accounting for the substantially larger size of the detector and some lessons learned from EXO-200.

The nEXO location is still under evaluation and multiple options are being investigated. For concreteness, here we assume a deep underground installation in the existing Cryopit at the Sudbury Neutrino Observatory Laboratory (SNOLAB), near Sudbury, Ontario, Canada, with an overburden of 6010 meters water equivalent (m.w.e.) [5].

## 3.2 EXO-200

EXO-200 is one of the world's largest $0\nu\beta\beta$ decay experiments currently in operation. Since the start of the physics data taking in 2011, the collaboration has published several high impact physics results, including the first observation of the $2\nu\beta\beta$ decay mode of $^{136}$Xe [6] in 2011, the most precise measurement of the $2\nu\beta\beta$ decay half-life for any isotope [3] in 2014, and three competitive $0\nu\beta\beta$ search results with successively improved sensitivities in 2012, 2014 and 2018 [7–9].

As a precursor to nEXO, EXO-200 has demonstrated the key performance parameters of the LXe detector technology required for a tonne scale $0\nu\beta\beta$ experiment, such as the detector energy resolution and the $\gamma/\beta$ discrimination capabilities. The measured background in the detector agrees well with the background estimate derived from the detector design and radioassay data [10]. It demonstrates that a rigorous material assay program coupled with a meticulous detector simulation can precisely model the detector background. Such an approach is used in the background and sensitivity prediction of nEXO. In this section, we briefly describe the EXO-200 detector, its performance and the physics results.

### 3.2.1 The EXO-200 Detector and Operation

The EXO-200 detector is located at a depth of $1624^{+22}_{-21}$ m.w.e. [11] at the Waste Isolation Pilot Plant (WIPP) near Carlsbad, New Mexico. The left panel of Figure 3.3 shows a cutaway view of the underground installation, including a class 1000 clean room, a double-walled copper cryostat with 25 cm of lead shielding, and an active muon veto system. The inner cryostat is filled with the ultra-clean heat transfer fluid (HFE-7000), providing both shielding and thermal uniformity. The front-end electronics, built with conventional surface mount components, are located outside of the lead shielding and connect to the detector through thin cables with Cu traces on a polyimide substrate.

The central component of the detector is a LXe time projection chamber (TPC) with $\sim$110 kg active mass. A cutaway view of the TPC is shown in the right panel of Figure 3.3. The cylindrically symmetric outer vessel is made of 1.3 mm thick ultra-low activity copper. The chamber is divided into two equal volumes by a photo-etched, perforated phosphor bronze cathode plane. Near each end of the chamber are copper platters, each housing an array of large-area avalanche photodiodes (LAAPDs) [12], behind two wire planes crossed at 60°. When ionizing radiation deposits energy in the TPC, the electrons produced are drifted towards the wire grids by a drift field applied between the wires and the cathode plane. The two wire plane potentials are set in such a way that electrons drift through the plane closer to the cathode, the so-called "V wires", providing an induced signal, and are collected on a second plane, denoted as "U wires". The wire signals provide two-dimensional position information for the event. The position in the drift direction is derived from the known electron drift velocity and the delay time after the scintillation pulse from the event is recorded by the LAAPDs. A more detailed description of the EXO-200 detector can be found in Ref. [13].

EXO-200 Phase-I running took place between June 2011 and January 2014 with a "golden data" fraction of 66% ("golden data" are defined as periods used for the physics analysis) and a calibration data fraction of 10%. Most of the data loss (24%) was due to underground power outages, as it often took a long time for the detector purity and Rn background to fully recover after an outage. The detector operation was stopped from Jan. 2014 to Oct. 2015, due to two independent underground incidents at WIPP, unrelated to EXO-200 [14]. Once underground access resumed, the detector was upgraded with new front end electronics and a Rn suppression system for the air immediately surrounding the cryostat. The Phase-II operation started in April 2016 and is expected to last until the end of 2018. Strategies to mitigate livetime loss from power outages have been implemented in Phase-II running, leading to an improved data collection efficiency, 76% of "golden data" and 10% of calibration data.

### 3.2.2 EXO-200 Detector Performance

The EXO-200 detector has met or exceeded its design goals in all major areas: electron lifetime, energy resolution, and background rejection capability using event topology. Its remarkable performance ensured the success of its physics program and serves as a benchmark for nEXO.

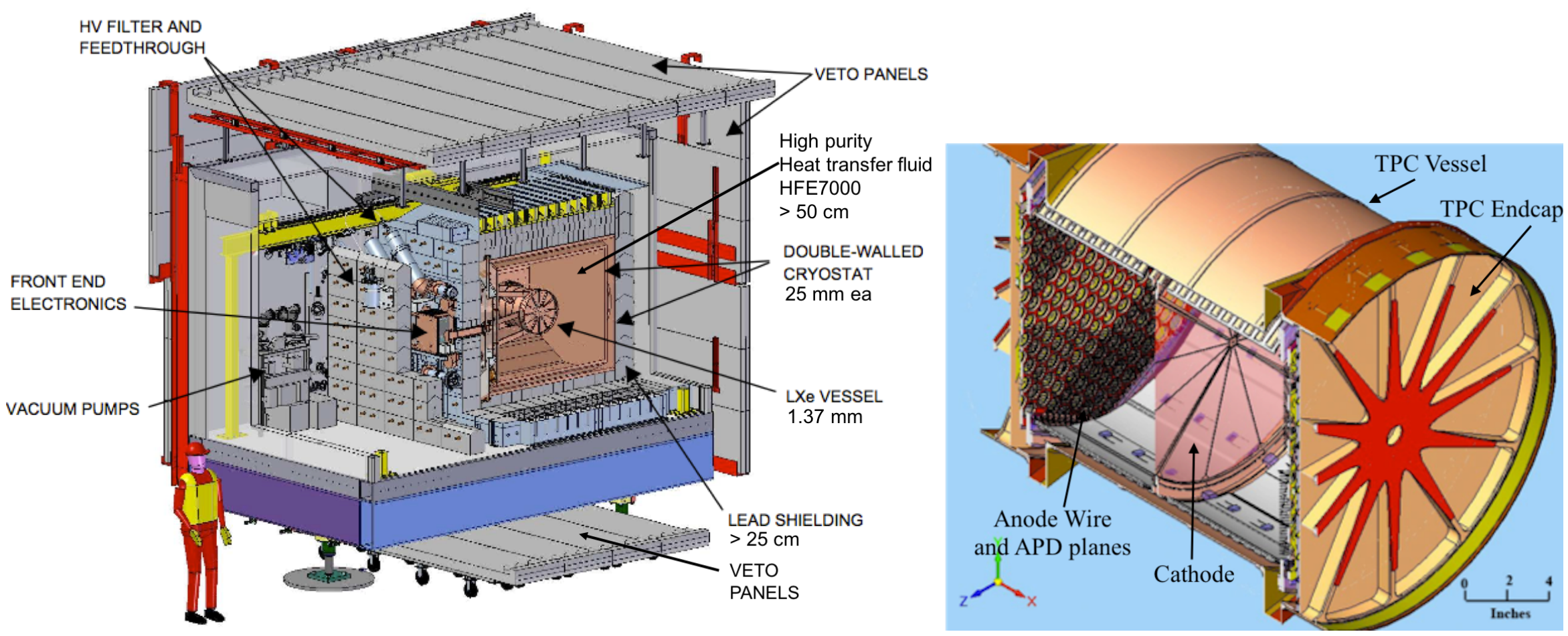

**Figure 3.3:** Cutaway view of the EXO-200 setup, with the primary sub-assemblies identified (left), and a cutaway view of the EXO-200 time projection chamber (right).

The xenon in the TPC is continuously recirculated and cleaned by a hot getter gas purifier [15]. At a flow rate of $>$ 15 slpm, electron lifetimes of 3–5 ms are routinely achieved in EXO-200. The electron lifetime is likely limited by outgassing of impurities from the large plastic surfaces (Teflon, acrylic, and Kapton) in direct contact with the LXe.

Detector energy resolution is a critical parameter for double beta decay experiments, as it is the only experimental handle for separating $0\nu\beta\beta$ events from the tail of $2\nu\beta\beta$ continuum. Due to energy conservation, the light and charge signals from LXe detectors are anti-correlated. By forming a linear combination of the two signals, the detector energy resolution can be improved considerably. This technique was first demonstrated in a small LXe test cell by the EXO-200 collaboration [4], and later confirmed by other groups [16]. The same anti-correlation is observed in the EXO-200 detector. Figure 3.4 shows events with a single reconstructed interaction location (i.e., "Single Site" events) induced by a $^{228}$Th calibration source. EXO-200 was designed to reach an energy resolution of 1.6% at $Q_{\beta\beta}$, but was able to achieve an energy resolution of 1.23% in its Phase-II running. Detailed simulation and analysis show that the energy resolution is limited by the APD front end electronic noise, APD avalanche gain fluctuations, and the overall light collection efficiency.

The EXO-200 detector demonstrates the power of LXe TPCs for rejecting background events. As a monolithic detector, surface $\alpha$s and $\beta$s can be rejected by a fiducial volume cut. Decays producing $\alpha$s inside the LXe bulk are rejected by a light to charge ratio cut, as $\alpha$ events create much denser charge clouds than electrons, and therefore a larger amount of scintillation light is produced from recombination. Coincidence techniques can also be used to identify backgrounds. For example, the delayed coincidence of $^{214}$Bi–$^{214}$Po is used to measure the $^{222}$Rn content of the LXe. In stable operation, the $^{222}$Rn in the detector is found to be 3.6 $\pm$ 0.4 $\mu$Bq/kg ($\sim$ 1 decay/hr) [17]. For $\gamma$ backgrounds, which preferentially Compton scatter leaving multiple energy depositions, the 3-D event reconstruction capability of a TPC and its monolithic active volume are critical for background rejection. TPC events can be separated into "Multi Site" (MS) and SS events based on the number of spatially separated energy depositions inside the chamber. Events in the MS spectrum are predominantly Compton scattering events and are excluded from the signal region. Furthermore, the MS spectrum measures the $\gamma$ backgrounds, providing constraints on the leakage of such background events into the SS spectrum. The observed SS/MS ratio for the EXO-200 detector is approximately 1:5 near the $Q_{\beta\beta}$ value. In the most recent analysis, topological information for each event was used to further discriminate between signal and $\gamma$ backgrounds in the SS event population. A multivariate discriminator was developed that combines the number of collection wires for each event, the signal rise time, and the event standoff distance (defined as the distance of the event vertex from the nearest detector surface) in a boosted decision tree (BDT) using the TMVA software package [18]. Figure 3.5 shows a comparison between data and Monte-Carlo simulation for the variables used in the BDT, as well as the BDT discriminator. The use of BDT variable provides a $\sim$ 15% increase in sensitivity compared to the SS/MS classification alone [9].

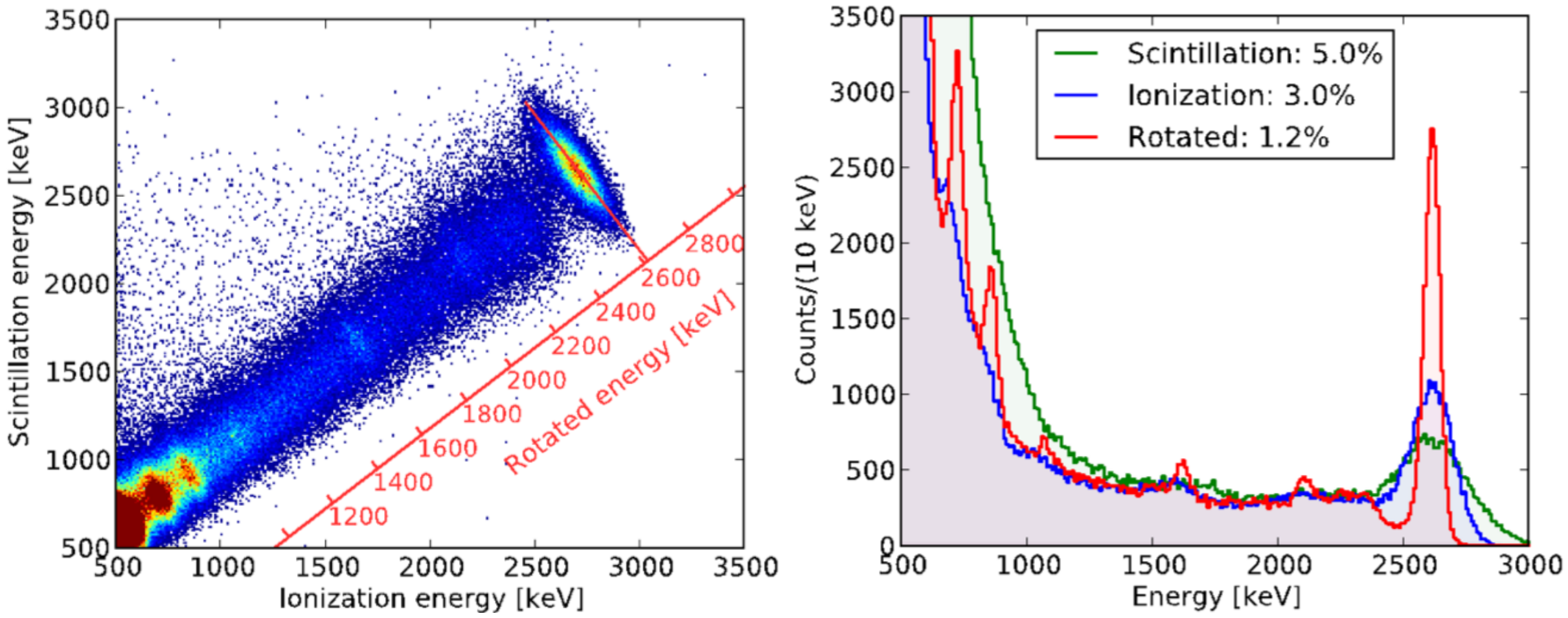

**Figure 3.4:** $^{228}$Th source calibration data for "Single Site" events in EXO-200. A clear anti-correlation between light and charge signals is observed (left). The energy resolution is greatly improved when the optimal linear combination of scintillation and ionization is used for the energy measurement (right).

### 3.2.3 Physics Results from EXO-200

Combining the entire Phase-I data and the first year of Phase-II data, the EXO-200 collaboration published a $0\nu\beta\beta$ search result with a total $^{136}$Xe exposure of 177.6 kg·yr [9]. Using a profile likelihood study, a lower limit on the $^{136}$Xe $0\nu\beta\beta$ decay half-life of $T_{1/2} > 1.8 \times 10^{25}$ yr was obtained at 90% confidence level. This value is lower than the experimental sensitivity of $3.7\times10^{25}$ yr, which is consistent with a positive fluctuation of events above the background expectation, since the excess is not sufficiently significant to warrant interpretation as a signal. This result corresponds to an upper limit on the Majorana neutrino mass, $\langle m_{\beta\beta}\rangle <$ (147 - 398) meV, assuming the unquenched value of the axial-vector coupling constant. The best-fit total background event rate near the $Q_{\beta\beta}$ value is $(1.5 \pm 0.3) \times 10^{-3}$ kg$^{-1}$ yr$^{-1}$ keV$^{-1}$ within the entire fiducial volume.

The EXO-200 results are competitive with other leading experiments, as shown in Table 3.2. The Majorana neutrino mass limit of EXO-200 is comparable to those of GERDA and CUORE. Although EXO-200 has a lower $0\nu\beta\beta$ decay half-life limit than KamLAND-Zen, the sensitivities of the two experiments are close and, providing a more specific signature, EXO-200 has better handles to validate a signal in the case of a discovery. EXO-200 will continue to take data until the end of 2018. With the entire data set, it is expected to reach a $0\nu\beta\beta$ decay half-life sensitivity of $5.7 \times 10^{25}$ yrs, as shown in Figure 3.6.

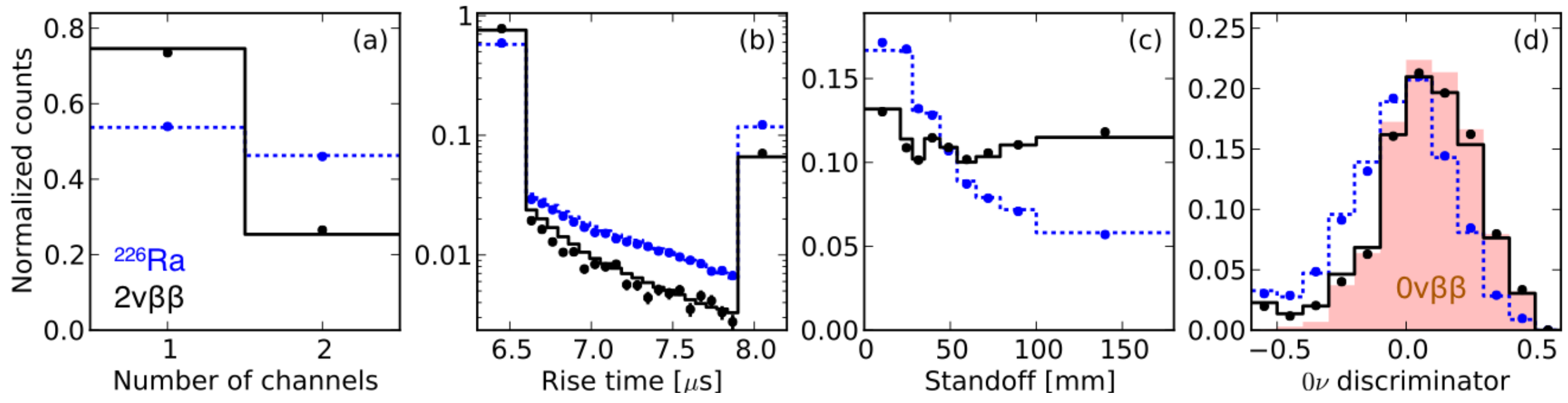

**Figure 3.5:** Comparison between data (dots) and MC (solid/dashed lines) for the individual variables used in the BDT and the overall discriminator distribution. Both source calibration data using a $^{226}$Ra source at the cathode (blue dashed) and the background-subtracted $2\nu\beta\beta$ spectrum from low background data (black solid) are shown. The expected BDT discriminator distribution for a $0\nu\beta\beta$ signal obtained from Monte Carlo data is indicated by the red filled region. [9]

Besides the search for $0\nu\beta\beta$ decay of $^{136}$Xe, EXO-200 has set some of the most stringent limits on the $0\nu\beta\beta$ and $2\nu\beta\beta$ decays of $^{134}$Xe [22], the Majoron-emitting $\beta\beta$ modes of $^{136}$Xe [23], the $2\nu\beta\beta$ decay of $^{136}$Xe to the $0^{+1}$ excited state of $^{136}$Ba [24], Lorentz and CPT violation in $2\nu\beta\beta$ decay [25], and triple nucleon decays of $^{136}$Xe [26]. Furthermore, EXO-200 has made several measurements that enhanced our understanding of LXe detector physics, including the drift velocity and transverse diffusion of electrons in LXe [27], and the ion fraction and mobility of $\alpha$ and $\beta$ decaying Rn progeny in LXe [28].

## 3.3 nEXO Background Model and Sensitivity

The following sections present the current estimates for the physics reach of nEXO. This is done in terms of a half-life limit, to be determined in case no effect is observed, and in the form of a "discovery potential," which is the magnitude of an effect that would be observable with a given degree of likelihood. While the $0\nu\beta\beta$ half-life sets the scale for discovery of physics beyond the standard model, the experimentally observable "event rate" is connected here to neutrino physics by reporting the constraints that nEXO expects to provide on $\langle m_{\beta\beta}\rangle$.

Many experimental parameters of the rather complex nEXO detector enter into the sensitivity estimation. The detector background is one of the key ingredients. In Section 3.3.1 we present the nEXO background model as it arises from an assessment of all reasonable background sources in nEXO. These estimates are based on a GEANT4 [30] detector Monte Carlo simulation and a separate event reconstruction. The latter step accounts for the expected detector performance. Section 3.3.6 reports the studies performed to convert the calculated distributions of signal and background into a quantitative estimate of the sensitivity in terms of $T_{1/2}$. The observation that the $\langle m_{\beta\beta}\rangle$-sensitivity meets the expectations for a next generation experiment serves as justification for the experiment size and all other boundary conditions defined in this proposal.

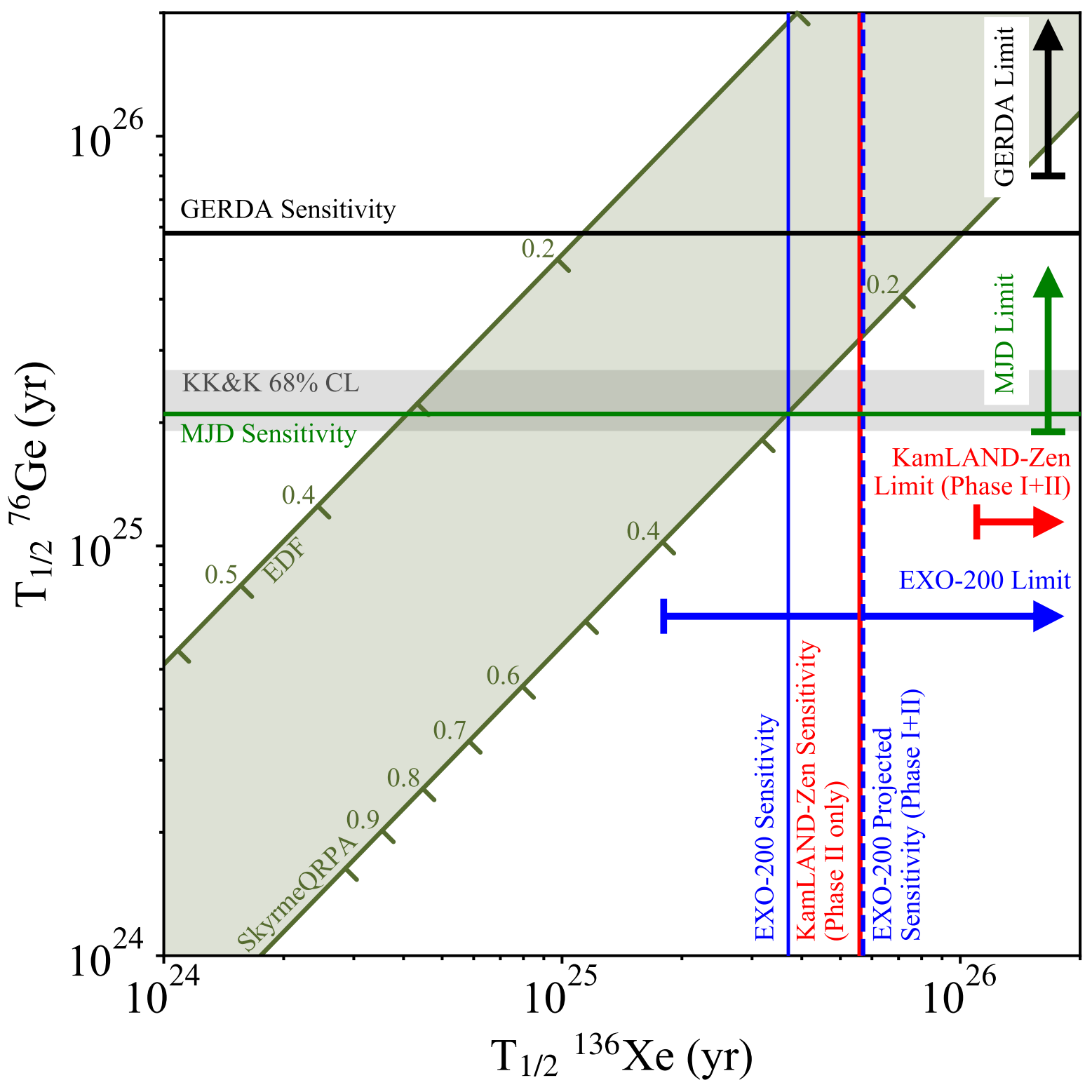

**Figure 3.6:** Comparison of $^{76}$Ge and $^{136}$Xe derived $0\nu\beta\beta$ half-lives. The horizontal and vertical lines represent the GERDA [19], MAJORANA DEMONSTRATOR (MJD) [20], KamLAND-Zen [21], and EXO-200 [9] sensitivities and limits. The shaded diagonal band indicates uncertainties due to different matrix element calculations. The marks on the diagonals denote model-specific effective Majorana neutrino masses.

| Experiment | Isotope | Exposure (kg·yr)[a] | $T_{1/2}^{0\nu\beta\beta}$ sensitivity (yrs) | $T_{1/2}^{0\nu\beta\beta}$ limit (yrs) | $\langle m_{\beta\beta}\rangle$ limit (eV) |
|---|---|---|---|---|---|
| EXO-200 [9] | $^{136}$Xe | 177.6 | $3.7 \times 10^{25}$ | $> 1.8 \times 10^{25}$ | $< 0.15 - 0.40$ |
| KamLAND-Zen [21] | $^{136}$Xe | 504 | $5.6 \times 10^{25}$ | $> 10.7 \times 10^{25}$ | $< 0.06 - 0.17$ |
| GERDA [19] | $^{76}$Ge | 46.7 | $5.8 \times 10^{25}$ | $> 8.0 \times 10^{25}$ | $< 0.12 - 0.26$ |
| MAJORANA DEMONSTRATOR [20] | $^{76}$Ge | 9.95 | $2.1 \times 10^{25}$ | $> 1.9 \times 10^{25}$ | $< 0.24 - 0.52$ |
| CUORE [29] | $^{130}$Te | 86.3 | $0.7 \times 10^{25}$ | $> 1.3 \times 10^{25}$ | $< 0.11 - 0.52$[b] |

[a] EXO-200 and KamLAND-Zen numbers are the isotope mass exposure, while the GERDA, MAJORANA and CUORE numbers denote the exposure of their active detector material.
[b] While the half-life limit refers to the search result with CUORE alone, the bounds on Majorana masses shown here are the combined results of CUORE with two earlier experiments, Cuoricino and CUORE-0.

**Table 3.2:** A list of $0\nu\beta\beta$ decay experiments and results (April 2018).

### 3.3.1 Background Sources

Building a background model involves the pre-selection of "reasonable" radio-nuclides of interest, since modeling all known radioactive nuclides would be prohibitively complex. The nEXO background model, as discussed here, includes components that can interfere with the detection of the $0\nu\beta\beta$ decay mode. In addition, the background model and radioassay are designed to include backgrounds at the lower energies relevant to the $2\nu\beta\beta$ decay. $2\nu\beta\beta$ decay is the only internal source of two-electron events and, therefore, is expected to be useful as a calibration tool, as it is in the EXO-200 data-analysis effort. For the nEXO background model the following selection criteria are used:

1. The decay must release sufficient energy to interfere with the detection of the $^{136}$Xe $0\nu\beta\beta$ mode.
2. The decaying nuclide must have sufficiently long half-life or be produced in steady-state in the detector. Otherwise, radioactive decay quickly diminishes its impact on the background. Nuclides with a half-life of less than half a year were considered "short lived".

A list of background sources that were considered during the development of nEXO's background model is provided in Table 3.3. Components resulting in background event rates $\leq 0.02$ SS

| Background Source | |
|---|---|
| Long-lived radionuclides ($\gamma$- and $\beta$-emitters) in detector materials | In model |
| Th and U in water shield and lab rock | Negligible |
| Surface radioactivity | Negligible |
| $\alpha$ radioactivity | Negligible |
| Aboveground cosmogenic activation products | Negligible |
| Underground cosmogenic activation products in LXe | In model |
| Underground cosmogenic activation products in other detector materials | Negligible |
| $^{136}$Xe $2\nu\beta\beta$ | In model |
| Activation products from ($\alpha$,n) reactions | Negligible/see text |
| Electron-neutrino elastic scattering | Negligible |
| Neutrino capture on $^{136}$Xe | Negligible |
| $^{222}$Rn steady-state presence in LXe | In model |
| $^{222}$Rn steady-state presence in water shield | Negligible |

**Table 3.3:** List of background sources that were considered during the development of nEXO's background model and whether they were included in the sensitivity calculations. Details are provided in the text.

events/(FWHM·y) in the inner 2000 kg LXe mass are not further considered (labeled "negligible" in Table 3.3). The choice of this particular Xe mass derives from the event-rate analysis depicted in Figure 3.15. The background and signal within this mass value drives $\sim$ 90% of the sensitivity reach of nEXO.

#### 3.3.1.1 Long-lived Radionuclides

Naturally occurring $^{232}$Th and $^{238}$U radioactivity fulfills the half-life selection criteria and, as a result, the decay chain daughters from each nuclide are included in the background model. Of particular interest is the $^{238}$U daughter $^{214}$Bi whose decay includes a $\gamma$-ray line at 2448 keV, an energy within 10 keV of the $0\nu\beta\beta$-decay peak.

Long-lived nuclides such as $^{137}$Cs, $^{60}$Co, and $^{40}$K are also considered. While they do not contribute to the $0\nu\beta\beta$ background due to energy and $\gamma$-ray multiplicity, they significantly affect the measurement of the $2\nu\beta\beta$ decay and are therefore systematically tracked as part of nEXO's materials analysis program. $^{40}$K was explicitly included in the model as representative of low-energy spectral features. The probability that a $^{60}$Co decay results in a SS event with energy within $Q_{\beta\beta}$ $\pm$FWHM/2 was estimated from Monte Carlo to be $< 2.3 \times 10^{-8}$ (at 90% CL) in the fiducial volume corresponding to a mass of 3740 kg. $^{60}$Co-induced background is therefore negligible for the $0\nu\beta\beta$ signal. The $2\nu\beta\beta$ decay of $^{136}$Xe is included in the model. $^{26}$Al, which decays by $\beta^+$ emission or electron capture with $Q$-value of 4.00 MeV and $T_{1/2} = 7.17 \times 10^5$ y [31] and is potentially present in sapphire components, is not currently included in the background model, but is planned for future study.

Only $\gamma$ and $\beta$ decays from the nuclides listed above create background events in nEXO. Decays emitting only $\alpha$ particles are rejected with high efficiency using a charge/light ratio analysis [28]. Secondary radionuclide production, e.g. through ($\alpha$,n) reactions, is discussed below.

Background radioactivity was then further sub-divided into bulk and surface activities. The materials analysis program described in Section 6.1 tests all materials of interest for their bulk radioactivity content. It is assumed that surface activities can be mitigated by an appropriate surface treatment, cleaning, clean machining, and/or etching. This strategy was effective in EXO-200. The list of bulk radionuclides included in the Monte Carlo detector model for each detector component in the nEXO background model is given in Table 3 of reference [32].

Several studies have demonstrated that a set of long-lived radionuclides from certain components of the experiment do not contribute significantly. In particular, natural radioactivity in the water shielding and underground laboratory walls have been evaluated (see Section 4.5) and found to be negligible.

#### 3.3.1.2 Cosmogenically-created Radionuclides

Radionuclides with a half-life of less than 0.5 years were only considered if they can be created by the interaction of cosmic radiation with a material of interest. The estimation of this background class had two components: activation while materials are stored, handled, or machined above ground and the steady-state production underground. The former results in guidelines for the exposure management, the latter defines the requirements for the overburden.

Above-ground radio-nuclide production is important for all passive detector materials. In particular, the production of radio-nuclides in copper (e.g. $^{56}$Co and $^{60}$Co) was estimated and found to be acceptable with proper management of the cosmic ray exposure. Because the xenon will be continuously purified during detector operation, long-lived spallation products created by the cosmic radiation while the xenon is above ground (e.g. $^{137}$Cs) are not a concern. Xenon has no long-lived cosmogenically-produced isotopes.

EXO-200 data were used to quantify a broad range of cosmogenic backgrounds [11] that would arise during underground operation. This was accomplished by testing GEANT4 and FLUKA Monte Carlo simulations against data, thus validating the models. These simulations were used, appropriately modified, to estimate cosmogenic backgrounds in nEXO, providing more confidence in the procedure. The simulations indicate that with sufficient overburden, for example that available at SNOLAB, all cosmogenic backgrounds except $^{137}$Xe are negligible. $^{137}$Xe, which $\beta$-decays with a Q-value of 4173 keV, is therefore the only cosmogenic activity contained in the nEXO background model.

At SNOLAB, the steady-state production of cosmogenic $^{137}$Xe in nEXO was estimated using FLUKA at $2.2 \times 10^{-3}$ atoms/(kg·y). Siting nEXO at locations with similar overburden, like China Jinping Underground Laboratory [33], would also result in an acceptable cosmogenic nuclide production, even without active vetoing. The depth of Laboratori Nazionali del Gran Sasso (LNGS) [34], Italy, was found to be marginally acceptable and would require the development of a more sophisticated active veto system. The Sanford Underground Research Facility (SURF), South Dakota, USA, [35, 36] at the 4850′ level is adequate assuming a simple active veto scheme.

#### 3.3.1.3 Neutrino-induced Backgrounds

Interactions of solar neutrinos in the detector are a potential source of background for $0\nu\beta\beta$ experiments, as discussed in [37–39].

Electron-neutrino elastic scattering ($\nu + e^- \rightarrow \nu + e^-$) in the detector volume results in the emission of energetic electrons that can mimic the signature of a $0\nu\beta\beta$ event. Using the background rate for this reaction in $^{136}$Xe from [39], ∼0.02 SS events/(FWHM·y) are expected in nEXO's inner 2000 kg of LXe. At this level, neutrino-induced backgrounds are small compared to other backgrounds, and thus we have chosen to set this rate as the level at which backgrounds are excluded from the sensitivity calculation.

The neutrino capture process via the charged-current reaction $\nu$+$^{136}$Xe→ $e^-$+$^{136}$Cs also contributes background events due to (1) the prompt $e^-$ combined with any $\gamma$-ray emitted from the $^{136}$Cs de-excitation, and (2) the delayed decay of $^{136}$Cs into $^{136}$Ba with a half-life of 13.16 days and $Q = 2548.2$ keV, which is approximately 90 keV higher than $Q_{\beta\beta}$.

The rate of events due to the neutrino capture process and falling near $Q_{\beta\beta}$ is expected to be very small. The dominant $^7$Be and other low energy solar neutrinos cannot produce enough visible energy to reach $Q_{\beta\beta}$ while almost all of the events due to the $^8$B flux release too much visible energy. Estimates of the true background rate have been made following [38, 40, 41], predicting background rates that are negligible for nEXO.

The total charged-current interaction rate for solar neutrinos on $^{136}$Xe has been shown [38, 40, 41] to be about 20 interactions/(2000 kg·y). In [38] it is estimated that the rate of (MS+SS) events within $Q_{\beta\beta}$ ±FWHM/2 would be about 0.6 events/(2000 kg·y) and a similar result is found in [40]. The decay typically proceeds with 3 $\gamma$-rays in cascade together with the electron with a total energy release of 2548 keV, and is rejected with very high efficiency by the single site selection cut. Indeed, a Monte Carlo simulation of $10^7$ decays of $^{136}$Cs in nEXO's LXe volume gave no events that satisfied both the energy and the single site criterion. Even assuming that $^{136}$Cs is not removed by the LXe purification system or does not freeze-out on metal surfaces, the decay of all $^{136}$Cs in the LXe would thus result in a negligible background rate in nEXO.

Finally, solar neutrinos can interact with $^{136}$Xe through an inelastic scattering neutral-current interaction. Such a process could excite a $1^+$ state which would have a high branching ratio back to the ground state, thus potentially resulting in an SS event. We are not aware of any $1^+$ state with energy near $Q_{\beta\beta}$ and therefore neglected this background source.

#### 3.3.1.4 Radionuclides from ($\alpha$,n) Reactions

Deposition of the $\alpha$-unstable $^{222}$Rn daughter $^{210}$Po can create background through ($\alpha$,n) reactions with low-Z detector materials such as F, C, O, Al, and Si which are contained in nEXO's HFE-7000, sapphire, and quartz. The emitted neutrons can subsequently produce $^{137}$Xe when captured by $^{136}$Xe.

A calculation was performed to determine the allowable exposure time of these materials to standard lab air (25 Bq/m$^3$ of $^{222}$Rn) resulting in no more than 0.01 events/(FWHM·y·3000 kg) of background. Neutron yields for the relevant ($\alpha$,n) reactions were calculated from tabulated stopping power and cross-section data, and used as input into a FLUKA simulation to determine the position distribution and probability of neutron captures on $^{136}$Xe in nEXO from neutrons generated on the surface of the relevant components. The neutron production yield calculations have been compared against results obtained from the ORNL code SOURCES-4C [42]. The nEXO GEANT4 Monte Carlo provided the fraction of $^{137}$Xe decays that result in an SS energy deposition within $Q_{\beta\beta}$ ±FWHM/2.

$^{210}$Po $\alpha$s emitted into the fluorine-rich HFE-7000 have the highest neutron production probability. Approximately 8 m$^2$ of TPC copper will be facing the HFE-7000. The reaction yield in HFE-7000 was estimated as $4.3 \cdot 10^{-6}$ neutrons per $^{210}$Po $\alpha$-emission. There will be ~4.5 m$^2$ of SiPMs inside the TPC whose surfaces will have some attached $^{210}$Po from air exposure. In silicon the ($\alpha$,n)-reaction yield is only $7.4 \cdot 10^{-8}$. It is therefore assumed that the TPC copper imposes the most stringent Po-constraint. An (assumed) constant Po growth rate from air deposition and the $^{210}$Po decay rate observed by the LUX experiment [43] yields an allowed exposure length of

340 days for the TPC copper. Using directly measured radon-daughter deposition rates in reference [44] and a more realistic Po-growth treatment yields even longer allowed exposure lengths. The deposition of radon daughters on surfaces is currently under review by the collaboration as new radon daughter deposition rates seem to contradict some of the older data [45].

#### 3.3.1.5 $^{222}$Rn

Contributions to the background rate from the $^{222}$Rn daughter $^{214}$Bi are particularly important because $^{214}$Bi emits a $\gamma$-ray with an energy only 10 keV lower than the $Q_{\beta\beta}$ value. As a result, any process that contributes a steady-state population of $^{222}$Rn inside the LXe is important. For the purpose of estimating the sensitivity, it is assumed that nEXO will have 600 $^{222}$Rn atoms continuously present in the LXe, a factor of 3 higher than observed in EXO-200 [10]. This factor is based on an estimate of the expected inner surface area of the xenon recirculation system in nEXO relative to EXO-200.

Events from the $^{222}$Rn decay chain can be tagged using a Bi-Po veto which identifies time- and space-correlated $\beta$ and $\alpha$ decays. An equivalent efficiency as that achieved by EXO-200 was assumed for nEXO to reject Bi-Po events from $^{214}$Bi decaying directly in the LXe volume inside the TPC field cage. Tagging or vetoing of $^{214}$Bi decays in the LXe outside the TPC field cage may be possible in nEXO by exploiting the light collected by the SiPMs on the barrel (not possible in EXO-200). The ability to identify $^{214}$Bi decays from $^{222}$Rn daughters that have drifted on to the cathode is the subject of ongoing studies, including special cathode designs and analysis techniques. The background model presented in this work combines $^{214}$Bi in the region outside the TPC field cage and on the cathode into one term, and assumes a tagging efficiency of $\sim$40% for these decays.

### 3.3.2 The Monte Carlo Detector Model

A GEANT4-based application [30] is the primary tool used to simulate energy depositions in the detector. A GEANT4 geometry model of the detector design described in Section 4 has been fully implemented. This geometry uses standard GEANT4 shapes to facilitate modifications allowing evaluation of design alternatives. While approximate geometries were used, care was taken to ensure all significant components are included, properly accounting for mass and materials. Visualizations of the simulated geometry are shown in Figure 3.7. A detailed list of components in the detector model and physics processes utilized by the simulation is provided in [32].

The GEANT4 tool kit is used to simulate generation and transport of particles from radioactive decay. In order to save computing time, a subset of daughters of the $^{238}$U and $^{232}$Th chains are simulated independently; their resulting energy deposits are subsequently merged with the appropriate branching ratios. Radionuclide selection is based on the emission of $\gamma$-radiation with energy $>$100 keV and with intensity $> 1\%$. In the $^{238}$U chain, decays from $^{234}$Pa, $^{226}$Ra, $^{214}$Pb, and $^{214}$Bi are simulated. $^{228}$Ac, $^{224}$Ra, $^{212}$Pb, $^{212}$Bi, and $^{208}$Tl are the only $^{232}$Th daughters considered. Generation of $2\nu\beta\beta$ decays is performed using the algorithm from [46], validated by the analysis of EXO-200 data. The simulations are weighted to ensure a minimum of $10^5$ events are generated with summed electron energies above 2250 keV. A FLUKA [47] model of nEXO has been developed for use in dedicated studies involving cosmogenic activation of detector components and neutron interactions.

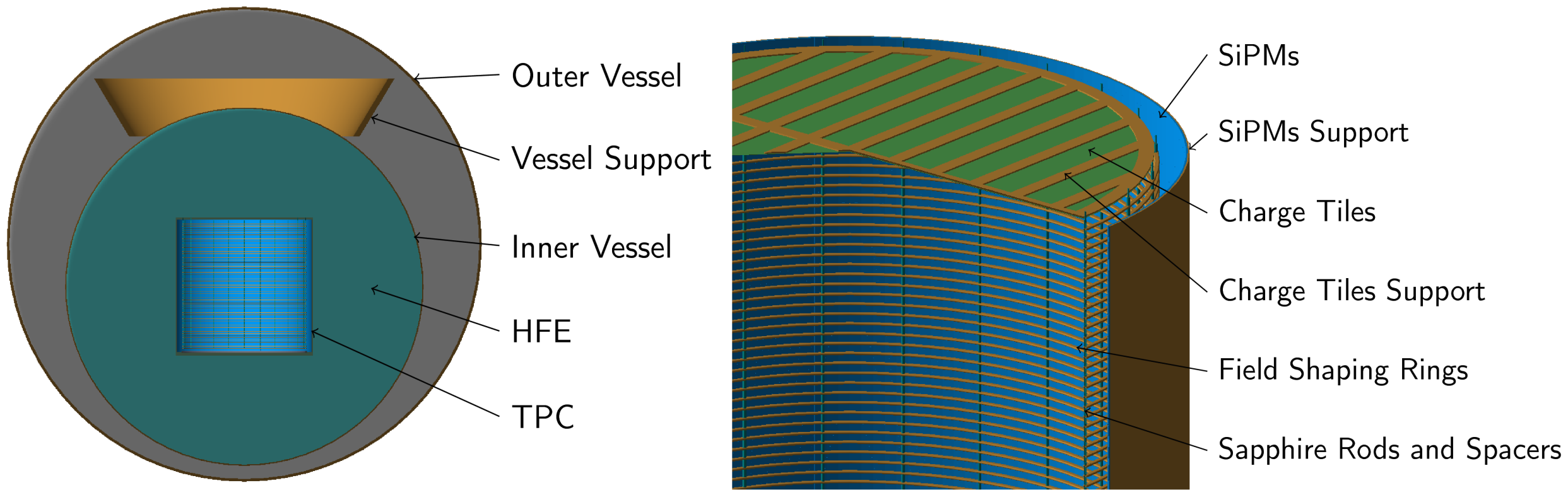

**Figure 3.7:** Visualization of the GEANT4 simulation geometry. A cross-section of the components within the outer vessel are shown (left) with a close-up of the TPC (right). The underground laboratory walls and the large water shield surrounding the outer vessel are not shown but are included in the full GEANT4 model geometry.

### 3.3.3 Detector Response

The experience of EXO-200 provides a basis for estimating the performance of nEXO, in particular of reconstructing energy, position, and multiplicity of each event.

The output of the GEANT4 simulation is reconstructed by software that applies detector effects from charge and light transport, based on MC energy depositions in the detector. Relevant event parameters are obtained by an event-based analysis.

In order to mimic the detector's ability to identify distinct interaction sites, the simulated energy deposits are aggregated by a clustering algorithm. This algorithm produces clusters of energy depositions by sequentially combining energy deposits within a radius $R = 3$ mm from the energy-weighted cluster center. Clusters below an energy threshold of 75 keV are discarded. The cluster multiplicity in each event allows the classification of events as either SS or MS. This choice of the cluster parameter $R$ results in a background $\gamma$-ray event SS fraction, within $Q_{\beta\beta}$ $\pm 1.7 \cdot$FHWM, that is about half ($\sim$10%) of that seen in EXO-200 ($\sim$20%). This extrapolation is justified by the projected hardware improvements (most notably the factor 3 reduction in charge channel pitch) coupled with the improvement generated by using information about the SS cluster size as obtained in EXO-200 Phase II [9]. A fiducial cut removes clusters that fall within 1.5 cm of the inner edges of the TPC field cage, including around the cathode and anode.

In the post-simulation reconstruction code, energy resolution is applied through a convolution with a Gaussian distribution of width obtained from a quadratic function analogous to that observed in EXO-200 [7] and coefficients that are scaled to achieve the expected nEXO resolution of $\sigma/Q_{\beta\beta} = 1\%$.

The collection efficiency of scintillation light is a key factor in determining the energy resolution in LXe TPCs. The GEANT4 nEXO simulation was used to propagate scintillation photons through the detector until they were absorbed by a SiPM depletion region, as discussed in Section 4.1.4.

The reconstruction algorithm employed for the present sensitivity estimate is simple and computationally inexpensive; it has been demonstrated that it can reproduce the shape of spectra observed by EXO-200, providing confidence in the methodology.

A more sophisticated approach, based on modeling and analysis of the waveform signal, induced by drifted electrons on each charge-collection strip, is under development. It also includes realistic effects from charge generation, charge drift, and readout electronics. Preliminary results indicate that the simple algorithm likely underestimates the SS-MS discrimination capability that is achievable.

Uncertainties due to systematic effects have not yet been investigated in detail. Such systematics could arise, for example, from biases in the energy reconstruction or other calibration effects, or from detector response non-uniformity. Such effects are not significant in EXO-200 and are therefore not expected to significantly impact the nEXO sensitivity calculation.

### 3.3.4 Background Budget

Figure 3.8 visualizes the contribution of different background components and their uncertainties. This figure shows the measured activities in Table 6.1 multiplied by the SS hit efficiency for events with energy within $Q_{\beta\beta}$ $\pm$FWHM/2 and within the inner 2000 kg of LXe. Contributions to the background budget are grouped by material, nuclide, and component. Contributions for which a non-zero contamination level was measured by radioassay are considered separately from those for which only an upper limit is available.

$\gamma$-ray interactions from the $^{238}$U decay chain constitute more than 70% of nEXO's background in the inner 2000 kg. A fraction of this component arises from materials for which only upper limits are currently available. Hence nEXO's expected background may fall as these radioassays are replaced by more precise measurements or higher purity materials are selected for use in the nEXO design.

Improvements in the data analysis are also expected to reduce the background arising from $^{137}$Xe. This is important because, as shown in Figure 3.10, this background is uniformly distributed in the detector volume. At a sufficiently deep location, a straightforward active muon veto could efficiently reduce the $^{137}$Xe with acceptable loss in livetime. As a conservative measure, no muon-based vetoing has been assumed in the analysis presented here.

The breakdown by component in Figure 3.8 shows that TPC elements dominate the background due to their vicinity to the central LXe region, while massive but distant components such as the cryostat vessels are subdominant. By material, the largest contribution arises from radio-impurities in the copper, primarily in the TPC vessel, which is the largest-mass component near the LXe. Cables and field rings (and their associated support equipment) are the next largest components. Overall, background counts are rather evenly distributed across various TPC internal components. This is indicative of a well-balanced design for the experiment.

### 3.3.5 Analysis of Trial Monte Carlo Data

The data analysis of nEXO will be based on EXO-200 experience and will integrate new features arising from nEXO's specific design and readout. For each event, the following quantities are extracted by the basic reconstruction code: energy, event position (X,Y, Z), and multiplicity (SS/MS). From the event position, the distance from the closest detector surface, labeled Standoff Distance (SD), is also computed.

The event reconstruction capability is utilized to categorize events into the SS and MS classes. The former is predominantly composed of $\beta$-induced signal-like events, the latter of $\gamma$-ray induced background-like events. Point-like $\alpha$-induced events are identified by their large scintillation to ionization signal ratio. These separations are analyzed and determined event-by-event. The nEXO analysis utilizes all event sets by performing a simultaneous fit of both the SS and MS event distributions. The approach of this coupled fit method offers the advantage that signal and background can be determined simultaneously. The energy resolution, which enables identification of multiple peaks within the decay series, provides important constraints for the background model. Further statistical signal and background discrimination is achieved by utilizing the event location. On average, $\gamma$-ray interactions occur preferentially near the detector surface. The SD parameter is used as a third independent fit variable. This additional analysis dimension helps refine the background model fit, which is dominated by $\gamma$-ray components.

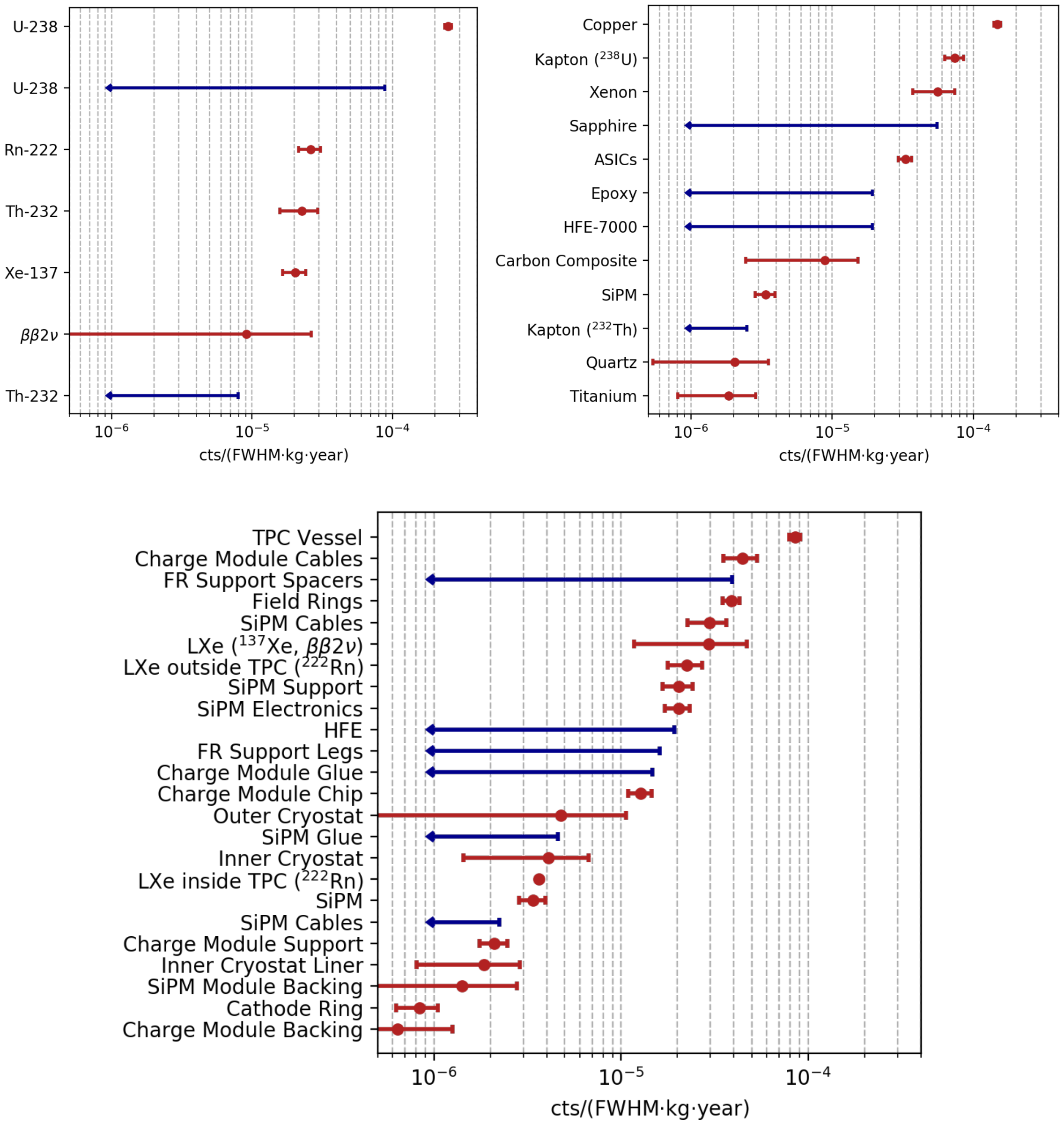

**Figure 3.8:** Histograms of the SS background contributions by nuclide (top left), material (top right), and detector component (bottom) for nEXO with energy within $Q_{\beta\beta}$ $\pm$FWHM/2 and in the inner 2000 kg. The blue arrows indicate 90% C.L. upper limits while the red circles indicate measured values with $1\sigma$ uncertainties. For the top left plot, upper limits and measured values are grouped separately, leading to multiple entries for certain nuclides. Systematic uncertainties and contributions smaller than $5 \times 10^{-7}$ cts/(FWHM·kg·y) are not shown.

For the purposes of determining sensitivity and discovery potential of nEXO, the data analysis is performed on an ensemble of simulated trial experiments, here called "toys."

Toy datasets are generated by randomly sampling the probability distributions functions (PDFs), $S_j^{\text{SS,MS}}$, describing the energy-standoff distribution in the detector arising from each background component $j$, with SS and MS PDFs considered separately. These PDFs are created from the distribution in energy-standoff space of simulated backgrounds generated by nEXO's GEANT4 and reconstruction simulations.

The overall normalization $n_j$ of each component $j$'s PDF is set using the formula:

$$n_j^{\text{SS,MS}} = M_j \cdot \varepsilon_j^{\text{SS,MS}} \cdot A_j \cdot T \tag{3.1}$$

where $M_j$ is either the mass or the surface area of the detector component $j$ (whichever is relevant), $\varepsilon_j$ is the hit efficiency, $A_j$ is the specific activity for the nuclide (normalized by mass or surface area), and $T$ is the observation time. Values for the $M_j$ and $A_j$ parameters can be found in [32]. The hit efficiency $\varepsilon^{\text{SS,MS}}$ is the probability that a decay in a specific detector component will produce an event of given multiplicity within the energy and standoff selection region. $\varepsilon^{\text{SS,MS}}$ is obtained from nEXO's GEANT4 MC.

Equation 3.1 and the $S_j^{\text{SS,MS}}$ are combined to produce a total background spectrum PDF in energy-standoff space. This PDF is then randomly sampled to produce a toy dataset, represented by two histograms for the SS and MS events.

Each toy dataset is fit by minimizing the negative log-likelihood (NLL, $\mathcal{L}$) constructed from the MC-generated PDFs of each component:

$$\mathcal{L}_{\text{nEXO}} = \mathcal{L}_{\text{SS}} + \mathcal{L}_{\text{MS}} - \ln(G_{\text{const}}) \tag{3.2}$$

where $\mathcal{L}_{\text{SS(MS)}}$ is the binned NLL built from the toy SS (MS) data compared to the corresponding PDFs $S_j^{\text{SS,MS}}$ generated by the nEXO MC for each background $j$ (and the $0\nu\beta\beta$ signal). The definition of the logarithmic likelihood function closely follows that outlined in Section 6.2 of [3]. The fit parameters are the expected number of counts in each component, $n_j^{SS+MS}$ (total, both SS and MS), and the fraction of SS events in that component $f_j^{\text{SS}}$. These parameters are fit, rather than being fixed, to accommodate the uncertainty the final nEXO experiment will have about background intensities and SS/MS discrimination. $G_{\text{const}}$ is a multivariate Gaussian function constraining some fit parameters. The NLL fit is implemented using ROOFIT [48] and MINUIT [49].

The choice and combinations of parameters used in the analysis can likely be further tuned to optimize the sensitivity reach of future analyses. EXO-200 showed a $\sim$15% sensitivity improvement [9] over the analysis described above when additional information about the spatial extent of SS events was incorporated with other topological variables in a BDT parameter, as described in Section 3.2.2.

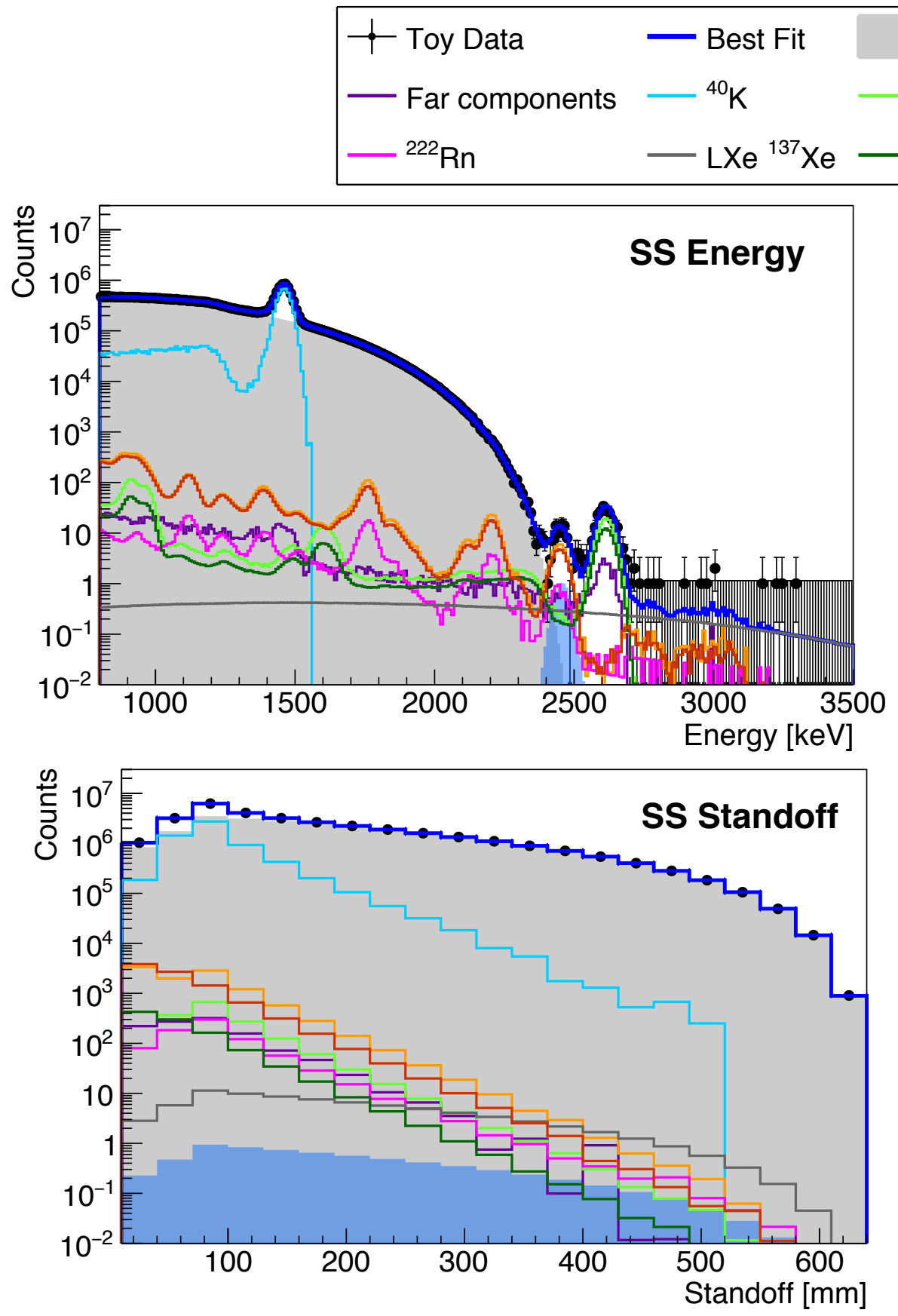

**Figure 3.9:** Result of the NLL fit to a representative nEXO toy dataset generated assuming a $0\nu\beta\beta$ signal corresponding to a half-life of $5.7 \times 10^{27}$y and 10 years of detector live time. The top plots are the energy distribution histograms while the bottom plots are the standoff distances; left (right) spectra are for SS (MS) events.

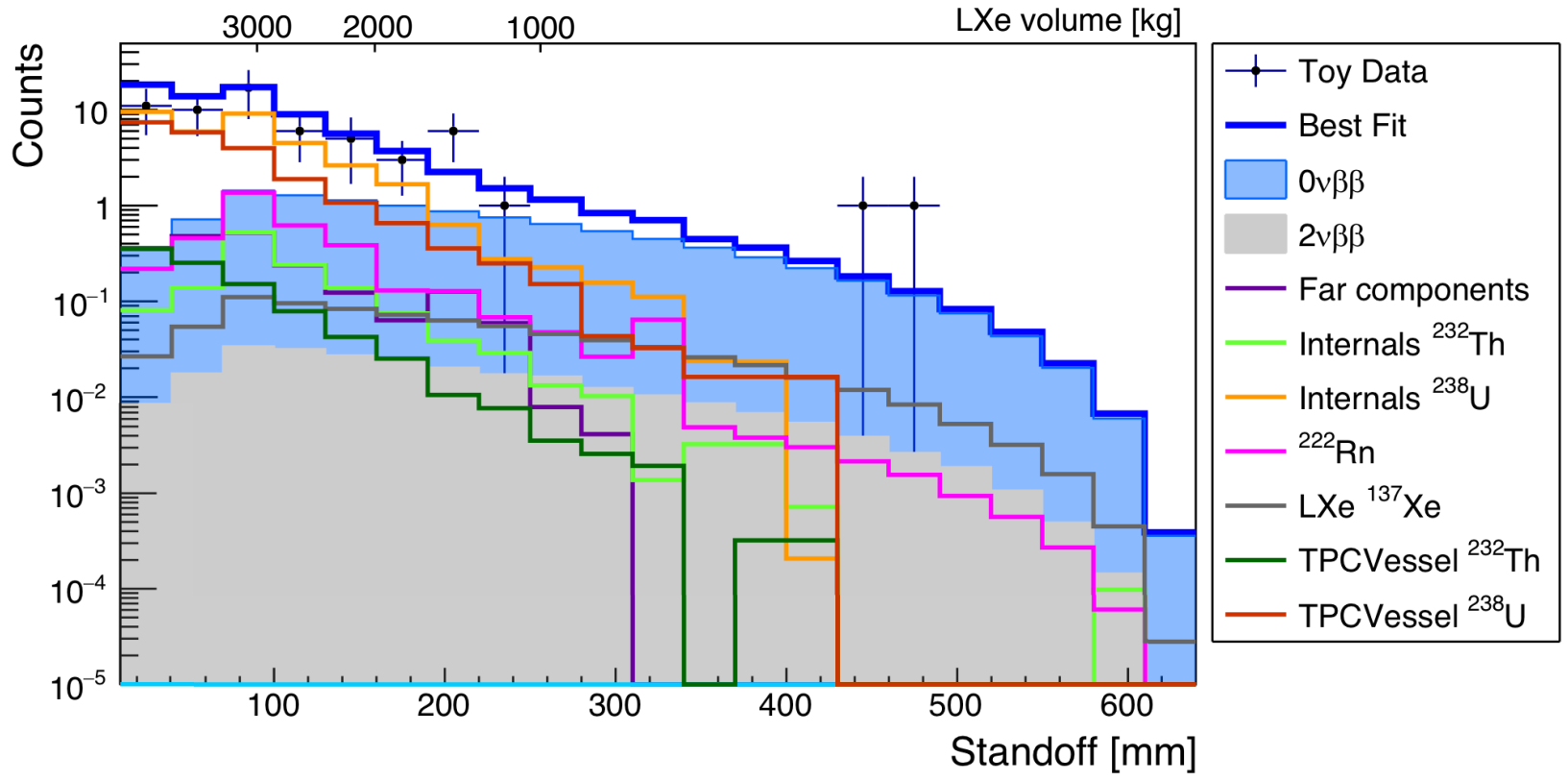

**Figure 3.10:** Standoff distribution for the fit results from a representative toy MC dataset with 10 years live time. Only SS events with energy within $Q_{\beta\beta}$ $\pm$FWHM/2 are included. The $0\nu\beta\beta$ signal corresponds to a half-life of $5.7 \times 10^{27}$y.

As an example of this multi-dimensional analysis, the SS and MS energy and SD distributions from a simulated toy experiment are shown in Figure 3.9. Best fit results for all radionuclides (arranged by component groups) are shown. Overall, the SS energy spectrum is dominated by $2\nu\beta\beta$ events, while the tail of the $^{214}$Bi photoelectric peak is the largest background contributor near $Q_{\beta\beta}$. As a function of standoff distance, the distribution of external radioactivity drops rapidly and is markedly different than the distribution of $0\nu\beta\beta$, $2\nu\beta\beta$, and $^{137}$Xe events, which are uniform throughout the LXe volume. This behavior adds additional resolving power beyond energy and event topology to the analysis, improving the ability to distinguish a $0\nu\beta\beta$ signal from background. In the central region of the detector, external gamma backgrounds are reduced by several orders of magnitude. Given the size of nEXO and the absence of any material other than LXe within the TPC volume, 2.5 MeV gamma-rays have to traverse more than 7 attenuation lengths, and likely scatter multiple times, before reaching the center of the LXe volume. To highlight the backgrounds of greatest concern, the standoff distribution is shown in Figure 3.10 with a cut selecting the $\pm$FWHM/2 energy region around $Q_{\beta\beta}$.

The power of nEXO's multi-parameter approach to discriminate the signal from backgrounds can be further appreciated by looking at Figure 3.11. While the energy resolution alone marginally resolves a $0\nu\beta\beta$-peak from $\gamma$-peaks caused by external radioactivity, the standoff distance variable provides additional resolving power in combination with the event-type variable (SS/MS). The large body of xenon is not simply used as a passive shield but actively measures external backgrounds and internal double beta decays simultaneously. The outer volumes effectively quantify external backgrounds, while the inner volumes determine the $\beta\beta$-signal. This combination of variables adds confidence in case of a discovery.

The background count rate as a function of the fiducial mass is shown in Figure 3.12 for events in an energy window $\pm$FWHM/2 around $Q_{\beta\beta}$. Shown is the median and 95% band of the distribution resulting from the random draw of activities values $A_j$. Clearly, a large homogeneous detector like nEXO cannot be characterized by a single background index value. Instead, its background rate is a position-dependent function. While for specific and circumscribed purposes it may be convenient to think in terms of a single background rate in a region of energy, one should always be aware that this point of view is not generally appropriate for nEXO. However, as a reference, nEXO is predicted to achieve a background rate of $3.6 \times 10^{-4}$ cts/(FWHM·kg·y) in the inner 2000 kg of LXe. This choice of mass value will become clear in the next section.

### 3.3.6 Sensitivity Calculations

In this section we discuss the technical details of the sensitivity calculation. The results of this calculation are presented in Section 3.3.7.

The sensitivity and discovery potential of nEXO are determined by applying the analysis described above to an ensemble of simulated "toy" experiments, and finding the confidence interval on the $0\nu\beta\beta$ rate using a profile likelihood method. This frequentist limit-setting methodology is discussed in detail in [32], and only its main traits are reviewed here.

Instead of relying on Wilks's theorem to describe the expected distribution of the NLL test statistic under the null hypothesis, the NLL ratio threshold $\lambda_c(\mu)$ is explicitly computed at all values of $\mu$ covered in this study. The EXO-200 analysis verified that EXO-200's data falls under the conditions of validity of Wilks's theorem [50], and therefore took advantage of the consequent statistical simplification. nEXO's data will include fewer background events than EXO-200, and thus the applicability of Wilks's theorem is not guaranteed, since it provides an accurate approximation only in the limit of a large number of events. As a result, when performing the profile likelihood analysis, nEXO must calculate a separate value for the test statistic at each hypothesis to include or exclude a given fit result from the confidence interval.

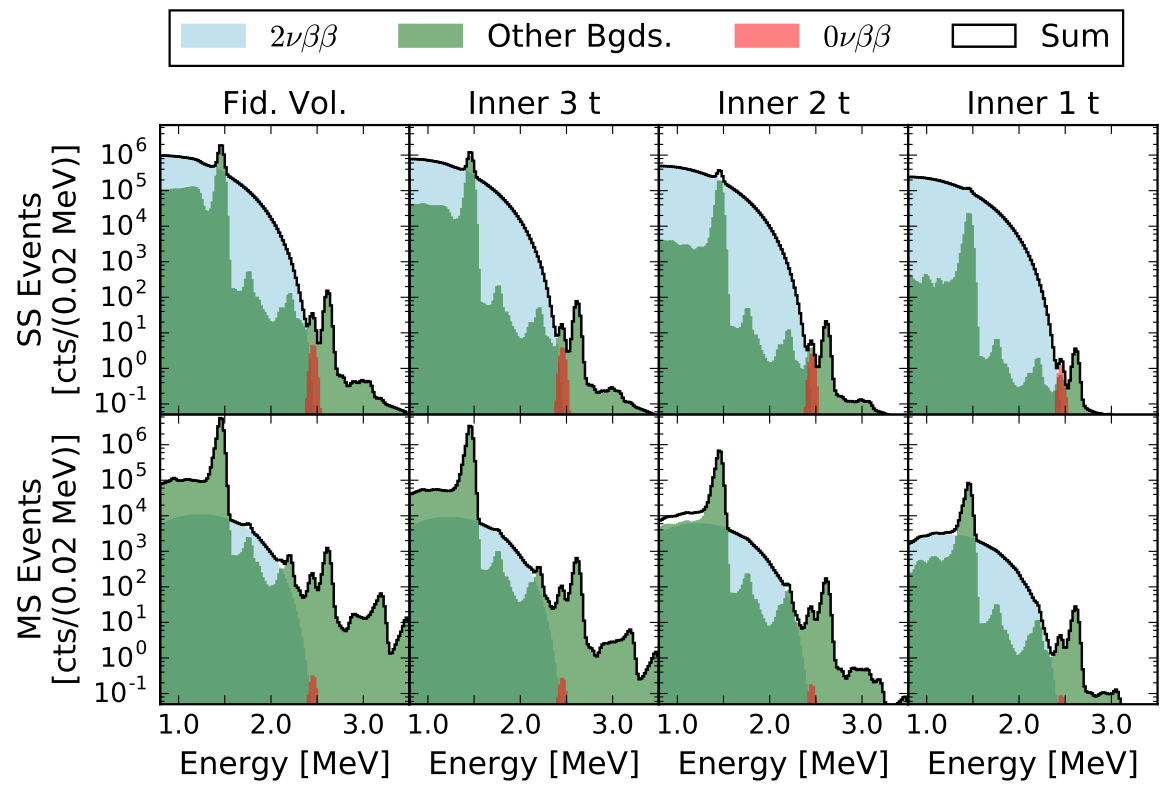

**Figure 3.11:** Energy spectra for SS and MS events as a function of the LXe mass. Spectra are evaluated for a detector live time of 10 years. The $0\nu\beta\beta$ signal corresponds to a half-life of $5.7 \times 10^{27}$ years.

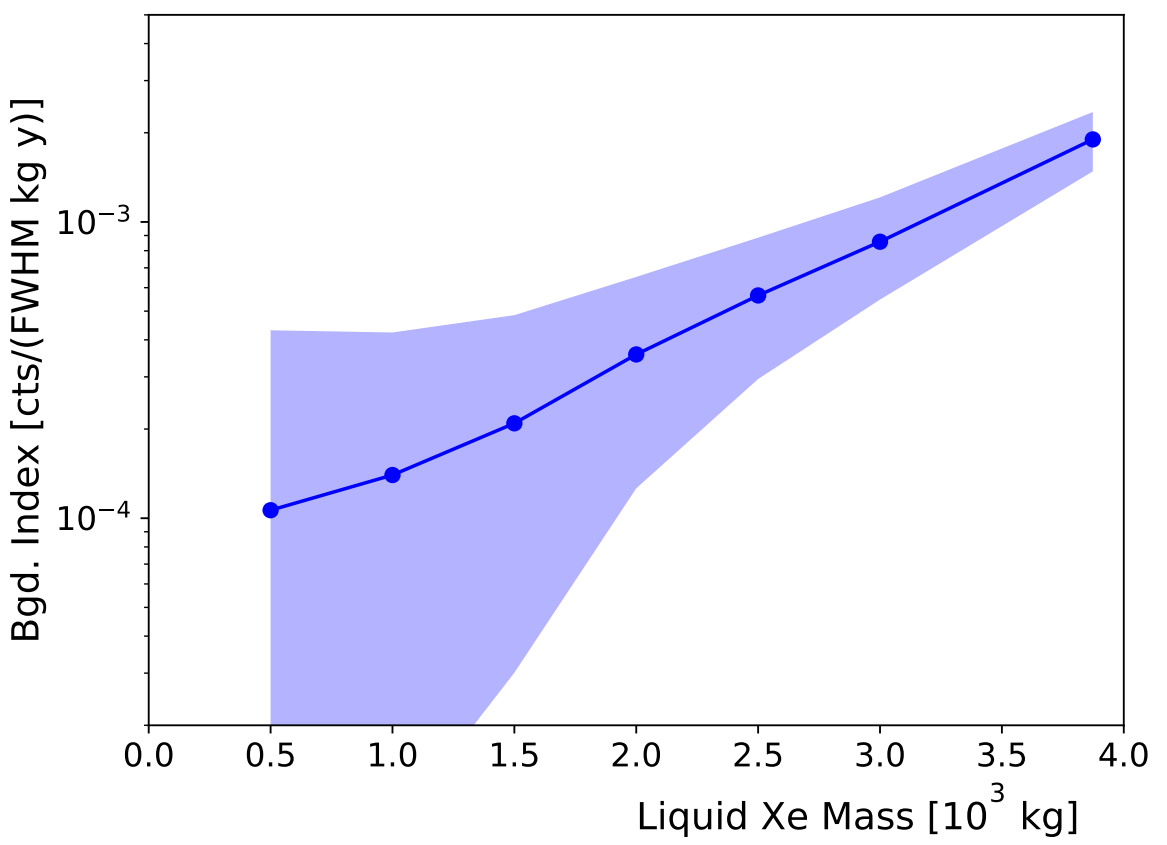

**Figure 3.12:** Median background as a function of the LXe fiducial mass derived from $10^4$ toy-MC simulations with detector live time of 10 years. The band corresponds to the 95% confidence belt of the distribution of background counts at each fiducial mass value.

Following the approach suggested in [51], the $\lambda_c(\mu)$ curve is obtained in a frequentist manner via MC generation of the distribution of the test statistic under each hypothesis. The NLL ratio test statistic is calculated as

$$\lambda(\mu) = 2\left(\mathcal{L}_{\mu} - \mathcal{L}_{\mu_{\text{best}}}\right) \tag{3.3}$$

where $\mathcal{L}_{\mu}$ is the log-likelihood fixing the signal expectation to $\mu$ and $\mathcal{L}_{\mu_{\text{best}}}$ is the log-likelihood letting the signal parameter assume its best-fit value $\mu_{\text{best}}$.

Over a range of hypotheses $\mu$, an ensemble of toy experiments is generated with a number of signal counts randomly drawn from the expectation $\mu$. The value of $\lambda(\mu)$ for each of these experiments is computed and the 90$^{\text{th}}$ percentile of the resulting distribution of $\lambda(\mu)$ defines the critical value $\lambda_c(\mu)$ for the 90% confidence interval used for sensitivity. The 99.7$^{\text{th}}$ percentile similarly defines $\lambda_c(\mu)$ for the 99.7% confidence interval used for the discovery potential.

Calculating $\lambda_c(\mu)$ requires generating and fitting an ensemble of many toys under a range of values for $\mu$. In order to reduce computing time, the calculation is performed at several discrete points, which are then fit with a third-order spline. This produces a smooth curve that interpolates between the calculated points and reduces the impact of the statistical uncertainty of the calculation of a quantile on a finite distribution.

The $\lambda_c(\mu)$ curve is obtained under a specified live time and expected distribution of backgrounds. Changing either of those assumptions requires the calculation of a new curve.

For any single toy data set, a given value of $\mu$ is included in the set of hypotheses that make up the confidence interval $C$ if $\lambda(\mu) < \lambda_c(\mu)$. A bisection algorithm is used to minimize the number of $\lambda(\mu)$ points that must be calculated for each toy experiment to determine $\mu_{90}$, the crossing point (or the greater of two crossing points, if two exist) between $\lambda_c(\mu)$ and $\lambda(\mu)$. This approach was validated by comparing the results obtained in the high-statistics regime, where Wilks's theorem holds, against those from MINUIT.

nEXO's sensitivity at a given background and live time is extracted as the median of the distribution of the upper limit $\mu_{90}$ from an ensemble of toy experiments generated with the null hypothesis $\mu = 0$ and under those livetime and background assumptions. An example of the distribution of $\mu_{90}$ is shown in Figure 3.13. The $0\nu\beta\beta$ half-life sensitivity is then inferred from the median $\mu_{90}$, the number of $^{136}$Xe nuclei, and the experiment's live time.

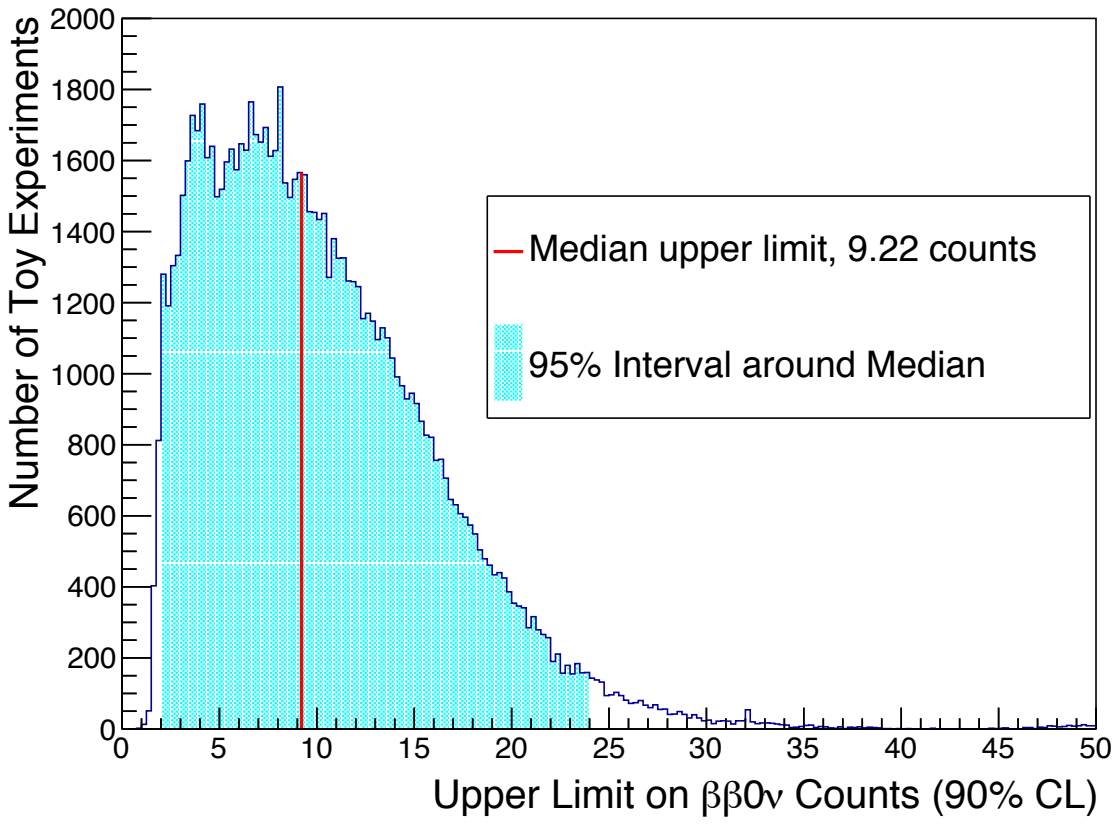

**Figure 3.13:** Distribution of $\mu_{90}$, the upper limit on the signal counts under the null hypothesis, obtained for several background realizations (toy experiments) at 10 years livetime.

In addition to the sensitivity, the median discovery potential at 3-sigma is also calculated. An experiment is evidence for a discovery if the 99.7% confidence interval, calculated as described above, does not include the null hypothesis $\mu = 0$. The median discovery potential is the $0\nu\beta\beta$ half life that produces discovery experiments 50% of the time. Determining the discovery potential entails a search over $0\nu\beta\beta$ rates to find which rate produces 50% discoveries.

### 3.3.7 Sensitivity Results

nEXO's size and extremely low background levels, coupled with an analysis that exploits the multi-parameter event signature provided by the TPC technique, result in a dramatic improvement in sensitivity compared to EXO-200. nEXO's median sensitivity to the $0\nu\beta\beta$ half-life for $^{136}$Xe at 90% C.L. is shown in Figure 3.14 as a function of the experiment's live time. After 10 years of data collection, the median 90% C.L. sensitivity reaches $9.2 \times 10^{27}$ years. A $3\sigma$ discovery potential of $5.7 \times 10^{27}$ years is predicted for the same live time.

The two-dimensional fit of energy and standoff distance allows nEXO to maximize its sensitivity by employing the largest possible fiducial volume, in contrast to a counting analysis which reaches maximum sensitivity only with a substantial fiducial volume cut. This is shown in Figure 3.15. Indeed, the full two-dimensional analysis shows an improvement of ~50% over a counting-style experiment. Figure 3.15 also motivates the earlier choice of presenting the background rate for the innermost 2000 kg of LXe where ~90% of the full sensitivity is achieved and a counting-style rate analysis reaches its maximal sensitivity.

The sensitivity to the $0\nu\beta\beta$ half-life of $^{136}$Xe can be converted into the corresponding sensitivity to the effective Majorana neutrino mass $\langle m_{\beta\beta}\rangle$ under the assumption of light Majorana neutrino exchange (Equation 2.9). Figure 3.16 shows the nEXO exclusion sensitivity to $\langle m_{\beta\beta}\rangle$ as a function of the lightest neutrino mass. The allowed neutrino mass bands are derived from neutrino oscillation parameters from [52, 53]. The $\langle m_{\beta\beta}\rangle$ exclusion band between 5.7 and 17.7 meV arises from the range of nuclear matrix elements, with EDF [54] and QRPA [55] at the minimum and maximum extreme respectively. Majorana neutrino masses are computed assuming the unquenched value of the axial-vector coupling constant of $g_A = 1.27$ [56], as discussed in Section 2.3.

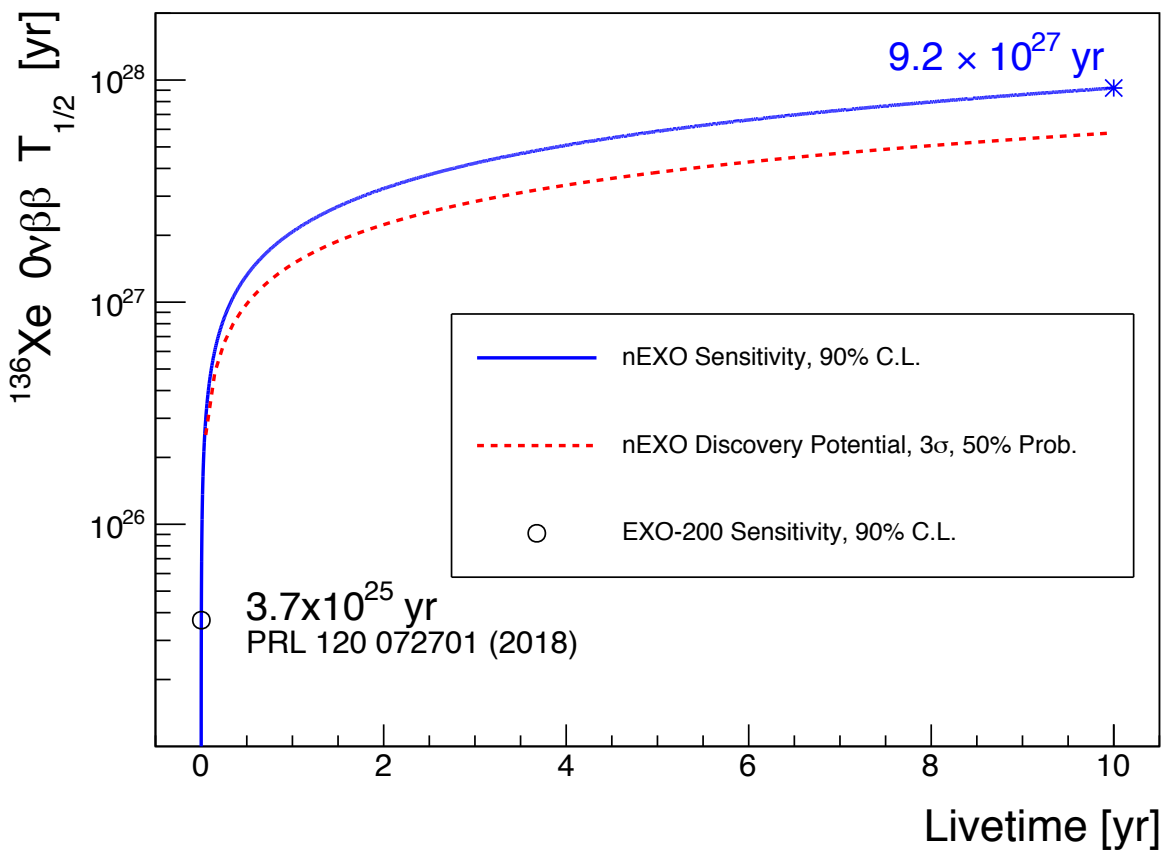

**Figure 3.14:** nEXO median sensitivity at 90% CL and 3$\sigma$ discovery potential as a function of the experiment livetime.

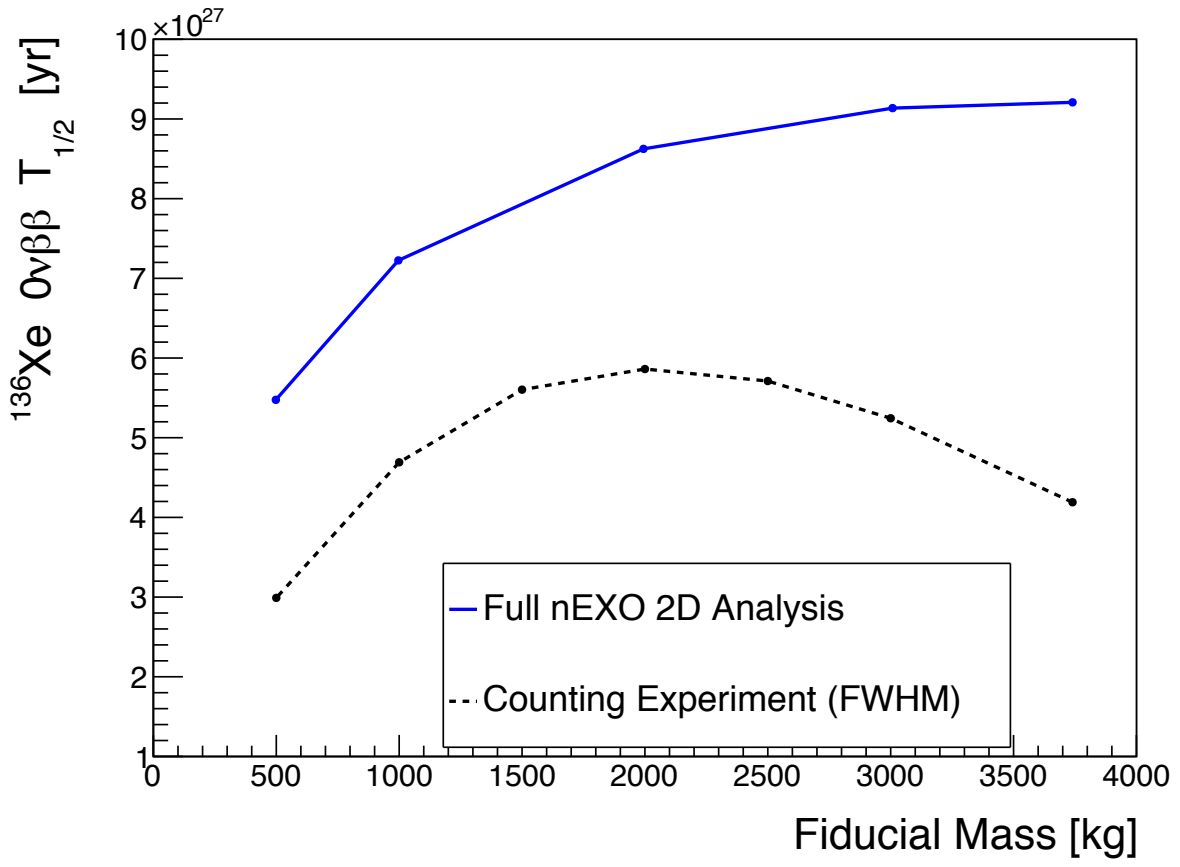

**Figure 3.15:** nEXO exclusion sensitivity at 90% C.L. as a function of fiducial LXe volume. The blue points (upper curve) are obtained from the full 2D energy-standoff fit, while the black points (lower curve) are the result of a pure counting experiment of events with energy in $Q_{\beta\beta}$ ±FWHM/2. Both analyses are performed using the method of [51].

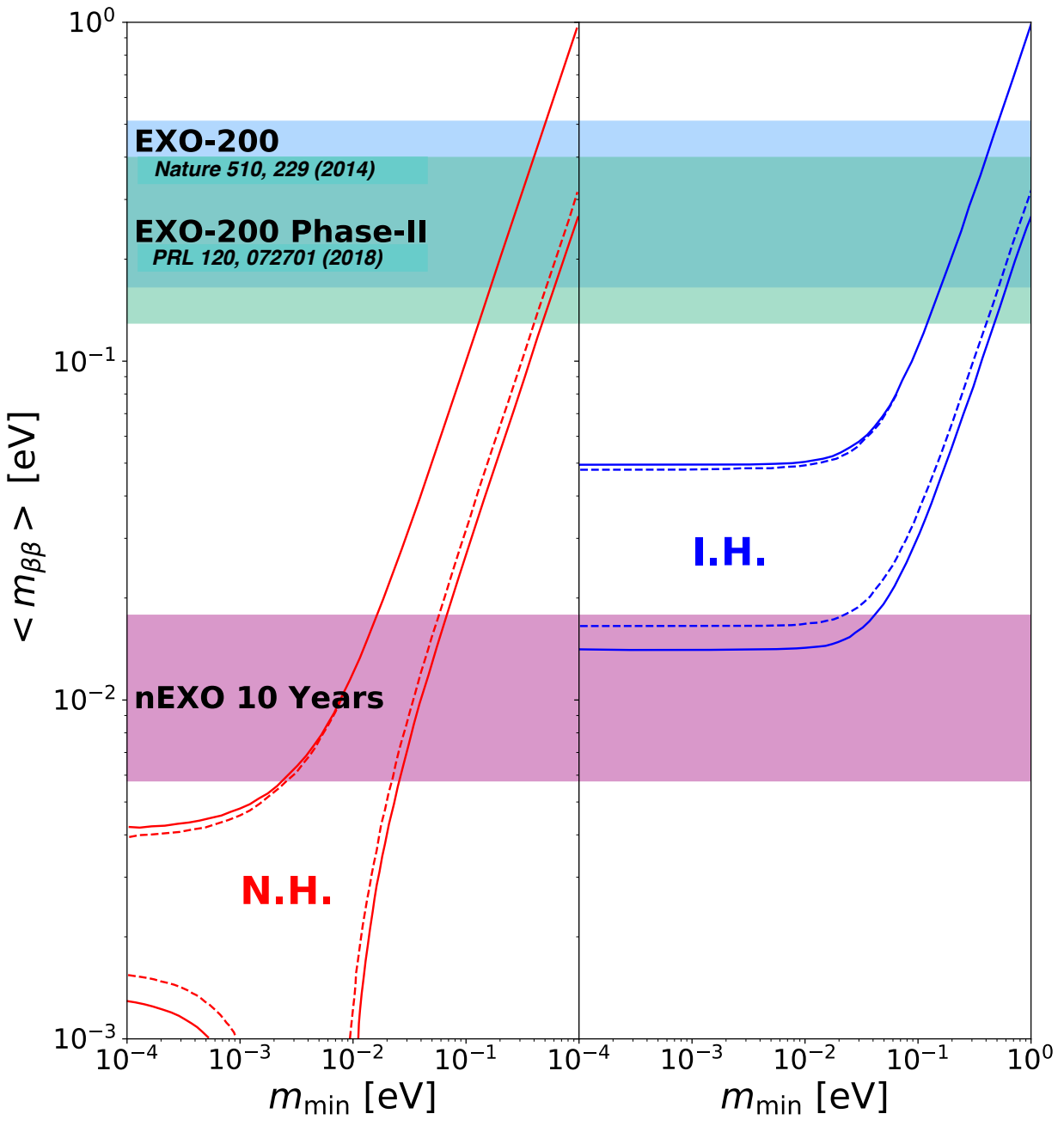

**Figure 3.16:** 90% C.L. exclusion sensitivity reach to the effective Majorana neutrino mass $\langle m_{\beta\beta}\rangle$ as a function of the lightest neutrino mass for normal (left) and inverted (right) neutrino mass hierarchy. The width of the horizontal bands derive from the uncertainty in nuclear matrix elements (see text) and it assumes that $g_A = 1.27$. The width of the inner dashed bands result from the unknown Majorana phases and is irreducible. The outer solid lines incorporate the 90% CL errors of the 3-flavor neutrino fit of reference [52, 53].

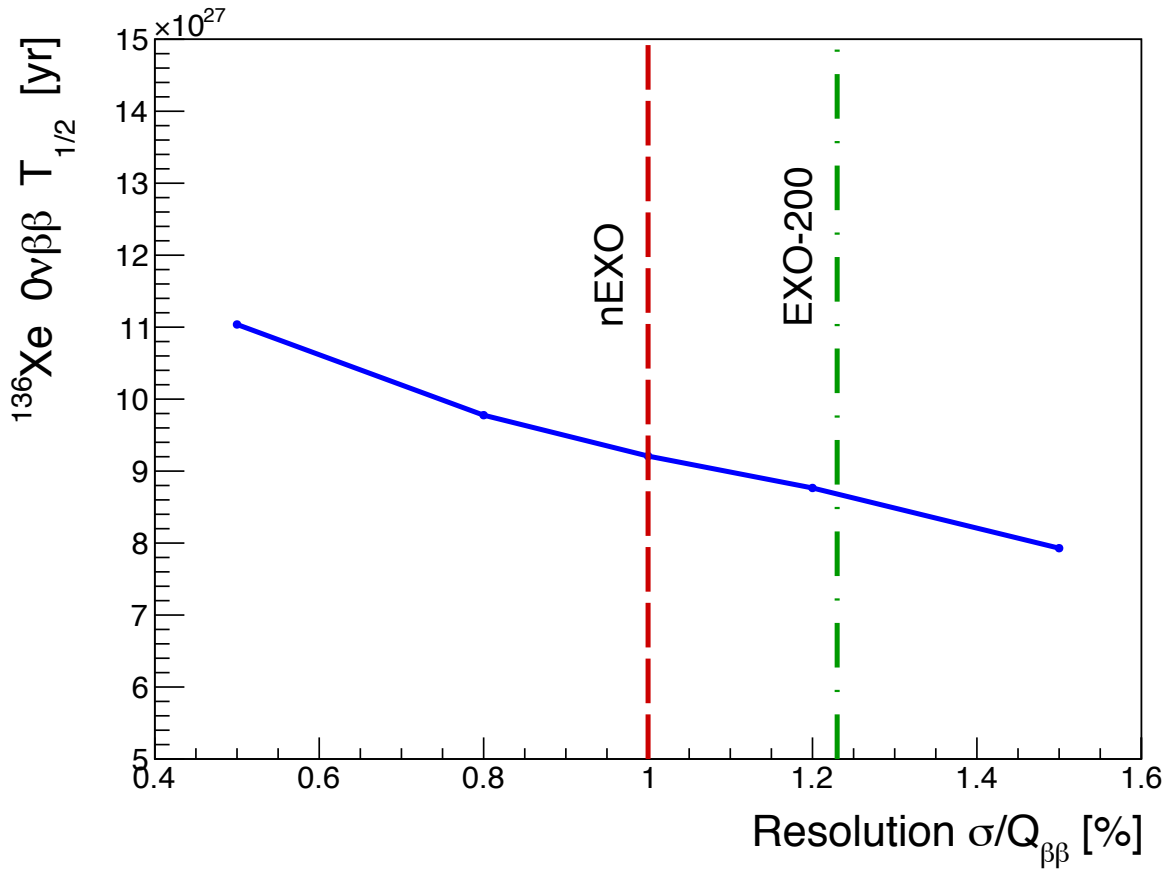

**Figure 3.17:** nEXO median exclusion sensitivity at 90% CL computed for different assumptions of the experiment's energy resolution.

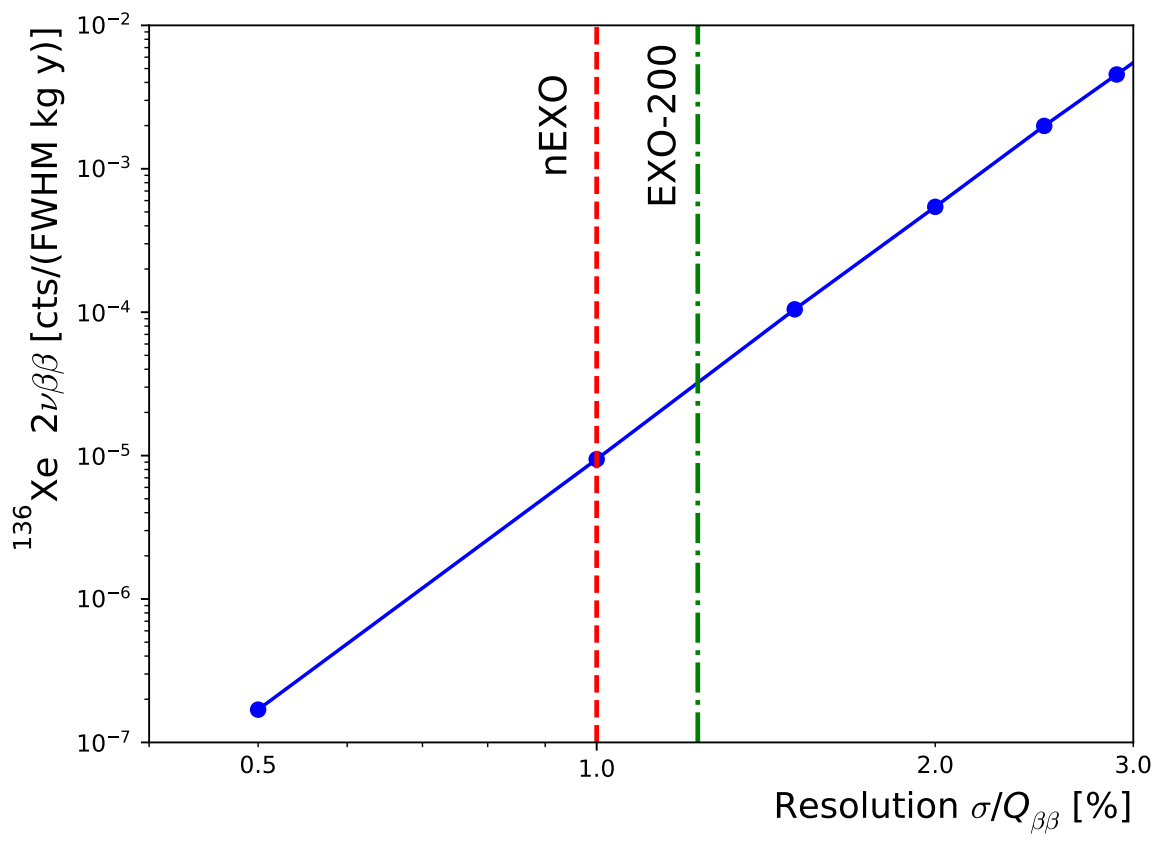

**Figure 3.18:** Calculated $^{136}$Xe $2\nu\beta\beta$ SS events falling within ± FWHM/2 of the $^{136}$Xe Q-value as a function of the energy resolution (assumed Gaussian). The expected and measured energy resolution of nEXO and EXO-200 respectively are shown for reference.

### 3.3.8 Sensitivity Variation Studies

In the following we explore how the nEXO sensitivity scales under a variation of important characteristic parameters: energy resolution and background. This section is meant to demonstrate that the nEXO sensitivity depends only weakly on these key parameters, counter to often stated community belief.

The results presented are based on a robust estimation of the backgrounds and realistic detector performance, extrapolated from EXO-200 and supported by nEXO-specific modeling results. We note that these results do not involve any extrapolation of materials radiopurity beyond what has been already measured. As the R&D continues, it is possible that better performance might be achieved, e.g. through improved material selection and engineering, improved analysis techniques, or hardware breakthroughs.

Traditionally, the analysis of energy spectra alone has been the workhorse of $0\nu\beta\beta$ searches, thus favoring calorimetric experiments with very high energy resolution. Over time, all $0\nu\beta\beta$ searches have started to introduce multiple parameters to measure and reject backgrounds (see e.g. [57–59]). By including standoff distance and event type (SS/MS), EXO-200 was able to provide outstanding physics results in spite of a somewhat limited energy resolution. nEXO's larger mass further enhances the utility of these additional variables in the multi-parameter analysis, leading to even less reliance on energy resolution. The size and homogeneous nature of the detector permits nEXO to take full advantage of this multi-parameter approach.

The impact of the energy resolution on nEXO's sensitivity is shown in Figure 3.17. The shallow slope of this curve is understood by considering the role of the photoelectric peak from $^{214}$Bi background, which falls only 10 keV away from the $^{136}$Xe $0\nu\beta\beta$ Q-value. In the range of energy resolutions considered here, only a small fraction of SS $^{214}$Bi background lies more than ±FWHM/2 away from $Q_{\beta\beta}$. Even at an energy resolution of $\sigma/Q_{\beta\beta} = 0.35\%$, 50% of the $^{214}$Bi SS background fall within $Q_{\beta\beta}$ ±FWHM/2. For this reason, the half-life sensitivity does not significantly change with the energy resolution. On the other hand, the sub-dominant contribution arising from the fraction of SS $^{208}$Tl decays that enter the same energy window is only $2.6 \times 10^{-5}$ at 1% resolution. This fraction increases rapidly as the resolution worsens, becoming $2.8 \times 10^{-2}$ at $\sigma/Q_{\beta\beta} \sim 1.5\%$. nEXO will utilize its position resolution and the characteristic difference in the spatial distributions of internal $\beta\beta$ and external $^{214}$Bi events to resolve these two signal components from each other. This position information adds resolving power that is independent of the energy observable.

The standoff and event type parameters in nEXO's multi-parameter analysis have no discriminating power against the unavoidable $2\nu\beta\beta$ background. As a result, energy resolution is the only proven method to suppress this background. Figure 3.18 shows the calculated $2\nu\beta\beta$ event rate in nEXO as a function of the energy resolution (assumed Gaussian). At nEXO's design energy resolution, the contribution of $2\nu\beta\beta$ decays at $Q_{\beta\beta}$ ±FWHM/2 amounts to only 0.34 counts over 10 years of data taking in the entire LXe volume, and is therefore negligible. This is also due to the fact that the $2\nu\beta\beta$ half-life for $^{136}$Xe has been found to be longer than that of all other common $0\nu\beta\beta$ candidates [60].

These results support our claim that nEXO's target energy resolution of $\sigma/Q_{\beta\beta} = 1\%$ will be sufficient to achieve the physics goal and suggest that further improvements, while beneficial, are not critical to achieving a compelling sensitivity.

Ongoing efforts focus on reducing the SS backgrounds through advancement in material screening and selection, optimization of the detector components (e.g. mass and location), and improved analysis. A parametric study was performed to evaluate the improvement in $0\nu\beta\beta$ sensitivity as a function of the total background. All materials activities from Table 6.1 were uniformly scaled down by a progressively larger fraction, with the exception of the $2\nu\beta\beta$ component which was held constant. New toy data sets were generated and then fit to obtain a median sensitivity estimate for different background scenarios. The resulting curve is shown in Figure 3.19, assuming 10 years of data taking. The $0\nu\beta\beta$ sensitivity increases by a factor 4 as the background rate is lowered by two orders of magnitude. The point labeled "baseline" refers to the case described in this report, while "aggressive" refers to a case in which plausible improvements are made.

It is interesting to observe that for nEXO, the common approximation that sensitivity $T^{0\nu}_{1/2}$ scales with background $B$ as $1/\sqrt{B}$ is not valid. Indeed, fitting the calculated sensitivity points in Figure 3.19 with a power law results in

$$T^{0\nu}_{1/2} \propto \frac{1}{B^{0.35}}$$

This finding is significant. First, it underlines the importance of using experiment-specific techniques to estimate sensitivities. Second, it shows that nEXO is less sensitive to background fluctuations than what might be inferred from a simple $1/\sqrt{B}$ scaling.

### 3.3.9 Why a 5-tonne Detector

The number of moles of $^{136}$Xe and hence the overall size of nEXO is clearly the most important parameter to set in the detector design. As discussed in Section 2, the relationship between the magnitude of Majorana masses and the half-life of the decay is complex and depends on many unknown factors. Here we recall that the particular decay mechanism (*e.g.*, which type of see-saw mechanism) assumed, the possible magnitude of $g_A$ quenching, and the amplitude of the nuclear matrix elements all contribute to a substantial uncertainty.

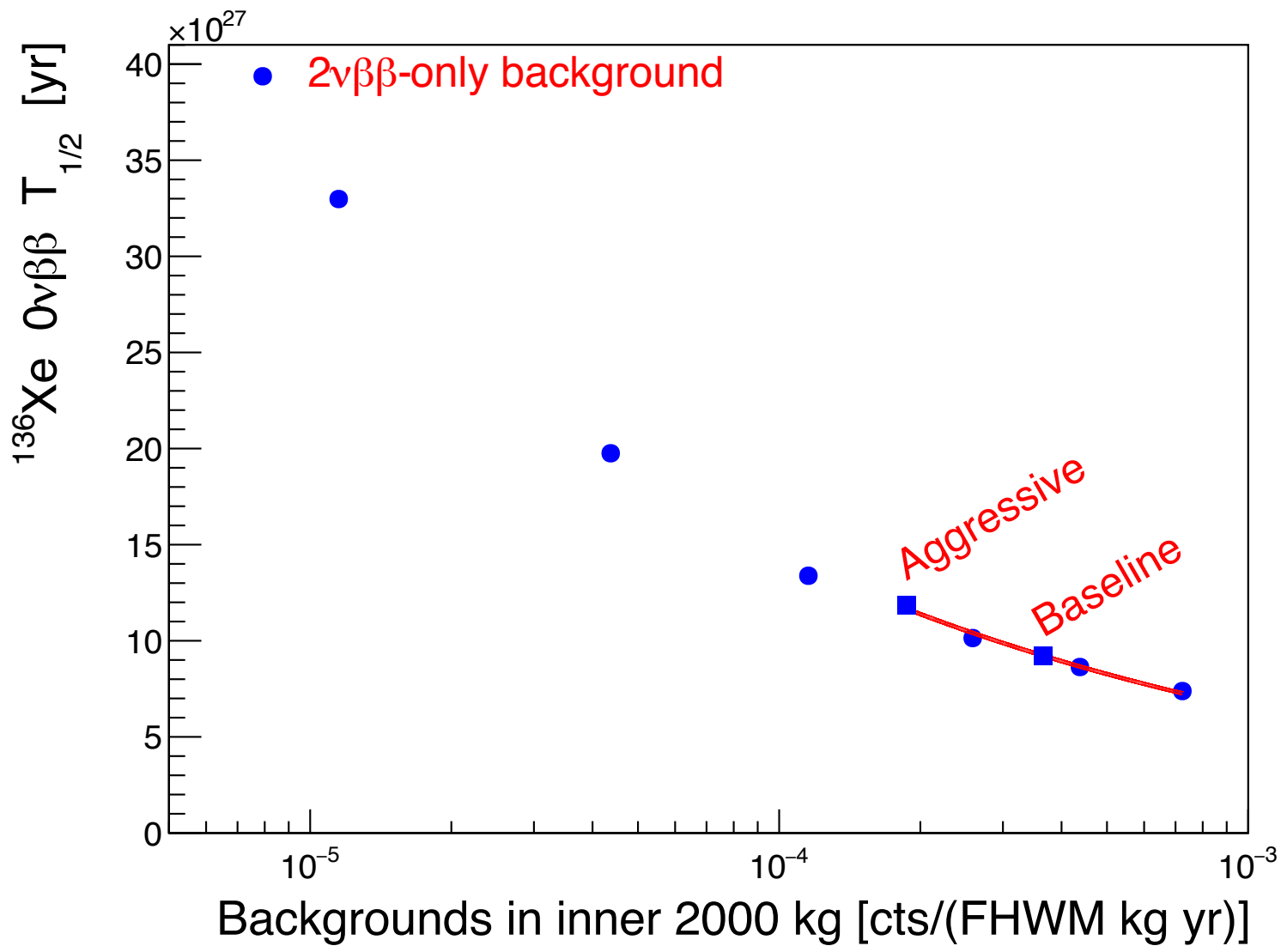

**Figure 3.19:** Sensitivity (blue circles) to the $0\nu\beta\beta$ half-life of a nEXO-like experiment as a function of total background in $Q_{\beta\beta}$ ±FWHM/2 in the inner 2000 kg. All components of nEXO's background model except for the $2\nu\beta\beta$ term are scaled to generate this curve. The red curve is the result of fitting the computed values with $T^{0\nu}_{1/2} \propto B^x$, giving $x = -0.35$ over the fitted region. The blue squares represent the sensitivity of the primary detector design discussed here, as well as an aggressive, but plausible improvement of the detector performance.

A valid, if somewhat generic, way to evaluate the projected impact of nEXO is to compare the discovery reach of $0\nu\beta\beta$ decay with that of a particle accelerator opening a new energy window on unexplored physics. In the accelerator case, increasing the energy by an order of magnitude is generally considered an appropriate and worthwhile step. In the case of the search for lepton number violation and Majorana neutrinos from $0\nu\beta\beta$ decay, the frontier is the decay half life, equivalent to the measurement of an ever smaller cross section unveiling exceedingly subtle physics. According to this admittedly loose metric, the increase in sensitivity from EXO-200 or KamLAND-Zen [21] to the nEXO sensitivity spans more than two orders of magnitude.

As discussed in Sections 3 and 3.3, the homogeneous LXe TPC offers advantages in terms of background identification and suppression, which scales with increasing mass, at least up to the size to which one can comfortably apply the technology successfully used in EXO-200. Again, $\sim 5000$ kg appear to be an appropriate choice, providing a very cost-effective discovery tool.

A sharper but model dependent argument can be derived from the goal set by the nuclear physics community for next generation $0\nu\beta\beta$ decay experiments. This argument relies on the Type-I see-saw mechanism, the traditional use of the unquenched value of $g_A$, and the consideration of all existing nuclear matrix element values for the relevant isotope, to compute a sensitivity in terms of the Majorana mass $\langle m_{\beta\beta}\rangle$. It is then considered a worthwhile—if somewhat arbitrary—goal to reach a sensitivity such that the horizontal band derived from neutrino oscillation experiments and displayed in the right panel of Figure 3.16 (inverted hierarchy) is entirely covered. The lower edge of such band corresponds to $\langle m_{\beta\beta}\rangle$=15 meV, as identified in the 2015 Long Range Plan for Nuclear Physics [2] and shown in Figure 2.3.

Utilizing the detailed sensitivity estimate laid out in Section 3.3.7 we derive the 90% CL sensitivity of nEXO to be of $T_{1/2} > 9.2 \cdot 10^{27}$ yr. Using the extremes of the matrix element calculations considered here [54, 55] this sensitivity translates into $\langle m_{\beta\beta}\rangle < 5.7 - 17.6$ meV. The upper end of this mass range is defined by a QRPA-type calculation that differs significantly from other models using this method. If one instead elects to use the nuclear shell model [61] as the reference point for the upper sensitivity value, the upper edge of the neutrino mass sensitivity range moves to 15.4 meV. Given the uncertainty in the nuclear matrix element calculations we consider nEXO with 5 tonnes of $^{\mathrm{enr}}$Xe to provide full coverage of the inverted mass hierarchy.

In order to provide a simple and transparent motivation for the source Xe mass of nEXO, we can consider only the event rate. The requirement of observing a certain number of events per unit time constitutes a simple criterion for the mass of enriched xenon $m_{\mathrm{enrXe}}$. Let $\varepsilon_e = 0.82$ be the $0\nu\beta\beta$ detection efficiency after all cuts, $\varepsilon_l = 0.9$ the detector live-time fraction, $M_{\mathrm{enrXe}}$ the molar mass of the enriched xenon, $a = 0.9$ the $^{136}$Xe isotope abundance (the enrichment), and $N_A$ the Avogadro constant. Assuming $\langle m_{\beta\beta}\rangle = 15$ meV and utilizing the NSM matrix element as benchmark [61], resulting in $\mathrm{T}_{1/2} > 9.7 \cdot 10^{27}$ yr, we calculate $m_{\mathrm{enrXe}}$ required to achieve a minimal $0\nu\beta\beta$ signal event rate of $R = 1\ \frac{event}{yr}$:

$$m_{\mathrm{enrXe}} = \frac{R \cdot T_{1/2} \cdot M_{\mathrm{enrXe}}}{\ln(2) \cdot N_A \cdot a \cdot \varepsilon_l \cdot \varepsilon_e} = 4758\ \mathrm{kg} \tag{3.4}$$

This simple estimate, depending on nothing but the radioactive decay law, shows that any low-mass detector will not have a sufficient number of events to make a sound case for discovery. This requirement for a "reasonable" event rate drives the need for a multi-ton detector.

It is also appropriate here to discuss the rationale behind the enrichment grade of the xenon. Natural xenon contains 8.9% of the isotope 136 [62]. The unit cost (\$/kg) of enrichment to $\sim 90\%$ grade in isotope 136 is roughly ten-fold the unit cost of the natural xenon feedstock. So, in general terms, the cost of 1 kg of $^{136}$Xe in the detector is roughly independent of whether this xenon is used "diluted" in natural xenon or enriched to 90%. On the other hand, a natural xenon detector would be ten times larger in volume, substantially more expensive, and present new technical challenges. We also note that the world production of (natural) xenon is $\sim 50,000$ kg/yr. So the extraction of $5,000$ kg of enriched material (in a few years, returning the light isotopes to the market) represents a reasonable, if challenging, endeavor. The use of $50,000$ kg of natural xenon, or the construction of a detector using substantially more than $5,000$ kg of enriched xenon, would provide a larger disruption to the world's xenon supply, and would likely require the development of a new technology to extract xenon from the air. A preliminary discussion of these xenon procurement issues is provided in Section 7.

# 4 The nEXO Detector

## 4.1 nEXO TPC

### 4.1.1 Overview

The design of the nEXO Time Projection Chamber (TPC) is rooted in that of its predecessor, EXO-200 [1]. It is a single-phase, LXe TPC filled with 5 tonnes of $^{enr}$Xe. The main design driver for the nEXO TPC is the realization of the largest possible, monolithic, instrumented LXe volume that allows to maximally exploit the background identification, suppression, and discrimination capabilities demonstrated by EXO-200. A general overview of the drivers for the design of nEXO are illustrated in Section 3.

The detector performance characteristics needed to maximize the physics reach of the larger detector define some scale-dependent, demanding engineering challenges and constraints. Among these are the efficient collection of scintillation light and ionization charge produced by interactions inside the detector. Another important technical hurdle is the robustness of the high voltage delivery to the TPC cathode in a detector where electric standoff is provided by a layer of $^{enr}$Xe, which needs to be kept to a minimum given its cost. Finally, the entire detector and its xenon-containing envelope needs to be built as lightweight as possible, using a small quantity of radiologically-certified materials, in order to keep the radioactivity load to a minimum.

The main features of the nEXO TPC design include:

- One single, 125 cm long drift volume, with anode and cathode at opposite ends of a cylinder having its base of 115 cm diameter (see Figure 3.2). This choice allows to define a large, all-xenon core of the detector where the radioactive background is essentially reduced to that intrinsic to the xenon source itself. The residual background at the cathode is optimally shielded from the core of the detector, an improvement over EXO-200, where the cathode was in the middle of the TPC and subtended a larger solid angle to the most sensitive volume of the detector. In addition to intrinsic $\gamma$-rays from impurities embedded in the cathode material, a relevant background is that from decays of $^{214}$Bi, a $^{222}$Rn daughter with a high probability of drifting to the cathode [2].
- In nEXO, the single, long drift volume requires excellent xenon purity from electronegative contaminants, with electron lifetimes better than 10 ms. This requirement derives from the expected quality of the electron lifetime correction needed to obtain a sufficient energy resolution and is discussed in details in Section 4.4. The nEXO TPC is designed with minimal use of large-area plastic components, which are known to out-gas impurities in vacuum-grade systems. In addition, the larger a monolithic detector is, the smaller the surface-to-volume ratio becomes, which reduces the load from out-gassing impurities. Finally, the diffusion of the ionization as it drifts the entire length of the TPC [3] is small enough not to degrade the topological information provided by the TPC (see Section 4.1.3.2). Xenon purification is discussed in Section 4.3.

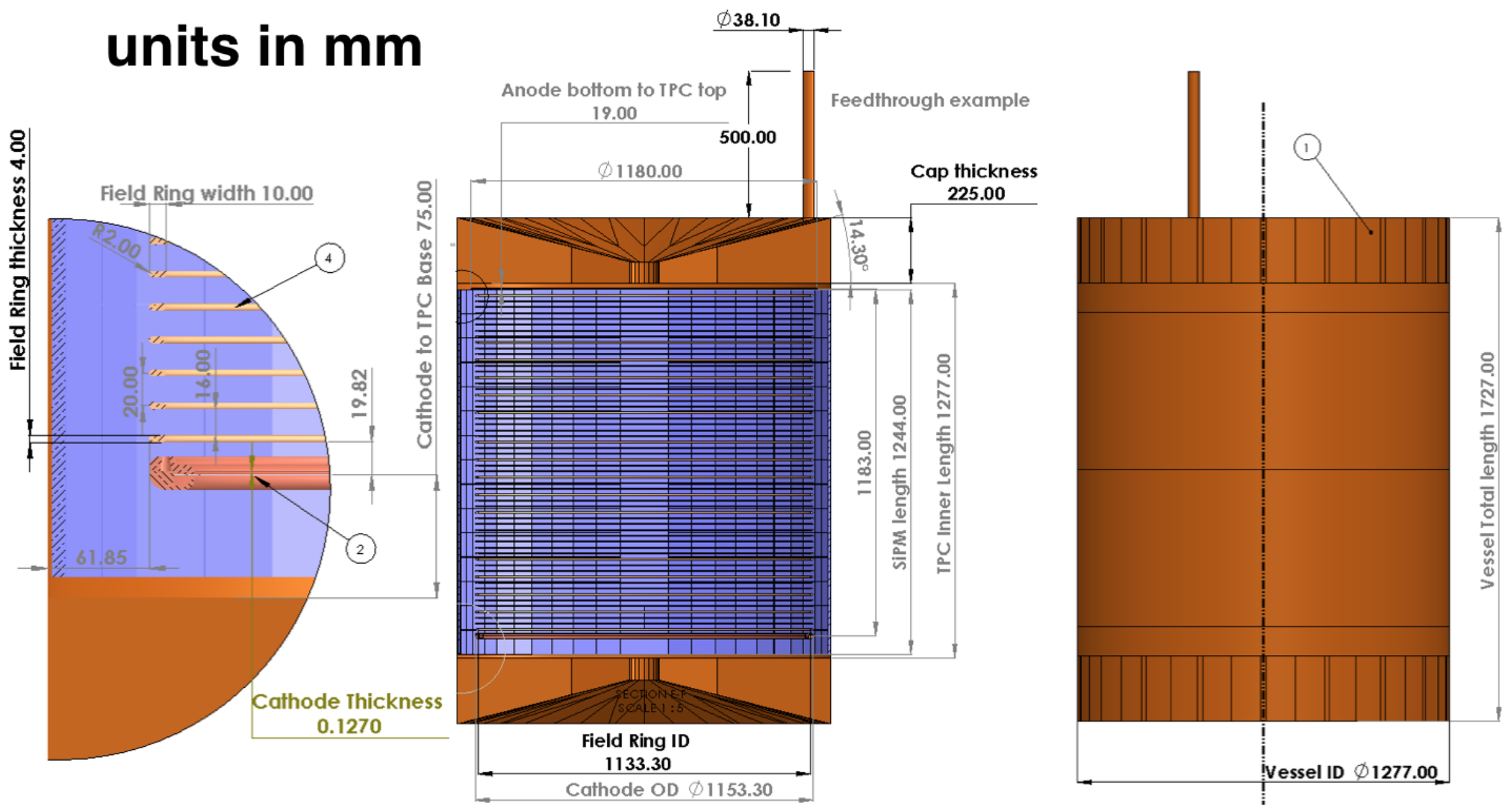

**Figure 4.1:** Schematic views with the basic dimensions of the nEXO TPC. The drawing to the right shows the projection of the copper vessel with the axis of the cylinder represented by the vertical dashed line. The central panel is the cross section of the TPC, in the same view and a detail, showing the cathode and the end of the field cage is represented to the left. Note that, while the LXe volume is a cylinder with diameter equal to its length, the external envelope of the vessel is longer because of the stiffening structure on the end-caps.

- The HV design parameters are derived from EXO-200, with appropriate scaling arguments. In the current, very conservative, configuration, the cathode needs to be safely biased to up to -100 kV, allowing for a maximum drift field of ~800 V/cm, twice that adopted in EXO-200 phase I [4]. Since in the case of nEXO the standoff insulating material is enriched LXe, a preliminary design that can safely handle the required high voltage was carefully studied, simultaneously optimizing the cathode profile while minimizing the cathode-to-vessel distance. The high voltage design for the nEXO TPC is discussed more extensively in Section 4.1.2, where the possibility of adopting a lower field is also considered.
- The nEXO TPC is designed to be as light as possible. It features 58 field shaping rings (FSRs), held together by 24, ~1.2 meter long sapphire single-crystal rods and spaced by 1.6 cm long cylindrical spacers (24 spacers between each pair of field rings). The sapphire rods are held in tension by springs above the anode region. The field is graded from anode to cathode by a set of custom-made, low-mass, low-radioactivity resistors, connecting to the FSRs.
- A particularly delicate component of the nEXO TPC is its cathode. The primary design calls for a thin, slightly tensioned, $\sim$ 125 $\mu$m-thick copper (or bronze) sheet, made VUV reflective by a thin layer of vacuum-evaporated aluminum protected by a final $MgF_2$ coating. At the edge, the sheet is welded to a copper support ring. The assembly is then captured in a profiled cathode rim annulus designed for proper field shaping at the edge, as shown in Fig-ure 4.2. Alternative designs include i) a tensioned 125 $\mu$m thick aluminized mylar film sandwiched at the cathode rim between a set of two support copper rings using cryogenically-rated, low background epoxy, and ii) a perforated, optically transparent, tensioned mesh inspired by the EXO-200 cathode.

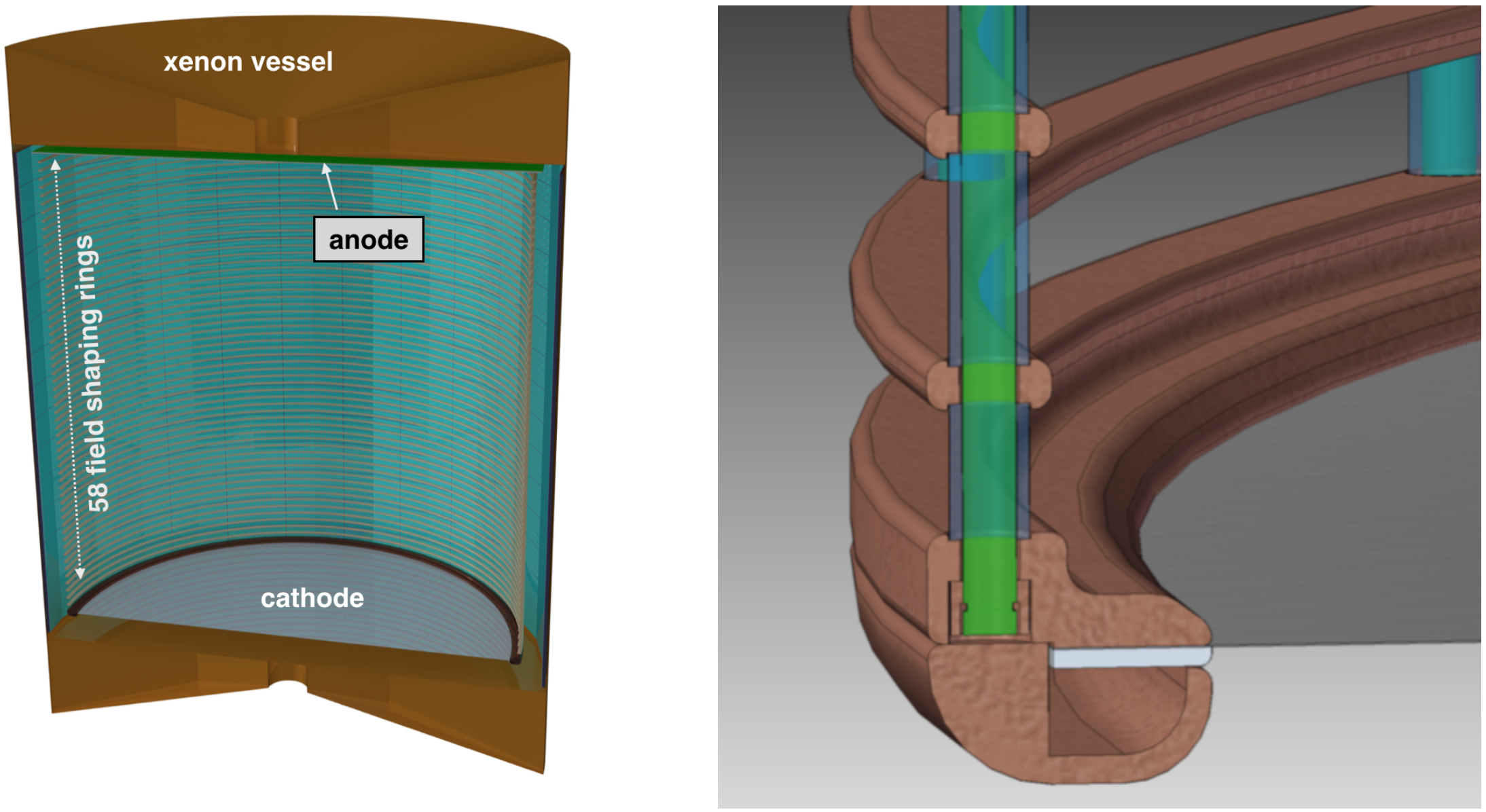

**Figure 4.2:** Left: the nEXO field cage mounted on the top end cap of the copper xenon vessel. Right: a view of the cathode profile, optimized for discharge protection.

For all designs, the tension required is rather modest, as some small degree of sagging, at the level of ∼0.5 cm deflection at the center, of the cathode membrane minimally distorts the drift electric field. In addition to withstanding the mechanical stresses and, more critically, complying with the HV requirements, the cathode design needs to address another important challenge. Rn progeny drifting to the cathode will generate a predictable amount of un-tagged $^{214}$Bi decays occurring at its surface [2]. While this background is predicted to be acceptable for the primary design, efforts are under way to better understand and, where possible, mitigate it.

- For nEXO, it was suggested [5–7] to adopt a modular charge collection scheme without a Frisch grid. Fused silica tiles of approximate dimensions $\sim 10 \times 10$ cm$^2$ are metalized with crossed strips of interleaved square pads, nominally 3 mm center-to-center. A tiled anode has the advantage of allowing for a modular assembly, with individual modules assembled, tested, installed, and integrated with the readout electronics, interconnections, and cables. The basic viability of this approach was recently demonstrated [8]. The optimization of the pad dimensions is currently being refined (see Section 4.1.3).

  In contrast, a crossed-wired design with a diameter $> 1$ m requires a rather substantial tensioning frame designed to withstand temperature cycling between room temperature and 165 K. In addition, the larger its diameter, the more vulnerable a crossed-wire design is to ambiguity in reconstructing the position of multiple energy deposits in the detector. Furthermore, wires are susceptible to microphonic pickup. This is particularly true for any wire plane held at voltage, as is the case for the EXO-200 ”induction” wire plane, operated at ∼1

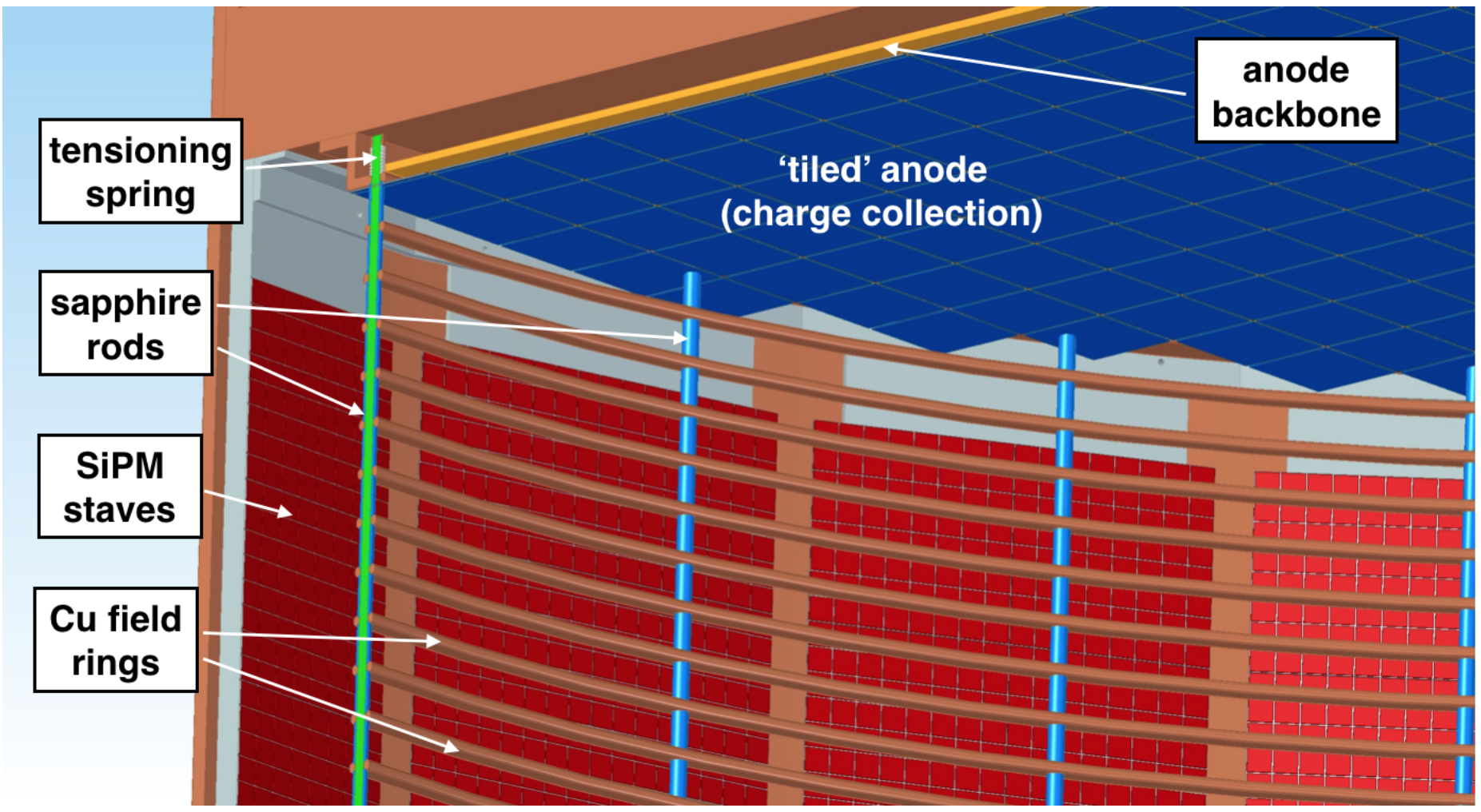

**Figure 4.3:** A close-up view of the anode region, showing the tiled charge-collecting anode, the sapphire tensioned rods, and the SiPM staves behind the field cage.

kV [1].

The fused silica charge tiles cover the anode plane as shown in Figure 4.3, and are mounted on a supporting copper backbone using cryogenically-rated silicone glue standoffs (four to eight per tile). These silicone glue standoffs are $\sim$ 1 mm long and absorb the differential thermal contraction between copper and fused silica, as demonstrated in a similar application for a silicon tracker detector in collider physics [9]. Through-Quartz-Vias (TQV) or metallized paths around the edges carry the charge collected by the strips to the back side of the anode, where they are processed by in-LXe front-end electronics and carried out of the detector on low radioactivity flat cables (see Section 4.1.5). The design of the readout for the ionization charge detector is discussed in Section 4.1.3.

- Efficient scintillation light collection is an essential requisite for nEXO to attain proper energy resolution and to provide the start time to localize events along the drift field (see Section 4.1.4.1). This challenge increases with the size of the detector, due to the larger surface area to be instrumented. Recent technological advances, however, make the task possible by using Silicon Photo-Multipliers (SiPMs). In nEXO, xenon scintillation light will be collected with a large-area array ($\sim$4.5 m$^2$) of SiPMs, a departure from EXO-200, where Large-Area Avalanche Photodiodes (LAAPDs) were used [10]. Vacuum Photo-multipliers are, to-date, too radioactive for use in nEXO. SiPM technology is rapidly evolving, and non-VUV-sensitive, cryogenic SiPMs have also recently been developed and tested for use in low-background liquid argon (LAr) detectors, both individually and read out in multiple cm$^2$ arrays [11–14]. VUV-sensitive SiPM devices from multiple manufacturers have been developed and successfully tested for nEXO [15]. The description of the nEXO scintillation light collection system is found in Section 4.1.4.
- The SiPM array for nEXO is installed on the xenon vessel barrel surface, behind the TPC field shaping rings. This is a departure from EXO-200, where the LAAPD photosensors are placed behind the anode crossed-wire planes. This arrangement is motivated by the larger surface of the TPC that can be covered to improve the overall light collection efficiency, and by the anode tiled design, which is no longer optically transparent (in the primary nEXO

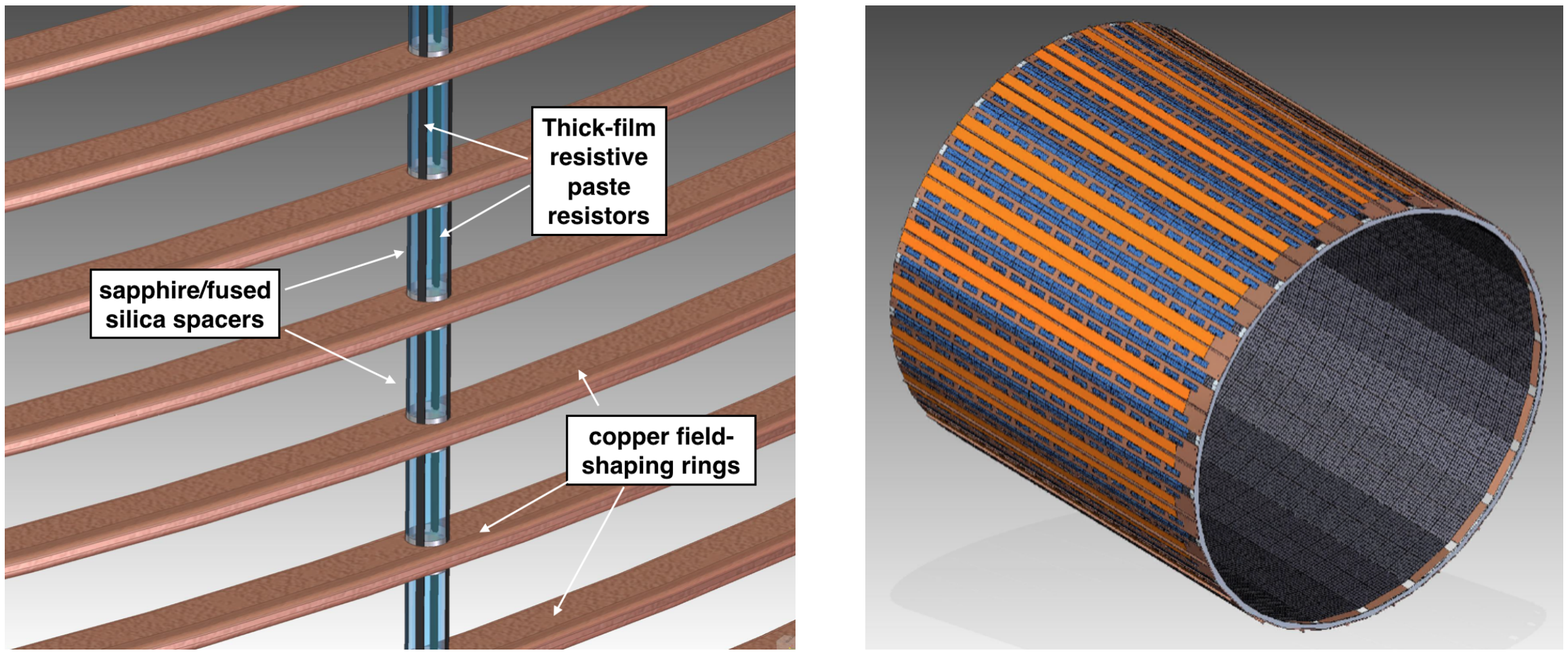

**Figure 4.4:** Left: detail of the primary design field cage voltage-dividing resistor chain. Right: the SiPM staves as a 24-sided polygon surrounding the nEXO field cage.

TPC concept the cathode is also opaque).
Individual SiPMs have an area of $\sim$1 cm$^2$ and are ganged in groups of several devices, read out as individual channels. Multiple channels are grouped in "tiles", assembled on a backing interposer, and mounted on 24 long staves surrounding the TPC field cage (see Figure 4.4). Section 4.1.5 puts forth the basic concepts for transporting signals from the SiPMs to the front-end ASICs (one per SiPM tile) and from there out of the detector, in digitized form.

- The placement of SiPMs on the barrel surface of the detector drives important aspects of the nEXO TPC:
  - The field cage must be optically open to allow scintillation light to reach the photosensors. In the primary design, the field shaping rings are flat rings, 1 cm wide and 4 mm thick, with rounded edges, on a 2 cm pitch, as shown in Figure 4.1. This layout was compared to many others and chosen as the best compromise between field cage transparency and electric field uniformity within the drift volume.
  - PTFE side reflectors, used in EXO-200 and most other noble liquid TPCs for rare event searches to concentrate the scintillation light towards the photodetectors, are no longer required. Eliminating PTFE reflectors addresses the complementary need to minimize impurity out-gassing from plastics into the xenon.
  - All passive detector surfaces need to be made as reflective as possible in the VUV. This includes the field shaping rings and the cathode. Simulations, reported on more extensively in Section 4.1.4, are being used to quantitatively estimate the requirements in this area. These requirements that can, to some extent, be traded against photodetectors performance. As successfully proven by EXO-200, vacuum deposited aluminum, protected against oxidation by a layer of $MgF_2$, can generally fulfill this role, although engineering details have not yet been defined.
- In nEXO, ionization and scintillation signals will be read out by cryogenic, in-xenon, low-radioactivity front-end electronics. The front-end readout electronics are discussed in detail in Section 4.2.
- Low radioactivity, flat signal cables (see Section 4.1.5) will be routed to the upper side of the

TPC and carried out through feedthroughs inspired by those developed for EXO-200 [1] (see Section 4.1.6).

- The nEXO TPC package will be mounted to a top flange and vertically lowered into the xenon vessel, as is standard for cryogenic setups. This is a departure from the cantilevered EXO-200 design, which was dictated by the limited vertical clearance available at the WIPP site.

### 4.1.2 High Voltage System

In nEXO, ionization electrons will be drifted to the anode strips held at virtual ground at the top end of the TPC under the influence of an electric field established by applying a negative potential to the cathode, located at the bottom end. The magnitude of the bulk drift field affects the detector performance in several ways, some of which are non-trivial.

- Since charge collection generally has higher efficiency than light collection, the energy resolution, deriving from a linear combination of the two, improves somewhat with increasing values of the field, as this shifts a larger fraction of the energy into the ionization channel. This is illustrated in Figure 4.5.
- While higher fields result in faster drift times and hence smaller spatial charge spread because of diffusion, for fields above 100 V/cm, this effect is too small to affect the electron-$\gamma$ discrimination appreciably. At lower electric fields, longer drift times as well as increased capture cross section for most contaminants [16], require higher electron lifetimes to obtain the same attenuation for electrons, although, also in this case, the effect is mild for fields above $\sim$ 100 V/cm. This is shown in Figure 4.6 for EXO-200 data.
- A lower value of the electric field further suppresses the charge yield for $\alpha$ events, but for $\alpha$ energies of a few MeV, such yield still produces easily detectable signals at 100 V/cm, guaranteeing the electron-$\alpha$ discrimination (that requires both light and charge collection).
- An increased sensitivity to stray fields is observed in many LXe detectors for small values of the bulk drift field. Figure 4.7 shows that, in the case of EXO-200, at low fields there is a depletion of events with long drift times and located near the detector edge. This is attributed to static charges accumulated (in a way which is difficult to assess quantitatively) on the Teflon VUV reflector placed inside the field cage. While in nEXO this kind of effect would be exacerbated by the longer drift, also in this case fields above 100 V/cm are probably safe, particularly as nEXO will have no dielectrics inside of the field cage.

On the other hand, running at lower fields relaxes many engineering constraints on HV standoff, with a potential effect on the sensitivity of the experiment. This is because, in the overall optimization, the lower voltage would result in a larger field cage and a larger fraction of the LXe being fully active (*i.e.* instrumented for charge collection in addition to scintillation light detection). However, also in this case, the trend is mild, because the extra LXe is located at the edge of the detector, where the background is largest.

A careful study of the trade-offs between all these factors will be carried out before the final definition of the size of the field cage. Such a study will have to be informed by more sophisticated simulation as well as by experimental work on HV performance of various components. For the purpose of this document, we assume a bulk drift field of 400 V/cm, a well-motivated value because it has been already utilized with success in EXO-200 Phase-I (although with different noise and photodetection efficiency values). Given the 125 cm length of the nEXO TPC, this corresponds to a cathode potential of $\simeq -50$ kV. Having set this value, and being mindful of the challenges reported by many experiments in this area [17], we then, conservatively, apply two independent safety measures: 1) the maximum cathode potential is assumed to be $-100$ kV and 2) the maximum design field anywhere in the detector is limited to 50 kV/cm, substantially lower than the 250 kV/cm breakdown field in LXe (see section 5.1 of [17]). These considerations result in the detailed shape and dimensions of the field cage, described in Section 4.1. In particular, this results in a clearance of $\simeq$ 7 cm around the cathode (the highest field region), computed using the COMSOL FEA software [18]. It is likely that further work will allow the relaxation of some of the safety margins, leading to a smaller clearance, even if the decision is made to design for the 400 V/cm drift field. The process is particularly complex in nEXO because of the cost of the $^{enr}$Xe that directly translates into the cost of a larger HV standoff in the TPC. This optimization will also have to take into account the role of mechanical tolerances, especially in the low field regime, when the nominal LXe standoff is the smallest. A final test of the full-size cathode and some field shaping rings will be performed in a large setup, capable of operating with up to 800 kg of LXe, currently under construction.

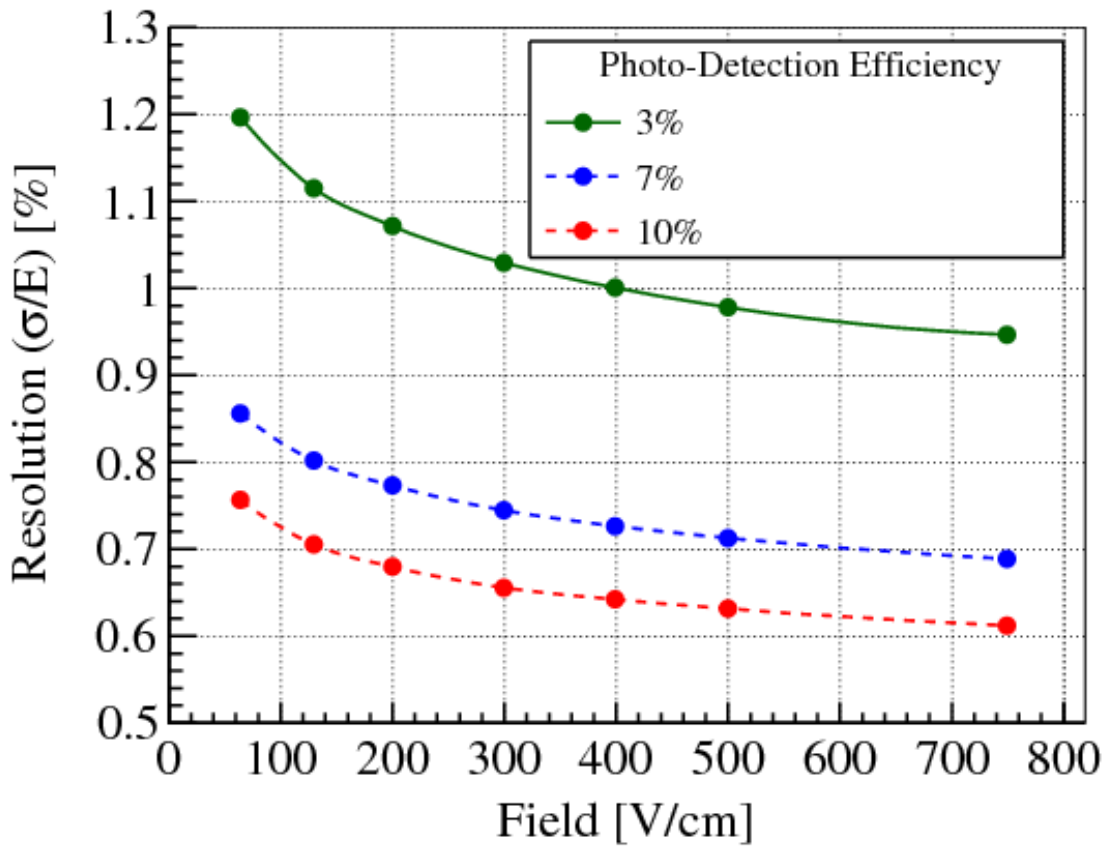

**Figure 4.5:** Calculated energy resolution in nEXO as a function of the drift electric field in the TPC. The three curves refer to different values of the photodetection efficiency, as indicated. The primary value of such efficiency, 3%, is shown by the solid line and is also expected to be a conservative estimate of the achievable performance.

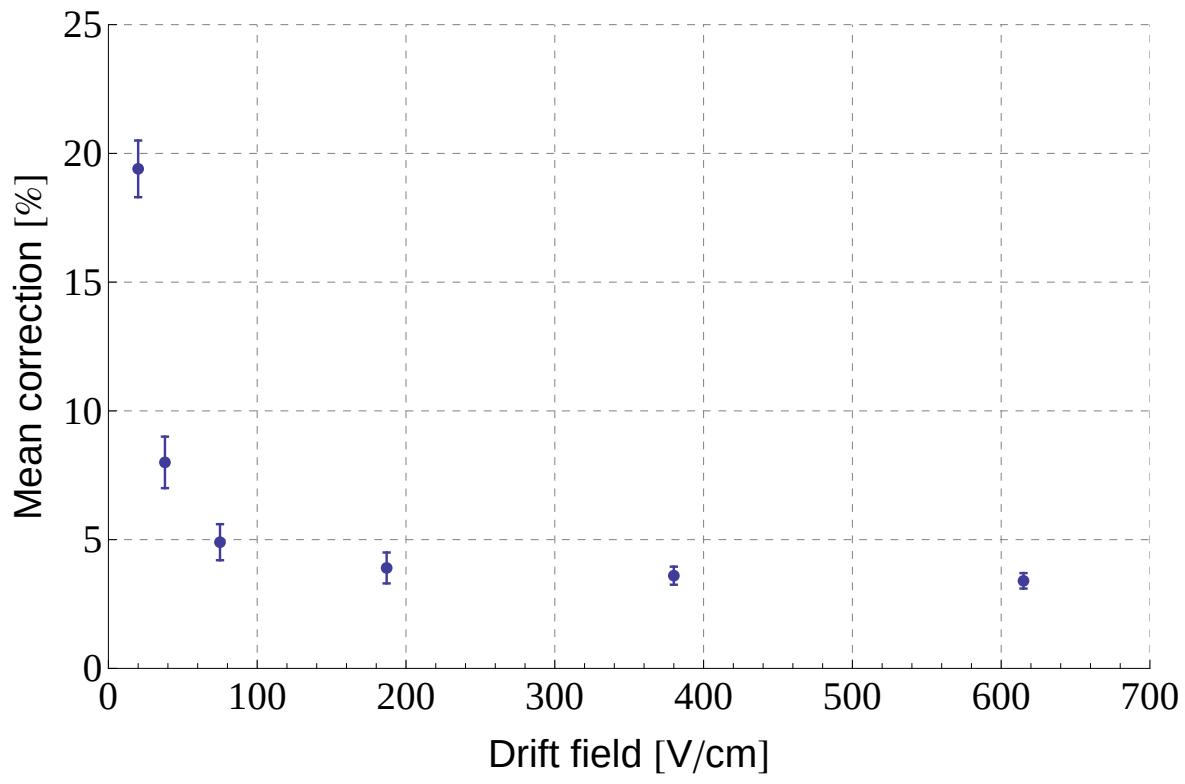

**Figure 4.6:** Magnitude (in percent) of the correction to the charge detected at the anode as a function of the drift electric field in EXO-200. The mean correction refers to the correction applied to an event midway between cathode and anode, and is determined from $^{228}$Th calibration data.

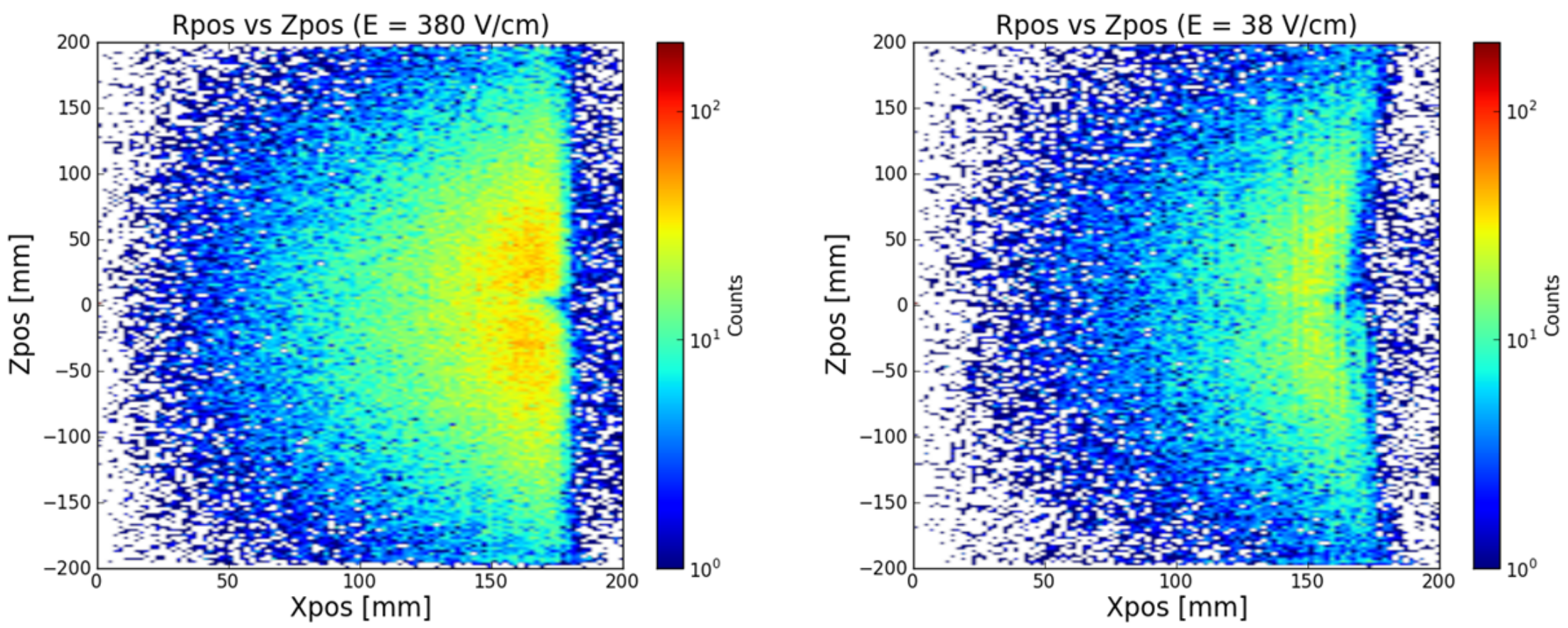

**Figure 4.7:** Apparent radial position ($X_{pos}$) of calibration events for different distances from the (central) cathode ($Z_{pos}$) in EXO-200 data for drift fields of 380 V/cm (left) and 38 V/cm (right). The $^{228}$Th calibration source is external to the detector, at a position corresponding to $Z_{pos} = 0$, where the cathode is located. While the main decrease in number of events moving away from the source is trivially due to the $1/R^2$ effect and the attenuation of $\gamma$s in the LXe, for the low field case, a deficit of events can be observed near the edge of the detector, particularly for positive values of $Z_{pos}$. This is attributed to static repulsion due to charges accumulating on the Teflon VUV reflectors placed just inside of the field cage. The smaller the bulk field in the TPC, the larger this effect is.

Finally, it is important to realize that the proper HV design requires a holistic approach as the details of dielectric layout, LXe purity and the surface finish of conductors can all affect the performance beyond what can be calculated for individual components. It is also essential that the cryogenics is sufficiently well engineered that no gas bubbles can possibly form in the system, including in proximity of in-LXe front-end electronic components.

In the rest of this section, we will review the primary configuration of the systems that are closely related to the transport and delivery of the high voltage (HV) from an external commercial 100 kV power supply to the TPC cathode. They include the HV cable, filter(s), feedthrough, and, in particular, the devices required to grade the field from ground to full potential.

#### 4.1.2.1 High Voltage Feed

The HV feed system transports the potential from the power supply to the cathode and provides filtering, current limiting and monitoring. The feedthrough has to accomplish three tasks: 1) transporting the HV to the cathode, 2) sealing the Xe in the detector and preventing air and radon, even in very minute amounts, from entering the detector, and 3) making the transition from cryogenic to room temperature. Conventional vacuum-rated feedthroughs for $\sim$ 100 kV are bulky and contain large amounts of materials, such as ceramics, which are known to have radioactive contaminations. The HV connection between power supply, filter and monitoring assembly(ies) and detector is achieved using a specially designed coaxial cable. The principle of the feedthrough design, schematically illustrated in Figure 4.8 and already successfully employed in EXO-200 [1] and other noble liquid detectors, is to directly insert a section of such a coax, with the outer screen and jacket removed, in a low-background copper conduit directly welded onto the TPC vessel on one side and to a flange in air and at room temperature on the other. In this way, the temperature transition occurs over a long distance (along a copper pipe, which is not ideal from the thermal standpoint but required by background considerations). The entire feedthrough assembly is light weight and progressively shielded by HFE-7000 with respect to the detector.

In the primary design, validated by EXO-200, the Xe-air seal occurs at room temperature, although a cold seal is also being investigated. The notional locations of the two styles of seal are shown in Figure 4.8 for illustration purposes. The basic tradeoff between the two include:

- *Cold seal:* the inherent risk of a seal that has to be temperature cycled through a range of over 100 K and the requirement of building such a seal near the detector, in the lowest background region and hard to access;

- *Warm seal:* the presence of a long region between the cable dielectric and the copper pipe which is open to the LXe environment and difficult to pump and the presence of the LXe-GXe interface at some location along the copper pipe that is not exactly known.

Mitigating strategies exist for both options and the warm seal performed well in EXO-200; further R&D and engineering will establish the lower risk solution for nEXO.

A dielectric component of generous size will be required at the bottom of the TPC to insulate the HV header connecting to the cathode from the surrounding copper vessel (at ground potential). The large field strength in and around this component will require a very careful design and testing. While in EXO-200 this component was made out of low background PTFE, outgassing constraints, driven by the higher electron lifetime specification of nEXO, are leading to the exploration, currently in progress, of low background fused silica or sapphire.

A special HV coax cable from Dielectric Sciences [19] has been tested for nEXO. This cable, shown in cross section in Figure 4.9, is rated to 150 kV and uses a polymer conductor, insulated by a polyethylene element from a conventional braided shield (that will be removed in the copper pipe region of the feedthrough). The resistivity of the polymer conductor, reported by the manufacturer as $\simeq$ 8.5 $\mathrm{k\Omega/m}$ at room temperature, is too small to be used as primary current limiter. However, this design provides much reduced differential coefficient of thermal expansion between the conductor and the insulator than in the case of a copper conductor. This is particularly important since a few meters of the cable will be in thermal contact with the cold HFE-7000 and, in the warm seal case, this all-polymer cable should reduce the risk of small leaks deriving from the loss of adhesion between the conductor and the dielectric. The cable also has a remarkably small overall diameter (14.0 mm including the outer jacket, 11.2 mm for the dielectric), resulting in a small mass of material near the low background part of the detector. Satisfactory radiopurity has already been demonstrated by $\gamma$-ray counting assays and extensive testing has been carried out up to 100 kV while at LXe temperatures.

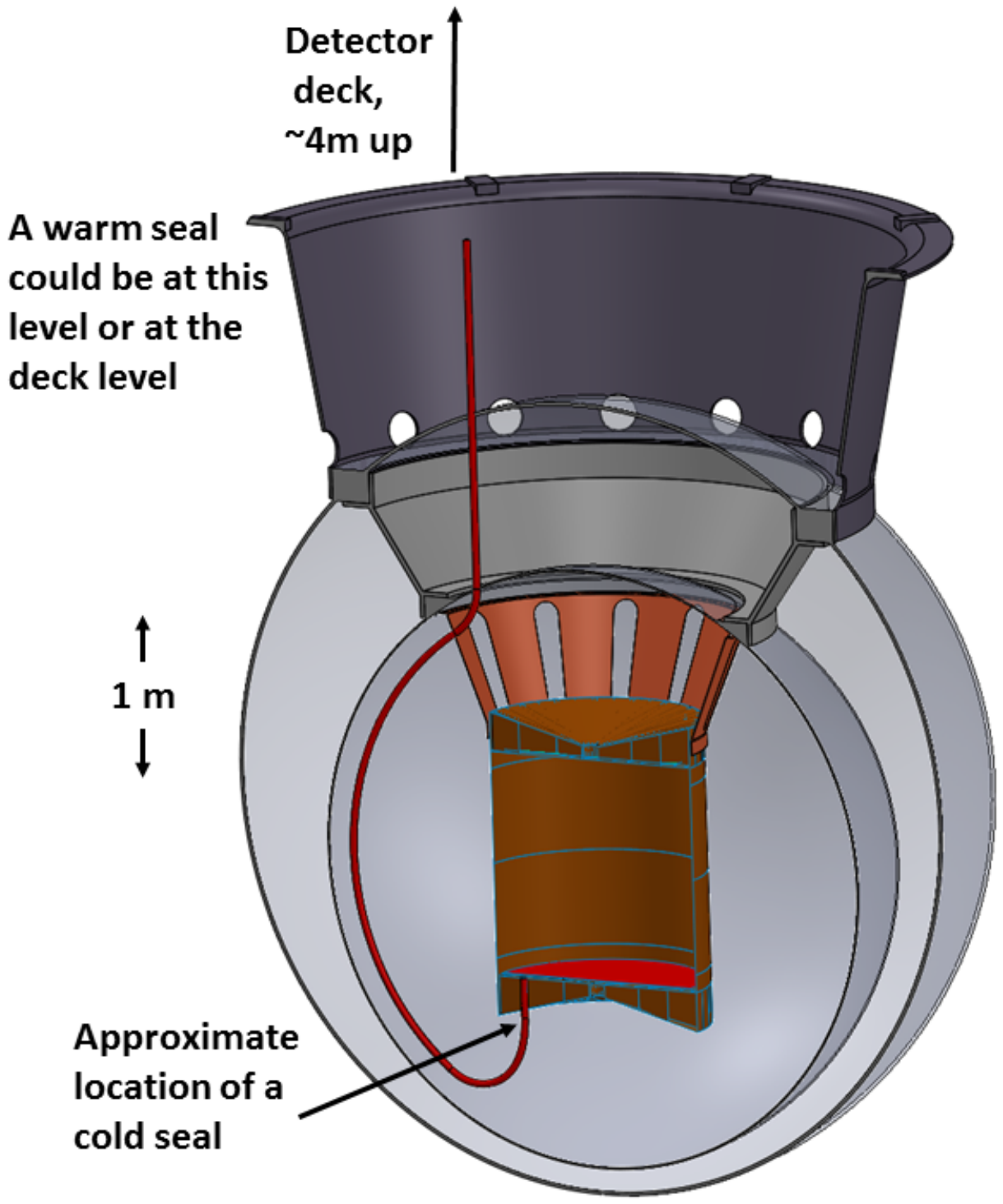

**Figure 4.8:** Schematic view of the HV feed system. The transition from cryogenic to room temperatures occurs along a low-background copper pipe welded to the TPC vessel on one side and to a flange in air on the other. The seal between Xe and air can be at cryogenic or room temperature, as discussed in the text.

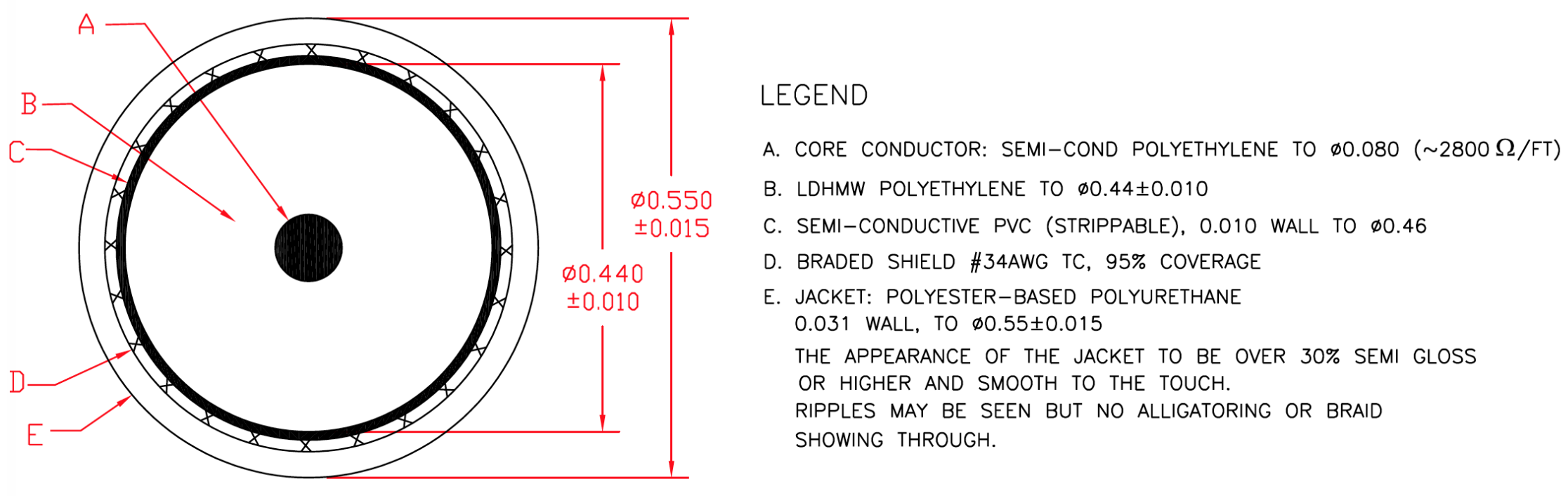

**Figure 4.9:** Cross section of the Dielectric Sciences polymer conductor HV cable being considered for nEXO. Dimensions are in inches [19].

A combination of aggressive low-pass filters and current limiters will be installed in-line with the HV feed, at the room temperature end of the HV cable. This R-C network will also include voltage dividers for monitoring purposes. Current monitoring will also be included, using an optically insulated meter. Details of the filter system will be engineered to protect the TPC from damage (although more on this will be discussed in the next section) and to stabilize the HV to a very tight tolerance. The filtering must extend over a broad range of frequencies, from under 50 Hz to over 500 kHz, due to the long drift time of the ionization and the lack of a Frisch grid on the ionization detector. Isolated multistage RC filters immersed in tanks of dielectric fluid, similar to those used in EXO-200, will be designed and tested with samples of the charge tile preamplifiers.

#### 4.1.2.2 Field Cage and Discharge Protection

The primary concept for the field cage of nEXO is shown in Figure 4.2. The field shaping rings are made out of low background copper and are held in place by dielectric (possibly sapphire) rods and spacers. Custom resistors, similar to the EXO-200 thick film devices, will be fabricated to comply with the tight radioactivity budget for the TPC. One or more of the stacks of dielectric spacers can be used as substrates on which to "paint" the resistive material. The two primary constraints for the design of the field cage geometry are the need for a uniform electric field in the largest possible fraction of the TPC volume, generally favoring a narrow spacing between rings, and the requirement that the scintillation light, to be collected by the SiPMs, be minimally obstructed (favoring the smallest possible number of rings). In addition, the rings have to be rigid enough to be assembled and support their own weight, and yet light enough to minimize the radioactive background and the light obstruction.

The current solution, to be further refined in a later stage of engineering, calls for 58 rings, each 4 mm high in the $Z$ direction and spaced with a 20 mm pitch (see Figure 4.1). This results in a field strength that is uniform to better than 1% over 90% of the field cage volume. The effect of the rings on the light transport is studied with ray-tracing software, as discussed in Section 4.1.4.3.

As already mentioned, the maximum field allowed by our requirements in LXe, near the outer edge of the cathode, is 50 kV/cm. While this is a conservative figure, the energy stored in the electric field is calculated to reach a few Joules, so that the concern still exists that occasional breakdowns may damage parts of the detector. Because the TPC is welded shut inside the LXe vessel, it is hard to access and very difficult and time consuming to repair. Thus care in minimizing the maximum field in the detector has to be paired with measures to mitigate unlikely breakdowns. The engineering of the field cage following the primary (copper) concept will have to be informed by the following considerations:

- The exact nature of these small discharges in LXe is not well understood, but it is dubious that very large currents can be produced, assuming some reasonable monitoring system providing advanced warning when ramping up the HV. In EXO-200 a system is installed that, by capacitively sensing microdischarges by the fast transients ("glitches") produced on the HV feed, provides the information required for a careful ramp up of the cathode potential and, in the extreme, trips the HV (although this last feature is not necessarily effective, because of the stored energy in the field cage). While the EXO-200 HV is unstable above $\sim 15$ kV, this limits the electric field in the TPC primarily because microdischarges affect the trigger rate. In five years of operation this system has never observed glitches of 100 mV or above on the HV feed. In EXO-200 such glitches correspond to $\sim 10^{-8}$ J of energy and $\sim 0.2$ mA of current, assuming a 10 ns duration, which is typical. The same 100 mV glitch amplitude in nEXO would correspond to a released energy of $\sim 5 \times 10^{-6}$ J and a current of 10 mA (assuming the same duration). These values appear unlikely to damage the readout electronics via inductive or capacitive coupling.

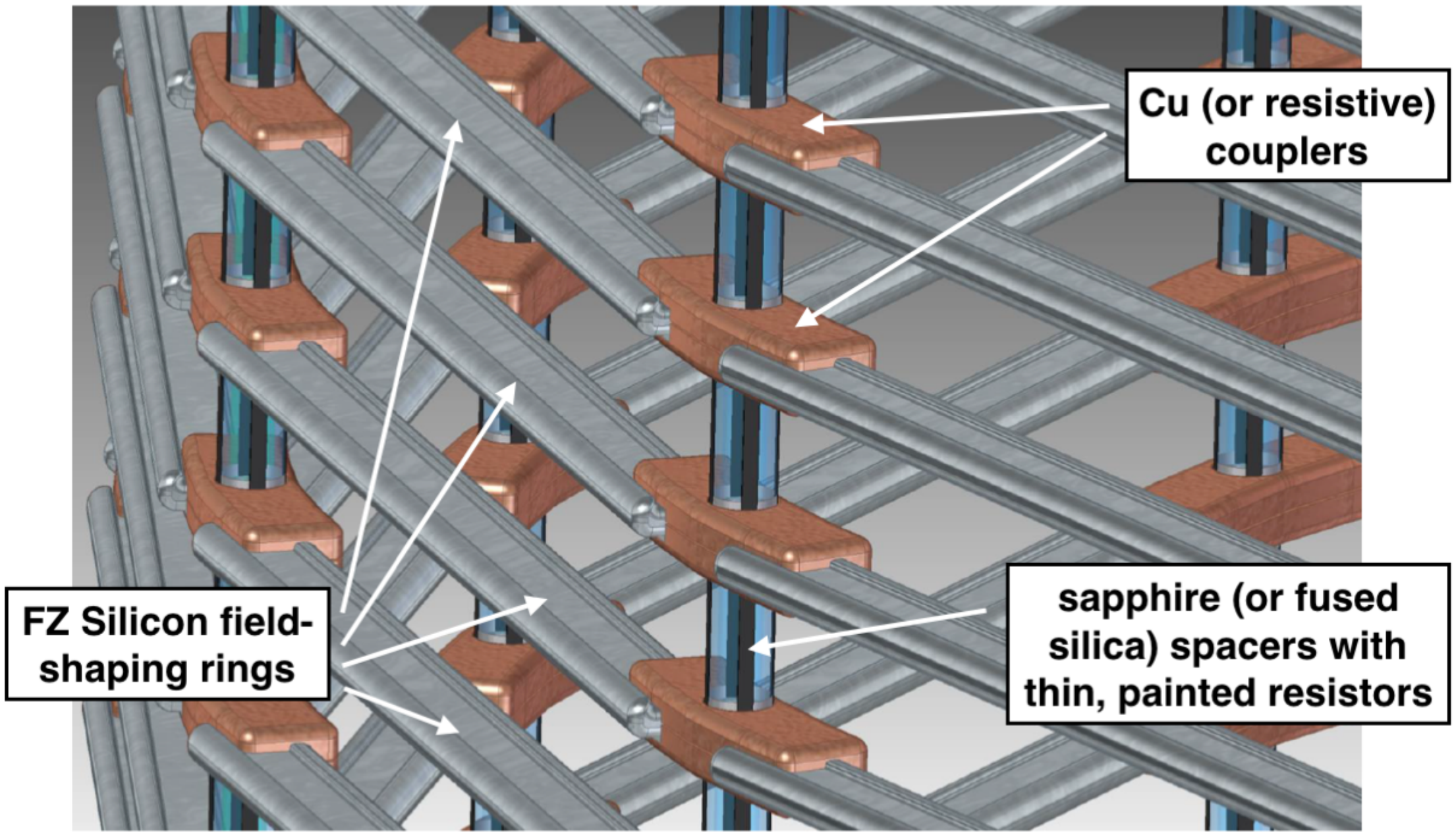

**Figure 4.10:** Notional view of a field cage built out of straight, high resistivity silicon members. The model shows the segmented nature of this arrangement.

- "Direct hits" from the relatively small discharges mentioned above onto key components are likely to be more dangerous. This could be the case should the discharges land on the SiPMs that are located behind the field cage, in some places where the field is close to the maximum, as in proximity of the cathode. To mitigate the risk, a conductive mesh could be installed in front of the SiPMs, at least in the regions where the field is particularly high. In this area, the challenges are in the detailed mechanical design, maintaining good optical transparency and low radioactive background. The performance of SiPMs in regions of space where substantial electric fields are present is not expected to change and this is currently being evaluated experimentally.

Since a substantial charge is stored in the field cage itself, the risk of breakdown cannot be fully eliminated unless all conductors are replaced by high resistivity material. A resistive field cage, also discussed by other experiments [20], would intrinsically limit the supply of charge to the site of a discharge. In the nEXO case, a substantial challenge is represented by the requirements of ultra-low radioactivity and outgassing (driven by the need for excellent electron lifetime). R&D is in progress on a scheme using float zone silicon (FZ Si) whose very high intrinsic purity is likely to result in sufficient radiopurity, as confirmed by an initial nEXO radioassay. It is plausible that a field cage could be built with linear segments of FZ Si, as notionally shown in Figure 4.10. Room temperature tests in air have shown that current suppressions of the order of 500 can be achieved with FZ Si of $80\ \mathrm{k}\Omega\cdot\mathrm{cm}$., as shown in Figure 4.11. Larger suppressions may be achieved at lower temperature. Substantial electrical experimentation in LXe, mechanical design work and radioactivity testing will be required to establish if this technology is practically usable in the nEXO detector.

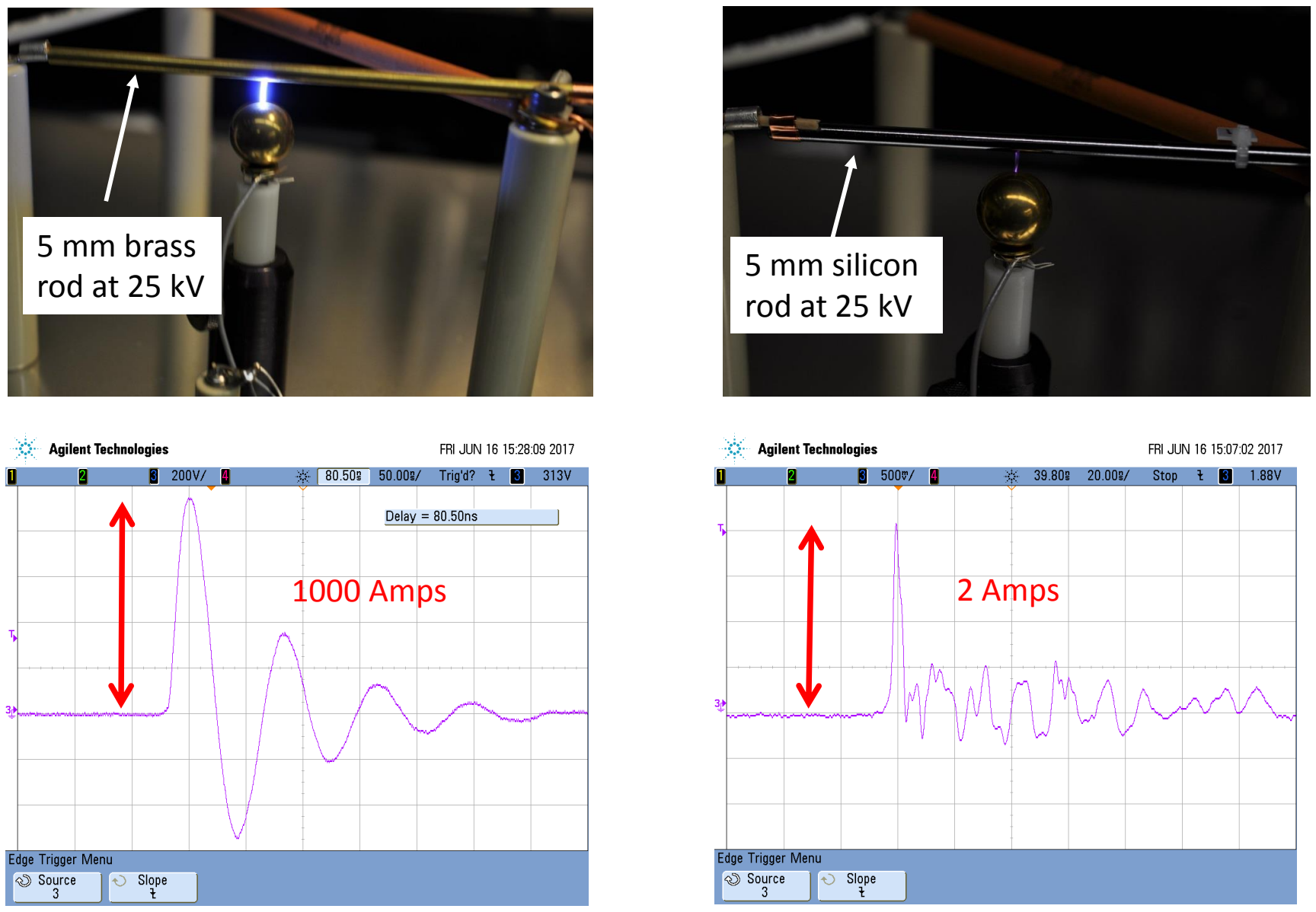

**Figure 4.11:** Breakdown test in air and at room temperature, using a conductive (brass) rod (left) and a resistive (silicon) one (right), shown in the photographs on top. Each rod represents a segment of the field cage and is connected to a 0.8 nF capacitor simulating the charge stored in the case of the full size detector. The potential applied (25 kV) is comparable to (but somewhat lower than) that of nEXO. The oscilloscope traces at the bottom represent the discharge current for the two cases, showing that the silicon rod supports about 500 times smaller current than the brass one.

The use of FZ Si for the cathode itself is more challenging, and other solutions involving a tensioned resistive film are also under study. This work needs to take into account the fact that it is desirable for the cathode and the rings to be reflective at 175 nm.

### 4.1.3 Charge Collection System

#### 4.1.3.1 Charge Detection Requirements and Conceptual Design

The nEXO primary design for charge collection consists of "tiles" made of a low radioactivity dielectric onto which crossed metallic strips are deposited. While, functionally, the metal artwork is arranged in orthogonal "X" and "Y" strips, at this early stage concerns about charge accumula-tion and small crossing capacitance suggest a structure whereby square pads are connected at one set of corners to form the strips. In the current primary design, shown in Figure 4.12, strips are 10 cm long and spaced 3 mm center-to-center, covering 96% of the fused silica surface. Each strip crossing has a metal surface area of 3600 $\mu m^2$, with the two orthogonal metal layers separated by a 1.5 $\mu$m thick $SiO_2$ dielectric layer, resulting in a capacitance of 0.56 pF/crossing. The capacitance between adjacent strips is 0.86 pF and the resistance of each strip is 4.4 $\Omega$.

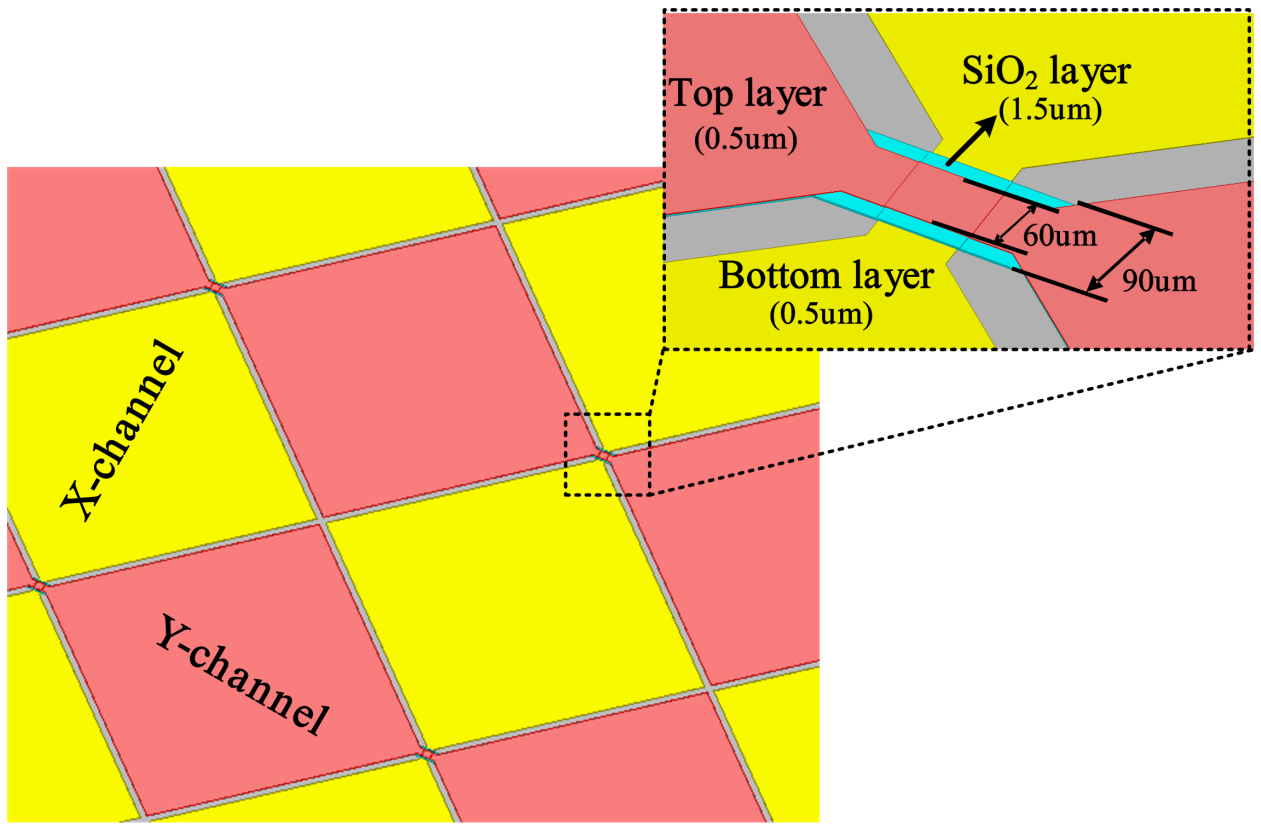

**Figure 4.12:** Details of a charge collection tile. The inset shows a crossing between strips. The metal X and Y strips cross each other at the pad junctions and are separated by a layer of $SiO_2$. This arrangement maximizes overall pad coverage while limiting the capacitance.

While initial prototypes have large bonding pads at the end of the strips for connection to different interface boards and readout electronics, production tiles are expected to bring the signals from the strips to the reverse side, using either through quartz vias or edge metallizations. In this way it will be possible to mount tiles very close to one another, possibly ganging together strips from different tiles using wirebonds, minimizing the number of independent channels to read out, consistent with the limitation on the total capacitance imposed by the electronics noise target. The optimal ganging (if any) is still to be determined. The fused silica substrates are currently 6-inch wafers, 300 $\mu$m thick. While silica samples with sufficiently low radioactive contamination have been tested by the collaboration, current tile prototypes are made from off the shelf silica that has not been directly certified for low radioactivity. Similarly, while the appropriate metals required for the tiles fabrications are known to be available with sufficiently small contaminations (in the small quantities required), no effort has been made for the prototypes to utilize low-background materials.

#### 4.1.3.2 Charge Detection Modeling

The charge collection tiles represent a departure from the more conventional readout with wire arrays, particularly because they operate without shielding grids and hence they record induced signals from the start of the drift, far away from the collection region. While the tiles are justified by their robustness, modularity and lack of structurally heavy supports to hold the tension of the wires, it is important to understand the signal shapes through simulation and validate models with actual measurements on prototypes.

From the point of view of the performance, the main difference between the system discussed here and a more conventional wire array with shielding grid is expected to be in the areas of 3-dimensional localization and single/multi-cluster identification from the signal shape and signal distribution between channels. Therefore, a GEANT4-based Monte Carlo simulation has been built, with the primary goal of analyzing the detector's ability to discriminate between $0\nu\beta\beta$ decays and $\gamma$-ray events. The Noble Element Simulation Technique (NEST) model [21] is used to handle the micro-physics of recombination in LXe (including the anti-correlation between ionization and scintillation). NEST free parameters are tuned to match the observed performance in EXO-200. Charge collection and induction are initially calculated for each square pad and then summed over strips. Thermal electrons produced during ionization are binned into small voxels with a side-length of 1/4 the strip pitch being analyzed. This choice represents a compromise between run time and fidelity in the simulation. The induced charge on each strip is calculated as each voxel drifts away from the interaction location towards the anode, driven by the electric field. To calculate waveforms, the Shockley-Ramo theorem is used [22, 23], summing over all voxels in LXe to compute the charge on each strip. This process is repeated at a frequency substantially above the digitization rate of 2 MS/s, following the charges as they drift in the LXe. EXO-200 measurements [3] are used for the electric-field-dependent transverse diffusion coefficient and the electric field-dependent drift velocity is from [24]. No longitudinal diffusion is implemented at present, since this is small compared to the transverse diffusion. The simulation also takes into account the electron lifetime, taken as the primary design value of 10 ms.

We use a multi-parameter method to discriminate between $0\nu\beta\beta$ and $\gamma$ events, using TMVA (Toolkit for Multivariate Data Analysis with ROOT). Quantities related to the shape and distribution of the collected and induced charge signals are combined using a boosted decision tree (BDT). Both the selection of input variables and the use of the BDT were based on their ability to optimally discriminate between signal and backgrounds in detailed Monte Carlo simulations of the charge measurement. Validation of the agreement for these input variables between the Monte Carlo and data from the prototype charge tiles is in progress.

To optimize detector characteristics such as the channel pitch, channel capacitance, and electronics noise, the discrimination between $0\nu\beta\beta$ signal events and background $\gamma$ events from the U and Th chains is under study using the simulation. This allows the strip pitch and electronics noise to be varied around the primary values of 3 mm and 200 electrons RMS, respectively. This work indicates that roughly a factor of two improvement in background rejection relative to EXO-200 [8] is possible with 3 mm channel pitch, consistent with the rejection assumed in the sensitivity calculation described in Section 3.3.4. A small degradation in the background discrimination is seen for 6 mm pitch, but future work will be required to balance this reduction of performance against the simplification and, ultimately, lower background, deriving from the reduced electronics channel count. For either pitch, the electronics noise of 200 electrons/channel RMS is sufficient to ensure that the charge noise is subdominant to the noise in the light channel, as assumed in the resolution calculation shown in Section 4.2.

#### 4.1.3.3 Prototyping and Bench Tests

Current prototype tiles have been made on 300 $\mu$m thick, 6-inch diameter fused silica wafers. The strips are obtained by physical vapor deposition of a 0.05 $\mu$m Ti adhesion layer, followed by a 0.5 $\mu$m Au layer. To limit the capacitance, strips thin-down to 60 $\mu$m-wide "bridges" at the crossings, where inter-layer dielectric (ILD) patches are used to insulate the strips. The 1.5 $\mu$m thick $SiO_2$ patch is wider than the bridges, to ensure a reliable standoff. Figure 4.13 shows scanning

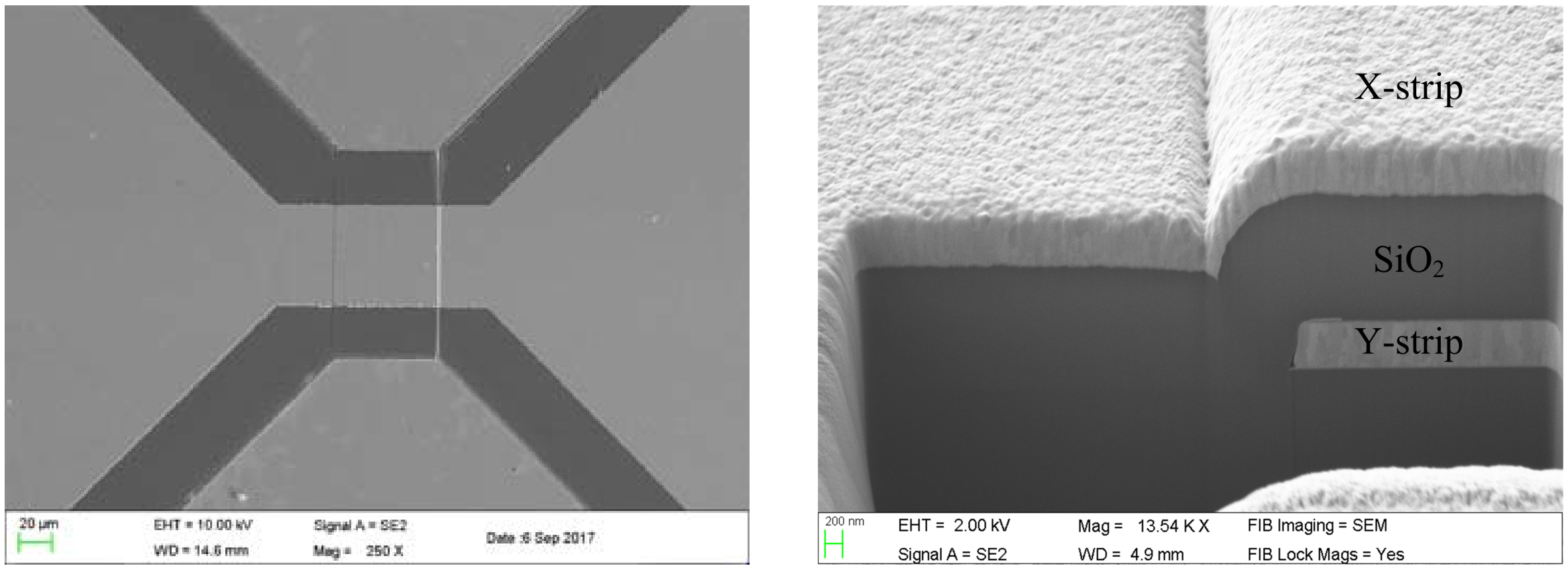

**Figure 4.13:** SEM photograph of the crossing between two orthogonal strips. Left: top view of the crossing region, where one of the bridges (the one running vertically in the image) is buried under the dielectric patch. Right: profile, showing the top bridge running over the inter-layer dielectric patch.

| Parameter | GND plane | No GND plane | |
|---|---|---|---|
| $C_{strip}$(pF) | 29.3 | 18.2 | 15.7 |
| $C_{adjacent}$(pF) | 0.40 | 0.86 | 0.64 |
| $C_{cross}$(pF) | 12.9 | 17.1 | 14.7 |
| $R_{strip}$($\Omega$) | 4.45 | 4.45 | 10.0 |
| $L_{strip}$(nH) | $< 110$ | $< 110$ | $< 110$ |
| | LXe (165 K) | | Vacuum (300 K) |

**Table 4.1:** Calculated electrical parameters for a prototype tile in LXe and vacuum (or air). For comparison, the calculation is also done for the case of a ground-plane on the reverse side of a 300 $\mu$m-thick tile, while, for the time being, only tiles without such a ground plane have been fabricated. $C_{adjacent}$ refers to the capacitance of one strip with respect to the two adjacent strips (one on each side).

electron microscope (SEM) photographs of a crossing.

The electrical properties of prototype tiles, calculated using a full 3D finite elements model, are shown in Table 4.1. In the table, $C_{strip}$ is the total capacitance of a single strip, $C_{adjacent}$ is the capacitance between a strip and the two adjacent parallel ones, $C_{cross}$ is the capacitance between a strip and all the crossing ones, $R_{strip}$ is the total resistance of a single strip, $L_{strip}$ is the total inductance of a single strip. Both the case of tiles with and without a ground plane on the reverse side are presented. The ground plane may be useful to shield the tile from the electronics, although only tiles without a ground plane have been produced until now. The substantially larger capacitance for the case with the ground plane suggests the use of thicker substrates or of a separate interposer, made out of a compatible material, to be used to support the readout chip.

#### 4.1.3.4 Charge Collection Studies in LXe

A prototype tile was tested in a TPC containing 9 kg of LXe [8], cooled down to an operating temperature of 168 K. Since the performance of the charge collection tile was the focus of this work, scintillation light was read out by an external PMT located behind a viewport and used for triggering only. Signals from the strips are fed to cold, discrete component preamplifiers located outside the LXe but still cooled to $\sim$ 170 K. Ceramic interface boards are wire-bonded to the strips on one side and support connections to individual leads connecting to the feedthrough extracting the signals from the LXe, as shown in Figure. 4.14. Signals are then digitized using a 125 MS/s VME ADC (Struck Model SIS3316-125-16). The 570 keV $\gamma$-rays from a $^{207}$Bi radioactive source, mounted directly on the cathode mesh, were utilized to calibrate the energy response of each individual channel and study the energy resolution. Figure 4.15 shows a sample event that passed all selection cuts and fell in the $^{207}$Bi peak region.

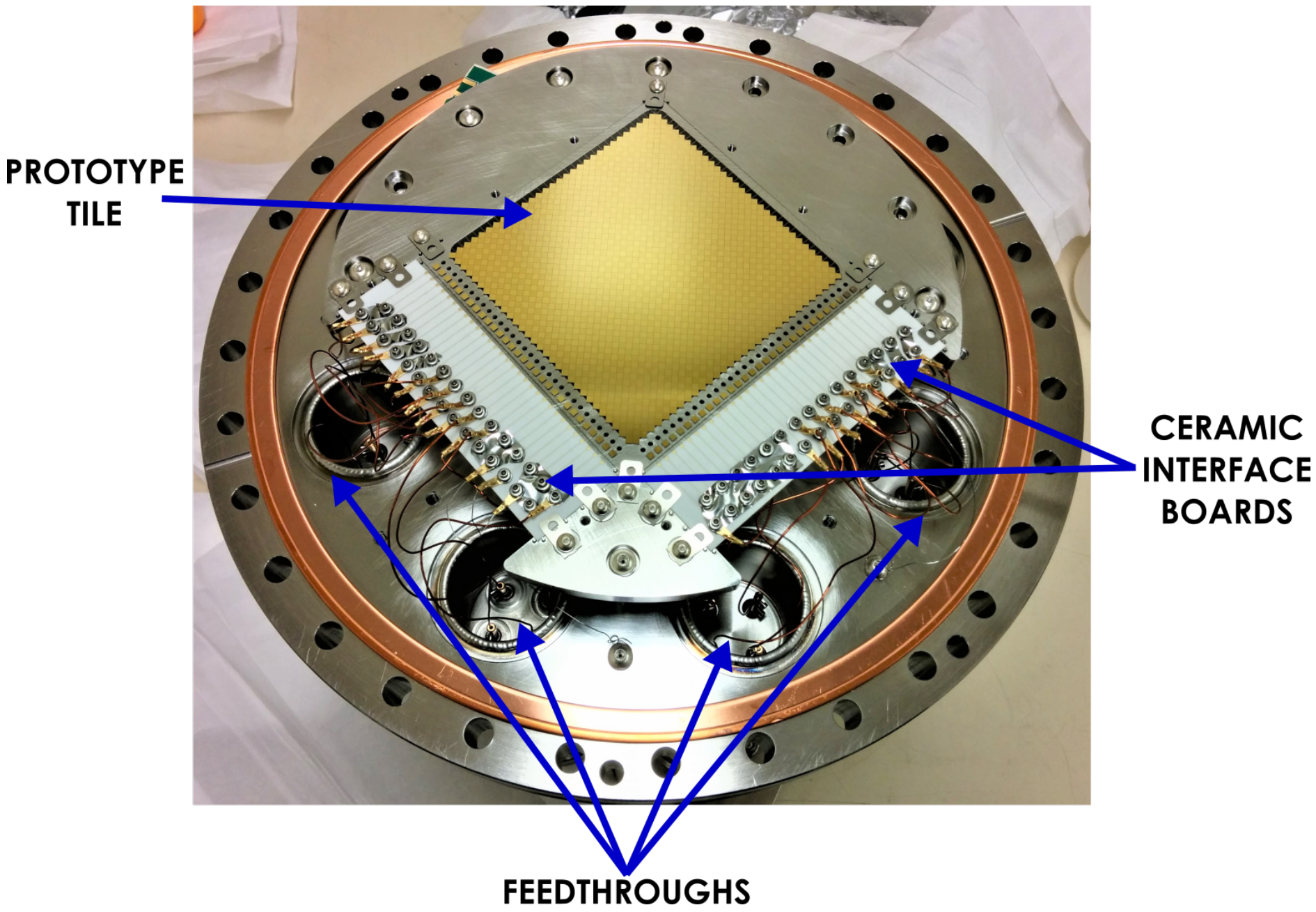

**Figure 4.14:** Photograph of the charge collection prototype tile mounted to the top flange of a test LXe TPC. Temporary ceramic interface boards are wire-bonded to the strips on one side and carry discrete leads on the other, connecting to feedthroughs and external, discrete component electronics.

The slow rise of the signals, due to the induction that starts displacing charge as soon as the ionization is produced, is visible, particularly in the expanded waveforms to the right in the figure. After calibration, the energy spectrum recorded from the source is shown in Figure 4.16, along with the result from the Monte Carlo simulation. The simulation is in good agreement with data, particularly above 200 keV, and the charge-only energy resolution of 5.6% at 570 keV is consistent with the intrinsic resolution of LXe measured by other experiments [25]. Such resolution is expected to improve once scintillation signals collected with high efficiency are available. In addition, the performance of the tile did not degrade after over ten cycles from room temperature to Xe liquefaction.

#### 4.1.3.5 3D-tile: Tile Integrated with ASIC

It is expected that a readout ASIC will be mounted on the reverse side of the tiles (possibly a subset of them, if strips are ganged together). Therefore signals from the strips need to be transported to the back. While existing prototype tiles are single-sided, double-sided tiles are being developed using two different techniques. In one case, signals are transferred to the back using TQVs. This technology is currently more advanced and thus adopted for the primary design. A demonstration of the TQV technique is shown in Figure. 4.17. Each TQV is drilled by a picosecond pulse-width laser beam, obtaining a slight taper angle with a sidewall roughness of about a few micrometers. Micro-cracks on the via's sidewall and surface debris near the via are etched away to guarantee a smooth surface quality for seed layer deposition. The TQVs are then electroplated with copper, as shown in the figure. The process is being optimized to mitigate the effects of differential thermal expansion between copper and fused silica.

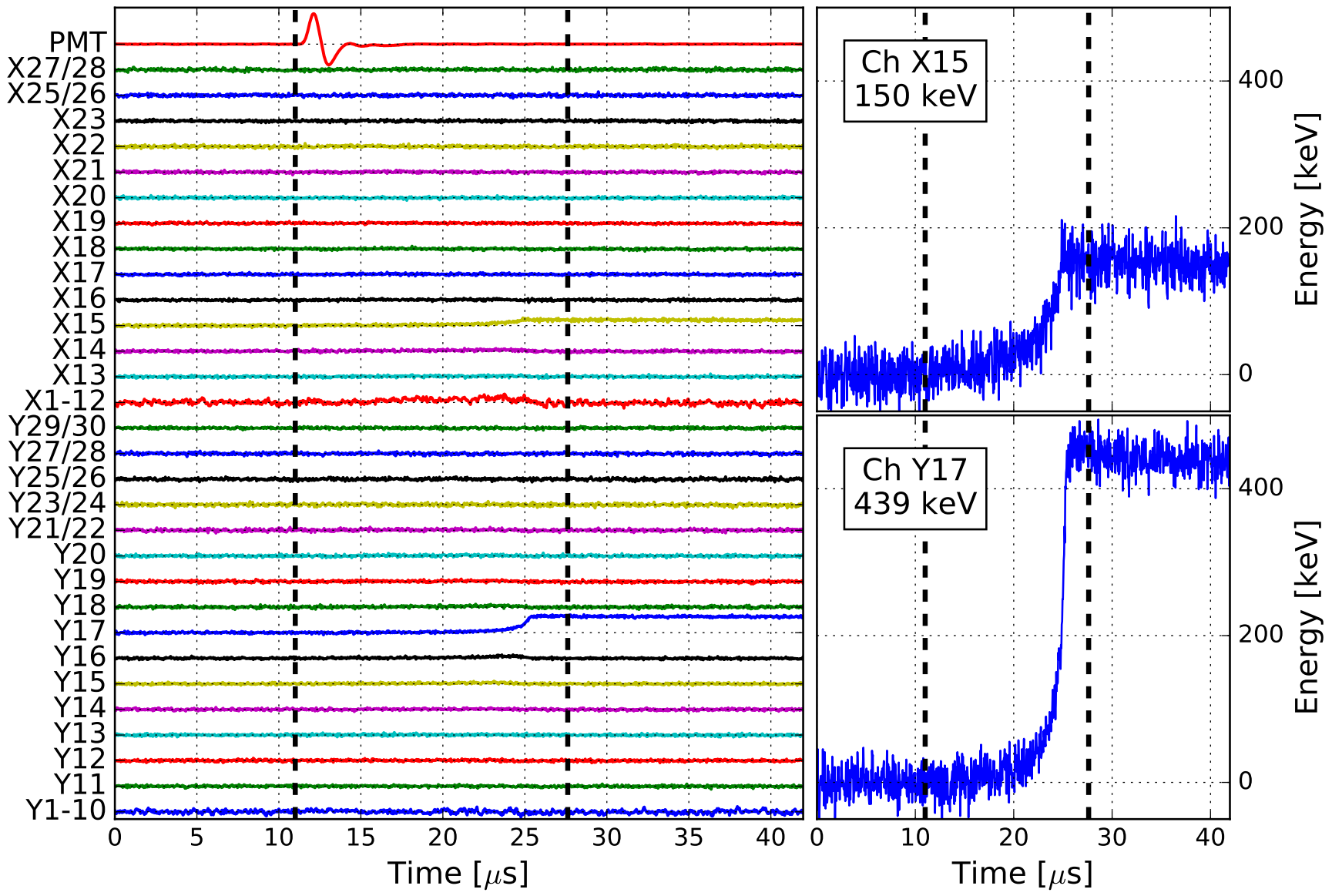

**Figure 4.15:** A sample event from the prototype tile tested in LXe. Traces to the left represent the digitized waveforms from the strips, except for the top-most that shows the (shaped) signal from the PMT reading out the scintillation light. The event passed all selection cuts and has energy near the $^{207}$Bi peak at 570 keV and the waveforms with the largest amplitude signals are shown magnified to the right. The sum energy in these two channels is 589 keV. The dashed vertical lines represents the trigger from the PMT and the maximum expected drift time for an event that started near the cathode.

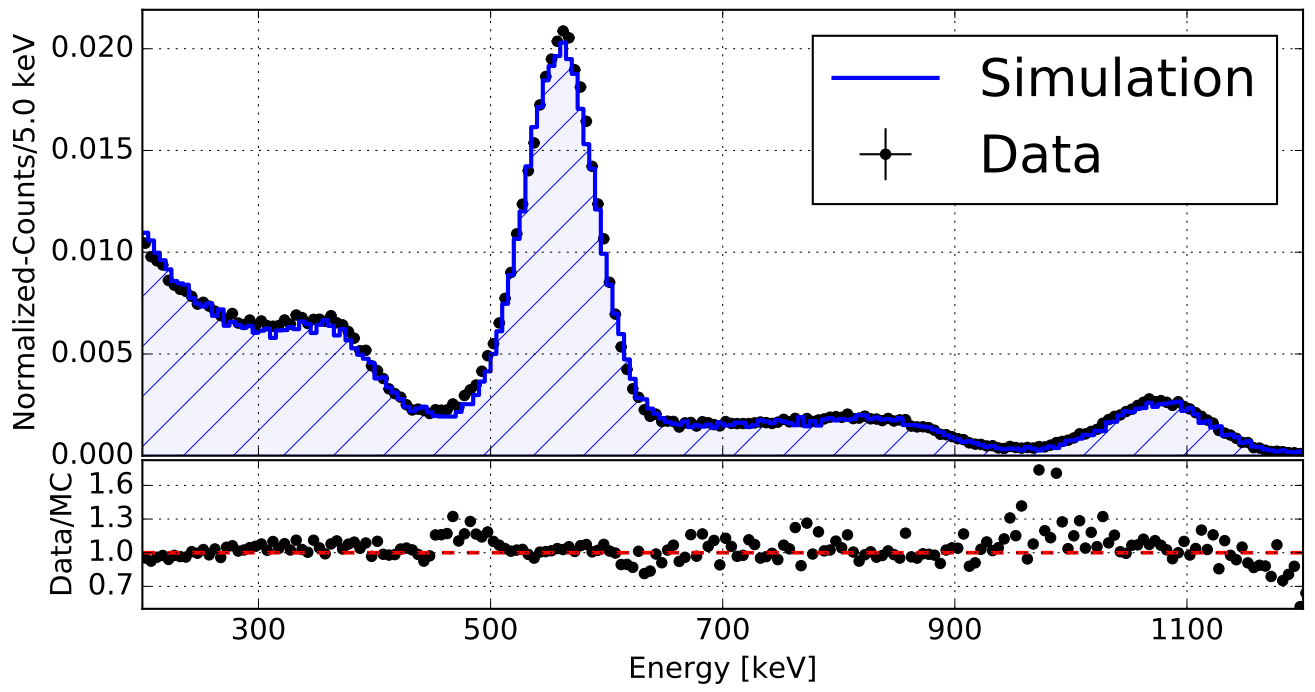

**Figure 4.16:** Top: Energy spectra from a charge collection tile in LXe, exposed to ionizing radiation from an internal $^{207}$Bi source. Bottom: Residuals between the two spectra. The agreement between data and Monte Carlo is excellent and the energy resolution is in line with the state-of-the-art for charge only.

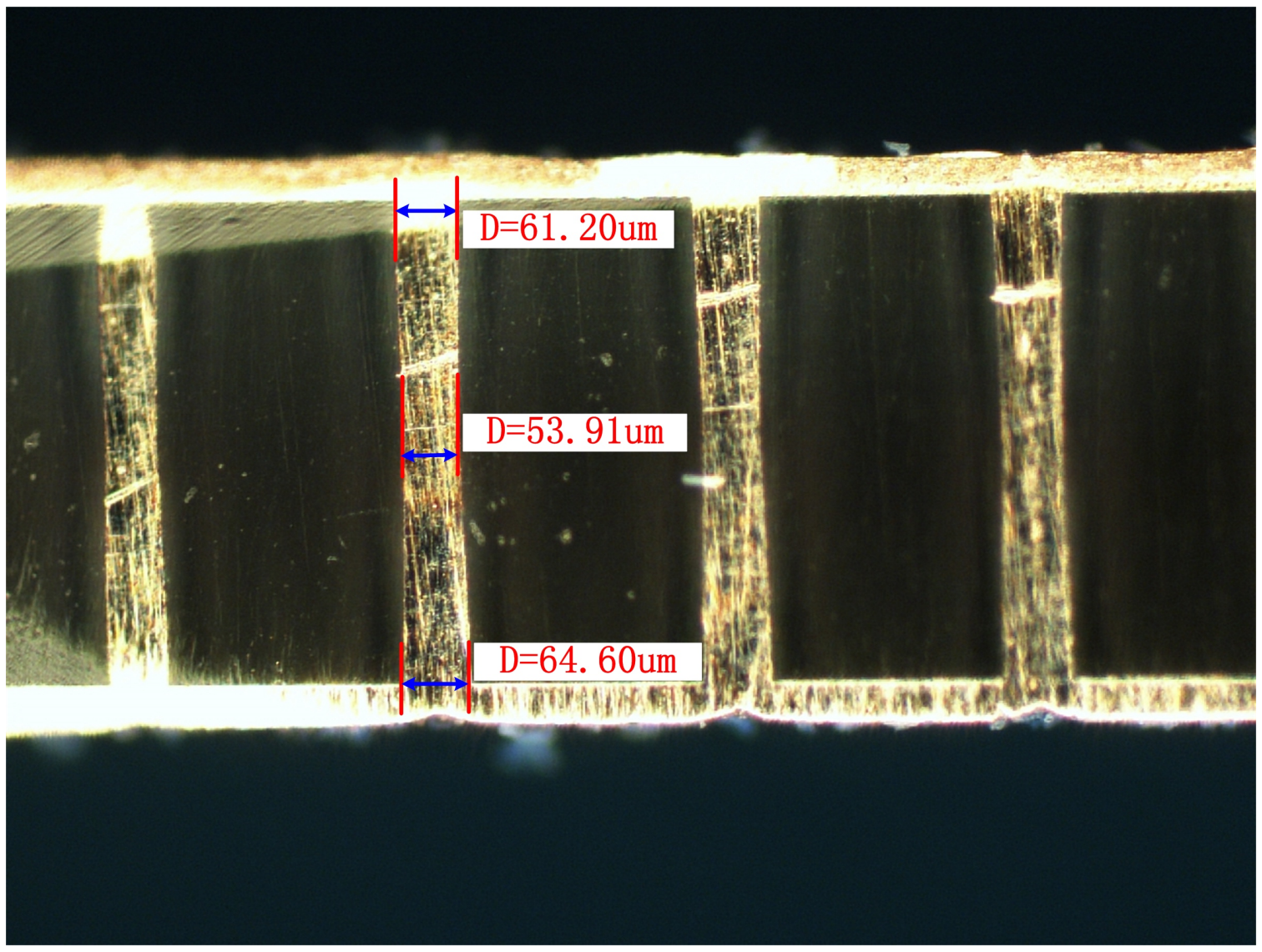

**Figure 4.17:** Demonstration of the TQV technology for nEXO. The image shows the cross-section of a fused silica wafer on which an array of TQVs has been obtained. Four vias, running vertically, are shown. The slightly conical shape of the TQVs is visible, along with the quality of the internal Cu coating, entirely filling the vias.

In a different scheme, also under development, signals are transferred to the back by metalizations on the edges of the tiles. This is practical because of the low density of the strips. In all cases, it is likely that an interposer will be required between the tile and the ASIC chip. This will carry the few discrete components required and provide the appropriate shielding layers to isolate the charge collection strips from pickup produced by the readout electronics.

### 4.1.4 Scintillation Light Detection

The readout of scintillation light from LXe is essential to provide a prompt signal at the interaction time, and to obtain the best possible energy resolution [4, 25]. The latter purpose requires a substantial light detection efficiency, to minimize the effect of fluctuations. Indeed, the resulting energy resolution tends to be dominated by the fluctuations in the scintillation light channel because, even under the best circumstances, its detection is less efficient than that of direct ionization (which generally approaches 100%). The efficient detection of the scintillation light is complicated by the wavelength emitted in LXe, $\lambda \simeq 175$ nm, in the vacuum ultraviolet (VUV), and by the radioactivity requirements, which makes vacuum photomultipliers unsuitable. This last require-ment is substantially stricter than for dark matter detectors, where, because of the lower energy regime, a thin layer of LXe is sufficient to shield radiation produced by the photomultipliers and their voltage dividers.

Silicon Photomultipliers (SiPMs) have been selected for the readout of the scintillation light in nEXO. SiPMs have gains of the order of $10^6$, comparable to those of vacuum photomultiplier tubes, in contrast to the Large Area Avalanche Photodiodes (LAAPDs) used in EXO-200 [10]. Indeed the energy resolution $\sigma_E/E \approx 1.2\%$ achieved in EXO-200 [4] is dominated by the LAAPD readout noise, in turn related to the large capacitance of the devices and their modest gain ($\simeq 200$). Modern SiPMs (not available at the time when EXO-200 was built) allow for better coverage (being square-shaped while LAAPDs are circular, with a substantial inactive "frame"), have a substantially lower mass and a simpler construction than LAAPDs, resulting in lower radioactivity background per unit area.

Until very recently, no SiPMs existed with sensitivity extending in the 175 nm region necessary for our purpose. R&D for a number of experiments, including nEXO, has led to devices with sufficient efficiency (albeit still substantially lower than that of LAAPDs) for the scintillation readout in nEXO [1]. While we expect further improvements before the beginning of nEXO construction, even with currently available devices the non-ideal SiPM efficiency is more than compensated by the other properties described below.

Since the energy resolution is dominated by the scintillation detection efficiency and the value of the electric field in the TPC can be used to adjust the ratio between scintillation and ionization signals, the optimization of the scintillation readout and the high voltage systems are closely related. In the final design, safety margins can be traded between these two areas.

#### 4.1.4.1 Photon Detection Requirements

In order to guide the discussion about the photo-detector specifications, it is useful to compute the required light detection efficiency using a simple model. Assuming that every electron-ion recombination yields a scintillation photon and that the electronics noise is sufficiently small, the optimal energy estimator, $\langle E\rangle$, for energy deposits in the LXe is: $\langle E\rangle = W(Q+S)$, where $Q$ is an estimate of the number of elementary charges deposited by the interaction, and $S$ is an estimate of the number of VUV photons produced. For this combination, $W = 13.7$ eV is the recombination-independent energy required to produce either a single electron or photon [21]. While the charge collection efficiency is nearly 100%, the light detection efficiency, $\epsilon_o$, can be substantially smaller. For $s$ detected photons, $S = s/\epsilon_o$. As described above, the electrons and photons are assumed to be perfectly anticorrelated, so that their intrinsic fluctuations do not cause variance in the total number of quanta deposited in the LXe. Due to the large number of quanta created, fluctuations in the total number of quanta (parameterized by a Fano factor in the LXe) are negligible [26]. Under these assumptions, the variance of $\langle E\rangle$ arises only from the imperfect measurement of the total quanta, and can be written:

$$\sigma^2_{\langle E\rangle} = W^2\left(\sigma^2_Q + \sigma^2_S\right) = W^2\left(\sigma^2_Q + \frac{S}{\epsilon_o}\left[(1-\epsilon_o) + \eta_N\right]\right) \tag{4.1}$$

[1]The use of visible light SiPMs coated with a wavelength shifting material has been proposed. In nEXO this option has not been pursued because most wavelength shifter materials have been found to be incompatible with the electron lifetime requirements in LXe.

Since the collection efficiency of the charge is approximately 100%, statistical fluctuations in the charge collection process itself are not significant, and the charge noise, $\sigma_Q$, is determined only by the charge electronics noise. As described in Section 4.1.3.2, the expected electronics noise is approximately 200 $\mathrm{e}^-$ per readout channel, which gives a total charge noise $\sigma_Q = \sqrt{N}(200\ \mathrm{e}^-) \approx 600\ \mathrm{e}^-$ for average channel multiplicity $N = 10$. The standard deviation of the light measurement, $\sigma_S$, has additional fluctuations due to the finite light collection efficiency, providing a binomial variance of $\sigma_S^2 = (\sigma_s/\epsilon_o)^2 \approx S(1-\epsilon_o)/\epsilon_o$. In addition, an excess noise factor, $\eta_N$, is included to represent fluctuations in the number of SiPM avalanches that are correlated in time with photon-induced avalanches. Assuming a Poisson distribution of such correlated avalanches with mean $\Lambda$, then $\eta_N = \Lambda/(1+\Lambda)^2$.

As described in Section 4.1.4.2, the gain of the SiPMs is sufficient to provide an electronics noise of $\sigma_p < 0.1$ spe (single photoelectron). The electronics noise of the light channel is negligible relative to the binomial variance when $\sigma_p \ll 1$ spe, so electronics noise is not included in Equation 4.1. In general, calculations show that the electronics noise requirement is met when the SiPM overvoltage is >3 V, which sets a lower bound on the required operating voltage.

The contribution of correlated avalanches to the resolution is subdominant when $\Lambda < 0.2$. This requirement sets a corresponding upper bound on the overvoltage since the correlated avalanche rate increases with voltage. The most stringent constraint on the dark rate of the SiPMs arises from the requirement to trigger on scintillation pulses with energy as low as 500 keV, for which a dark rate $< 50\ \mathrm{Hz/mm^2}$ is required. At this rate, easily achieved in the devices tested, the dark noise in the 1 $\mu$s window following an event is much smaller than the rate of correlated avalanches, and is neglected in the energy resolution calculation. Dark rate and correlated avalanche effects will be described in Section 4.1.4.2.

For the nEXO drift field of 400 V/cm, an electron-like 2.5 MeV energy deposit produces approximately $1.2 \times 10^5\ \mathrm{e}^-$ and $7 \times 10^4$ photons, which gives total quanta of (2.5 MeV)/$W \sim 1.9 \times 10^5$ [21]. The relative variance contributed by the charge noise is $(600\ \mathrm{e}^-)/(1.9 \times 10^5) \sim 0.3\%$, which is typically subdominant to the light noise. From Equation 4.1, achieving a resolution $\lesssim$1% requires a light collection efficiency of $\epsilon_o \gtrsim 3\%$. For this efficiency, the contribution of the light channel resolution to the relative energy resolution is $\sigma_S/(Q+S) = \sqrt{(0.8\%)^2 + (0.3\%)^2}$, where the two terms denote the binomial noise and the correlated avalanches ($\Lambda = 0.2$), respectively. Thus, the statistical noise in the light collection dominates the resolution at low collection efficiency, and the charge noise only becomes significant when the overall light efficiency approaches 10%. While this simple model provides a description of the dominant noise sources, a more complete model using the NEST simulation package gives consistent results. The results of this simulation are illustrated in Figure 4.18, where the energy resolution of the detector is given as a function of the light detection efficiency for three values of the drift field. This model is used to fine tune the detector configuration.

The primary specifications for the nEXO scintillation light detection system are listed in Table 4.2 and are described in the following sections. Based on the calculations above and Figure 4.18, the table provides a requirement on the overall light detection efficiency of $\epsilon_o > 3$ %. It is useful to break down $\epsilon_o$ as the product of the photon transport efficiency, $\epsilon_t$, and the probability that a scintillation photon hitting a SiPM is absorbed and triggers an avalanche, $\epsilon_a$. Since the reflectivity $R$ of silicon photo-detectors can be as large as 60% at 175 nm, photons may be reflected several times from the photo-detector surfaces before being absorbed, reducing the value of $\epsilon_o$. When measuring the SiPM photo-detection efficiency (PDE, $\epsilon_{PD}$), reflected photons are dis-carded, thus $\epsilon_{PD} = (1-R)\epsilon_a$. It follows that $\epsilon_o = \epsilon_t \cdot \epsilon_{PD}/(1-R)$. In general, the value of $\epsilon_a$ quoted by manufacturers assumes a reflectivity of $R = 50\%$. Thus, for nEXO, it is necessary to measure $R$ along with $\epsilon_{PD}$ in order to assess the performance of the devices. The photon transport efficiency $\epsilon_t$ is investigated using simulations as well as reflectivity measurements and is described in Section 4.1.4.3.

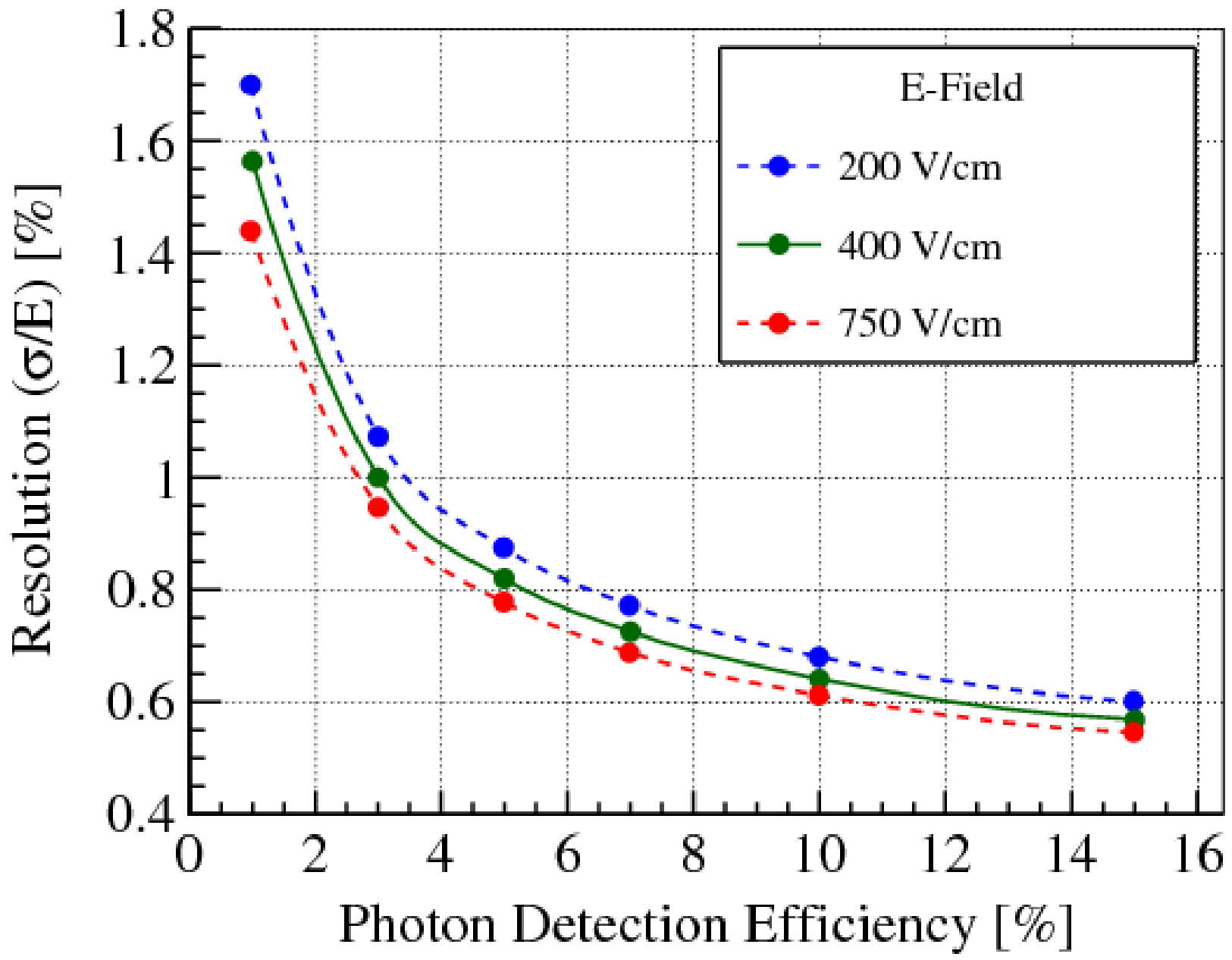

**Figure 4.18:** Energy resolution as a function of the light detection efficiency. The three curves refer to three values of the drift electric field. The primary value of the field is represented by the solid line.

| Parameter | Value |
|---|---|
| Total instrumented area | $\simeq 4.5$ m$^2$ |
| Overall light detection efficiency | $\epsilon_o > 3$ % |
| SiPM PDE (175 nm, normal incidence) | $\epsilon_{PD} > 15$ % |
| Overvoltage | $> 3$ V |
| Dark noise rate | $< 50$ Hz/mm$^2$ |
| Correlated avalanche rate | $< 0.2$ |

**Table 4.2:** List of key parameters for the nEXO light detection system and the corresponding SiPM device requirements.

#### 4.1.4.2 Silicon Photo-Multipliers

SiPMs are generally constructed as arrays of very small Geiger-mode avalanche diodes connected in parallel through appropriate resistors and read out by electronics which are external to the device. This is equivalent to making an analog sum of all the elementary charges generated by each firing avalanche diode. Conventional devices of this type ("analog SiPMs") are the primary choice for nEXO. However each diode can be considered a digital device with two possible states: "fired", indicating the presence of a photon and "idle" indicating that no photon has been detected. Electronics could be built directly on the devices to simply count the number of fired diodes. This scheme ("digital SiPMs") is being developed as a backup by the collaboration.

The nEXO collaboration has been driving the development of analog SiPMs sensitive at 175 nm for a number of years. Fondazione Bruno Kessler (FBK) [2] has produced several batches of SiPMs dedicated to nEXO. Hamamatsu Photonics Inc. [3] has also developed four generations of SiPMs for applications in LXe.

For SiPMs, the operating range is defined by two voltages: breakdown, where Geiger mode avalanches start to occur, and runaway, where dark noise and the correlated avalanche rate becomes so high that the SiPM continuously avalanches. The over-voltage is defined as the difference between the chosen operating voltage and the breakdown voltage. Dark noise and correlated avalanches adversely impact the detector performance and, in particular, the energy resolution. Hence care is required in finding the optimal operation point and in understanding whether such an optimal point is sufficient for the resolution requirements of the experiment. Dark noise is the production of Geiger-mode avalanches due to thermally generated charge carriers that are indistinguishable from photon-triggered avalanches. Correlated avalanches are due to two processes: after-pulsing and cross-talk. The former is generated by charges trapped and released at a subsequent time; the latter is a charge generation process due to the production of longer wavelength photons during the avalanche process within the avalanche region that travel across the device and set off more avalanches. Both phenomena generate additional avalanches whenever a primary avalanche occurs, and both are associated with statistical fluctuations.

The over-voltage affects the SiPM gain, $\epsilon_{PD}$, the dark noise rate and the correlated avalanche rate. $\epsilon_{PD}$ tends to saturate above a few volts of overvoltage, while the correlated avalanche rate continues to grow, eventually reaching the runaway condition. The saturation of $\epsilon_{PD}$, around 20%, is clearly visible in Figure 4.19 for a number of FBK device types measured by nEXO. However, since the rate of correlated avalanches continues to increase beyond the onset of this saturation, a better figure of merit can be derived by plotting $\epsilon_{PD}$ against the probability of correlated avalanches, instead of the overvoltage. This is shown in Figure 4.20 for the same FBK devices.

Preliminary models accounting for the gain, $\epsilon_{PD}$, dark noise, correlated avalanche rate, electronics noise and signal observation time produce the device specifications listed in Table 4.2.

[2] https://www.fbk.eu/en/
[3] http://www.hamamatsu.com

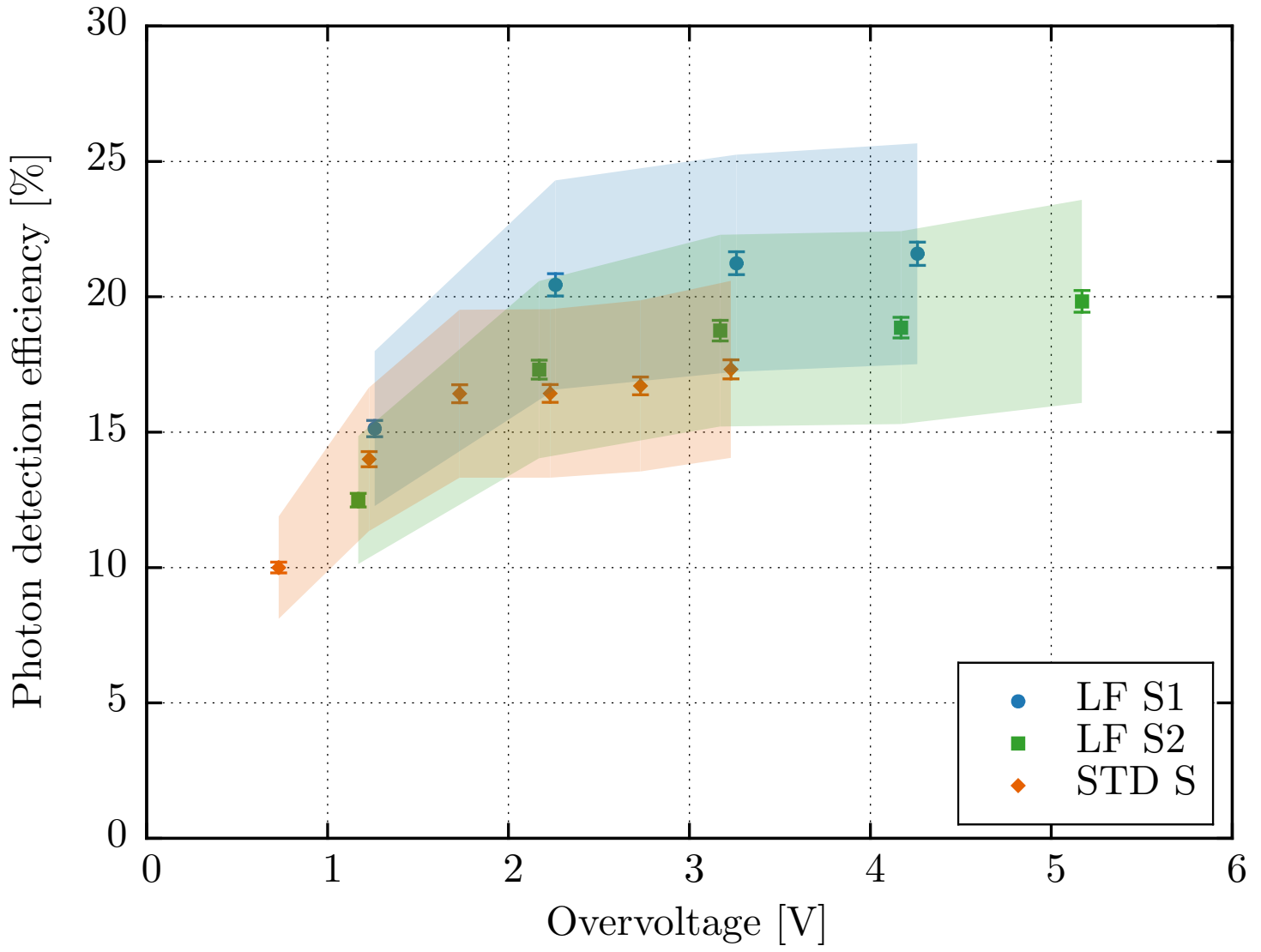

**Figure 4.19:** SiPM photo-detection efficiency, $\epsilon_{PD}$, as a function of the over-voltage for several FBK devices. A saturation around 20% is evident. The colored bands represent the systematic uncertainties of the measurements.

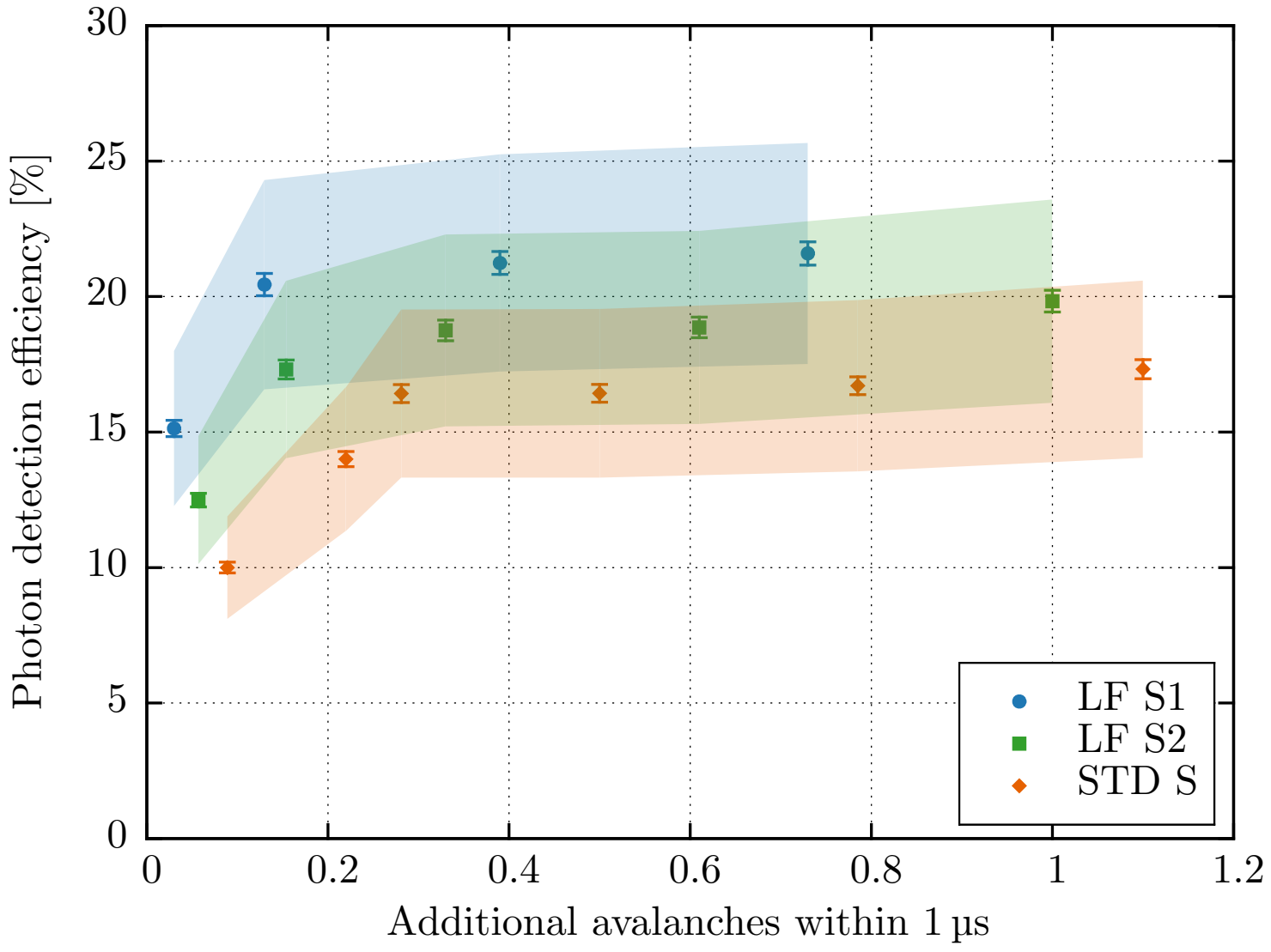

**Figure 4.20:** SiPM photo-detection efficiency, $\epsilon_{PD}$, as a function of the probability of correlated avalanches, for the same devices as in Figure 4.19. The efficiency of the "LF" devices exceeds the minimum requirements of nEXO and the optimal set point is near the beginning of the plateau. The colored bands represent the systematic uncertainties of the measurements.

Therefore it appears that at least the “Low Field” (LF) FBK devices tested meet the nEXO requirements. It should be noted, however, that the investigations done until now focused on ascertaining feasibility, and a new phase of work is required to optimize the system design, combining together the intrinsic performance of the devices, the light propagation in the detector and the performance of the electronics (that, through the SiPM capacitance, are also related to the physics of the devices).

Further work aims to improve $\epsilon_{PD}$ without adversely affecting other parameters. This is done in collaboration with manufacturers (a new, nEXO commissioned run with FBK is in progress) as well as by exploring post-processing options, such as the development of VUV anti-reflective coating for the SiPMs in LXe. We are also exploring the possibility of using “delta doping”, a process pioneered by JPL [27] that permits a very shallow doping layer on the surface of the silicon in order to maximize photon collection very close to the surface, where VUV photons are absorbed.

A batch of FBK VUV LF devices was tested for their radioactivity content by means of ICPMS (see Section 6). The results show that SiPMs of this type would contribute about 0.5% of the SS background rate around the Q-value for the inner 2000 kg of the detector. Hence, also in this case, measurements establish feasibility. The final choice of device will need to take into account electro-optical properties as well as issues related to radioactive contamination. These include the intrinsic contamination of the devices tested by nEXO, as well as the arrangements that manufacturers are willing to entertain to ensure quality and simplify routine testing by nEXO during production.

“Digital” SiPMs, simply reporting the number of photons detected on a certain surface, are currently being pursued for medical imaging cameras but not for large areas like those found in nEXO. The collaboration is investigating a specific architecture solution called “three dimensionally integrated digital SiPM” (3DdSiPM). In 3DdSiPM every Geiger-mode avalanche diode (also called “single photon avalanche diode”, SPAD) is individually connected to an active circuit used to quench and tag each avalanche. The end result of the process is a digital number corresponding to how many SPADs have fired. While still unproven at the scale required here, the main advantage of this solution is the much lower power dissipation in the electronics, because the technique avoids the requirement of low-noise analog front-ends reading out the SiPM’s large capacitance. 3DdSiPM are being developed as a backup for nEXO by the Université de Sherbrooke (Quebec, Canada) in partnership with Teledyne-DALSA Semiconductor with whom they have a close relationship. It is expected that a decision on the technology will occur at the time the detector design will need to be frozen.

#### 4.1.4.3 Light Transport in the TPC

Optimizing the photon transport in the TPC is very important for the performance of the detector and complements the development of the actual SiPM devices. This work involves simulation of light transport, using GEANT4, as well as the development of coatings with specific optical properties.

Three methods can be used to optimize the light transport to the active SiPMs. Since the SiPMs are installed in the barrel region of the detector, behind the FSRs, the shape and pitch of the rings have to be optimized for light transmission, along with drift field uniformity and minimum electric field between the SiPMs and TPC vessel. The reflectivity at 175 nm wavelength of various components can also be increased. Finally, antireflective coatings can be applied to the surface of the SiPMs. The $\sim 2\pi$ sr incidence angles, along with the relatively broad emission spectrum of the LXe scintillation light, make the use of multi-layer dielectric coatings challenging.

A simulation of the light transport efficiency versus the overall reflectivity of the field shaping rings and cathode is shown in Figure 4.21. With the current design of the field cage, at one extreme of the parameter space, all electrodes can be considered 100% absorbent, resulting, according to the simulation, in $\epsilon_t = 14$ %. This, in turn, requires $\epsilon_a > 20$ %. The target for the nEXO design is to achieve about 70% electrode reflectivity (FSRs and cathode), yielding $\epsilon_t = 25$ % and, in turn, the nominal $\epsilon_a > 15$ %, which would exceed the required $\epsilon_o$ in Table 4.2. Such reflectivity at 175 nm wavelength can be obtained by depositing a thin layer of aluminum covered by a layer of magnesium fluoride ($MgF_2$), protecting the aluminum from oxidation. This technique can routinely achieve 90% reflectivity for small surfaces and was successfully employed on the LAAPD platters of EXO-200. In nEXO one challenge is the size of the components to coat, although large evaporators for optical depositions are becoming more common because of the processing of large displays and parallel processing of eyeglasses. Very large size (3 m diameter and $5 \times 5 \times 3$ m$^3$) evaporators are available at the Changchun Institute of Optics, Fine Mechanics and Physics of the Chinese Academy of Sciences (CIOMP), associated to nEXO institutions in China. The presence of a thin ($MgF_2$) dielectric coating on various HV electrodes will need to be investigated, although this appears less problematic than using Teflon sheets inside the field cage of EXO-200.

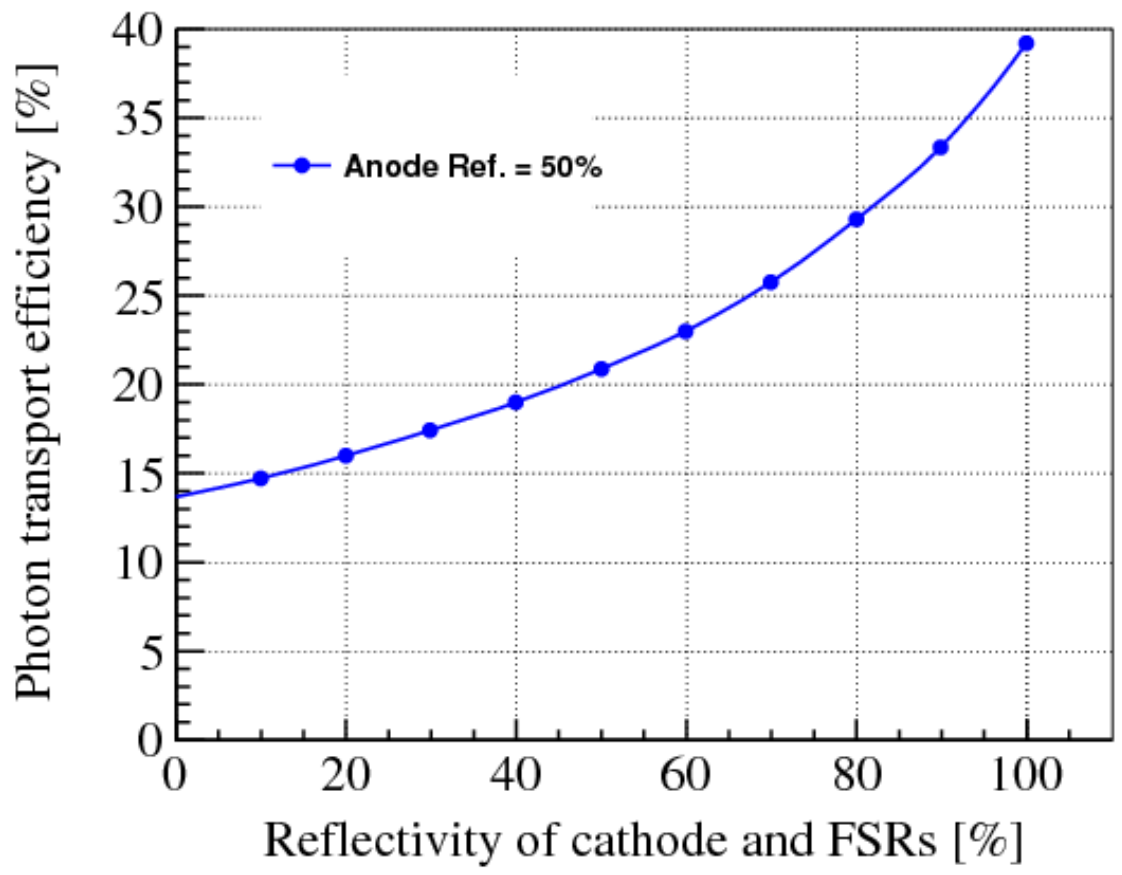

**Figure 4.21:** Simulation of the photon transport efficiency, $\epsilon_t$, versus reflectivity of the field shaping rings and cathode, incorporating measured reflectivity and interference effects on the SiPM surfaces.

The possibility of using bare, high resistivity silicon (Si) for the various electrodes has been already discussed for reasons related to minimizing discharge currents. Interestingly, Si exhibits good VUV reflectivity, especially at large incidence angles, as shown in Figure 4.22. In the figure, the oscillations are due to interference between the optical interfaces Si-$SiO_2$ and $SiO_2$-vacuum. These oscillations are expected to disappear in LXe because the indexes of refraction of $SiO_2$ and LXe are closely matched, and such interference is, consequently, minimal.

#### 4.1.4.4 SiPM Integration

The SiPMs are expected to be mounted on staves, behind the field shaping rings as discussed in Section 4.1.1, possibly first grouping them on dielectric tiles also hosting the readout electronics. While the engineering of this system has not been performed, some work has begun in building progressively larger SiPM arrays, in this initial stage without attention to the radiopurity of the arrangement. A first, 24 cm$^2$ prototype array, using FBK LF SiPMs, was recently assembled and is shown in Figure 4.23. This array, currently used in LXe in combination with a charge collection tile, is not yet optimized for density as SiPMs are held in place and contacted by dismountable springs. This is desirable at this early stage as SiPMs are still in short supply.

Since the SiPMs are to be installed behind the field cage, at the ground potential of the TPC vessel, tests are in progress to confirm that electric fields up to 50 kV/cm, expected at the bottom of the detector near the cathode, do not affect their operation. Initial measurements appear to exclude problems in this area. As mentioned in Section 4.1.2, tests are in progress to understand and mitigate possible HV breakdowns, particularly landing on or near the SiPMs.

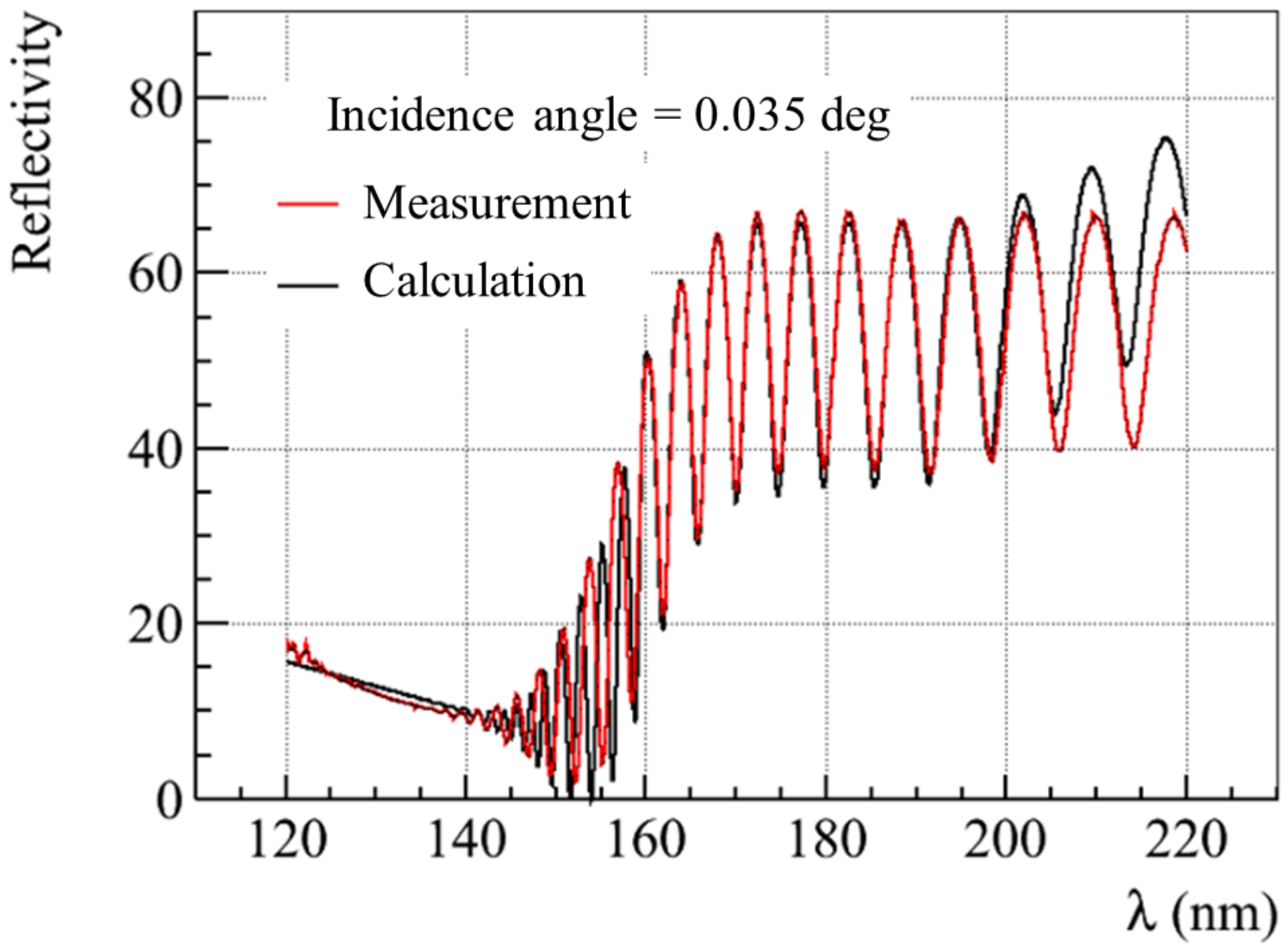

**Figure 4.22:** Simulated specular reflectivity of a silicon wafer coated with 1.5$\mu$m of $SiO_2$ (black) and measurements of a SiPM manufactured by FBK (red). The measurements were done in vacuum.

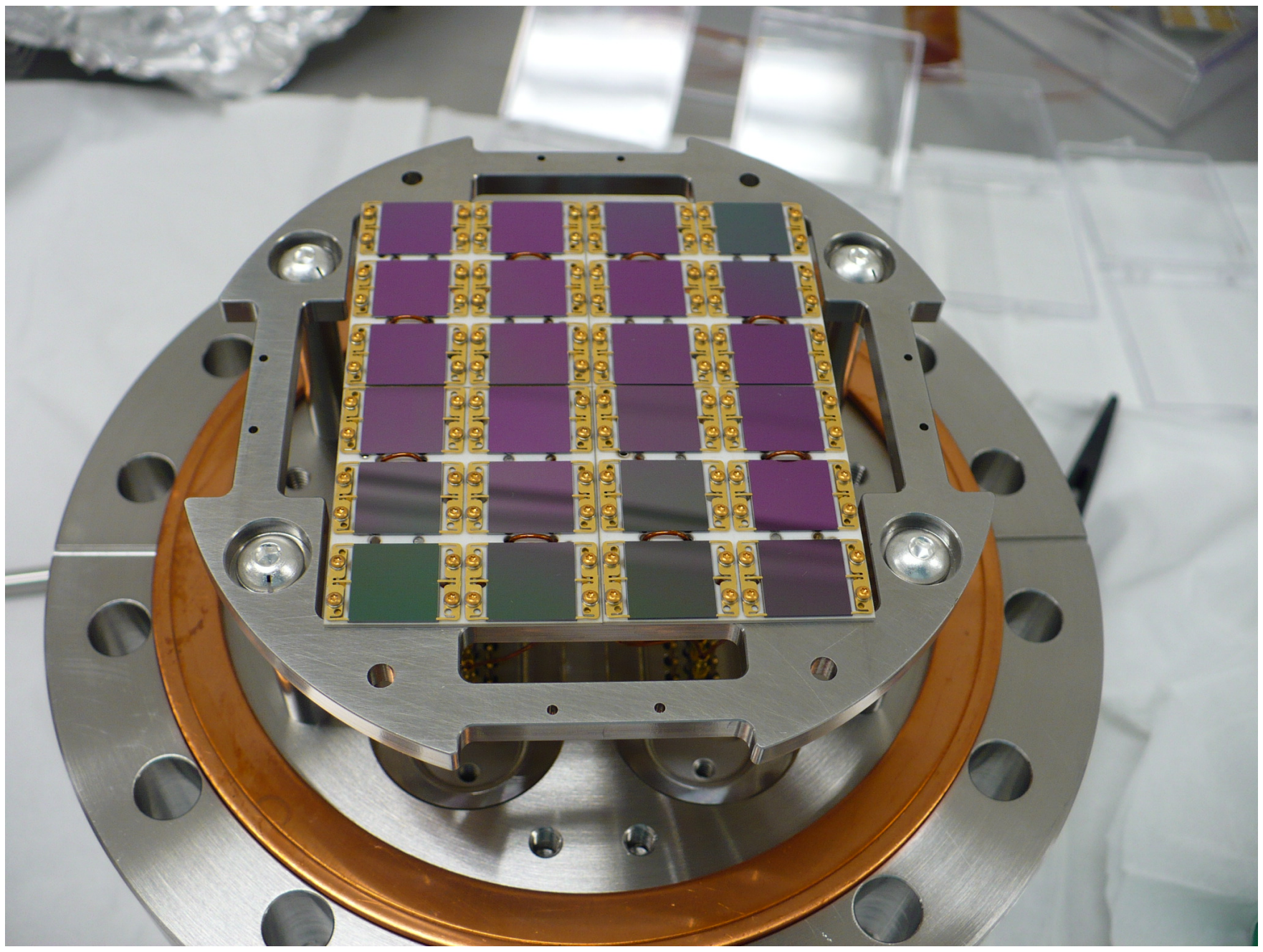

**Figure 4.23:** Test array of 24 SiPMs used in conjunction with a charge collection tile in LXe. Each SiPM is 1 $cm^2$. The mounting system is designed to permit the reconfiguration of the devices and not for optimal collection efficiency.

### 4.1.5 Interconnections

Several types of electrical connections are required inside the nEXO TPC. Charge collection strips have to be connected together and to the readout ASIC chips. Likewise, SiPMs have to be appropriately grouped and connected to the corresponding ASIC. These connections generally carry low-level analog signals over short distances in close proximity to the active LXe. In these cases, the electrical interconnections are intimately related with the mechanical structures of charge tiles and SiPM staves. Fast digital signals have to be transported from the ASIC chips out of the TPC, for logging on the data acquisition computer(s) and power and control signals have to be transported along the same path to the ASICs in the TPC. These connections need to be in the form of long, flexible cables. Since the SiPMs are arranged in staves spanning the entire length of the TPC, there will also need to be cables along staves, and these may be rigid or flexible.

While many commercial solutions for high density electrical interconnections between components and subsystems exist, nEXO presents unique constraints in terms of cryogenic reliability, radiopurity, and low outgassing into the LXe, so that specially engineered solutions have to be developed.

#### 4.1.5.1 Charge Modules

Due to fabrication yield constraints, the charge collection tiles are assumed to have maximum size of 10 cm $\times$ 10 cm, consistent with the use of 6-inch wafers. Taking into account the channel capacitance and constraints from the readout electronics, it may be optimal to connect multiple tiles into a single module to be read out by a single charge ASIC as described in Section 4.2.1. The readout ASIC with a few accompanying discrete components, will be mounted on the reverse side of one of the tiles for each module, most likely on a separate substrate to allow the inclusion of a shielding ground plane between the digital and analog lines while minimizing the capacitance of the strips to ground.

To connect multiple tiles, if needed, wire traces will be brought to wirebonding pads on the backside of the tile using through quartz vias (TQV) or edge metalization. Strips on adjoining tiles will be connected by wirebonds, and the tiles within each module will be mechanically supported by a rigid backing. A sketch of wirebond connections between multiple tiles in a module is shown in Figure 4.24. For 3 mm channel pitch, the total number of wirebonded connections between pads will be 128 per 2$\times$2 tile module, with an equal number of connections to the electronics daughterboard for each module. The total number of connections for the anode plane is then $\sim 10^4$. To ensure less than $< 1\%$ chance of a connection failure due to the loss of a wirebond during thermal cycling or assembly requires a total failure rate per connection of $< 10^{-6}$. Assuming three wirebonds per connection, and an uncorrelated rate of failures within a connection, this total failure rate can be achieved with a individual bond failure rate of $\lesssim 1\%$. Such a failure rate is expected to be straightforward, but will be confirmed with thermal cycling tests of prototype modules.

Traces will be routed to the ASIC substrate along the mechanical supports for the module. For a Cu backing structure, such traces can be routed along Kapton strips (while minimizing capac-itance to ground). As an alternative, large area quartz supports are being investigated on which routing traces can be directly patterned and the readout ASIC directly mounted. This solution may provide lower radioactivity and less outgassing of impurities into the LXe than the Kapton design. Work to characterize the background contribution and mechanical reliability of the various methods is in progress.

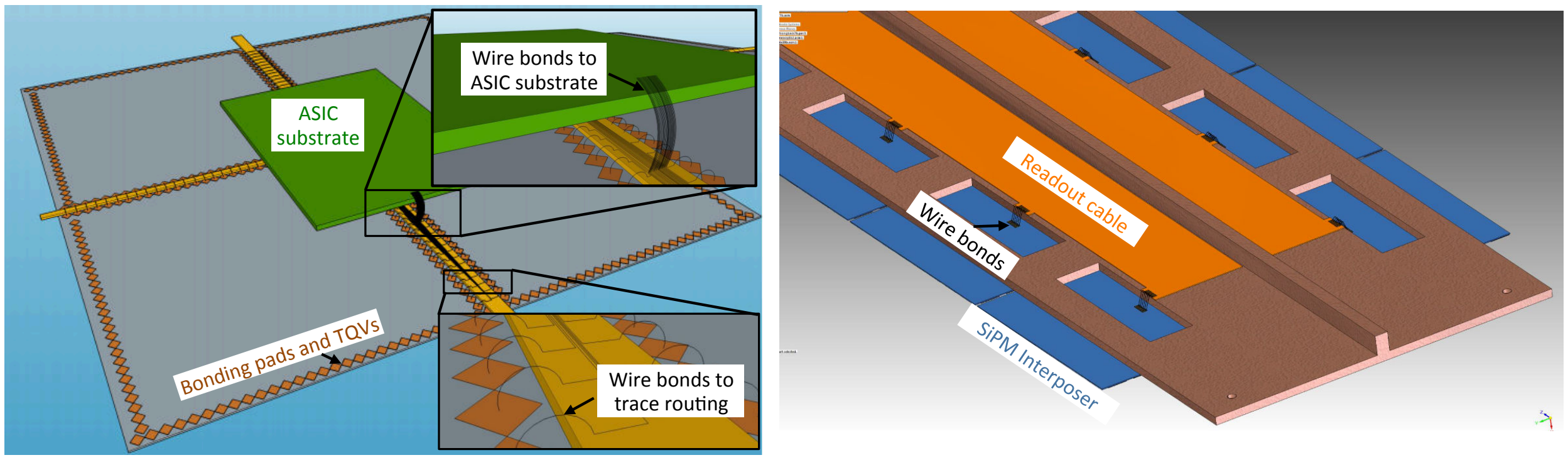

**Figure 4.24:** (left) Schematic of backside connections between a charge module consisting of 2×2 individual tiles. Signals on the opposite side of the charge tile are brought to the backside through TQVs in the center of bonding pads along the edges of the tiles. The bonding pads within the module are connected with wire bonds, as shown in the lower inset and described in the text. A separate substrate (upper inset) is used to readout the module. (right) Schematic of connections on a SiPM stave. The backside of an interposer on which the individual SiPMs are mounted and connected is wire bonded to readout cables running along the backside of the stave.

#### 4.1.5.2 SiPM Modules

Multiple SiPMs must be combined into a larger readout module as described in Section 4.2.2. Several options for connecting multiple devices into a larger module are currently under development. In general, the integration of multiple SiPMs is expected to require an "interposer," which provides a ∼10 cm ×10 cm mechanical support and electrical routing of signals to each individual SiPM.

Possible materials under investigation for the interposer include silicon, quartz, or Kapton. Silicon and quartz have the advantage of demonstrated high radiopurity, low outgassing rates and thermal stability, but fewer existing commercial fabrication options exist for multilayer boards of the required size with through substrate vias (TSVs or TQVs). Multilayer Kapton boards of the size and complexity required are trivial to obtain, but radiopurity, outgassing and thermal contraction need to be better understood. While quartz tiles with TQVs sufficient for such an interposer have been demonstrated for the charge tiles described in Section 4.1.2.2 silicon interposers, developed in parallel, have the additional advantage of a perfect match of coefficient of thermal expansion with respect to the SiPMs. Techniques for electrical connection between the SiPMs and interposer are under investigation, and the optimal solution will depend on the substrate material. Reliability tests under thermal cycling, similar to those to be done on the charge modules, will be performed to ensure negligible probability of a loss of connection to a SiPM channel. Connections to the interposer can be made with wirebonds that bridge the step between bonding pads on the back of the interposer and the readout cables, as shown in Figure 4.24.

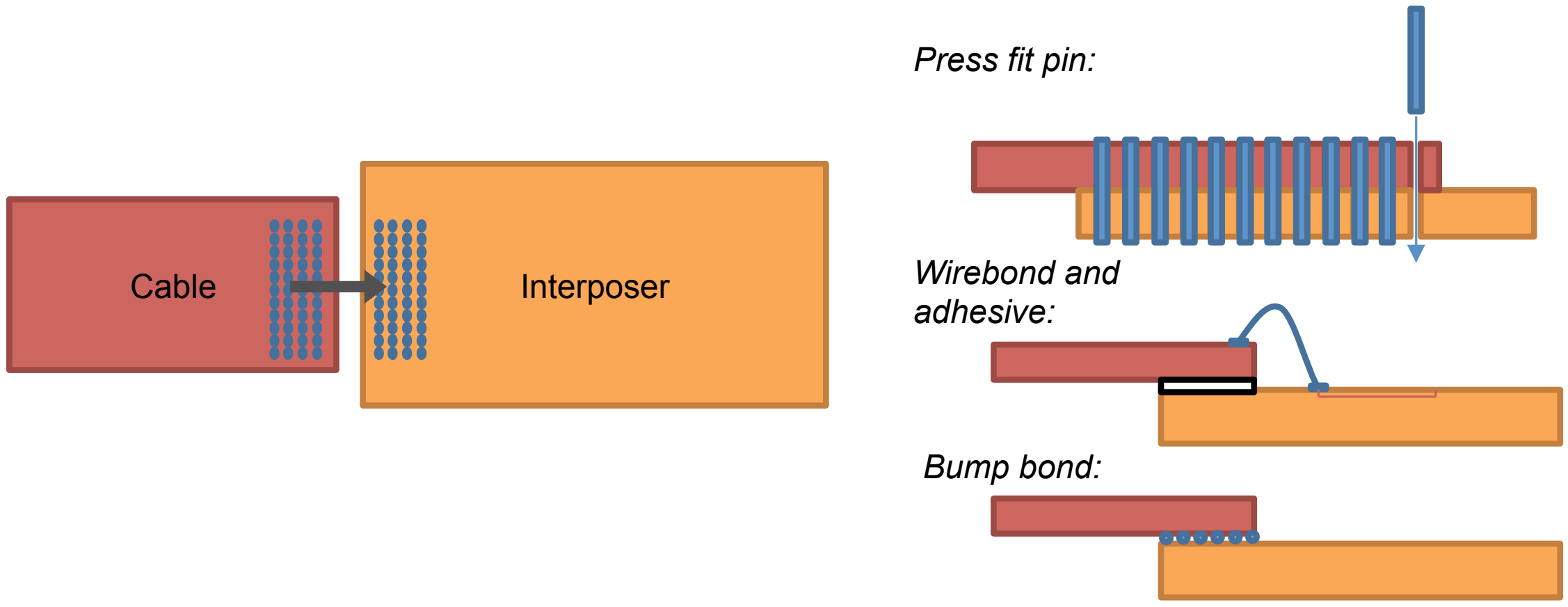

**Figure 4.25:** Top view (left) and cross-section (right) for various connection techniques between the readout cables and interposers, as described in the text. Connections using press-fit pins, wirebonds, or bump bonds will be investigated.

#### 4.1.5.3 High-speed Flex Cables

After digitization by the electronics for the charge and SiPM modules, signals will be routed out of the cryostat using 1 to 2 m-long high-speed flexible cables. The electrical properties of these cables are described in Section 4.2.3, while here we describe their mechanical design and layout. The time multiplexing of the digitized signals substantially reduces the required number of cables with respect to the case in which digitizers are external to the TPC (e.g. EXO-200). This is important, as it is often the case that cables contribute substantially to the background budget (see Section 6).

Each detector module will require several high-speed digital transmission lines as well as control and power lines to be routed to the daughter board or interposer containing the readout ASIC. In general, such cables must be flexible (or transition to a flexible cable) to enable their routing from the anode plane and SiPM staves to feedthroughs at the top of the vessel. In the primary design, such cables are routed on the backing structure supporting the charge tile modules and along the SiPM staves, providing a fan-out pattern to each module. Due to background considerations and the proximity to the LXe volume, it is planned to limit the total mass to $\lesssim 100$ g. This permits the use of up to 20 cables, each 50 $\mu$m-thick [4], 5 cm wide and with an average length of 1 m along the anode plane and/or the SiPM staves. This is sufficient to allow the expected data, power, and control required for each readout module. Cables will be tapered as they fan out among modules to minimize the total mass of Kapton.

Along the flat portions of the anode plane and SiPM staves, fabrication of readout traces on rigid materials (e.g. a large area quartz support) may be possible. Such a solution could minimize the amount of flexible cables required and further reduce the contribution to radioactivity and outgassing. In addition, the readout ASICs could be designed to receive and repeat signals from neighboring ASICs to allow interconnection only between modules rather than a full fan-out of traces to all modules individually. While not required in the primary design, future work will assess the feasibility of these techniques.

Several options will be investigated for connection of the cables to the readout modules, as shown in Figure 4.25. These include 1) press fit pins that provide both mechanical and electrical connection, 2) an adhesive mechanical connection with wirebonds, and 3) solder or gold bump bonds, with or without additional mechanical support. The number of cable connections is more than an order of magnitude smaller than the number of connections within the readout module itself, due to the digital multiplexing of signals onto high speed transmission lines. Nonetheless, high reliability connections are required since a single failure would remove an entire readout module, requiring similar failure rate testing to that described above for the individual module interconnections.

[4] EXO-200 Kapton signal cables are 25 $\mu$m thick but during construction were found to be exceedingly delicate and unsuitable in an application where longer sections are required

The flexible cables will be routed to about five wiring feedthroughs at the top of the LXe vessel, each of which can contain approximately 5 data or power cables. The primary design for the feedthroughs will follow the successful design of EXO-200 [1], in which continuous cables are routed through an epoxy filled plug. Unlike EXO-200, the use of electronics in the LXe will require the feedthrough to transmit high-speed digital signals without substantial impedance mismatch or loss. Impedance matching will be performed by appropriate tapering of the microstrip line geometry within the potted region, and the dielectric loss of the feedthrough will be characterized using the same techniques described in 4.2.3.

### 4.1.6 TPC Vessel Mechanics and Feedthroughs

The nEXO TPC vessel has the primary function of containing the LXe, field cage assembly, charge collection tiles, SiPM arrays and the front-end electronics necessary for the operation of the TPC.

To meet requirements, the vessel design has to address the following two key design challenges:

- Minimize its total mass in order to reduce the radioactive background contribution as much as possible while meeting structural requirements. This is important because the vessel is the most massive component in contact with the LXe.
- Fit around the TPC detector as snugly as possible, minimizing the LXe volume outside the TPC field cage, which is only partially active, although useful for background reduction. This, in turn, maximizes the LXe volume that is fully sensitive for physics.

#### 4.1.6.1 nEXO TPC and Xenon Vessel Concept

The vessel is envisaged as a right cylinder constructed of special low background copper [28] and assembled primarily via electron beam welding. The use of a vessel with very thin, electron-beam welded walls was pioneered in EXO-200 and is made possible by the HFE-7000 bath that, at the same time, transfers the pressure to the cryostat and reduces temperature gradients. The use of electron beam welding of a thin LXe-containing copper vessel was proven by EXO-200 to largely retain the radiopurity of the original copper stock.

The TPC vessel is supported from the Inner Cryostat by a copper structure as shown in Figure 4.39. Models of the TPC vessel are shown in Figure 4.26. On the left the stiffening structure of the end-plates is emphasized; we note that for reasons of economy of LXe these cannot be dished (that would be the rational design from a structural point of view). On the right a cutout view of the vessel with internal components is shown. The main field cage with the anode, FSRs, cathode, and SiPM arrays hang from the top end plate. The High Voltage feedthrough is connected via a weldment to the lower end-plate. Fluid and signal ports are also located on the top end-plate.

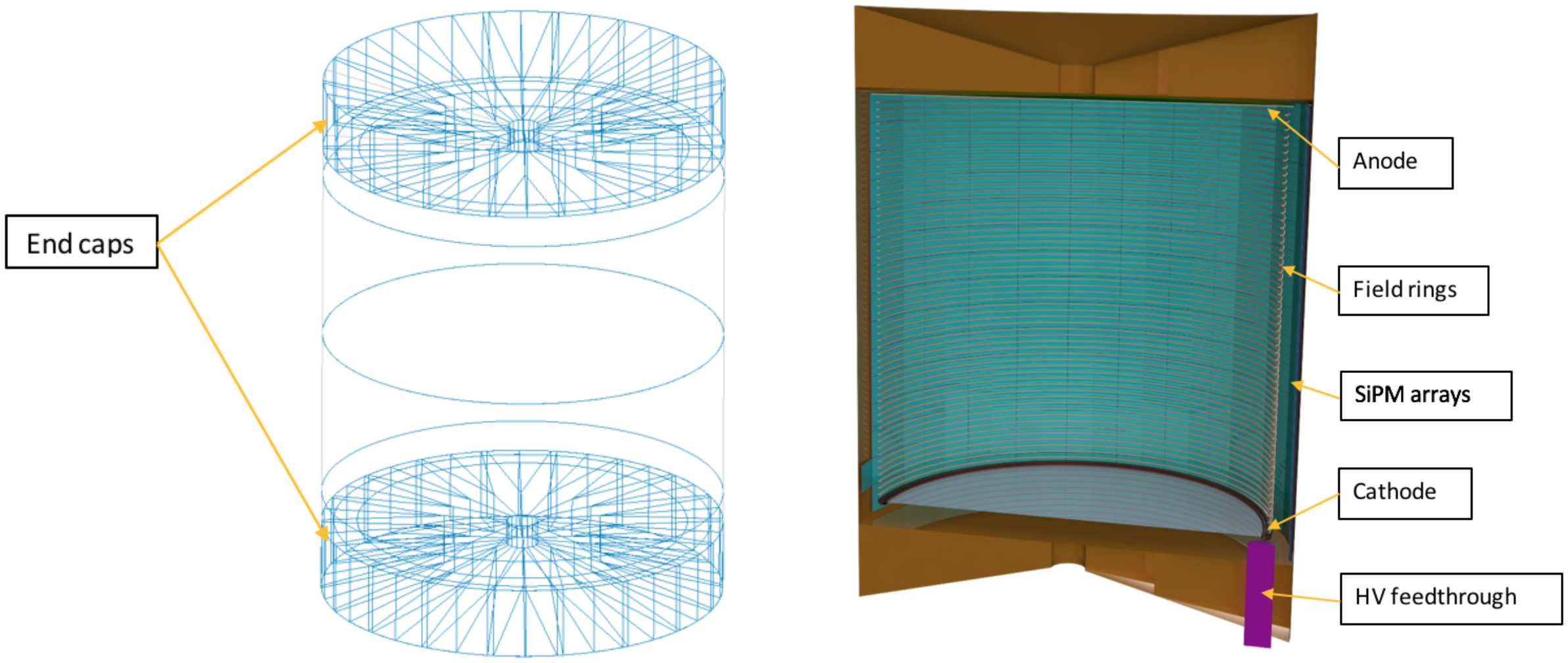

**Figure 4.26:** Left: wire-frame model of the TPC vessel, emphasizing the stiffening structure of the end-plates. Right: cutout showing the nEXO TPC inside the vessel, snugly shaped around it.

#### 4.1.6.2 Vessel Fabrication and Packaging Concept

The vessel fabrication will likely be done at the underground location to minimize cosmogenic activation of the copper and other materials. The majority of the welds will be made with a modular electron beam welding system in which the vacuum chamber can be custom built, so that it can be assembled underground, coupled to a standard electron gun and controls (as an example see [29]). The approximate assembly sequence is as follows:

- Raw copper material is acquired and stored underground, after passing the radiological assays. Concurrently, the appropriate e-beam welder is procured and installed in an underground cleanroom (this may be the water tank of the experiment, described in Section 4.5, specially outfitted for the purpose).
- The main LXe vessel barrel and the end-plates are machined, welded, surface-etched and cleaned.
- The bottom end-plate is welded to the main barrel (bottom assembly)
- The TPC package (field cage, anode structure, SiPM detector, front-end electronics, interconnections and cables and feedthroughs) is mounted under the top end-plate (top assembly).
- The top assembly is lowered into the bottom assembly and the final weld is performed, with a special TIG process, like already successfully done in EXO-200, should the use of the electron-beam welder be unpractical at this stage.

At every step of the assembly, parts are kept, as much as possible, under inert atmosphere, both to reduce Rn-daughter deposition onto exposed surfaces and, for the few plastic parts, to limit the diffusion of oxygen and other electronegative gases, limiting the impact on the electron lifetime in the TPC.

#### 4.1.6.3 Cable and Fluid Feedthroughs

The signal flat cable feedthroughs are modeled after those designed for EXO-200, in which stacks of flat cables are routed, uninterrupted, through cryogenic epoxy plugs on copper flanges. The mismatch in temperature expansion coefficient between copper and epoxy is absorbed by adequately thin "lip" along which the epoxy adheres. Signal and LXe feedthroughs will be in the top end-plate of the TPC vessel.

## 4.2 Readout Electronics

This section describes the electronics required to amplify, condition and digitize the charge and scintillation signals produced in the nEXO TPC. Locating these functions near the origin of the signals, inside the TPC at cryogenic temperature, has a number of advantages. The capacitance from the connections of the charge collecting devices to the preamplifiers is minimized with respect to more conventional room temperature electronics that would have to be located over 2 m away, outside of the cryostat. In addition, the long cables transmitting analog signals would be more susceptible to electromagnetic pickup. The low temperature environment results in lower noise for the input transistors. Finally, digitizing all signals inside the TPC reduces the number and/or size of cables and feedthroughs, simplifying the assembly and increasing the reliability. While many experiments have found it difficult to produce ultra-low background cables, in several cases silicon devices have been measured to have extremely low radioactive contaminations. Thus, the replacement of long cable runs with silicon chips in proximity of the fiducial volume is likely to reduce the overall radioactivity load of the detector.

To demonstrate one of these points, Figure 4.27 shows the result of a simulation deriving the overall energy resolution at the $Q$-value ($\sim 2.5$ MeV) as a function of the photodetection efficiency with a drift field of 400 V/cm, for cable capacitances expected in the case of external electronics and for electronics located in the LXe at the source of the signals. Under those assumptions, and using nominal nEXO parameters, the internal cryogenic electronics results in a substantial resolution improvement.

Of course, the design of electronics to be located in the TPC presents a number of challenges that have to be met in order to achieve the advantages described. In particular, the electrical specifications need to be fulfilled while complying with the ultra-low outgassing and radioactivity requirements. This leads to limits in both the total amount of materials and the number of different components, as each component needs to be carefully tested. In the case of electronics, these issues generally preclude the use of discrete components and naturally lead to the use of integrated techniques, where several channels can be integrated onto a very thin piece of silicon measuring only a few millimeters on each side. Initial R&D within the nEXO collaboration has indicated the feasibility of the concept and provided an initial system architecture leveraging the virtues of a monolithic design.

Another important factor in the design of the nEXO electronics is the limited power budget, needed to ensure that the power dissipation within the cryostat does not cause unwanted temperature gradients which, in turn, may lead to excessive convection and, in the extreme, formation of gas bubbles within the LXe. Thermal analysis indicates that a reasonable goal is to limit the power dissipation of each of the charge and light readout systems to 100 W, with more limitations on the power density at the chip level that still needs to be analyzed in detail. The use of integrated technologies also aids in containing the power budget, although even in this case, extreme care needs to be applied to reach the goal. The implications of a low power budget have repercussions on the ability to meet a number of requirements, since high resolution, data rate, and analog bandwidth all directly impact the power consumption and heat dissipation.

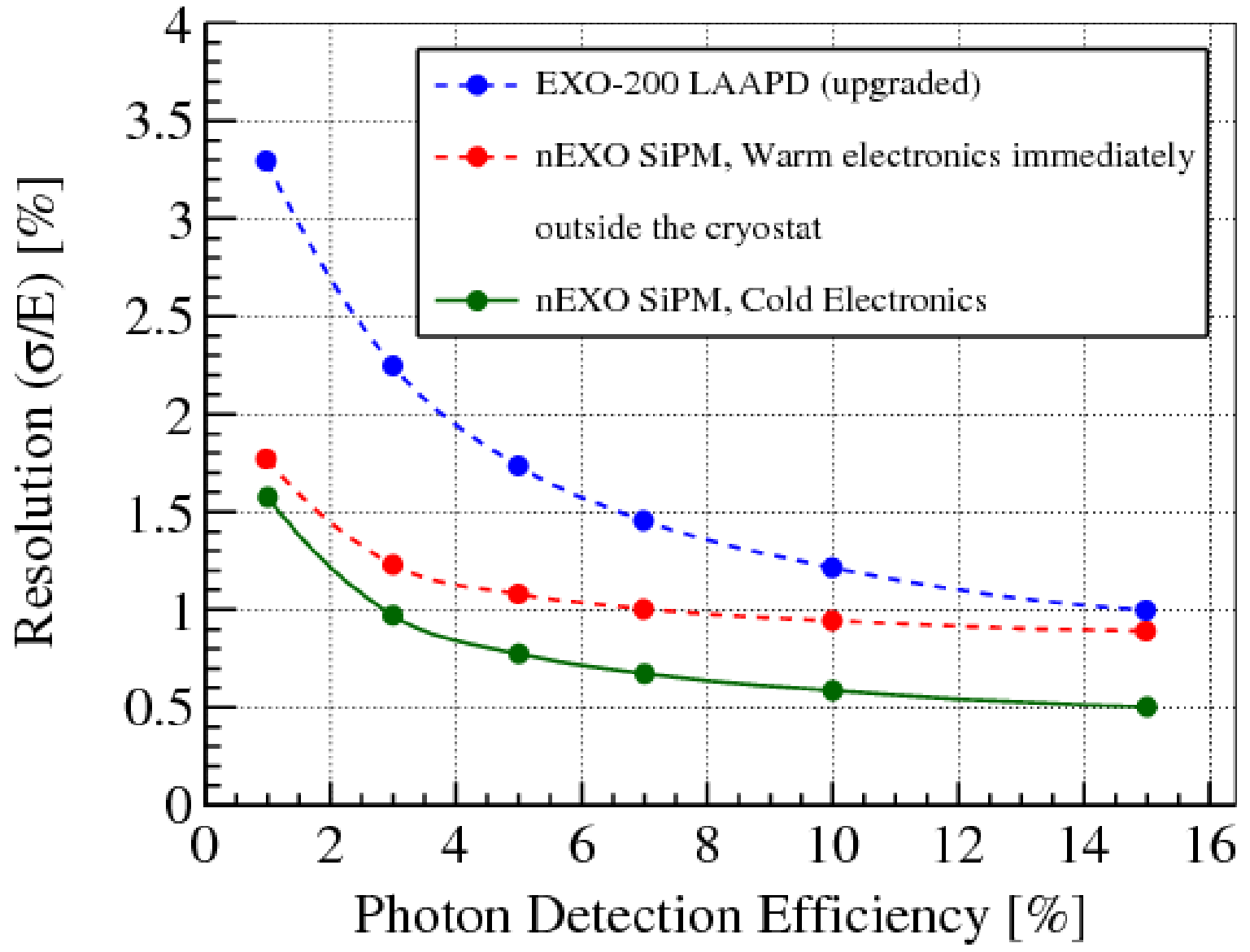

**Figure 4.27:** Energy resolution as a function of the overall scintillation light detection efficiency at a 400 V/cm drift field and for electronics noise in the charge channel appropriate for internal (cryogenic) and external (room temperature and immediately outside of the cryostat) front-end electronics. The curve for the upgraded (Phase-II) EXO-200 detector is also shown, for reference.

| Parameter | Charge channel | Scintillation channel |
|---|---|---|
| Operating temperature | 165 K | 165 K |
| Number of channels | $<$6000 | $<$6000 |
| Front end type | current amplifier with anti-alias filters | current amplifier |
| RMS noise floor | 200 $e^-$ | 0.1 single photoelectrons |
| On chip calibration | 0.2% | N/A |
| ADC resolution | 12 bits | 12 bits |
| ADC INL, DNL | $<$ 1 bit | N/A |
| Sampling rate | 2 MS/s | N/A |
| Processing type | continuous time waveform sampling | photon counting with spatial trigger |
| Power | $<$ 10 mW/ch | $<$ 10 mW/ch |

**Table 4.3:** List of key specifications of the nEXO front end readout. INL (DNL) refer to integral (differential) non-linearity.

The main requirements for the nEXO electronics are listed in Table 4.3, and in the following paragraphs, an overview of how such requirements can be met is presented for each system.

### 4.2.1 Charge Readout

#### 4.2.1.1 Charge Readout Specifications

The specifications of the nEXO charge readout are driven by the detector energy resolution ($< 1\%$) and background discrimination requirements: the charge noise has to be low enough to have a minimal impact on the overall energy resolution; the front end anti-aliasing filter and the digitization rate are chosen to maximally preserve the waveform shape information for background rejection; and the precision and dynamic range of the readout are determined based on simulated $\beta\beta$ decay and background events. The preliminary set of key specifications, shown in Table 4.3, have been developed using simplified models that have been validated with test data and other means. Further R&D and refined simulation will gradually lead to the maturation of the design.

#### 4.2.1.2 CMOS Cryogenic Performance and Characterization

While it is well known that at the moderately low temperatures of interest, the charge carrier mobility in silicon increases while thermal fluctuations decrease, resulting in higher gain and lower noise, the exact transistor parameters at 160 K had to be directly measured for nEXO and other cryogenic experiments. Measurements on test structures by the BNL and SLAC groups have confirmed the decrease of white noise for both PMOS and NMOS and of the 1/f noise for PMOS [30], while supplying the device-level parameters required for the design. The measurements were conducted with TSMC [31] 180 nm technology at 77 K and TSMC 130 nm technology at 160 K.

These measurements have been used to produce models on which new designs can be reliably based. At SLAC, the DC and noise behaviors of 8 different types of TSMC 130 nm transistors were characterized at 160 K. The transistors are a combination of different channel types (PMOS/NMOS), operating voltages (1.2 V/2.5 V) and threshold voltages ($V_{gs}$). Each transistor type has a subgroup of 16 devices with different width/length ratio. Data are fitted simultaneously to extract the DC and noise parameters. The end results are device models that work well for all device types, with typical deviations of less than $< 1$ %. As an example, Figure 4.28 shows the comparison between the model prediction and the measured $I_D$-$V_{DS}$ curves at several gate-source voltages for a NMOS 2.5 V device.

Impact ionization can adversely affect CMOS circuits at cryogenic temperatures and produce aging. This causes charge trapping in the MOSFET gate oxide at large drain current densities (the so-called "hot carrier effect" [32]). This effect can be avoided by following certain design rules, e.g. operating analog devices at moderate to low drain current densities. The cryogenic ASIC designed by the BNL group for the LAr detectors has gone through a series of accelerated aging tests that demonstrated that the lifetime of these circuits can significantly exceed 20 years at 77 K [33]. Similar aging tests will be performed for the nEXO readout ASICs in the future to guarantee a long operating lifetime at 165 K.

The possibility of building all functions on a single chip is, of course, extremely attractive. Several batches of chips from the TSMC process(es) have been tested for radioactivity and have been generally found encouraging in the quantities required. Some external high-capacitance bypass capacitors will still be required, but commercial components made entirely from silicon (and silicon dioxide) with capacitance up to 4.7 $\mu$F and voltage rating up to 11 V (e.g. Murata Inc.) have been identified and successfully screened for radioactivity.

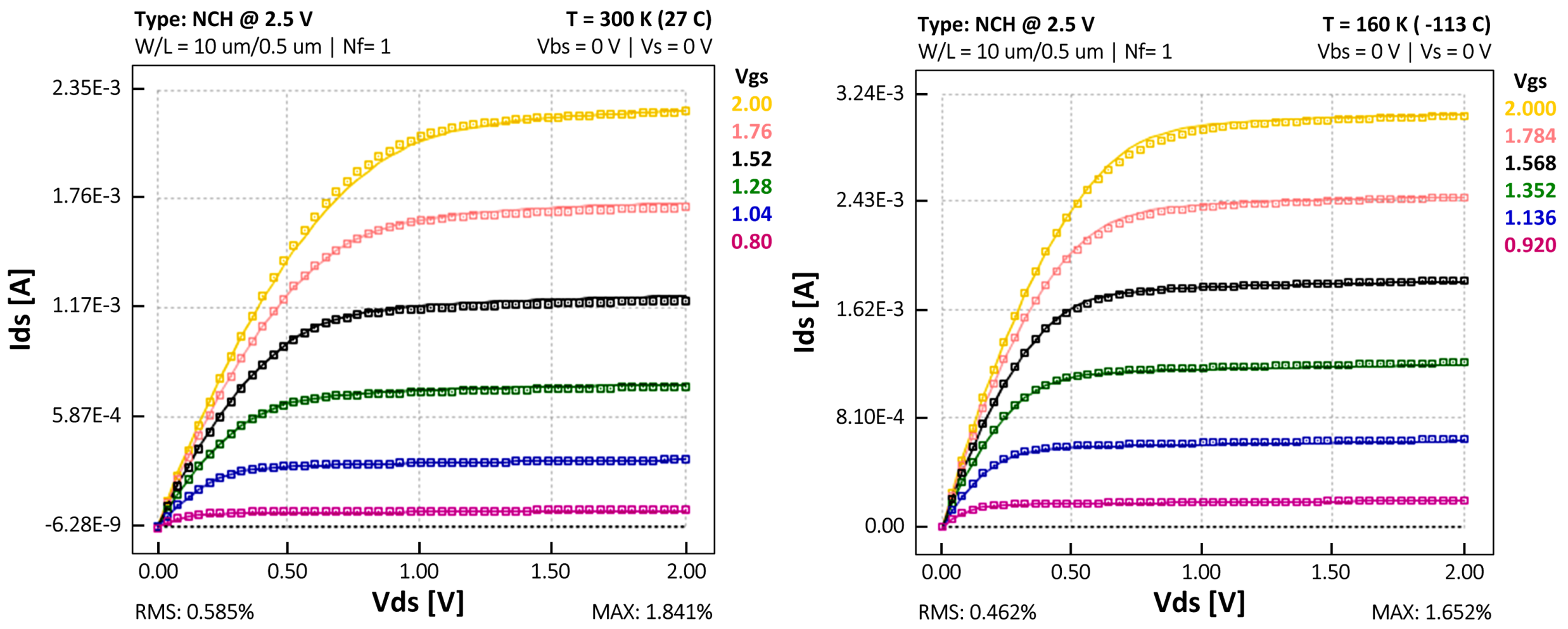

**Figure 4.28:** An example of the cryogenic measurement and modeling of CMOS transistors. The $I_{ds}$ vs $V_{ds}$ curves were measured for several $V_{gs}$ voltages at 300 K (left) and 160 K (right). The DC response predicted by the models (solid lines) reproduces the data very well.

#### 4.2.1.3 nEXO Charge Readout ASIC Development

As already mentioned, nEXO requirements are optimally met by designing a specific ASIC, maximally integrating all front-end and conversion functions on a single chip. SLAC has been designing such a prototype nEXO chip (CRYO), keeping in mind all parameters from Table 4.3.

The resulting architecture is shown in Figure 4.29 for 64 channels, as appropriate to read out a charge collection tile with the maximum density under consideration (3 mm pitch). This chip combines both analog and digital functionalities including signal preamplification, waveform digitization, and channel multiplexing with minimal numbers of I/Os. It has built-in low dropout (LDO) voltage regulators for analog and digital sections, requiring only one external supply at 2.5 V. TSMC's 130 nm CMOS process is chosen as a good compromise between the front-end performance and the speed of the back-end section. 2.5 V devices are used for the front-end, while 1.2 V devices are used for the digital back-end. The digital domain is isolated in Deep N-well (DNW) to suppress the coupling of digital noise into the analog front end [34].

The 64 channels on the ASIC are divided into two 32-channel banks with a single data output. Each bank is further divided into eight 4-channel sub-banks. The front end of each channel consists of a current preamplifier, an anti-aliasing filter and a sample-and-hold (S/H) stage. The preamplifier also has pole-zero cancellation [35]. Because of the long signal rise time, the lowest charge noise is achieved by sampling the current waveform over its slow evolution. The digitized data is then processed offline to obtain the charge measurement. The preamplifier's noise is designed to be $< 200$ e$^-$ with 20 pF input capacitance, appropriate for the charge collection strips. A Bessel filter is implemented as the anti-aliasing filter, because it appropriately limits the noise bandwidth while minimizing waveform shape distortions. A sample and hold (S/H) stage allows the multiplexing of several channels onto a single analog to digital converter (ADC). In this case, four front end channels share one 8 MS/s, 12-bit SAR (successive approximation register) ADC in fully differential mode. The 12-bit data from the ADCs are encoded onto 14-bit frames (12b-to-14b) by a custom encoder through a sequential 8-to-1 digital multiplexer. This ensures sufficient redundancy for the transmission over a substantial distance using imperfect cables. Finally, the digital data is serialized and transmitted off-chip and out of the TPC at 896 Mbps using the LVDS (Low-Voltage Differential Signaling) protocol. The CRYO chip is controlled by a dedicated slow control unit section, SACI (SLAC ASIC Control Interface) and global registers. The power consumption is estimated to be $\sim 4$ mW/ch.

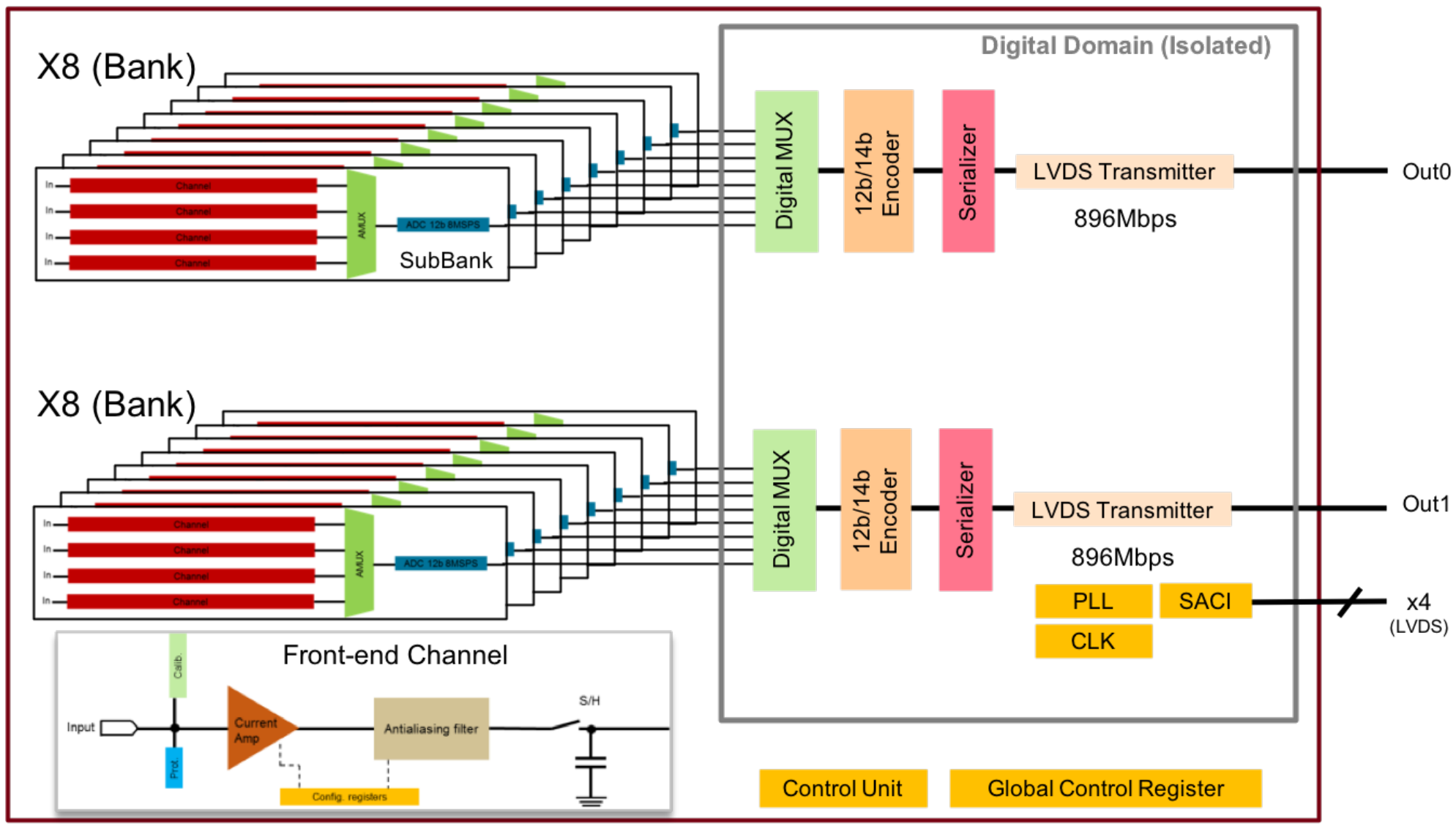

**Figure 4.29:** Architecture of the CRYO ASIC designed for the nEXO charge readout.

The individual blocks of the CRYO ASIC were first designed using existing foundry models for 220 K, then optimized using the models extracted from the measurements at 160 K. Simulations show that each subsystem on the chip can meet the design goals. The analog and digital blocks are currently being integrated together. Further optimization may be needed after the full system simulation. The prototype chip design is expected to be submitted to TSMC in the second quarter of 2018 and testing will begin when the chips have been received. We plan for two more submissions of the prototype chip in 2019 and 2020 before finalizing the ASIC design for production.

#### 4.2.1.4 Initial Demonstrator Using LAr ASICs

While the nEXO-specific CRYO ASIC is under design and prototyping, tests with a charge collection tile in LXe are being performed taking advantage of systems already designed for LAr experiments (such as MicroBooNE [36] and ProtoDUNE [20]). These test systems, of course, do not comply with the low outgassing and radioactivity requirements of the nEXO detector and are substantially bulkier that the CRYO ASIC implementation.

The LAr ASIC was adapted to a nEXO specific prototype readout board for the purpose of initial testing in LXe. The board material (Rogers RO4003) and components inside the LXe vessel are selected to have relatively low outgassing properties. For example, signal and power connec-tors are replaced by a custom Kapton rigid-flex cable directly soldered to the board. The signals are brought outside of the LXe vessel through a ceramic micro-D feedthrough, then carried to an FPGA at room temperature by a custom high speed digital cable. The prototype board, shown in Figure 4.30, is mounted directly on the back of a charge collection tile and can read out 64 channels simultaneously. The full readout chain has been tested successfully and reached an equivalent noise charge (ENC) of 240 $e^-$ at liquid nitrogen temperature. Data from a charge collection tile in LXe was recently collected and is being analyzed.

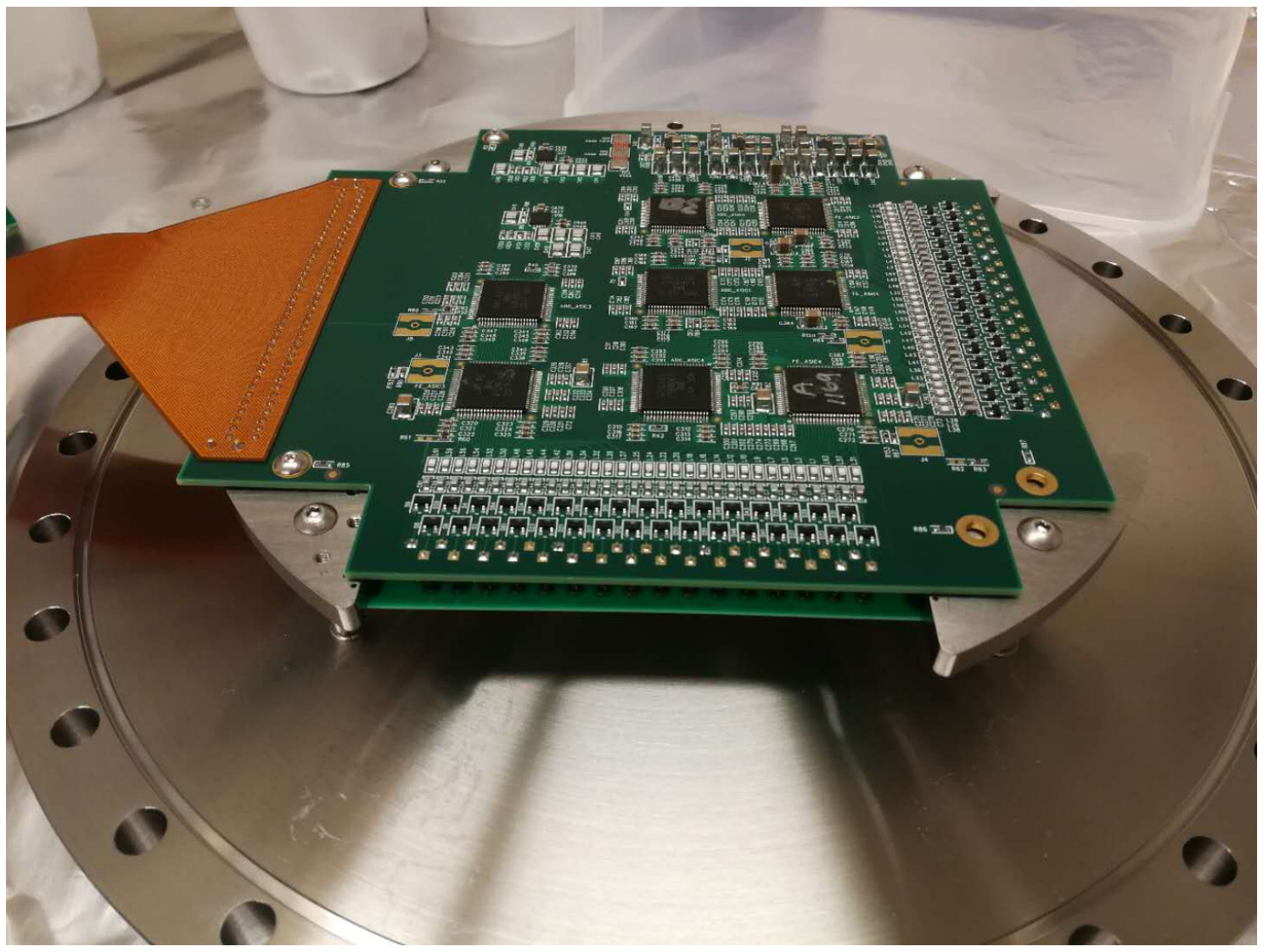

**Figure 4.30:** Overview of the prototype charge readout system installed in the LXe cell and connected to a charge collection tile (on the reverse side and not visible). The board and the digital connection with the outside world through a ceramic feedthrough are clearly not low-background and are meant only as a first demonstration of a cryogenic electronics system.

#### 4.2.1.5 An Alternative Charge ASIC Design

One risk of the primary design described in Section 4.2.1.3 is noise pick-up by the sensor strips due to their close proximity to the digital front end electronics. This is particularly insidious for an experiment where signals are random in time and the digital activity cannot be de-synchronized from the physics events. While the SLAC and BNL groups have experience with mixed signals systems and appropriate shielding techniques, it appears prudent, at this early stage, to study mitigating alternatives that do not involve fast digital signals inside the TPC. The IHEP group has pursued such an architecture, whereby analog waveforms are multiplexed inside an ASIC and transmitted outside of the TPC, with the architecture illustrated in Figure 4.31. The digital conversion and processing then occurs with more conventional circuitry at room temperature. This approach also reduces the complexity of the ASIC design and the power consumption. The drawback is that long distance transmission of precision and fast analog signals is non-trivial and high quality coaxial cables are needed to preserve fidelity. Finding coaxial cables that meet the radiopurity and low outgassing requirements of nEXO will be challenging.

The development of the IHEP analog ASIC began in 2014 and is proceeding in stages. Version 0.1 only included the analog preamplifiers with no multiplexing. As shown in Figure 4.31 the noise decreases with temperature, reaching a floor of 270 $e^-$ at 165 K. Version 1, developed in 2015, included both analog front end and multiplexing. Analog signals have been successfully multiplexed and transmitted through a 2 m long micro-coaxial cable at cryogenic conditions, though input charge-dependent baseline drifts were observed in the waveforms due to parasitic capacitances. The ASIC version 2, designed in 2017, aims both at reducing the front end noise below 200 $e^-$ and eliminating the baseline overshoots. The performance of the chip will be measured during tile integration tests in 2018.

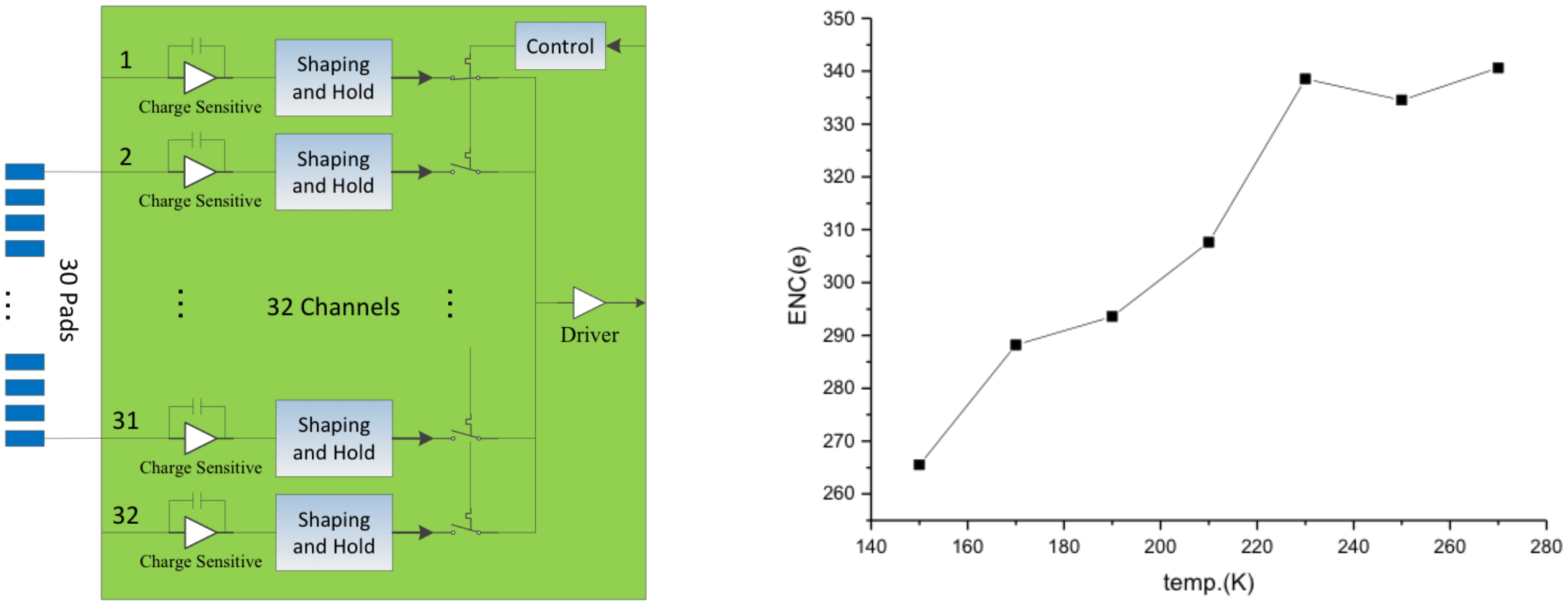

**Figure 4.31:** Architecture of the IHEP analog multiplexing ASIC (left), and the measured temperature dependence of the preamplifier noise (right).

The final choice of ASIC architecture and design will be informed by the results of the various test devices and prototypes described.

### 4.2.2 Scintillation Readout

The readout electronics for the scintillation light is logically similar to that, already described, used for the charge. However, unlike for the case of the charge, topological information about the sources of photons is virtually impossible to reconstruct, particularly at the cm-scale of the SiPMs granularity. Therefore the channel count is simply driven by noise considerations related to the large capacitance of the devices. The specific capacitance of the SiPMs under consideration ranges from 35 $\mathrm{pF/mm^2}$ (Hamamatsu devices) to 85 $\mathrm{pF/mm^2}$ (FBK devices). The most important question is to ascertain whether the total capacitance of 4.5 $\mathrm{m^2}$ of SiPMs can be readout with the required noise within the total power budget of 100 W, corresponding to 2.2 $\mathrm{mW/cm^2}$.

Furthermore, whereas the temporal structure of the signal collected in the charge channel contains useful information, for the case of the scintillation light readout, it is assumed that no useful information can be extracted in this way. [5]

Because of the challenge of extracting signals out of large capacitance devices and due to the fact that SiPMs compatible with nEXO requirements have been under development and are be-coming available only now, the work on electronics for the scintillation channel has concentrated on assessing the limits of performance and the feasibility, as described below. The design of an ASIC chip will emerge in the near future from the information gathered during the development of the charge readout ASIC and testing discrete-components readout schemes with now available SiPMs.

[5] While nanosecond timing may allow for the discrimination of Cherenkov light from scintillation and may access further information on the type of energy deposition, this regime may be studied for future upgrades but is not part of the nEXO design.

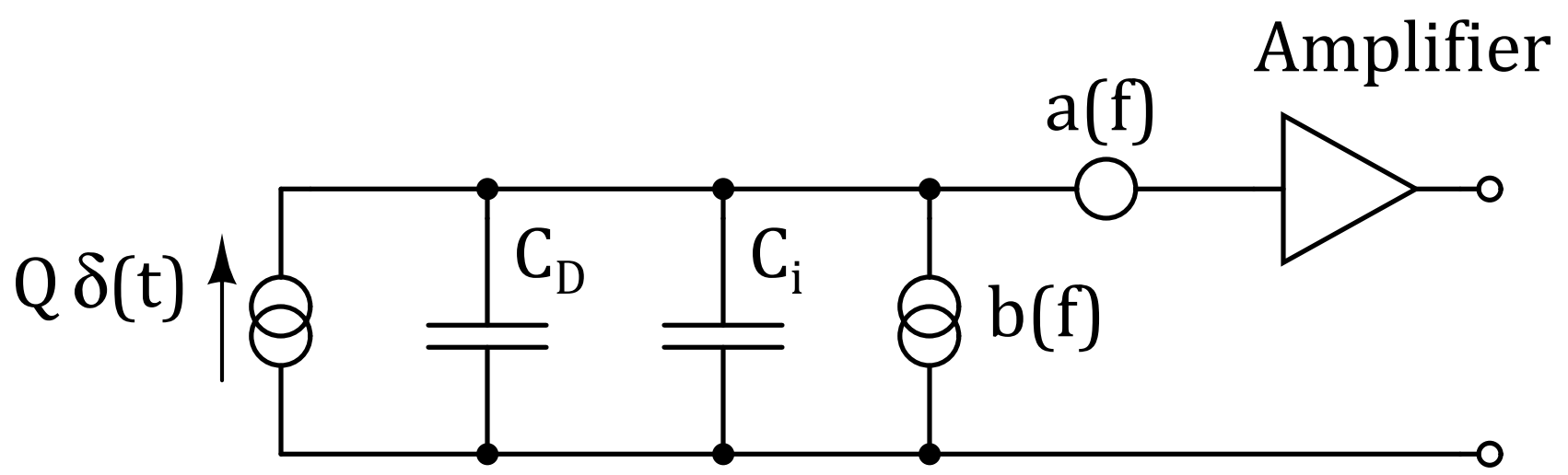

**Figure 4.32:** Simplified schematic of the readout electronics for a photodetector channel. The symbols are explained in the text.

In developing the scintillation design, it is useful to start with the observation that an event at the $Q$-value ($\sim 2.5$ MeV) on the TPC axis will produce in the LXe approximately $7 \times 10^4$ photons over a $4\pi$ solid angle. For the canonical 3% photodetection efficiency this corresponds to the signal from 2100 photons, distributed over the entire $\sim 4.5\text{m}^2$ surface of the SiPMs. This means that, on average, a 1 cm$^2$ SiPM device will detect less than one photon, hence, the readout system must be able to efficiently detect single photon events. This requirement also applies to realistic clusters of SiPMs, hardwired together, as discussed later, and defines the signal-to-noise ratio (SNR).

Any charge readout system can be modeled as indicated in Figure 4.32, where the current source is a noiseless source producing a Dirac's delta of charge $Q$. $C_D$ and $C_i$ are respectively the detector capacitance and the amplifier's input capacitance. The amplifier is modeled as ideal with a response to a delta function $A(t)$. The sources $a(f)$ and $b(f)$ represent the series noise and the parallel noise power densities of the readout [37]. If we neglect the $b(f)$ component (appropriately in our case, as at the operating temperatures all leakage current and dark noise are negligible), we can, after some considerations [38], write an expression for the resolution as a function of the SiPM parameters and the electronics noise:

$$\overline{E_{eq}^2} = (0.1 \cdot \text{SPE})^2 = \frac{\epsilon^2 \left(C_D + C_i\right)^2}{G^2 q^2} a R_D. \tag{4.2}$$

In the equation, $G$ is the SiPM gain at a set value of over-voltage, $a(f)$ is assumed constant, such as is the case for white noise, $q$ is the electron's charge, and $R_D$ is known as "delta noise residual function", $R_D = \frac{\int_0^\infty |A(f)|^2 df}{\max_t(A(t))}$ [39], and is a property of the amplifier. The value of $0.1$ SPE was chosen from the requirement that the system be able to detect a single photoelectron.

It is interesting to observe that once the preamplifier bandwidth, affecting $R_D$, is set and a photodetector is chosen, providing the value of $C_D$, the only parameter one can control to attain the desired SNR is the amplifier's noise $a$. This is of fundamental importance in designing the proper readout system. In the case of nEXO, the power budget is set by external requirements and, as just observed, the noise is determined by resolution requirements. Consequently, Equation 4.2 supplies the parameter space within which the requirements can be satisfied. By writing the expression for the amplifier's noise, it is possible to write an explicit form for $a$:

$$a = \frac{4kT\Gamma}{g_m},$$

where $k$ is Boltzmann's constant, $T$ the temperature and $\Gamma$ a parameter intrinsic to the amplifier. Only $g_m$ depends on design factors such as power and type of technology used for the amplifier. In the case of bipolar transistors, for example, $g_m = \frac{qI}{kT}$, with $I$ being the current flowing through the device. Similar results with different dependencies on the current can be shown for JFET as well as CMOS technologies. Because $C_D$ is directly proportional to the area of the photosensor, the explicit dependence of such area on the power is now trivial.

Since the constraints are all known, a specific readout architecture can be selected. A classic charge preamplifier will not work because of the high detector capacitance and the long response introduced by the charge-reset mechanism. A proven solution for this type of detectors has been designed following the principles illustrated in [40]. The circuit, shown in Figure 4.33 provides the low impedance needed for a high-capacitance detector with minimal power, thanks to the local feedback provided on the input transistor. Also, it has the benefit of being technology-independent, in that it can be implemented in bipolar or CMOS technology without much additional effort. This also means that results obtained from a preamplifier version made from discrete components will be useful in predicting the behavior of the same circuit when implemented in an ASIC.

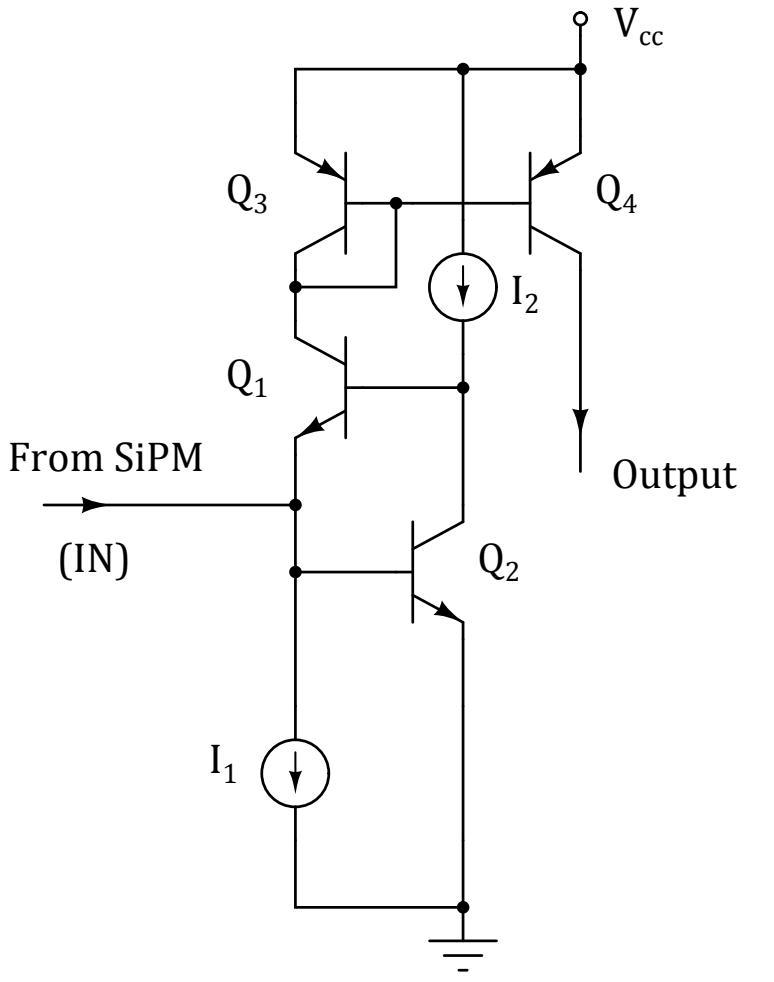

**Figure 4.33:** Optimum large area SiPM front-end in its BJT implementation.

Calculations using the noise formula discussed earlier, with the appropriate choice for the parameter $R_D$, and assuming a SiPM gain of $3{\cdot}10^6$, shows that this circuit is able to instrument a single channel up to a total detector capacitance of 54 nF, corresponding to either 6.3 cm$^2$ (FBK devices) or 18 cm$^2$ (Hamamatsu devices). The power dissipation of such a front-end is six times lower than the 100 W total, providing generous overhead for the rest of the readout.

This prediction has been recently confirmed using a discrete-component two-channel system, each reading out a total of 6 cm$^2$ of (FBK) SiPMs (six 1 cm$^2$ devices in parallel). The prototype dissipates 2.5 mW for 6 cm$^2$. The pulse height spectrum, shown in Figure 4.34, can be fit to extract a 0.27 photoelectron noise RMS broadband and 0.13 photoelectron noise RMS with an appropriate bandwidth limit, and SiPM gain of $1.8{\cdot}10^6$.

Before developing an ASIC version of the scintillation readout electronics more tests will be required to further refine the design, particularly in connection with the multiplicity of SiPMs in one electronics channel and their connection scheme. In particular, the capacitance seen by the preamplifier stage can be lowered by connecting the SiPMs in series. This scheme, however, requires careful matching of the leakage currents so that each device is biased in the same way. Parallel resistors, lowering the impedance of the network and hence fixing the potentials across each SiPM, need to be investigated.

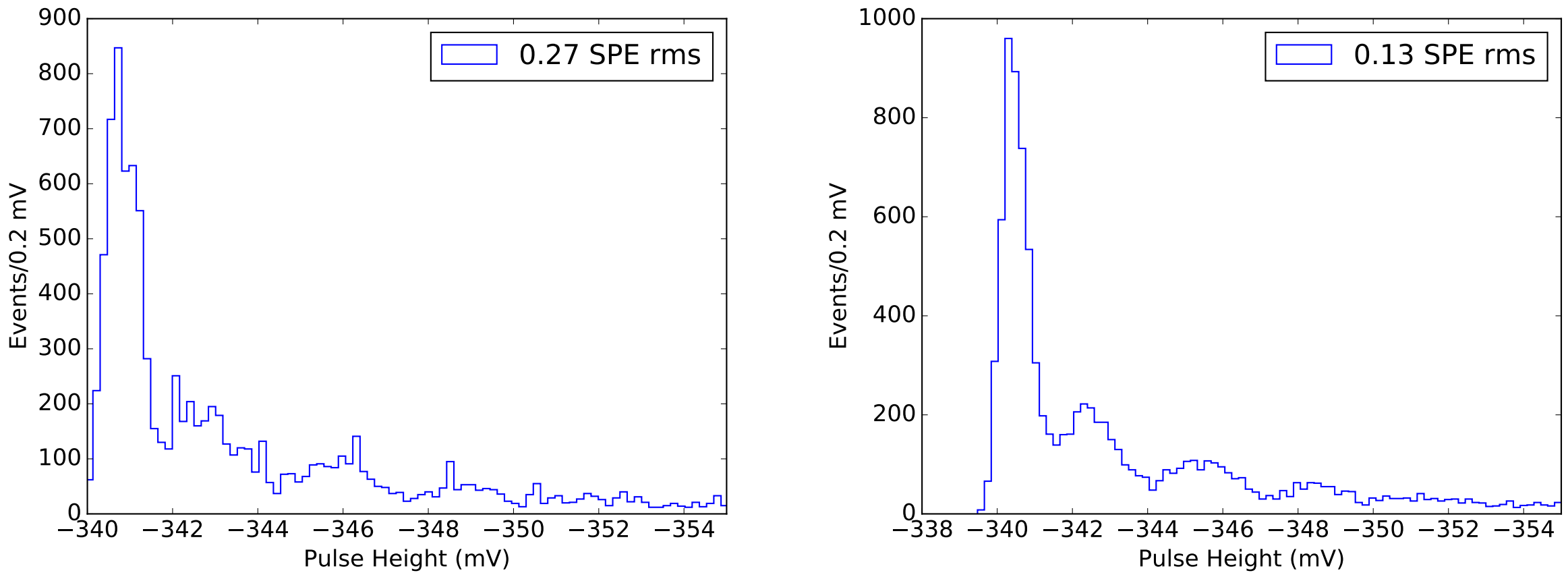

**Figure 4.34:** Pulse height distributions from six 1 cm$^2$ (FBK) SiPM read out in parallel using the discrete components system described in the text. On the left is the spectrum obtained in a broad-band situation, while the spectrum on the right has been obtained limiting the bandwidth of the signals. The resulting RMS noise is indicated in each panel and peaks from different number of photoelectrons are clearly visible.

#### 4.2.2.1 Light Readout Architectures

The light readout architecture is mostly dictated by a different set of constraints than those highlighted in Section 4.2.2. In that section, we considered the attainable area vs. SNR vs. power as one of the most important trade-off studies. For the design of an architecture, it is assumed the trade-off has been resolved, and system-level constraints must come into play. One important consideration in this case is the amount of dark counts from the SiPM that the readout will experience. At the target SiPM dark count rate of 50 Hz/mm$^2$, the light readout system would see a total dark rate of 225 MHz. A serial readout would have to be capable of a few Gb/sec sustained rate, with important implications on power consumption. In such a case, a level-0 trigger capable of filtering signals from dark counts based on spatial and temporal characteristics would be necessary. The rate could also be broken down by using a number of parallel links rather than a single serial link. It is possible that both solutions may be used at the same time. On the other hand, should the rate be even just a factor five lower, some mitigation components, such as the level-0 trigger, could be avoided with consequent savings in power and complexity, and increased reliability. Fortunately, this seems to be the case, according to our recent SiPM measurements.

By taking into account these considerations, the architectural block diagram in Figure 4.35 can be drawn. The front-end stages are based on the work of Section 4.2.2. A current preamplifier buffers the charge signal into a gain stage followed by a simple integration to optimize the SNR for the information. The amplitude of the integrated signal is sampled and sent into an analog-to-digital converter for digitization. The digital data is multiplexed and sent on appropriate lines to the external data acquisition system. Each event has a time stamp to allow for sorting and triggering by the DAQ.

### 4.2.3 Digital Data Transmission

Digitized signals from the charge module ASICs in Section 4.2.1 and SiPM module ASICs in Section 4.2.2 must be transmitted from the cold electronics to the room temperature data acquisition hardware. For digital data transmission within the LXe, this requires high-bandwidth (Gbit/s) and radiopure data transmission lines to avoid significant contribution to the backgrounds. Flexible polyimide (Kapton) differential microstrip cables, described in the following sections, can meet these requirements as long as great care is used in material selection and processing. The mechanical layout and total mass of these cables are summarized in Section 4.1.5.

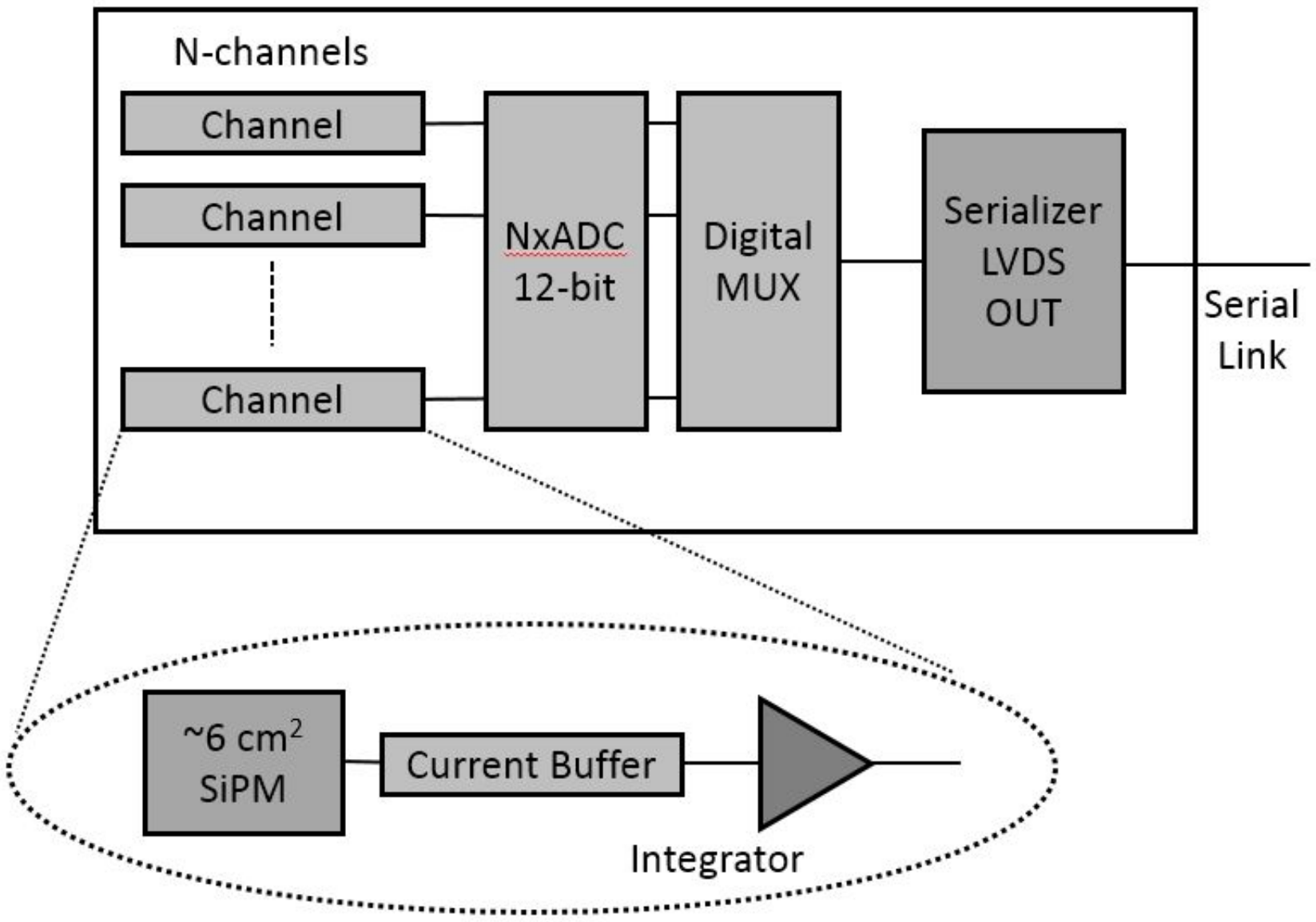

**Figure 4.35:** Silicon photomultiplier readout architecture.

#### 4.2.3.1 Data Rate

For 10 cm $\times$ 10 cm charge tile modules, 120 tiles are required to instrument the anode. In the primary design, each tile contains 64 readout channels with 3 mm pitch. For continuous streaming of data at 2 MS/s, the total data rate would be (12 bits)$\times$(2 MS/s)$\times$(64 channels/tile)$\times$(120 tiles) $\sim$ 200 Gb/s. This data rate can be reduced by triggering, data compression, or zero suppression, as well as by stitching multiple 10 cm $\times$ 10 cm tiles into larger readout modules. However, at this early stage and considering high event rate calibrations, it appears prudent to plan for total rate out of the cryostat close to 200 Gb/s.

The current prototype ASICs require 2 differential lines for each tile to provide this data transmission rate, using low voltage differential signaling (LVDS) at 1 Gbps. In addition, 3 digital control lines as well as power lines are required. The differential lines can be implemented as edge coupled microstrip lines on a Kapton substrate, as described in the following section. The microstrip pairs can be fabricated with a line pitch of $\sim$1 mm allowing approximately 40 differential lines per 5 cm width cable.

The data rate for the SiPM tiles can be substantially reduced relative to the charge channels by using zero suppression due to the small fraction of channels recording a photon. However, even in the most conservative case that all SiPM channels are digitized in the same way as the charge channels (2 MS/s), and a channel density of 16 readout channels per 10 cm $\times$ 10 cm module is assumed, the total channel count is $\sim$6400. Hence the data rate and required number of transmission

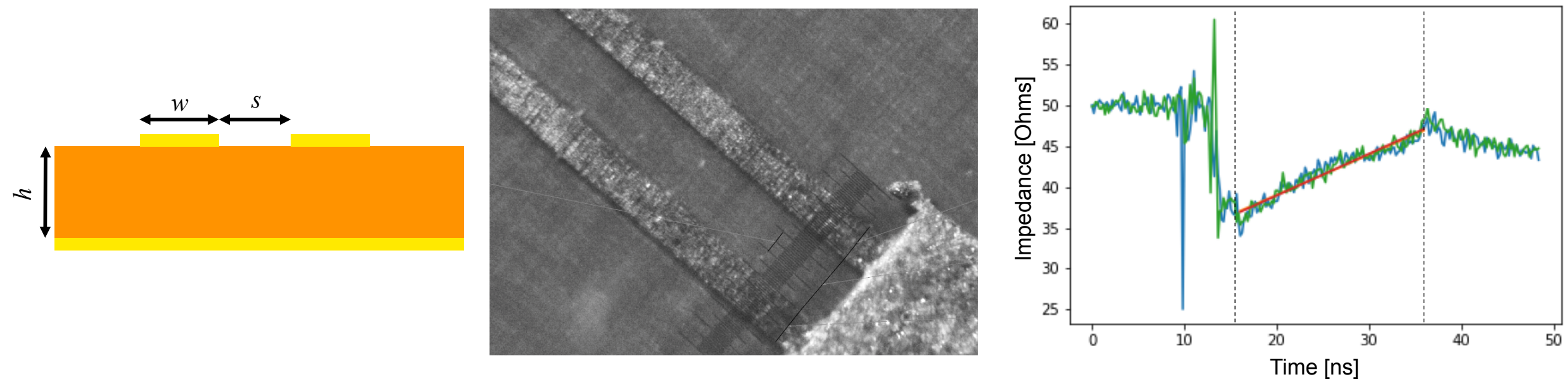

**Figure 4.36:** (left) Cross-sectional schematic of single edge-coupled differential microstrip. The prototype cables described in the text have $h = 50$ $\mu$m, $w = 165$ $\mu$m, and $s = 277$ $\mu$m. (center) Microscope image of short portion of a fabricated microstrip. (right) Impedance measured using the reflected voltage from a 2 m long microstrip with TDR. The center region between the dashed lines indicates the time period in which the TDR pulse traverses the cable. The observed slope (red fit) is consistent with simulations of the resistive loss in the cable.

lines is comparable to that of the charge tiles.

#### 4.2.3.2 High-bandwidth Cables

Commercial Gb/s data transmission cables do not have sufficient radiopurity to meet nEXO requirements. Instead, custom low radioactivity cables will be used to transfer digital signals from the in-LXe electronics to outside the cryostat. Initial tests have characterized prototype cables to validate their electrical properties and determine their performance for use as high-bandwidth differential transmission lines. The primary design for the high-speed cable consists of microstrip lines patterned on one surface of a double-sided copper clad polyimide laminate.

A cross-sectional schematic of a single microstrip transmission line is shown in Figure 4.36. Prototype transmission lines of this geometry have been fabricated and characterized. The initial prototype cables consist of microstrips with total length 1 m, 1.5 m, and 2 m, with copper thickness of 25 $\mu$m and substrate thickness of 50 $\mu$m. The microstrips are each 165 $\mu$m wide with a spacing of 277 $\mu$m. The dimensions of the fabricated microstrips were measured along their length with an optical microscope and found to agree with the design dimensions within 5% over their full length. A microscope image of a short portion of the transmission line is shown in the center of Figure 4.36.

The electrical properties of the cables were characterized using high-bandwidth time domain reflectrometry (TDR). Figure 4.36 (right) shows the reflected voltage versus time from the 2 m long cable for a simultaneous differential step function pulse (with rise time $< 100$ ps) with opposite polarity on each microstrip. No large voltage reflections resulting from impedance mismatches or cable defects are seen for the time period where the pulse traverses the cable, indicated by the vertical dashed lines in the center of the plot. Comparison of the reflected voltage to known terminations (short circuit, 50 $\Omega$, and open circuit) allows the conversion of the reflected voltage into the impedance for the differential microstrip mode, which is found to be roughly 40 $\Omega$, in good agreement with the theoretical impedance for this geometry of 39.2 $\Omega$. A rising slope is seen in the reflected voltage indicating loss as the signal propagates along the cable. A finite element simulation of the cable geometry using COMSOL that included the measured trace resistance and nominal value of the dielectric loss tangent of the Kapton substrate was found to reproduce the observed slope within 5%. The COMSOL simulation indicates that this loss is dominated by the trace resistance, and should improve at the lower temperature of operation in LXe. Future work will characterize the effect of cable loss on the bit error rate for digital data transmission, although the observed loss appears to be sufficiently low to permit the $< 3$ m cable lengths required for nEXO.

| | |
|---|---|
| Data rate (physics) | $< 1$ Hz |
| Data rate (calibration) | $\sim 1.6$ kHz |
| Trigger types | Charge or Light |
| Single event size | 10 MB |

**Table 4.4:** List of parameters of the nEXO data acquisition system.

#### 4.2.3.3 Room-temperature Transmission

Upon exiting the vacuum cryostat, digital signals from the custom, low-radioactivity transmission lines will be converted to optical signals and transmitted to the data acquisition electronics. This electric-to-optical signal conversion and transmission can be accomplished using commercial, room temperature electronics due to the less stringent radioactivity and cryogenic requirements outside the cryostat.

### 4.2.4 Data Acquisition System

The requirements of the nEXO data acquisition system are driven by the physics goals. Table 4.4 lists some of the parameters for the DAQ system. The nEXO DAQ needs to be able to handle the calibration event rate, which is more than three orders of magnitude higher ($\sim 1.6$ kHz) than that of low background data taking. It also needs to provide flexible trigger conditions in order to maximize the physics output. A schematic of the DAQ system is shown in Figure 4.37. As described in Section 4.2.3, digitized charge and light signals are extracted from the TPC and to room temperature on copper conductors. A transition to optical fibers is then provided, for long distance signal transmission. The location of this transition is to be determined, as on the one hand a location immediately outside of the warm vessel of the cryostat would minimize the length of the custom-built copper conductors, while a location on the deck capping the water tank would simplify the access for installation and repairs. The event building process takes place in a custom "Event Builder" module, potentially installed quite far from the electro-optical conversion. Triggered data are compressed, then stored on disks for offline analysis. Here, we assume a lossless data compression ratio of 5:1. For comparison, EXO-200 achieved a compression ratio of 8:1. A separate "Timing and FE control Board" module will provide timing and control signals to the front end readout.

nEXO's readout system is similar, in principle, to those of some existing LAr TPCs. The ProtoDUNE experiment, for example, uses cryogenic ASIC readout and has similar channel numbers and data rate. The ProtoDUNE DAQ uses RCEs (Reconfigurable Cluster Element) housed in industry-standard ATCA shelves on COB (cluster-on-board) motherboards designed at SLAC for high energy experiments [41]. This DAQ system can handle data rates of 480 Gb/s and triggers at 25 Hz [20]. This would be sufficient for nEXO's low background data taking, but not for source calibrations. As discussed in Section 4.2.5, data compression is necessary during source calibra-tion. A simple scheme of zero-suppression of channels with no signals would bring the stored data to a reasonable rate. At this early stage, however, there is no need for nEXO to lock into a specific DAQ hardware choice as DAQ technologies evolve rapidly. The ProtoDUNE example illustrates that hardware solutions already exist that can at least partially meet the nEXO requirements. The specific choice of the nEXO DAQ hardware and architecture will be made when the front end designs become more mature.

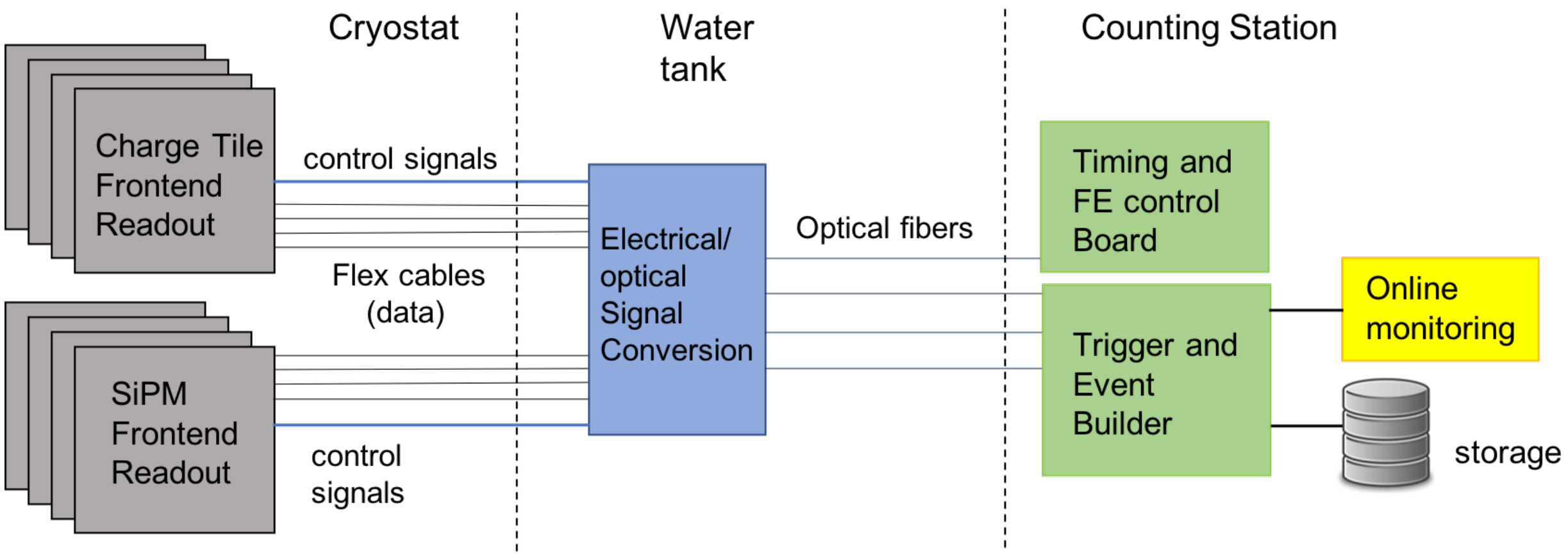

**Figure 4.37:** Block diagram of the nEXO data acquisition system.

### 4.2.5 Data Reduction and Storage

As mentioned in Section 4.2.3.1, a continuous stream of all charge channels for nEXO would result in a data rate of ~200 Gb/s. However, based on simulations, nEXO expects about $1.5 \times 10^6$ events/yr above 700 keV, including daily calibrations. Even assuming that all data from all channels are recorded for each event, nEXO will only produce an average of 75 TB/yr. Data compression will allow to further reduce the data volume.

The calculations above assume a data reduction technique is employed on calibration data. The primary plan for the external source calibration (Section 4.4.3) involves exposing the detector to about 1.6 kHz of calibration gammas. This rate would overwhelm nEXO's ability to store the resulting data. To manage this, a hardware or software trigger system will be implemented to reduce the stored data to a reasonable rate.

A storage array located onsite near the experimental area will receive data from the DAQ and provide temporary storage until they are transferred to the above ground facility and then offsite. The onsite storage will also provide buffer storage, to prevent the possibility of data loss during long distance transmission. The primary data storage facility for the experiment has not been chosen, but the data load indicated above under the most conservative assumptions is not particularly challenging by modern standards. Additional copies will be distributed to collaborating institutions, both for security of the data and for ease of analysis.

## 4.3 Cryogenics

### 4.3.1 Overview

The cryogenic system includes the cryostat, the refrigeration infrastructure and the xenon and HFE-7000 handling systems. The tasks of the cryogenic system are to support stable conditions for optimal detector operation and to protect both the delicate apparatus and $^{enr}$Xe stockpile in liquid phase. These goals must be accomplished while also meeting low background requirements, including the requirement on dissolved Rn, and guaranteeing a sufficient electron lifetime in the TPC. It is useful to consider that the components physically located inside the cryostat (including the cryostat) have to be built out of low background materials, while for elements in the cryogenic fluid systems which are outside the cryostat the concern is only that of outgassing and Rn emanation.

The LXe vessel is cooled and shielded by a vacuum-insulated cryostat containing 32,000 kg of HFE-7000 fluid (sometimes referred to as Novec 7000 [42]). Both the inner and outer cryostat vessels will be spherical in order to optimize the strength/mass ratio. In the primary concept, the vessels will be made of carbon composite material and lined with a thin, low background liquid-resistant material such as titanium.

HFE-7000 is a dense, radiopure fluid that is liquid both at room temperature and at the LXe operating temperature of 165 K. This fluid provides ultra-low background shielding, thermal uniformity to the TPC and the ability of transferring pressure loads from the TPC vessel to the cryostat. This application of HFE-7000 was pioneered by EXO-200 [1] and is essential for a detector requiring the lowest background at MeV energies, where $\gamma$ radiation is hardest to shield, since it allows for the construction of the lightest possible LXe container. The large HFE-7000 mass, required for shielding, also provides a substantial thermal inertia, making the system intrinsically fault tolerant. The cryogenic system will be designed to optimally take advantage of this feature.

The “system pressure”, that is the pressure of the LXe and, by design, the HFE-7000, during operations, is a free parameter that can be chosen at the design phase. Figure 4.38 shows the liquid-gas saturation curve for xenon, along with the same curve for the solid-liquid phase transition. As the pressure increases, the range of temperature over which xenon is in liquid phase also increases, going from 3.6 K at 1 bar to 11 K at 1.5 bar. Hence, running at a higher system pressure reduces the risk of boiling or freezing of the LXe. In addition, also shown in Figure 4.38, the viscosity of the HFE-7000 fluid decreases quite substantially with increasing temperature, around the region of temperatures of interest. Since lower viscosity corresponds to better HFE-7000 convection and mixing, running at higher pressure also simplifies the problems of temperature uniformity and heat removal. Of course, the system pressure has to be held by the inner cryostat vessel and this imposes a structural constraint, ultimately tied to background considerations (thicker cryostat vessel). While this fine optimization has not been done yet, here we conservatively assume that the system pressure will be 1.5 bar, probably at the upper limit of what is reasonable, and an operating temperature of 165 K, towards the lower limit, keeping in mind that these two parameters will not be applied to the final system simultaneously.

The refrigeration system must provide sufficient cooling power to cool down the large thermal mass from room temperature to 165 K in less than one month. It must also be capable of maintaining the system at low temperature with temporal variations of $< 0.1$ K over the course of many years of data taking. The projected power to maintain the detector at base temperature is only 500 W, a power that can be comfortably transferred to the cold mass by a system of ther-mosyphons. On the other hand, in order to complete initial cooldown in the required timescale, much higher cooling power must be provided. Circulating the HFE-7000 fluid to external heat exchangers can provide the required cooling power while keeping the heat exchangers outside of the low background region. The same system can also be used to apply heat to the HFE-7000 in the (rare) warm-up cases. During operations, the HFE-7000 recirculation system will be turned off, excluding the possibility of Rn intake from external equipment.

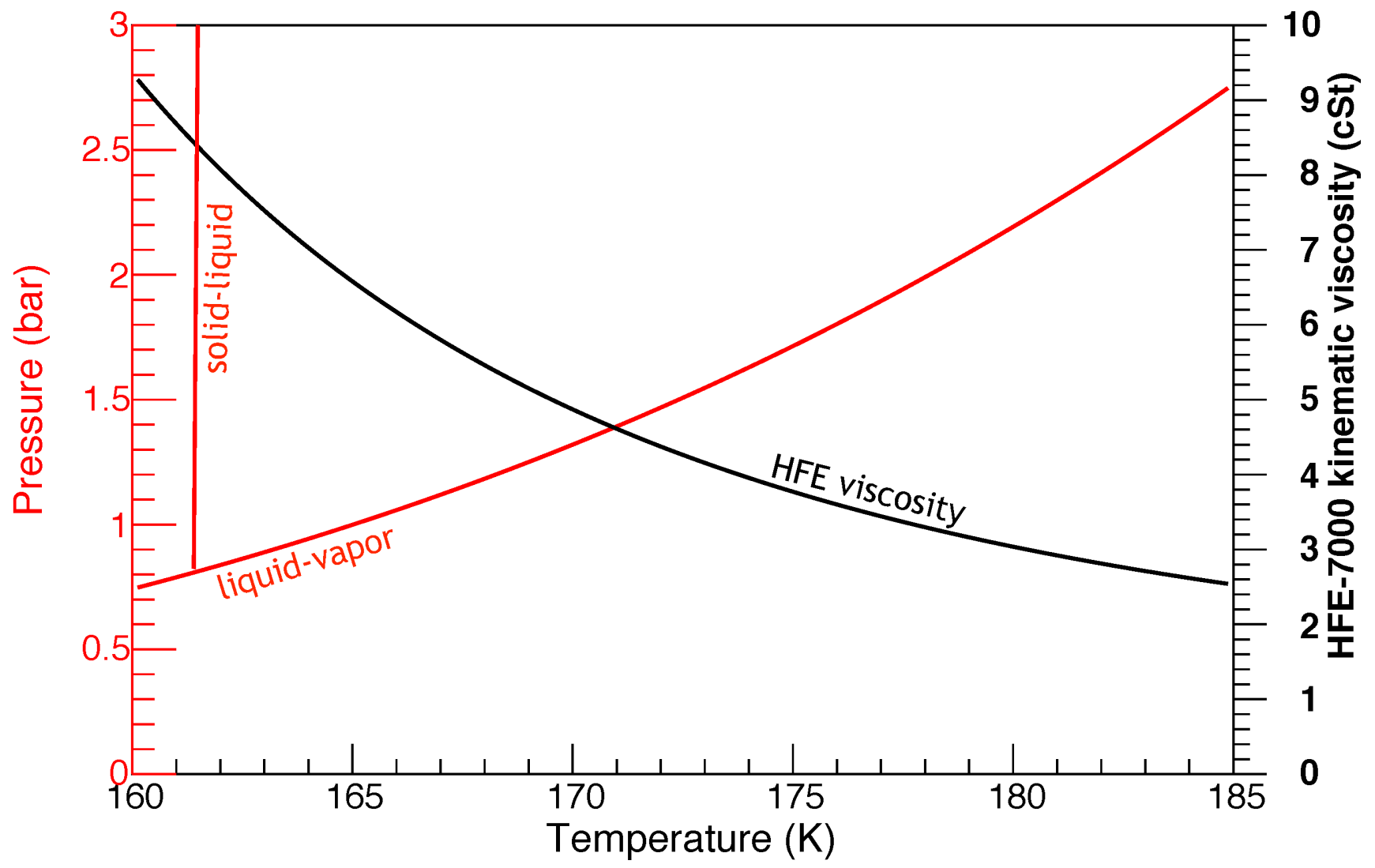

**Figure 4.38:** Xenon phase diagram showing saturation curves for both the solid-liquid and liquid-gas phase transitions (pressure shown on the left ordinate axis). Also shown (right ordinate axis) is the viscosity of HFE-7000 in centistokes, as a function of the temperature. From NIST thermophysical properties of fluid systems [43]

In order to be useful as a detection medium, the LXe must be free from electronegative impurities such as oxygen, nitrogen, and water vapor. Xenon purification is accomplished in the gas phase through the use of heated zirconium getters, a technique already employed by EXO-200 [1], XENON [44, 45], and LUX/LZ [46, 47]. Purification is necessary during the initial detector filling procedure as well as continuously throughout detector operation. As discussed in Section 4.4, to achieve an energy resolution $\sigma/E \simeq 1\%$ at 2.5 MeV, an electron lifetime of 10 ms or better is required. In order to engineer the flow of purified liquid xenon for optimal purity, a sophisticated campaign of computational fluid-dynamic simulations is underway using both 3D modeling in SolidWorks [48] and 2D modeling in COMSOL [18].

Xenon is continuously circulated through a purification loop in gas phase at a rate of 350 slpm (standard liter per minute) in order to achieve the electron lifetime goal, as discussed in Section 4.3.4. Unique components of the recirculation system that are currently in the R&D phase include a sealed xenon pump with a magnetically coupled piston, currently under joint development for the XENON and nEXO Collaborations [49], a counter-flow heat exchanger, and various xenon purity monitoring devices.

The location of the various components of the xenon and HFE-7000 systems are currently under study. The “natural” location on the veto water tank deck simplifies access and construction, but it involves substantial liquid heads (about 1.5 bar additional pressure for LXe). This can be mitigated by the installation on the side of the water tank, at an arbitrary height, presumably close to the height of the cryostat and TPC. The process of optimization has to take into account three different fluids with different densities: liquid HFE-7000 ($\simeq 1.75\ \mathrm{g/cm^3}$ at 165 K), LXe ($\simeq 3\ \mathrm{g/cm^3}$) and gaseous xenon ($\simeq 0\ \mathrm{g/cm^3}$ for this purpose).

The xenon recovery system is a critical component, meant to safely recover the xenon in gas form when the detector is emptied, either for planned operational reasons or in case of emergency. While the large thermal mass at base temperature makes the system resilient to short-term power outages or other interruptions of cooling, a recovery system is being developed to safely return the xenon to high pressure storage.

At this time, it appears convenient to provide all cooling needs with liquid nitrogen ($LN_2$). The $LN_2$ can be used to cool the heads of the thermosyphons, the recirculating HFE-7000 heat exchangers, and the high-pressure-capable cold traps to be used for xenon recovery. The $LN_2$ can be produced by a (possibly redundant) plant located underground, with storage dimensioned to provide sufficient endurance in steady state and/or the capability of recovering the xenon without the need for substantial amounts of electric energy.

The slow controls system must be reliable and robust. Several commercial solutions exist for programmable logic controllers (PLCs) with associated software platforms. Redundant power will be required for the slow control system (along with some other auxiliary devices).

### 4.3.2 Cryostat

The nEXO cryostat is envisaged as a two-vessel, vacuum insulated system, as shown in Figure 4.39. The current plan calls for both vessels being fabricated from carbon composite material. Due to its high strength this material allows the design of low-mass vessels, promoting low radioactivity. Preliminary tests, performed with Mitsubishi Rayon Grafil 34-12k (Pyrofil) carbon filaments and cured Hexion 862/81k resin, show that a carbon composite cryostat would contribute about 11% of the SS event background at the Q-value in the inner 2 tonnes of the detector. In addition, carbon composite lends itself to underground fabrication, only requiring a winding machine that can be assembled in a clean room underground. This is important, because the cryostat vessels are too large to be lowered in the SNOLAB conveyance. As mentioned in Section 4.5.1 it is plausible to utilize the tank as a temporary clean room for the winding of the cryostat vessels. It should be mentioned that in a final tradeoff study other, more conventional, cryostat materials could be considered. A higher radioactivity content of such material could be remediated by a larger HFE-7000 thickness (and larger mass), resulting in equal or even reduced background. Background risk could, therefore, be traded with higher cost.

As shown in Figure 4.39, the outer vessel of the cryostat interfaces with the water tank, which is suspended from the deck structure. The inner vessel is completely inside the outer shell and is connected via a low thermal conductivity carbon structure. The two vessels are offset, to provide a wider gap at the top, where all the services are routed. The space between the two vessels is kept at a residual pressure of $\sim 10^{-5}$ mbar by active pumping and specially designed superinsulation will keep the radiation losses below 100 W. The EXO-200 experience shows that low radioactivity superinsulation is available and, in addition, the number of layers can be carefully tuned for the modest temperature difference required, resulting in assemblies that are substantially lighter than commonly used for $LN_2$ or LHe. The cryostat pumping system will need to be redundant and fed by uninterruptible power.

The inner cryostat vessel will be filled with HFE-7000 and designed to support the TPC via a structure probably made out of low background copper. The inside of the vessel will likely have a liquid resistant liner, perhaps made of thin titanium. This liner could also serve as a mandrel for the fabrication winding.

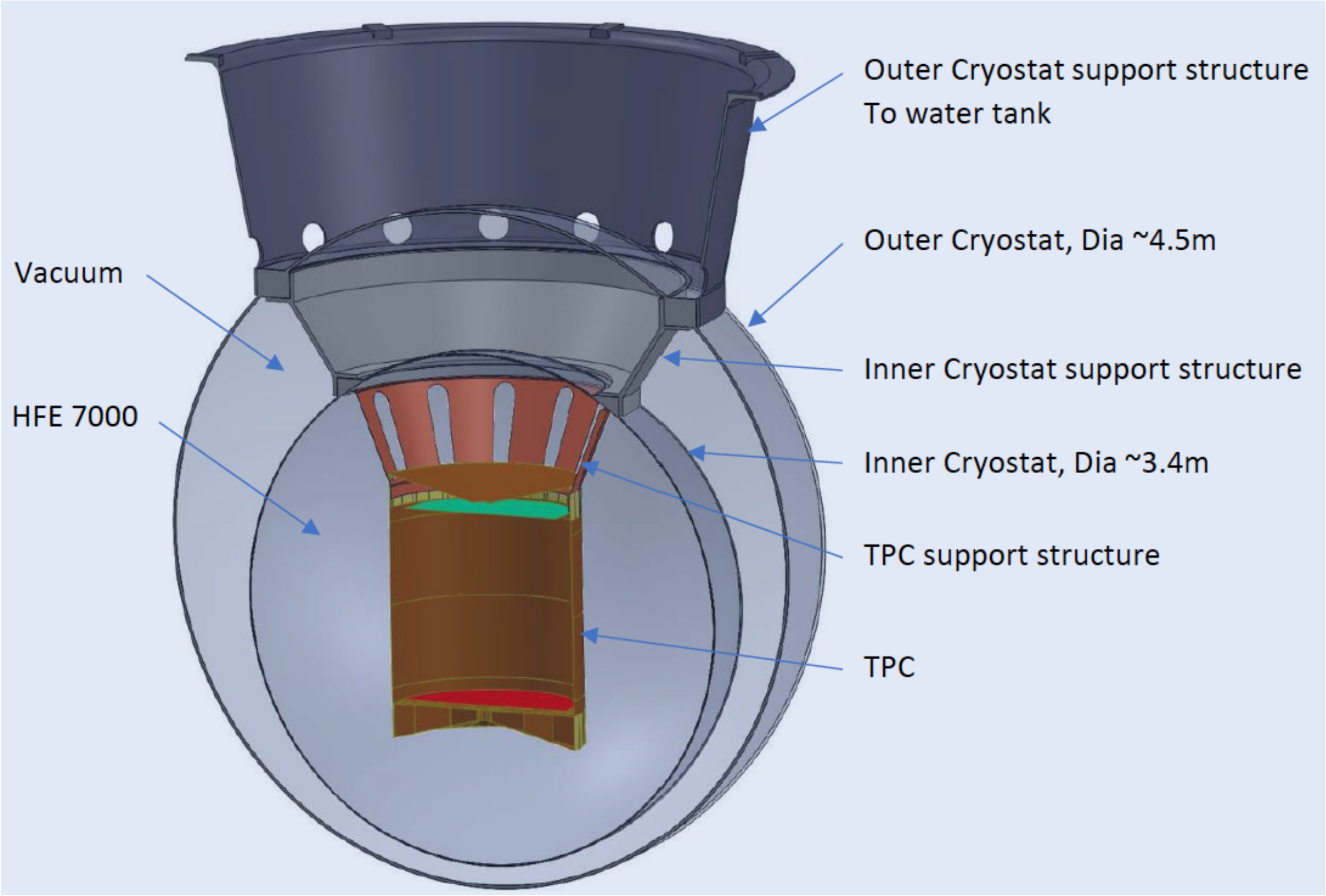

**Figure 4.39:** Cutout model of the nEXO cryostat system.

The 1.5 bar system pressure is also the internal load on the inner vessel of the cryostat. The outer vessel, on the other hand, is only externally loaded since it is evacuated, keeping in mind that the atmospheric pressure 2,000 m underground is 20% higher than at sea level. The design of the outer vessel must also take into account the hydrostatic load from the water in the veto detector. Preliminary modeling of the two vessels results in carbon composite thicknesses of 10 mm for the inner vessel and 20 mm for the outer one, although substantial engineering is still required to better understand safety margins and further refine the design. Penetrations, concentrated on the two top hatches have to be designed and their effects on the structure understood. The input of this preliminary design was required to set the minimum HFE-7000 thickness to 80 cm and, in turn, to estimate the background and sensitivity presented in Section 3.3.

The assembly is envisioned as a vertical sequence with the TPC with its services being mounted on the two cryostat hatches in the final position, with full access to route cables and piping. The two cryostat vessels will then be assembled in sequence, nested inside one another. The hatches will be bonded for sealing and additional structural members added, as needed.

A breakdown of the heat loads in the cryostat is presented in Table 4.5. Future engineering will provide a more accurate assessment along with a proper safety margin. At this stage, it is assumed that the cooling ends of the thermosyphons are in thermal contact with the HFE-7000 near the inner vessel of the cryostat, and remove the heat from the LXe through conduction on the copper TPC vessel and convection of the HFE-7000, as explained in Section 4.3.3.1.

| Item | Power (W) |
|---|---|
| Scintillation light readout electronics | 100 |
| Charge readout electronics | 100 |
| Radiative loss in the cryostat | 100 |
| Purification and recirculation system | 55 |
| Thermal conduction from the inner vessel attachment | 135 |
| Total | 490 |

**Table 4.5:** Breakdown of the heat losses in nEXO in steady state. The radiative loss is based on a 5-layer superinsulation blanket analogous to the one optimized, for the same temperature difference, in EXO-200. The heat loss in the recirculation system include inefficiencies of the counterflow heat exchanger system and heat produced in the purifier(s). The conductive loss is based on a carbon composite "skirt", like the one shown in Figure 4.39. The heat load from the purification and recirculation system may or may not have to be absorbed by the cryostat, but is conservatively listed here (and is subdominant).

| Component | Mass (kg) | Heat from 165 to 300 K (J) |
|---|---|---|
| LXe | 5000 | $4.8 \times 10^8$ |
| Copper TPC vessel | 566 | $2.8 \times 10^7$ |
| Inner cryostat carbon composite vessel | 644 | $5.7 \times 10^7$ |
| HFE-7000 | 32,000 | $6.7 \times 10^9$ |
| Total | – | $7.3 \times 10^9$ |

**Table 4.6:** Cold masses and energy to be extracted to cool from 300 K to 165 K for different components in the nEXO cryostat. The values for "Copper TPC vessel" include some allowance for other copper components auxiliary to the vessel itself.

Table 4.6 provides a breakdown of the cold masses, using the current understanding of the cryostat and TPC. The table also provides the heat to be removed when cooling to the operating temperature. From the table is clear that most of thermal inertia is due to the HFE-7000.

### 4.3.3 Refrigeration and HFE-7000 Systems

The nEXO refrigeration system can be separated into two largely independent functions: a) produce cooling power to be used by various systems and b) transfer the heat from the element that needs to be cooled to the source of cooling. Typically the source of cooling is a complex but commercially available electromechanical device located outside of the low background volume, possibly far from the detector. Devices that transfer the heat have to be custom made for nEXO and, at least in part, are connected to the low radioactivity environment inside the cryostat.

Substantial engineering is required for a final decision on the source of cooling power, but the general properties should be those of simplicity, reliability (including during power outages) and cost. At this early stage it is assumed that nEXO will install a $LN_2$ reservoir, possibly supplied by two liquefiers for redundancy, servicing all cooling needs of the experiment. Using the vaporization heat of $LN_2$ ($\simeq 1.5 \times 10^5$ J/l) a full cool-down requires 43 $m^3$ of $LN_2$, of which 3 $m^3$ is required for the xenon only. This is also the amount required to recover the xenon in a cryo-trap, should this technique be selected for the recovery. A modest size nitrogen liquefier [50] provides a cooling power of about 3 kW at $LN_2$ temperatures (with a $\sim$ 10% efficiency from electrical power) and can easily keep up with the need to maintain steady state, while requiring about a month of operation to build up the $LN_2$ required for a cool-down. Two units would provide redundancy and power to further reduce the cool-down preparation time. The $LN_2$ store contemplated here is compatible with SNOLAB operations, although a full safety analysis will be required. Following more detailed engineering and optimization, it is possible that, for the cool-down phase, mechanical refrigerators (instead of $LN_2$) will be directly coupled to the HFE-7000 heat exchanger, as shown in Figure 4.41.

As already mentioned, the heat transfer functionality is of two kinds. In the cool-down phase, which is expected to occur only a few times in the lifetime of the experiment, substantial power is required to remove thermal energy from the cold mass in order to reach the operating temperature in a reasonable amount of time, here set to 30 days, as a requirement. Using the total heat from Table 4.6 this corresponds to an average cooling power of about 3 kW. At steady state, on the other hand, only about 500 W need to be removed from the cryostat, according to Table 4.5.

#### 4.3.3.1 Steady-state Cooling with Thermosyphons

At this early stage, thermosyphons are envisaged to transfer the steady state heat out of the detector and to an external $LN_2$ heat exchanger. Thermosyphons are passive devices, proven to be very compact and efficient by other large scale LXe experiments. They are basically gravity-assisted heat pipes, consisting of three sections: a condenser, an evaporator (located below the condenser), and a passive adiabatic section connecting the two ends. The phase transition of a fluid that is sealed in the device is used for the heat transfer process. The thermal performance is scalable from a few W to kW, over a wide range of temperatures, depending on the choice of the fluid.

An initial set of 3D thermofluid mechanics calculations have been performed using ANSYS CFX [51] and a thermosyphon using nitrogen as working fluid. In the model, 100 W are generated on the copper top flange of the TPC vessel and a copper plate, conductively coupled to the bottom-end of the thermosyphon, is submerged in HFE-7000 at the top of the cryostat, as shown in Figure 4.40. The result of this preliminary analysis shows that this kind of arrangement is sufficient to maintain temperature gradients well below 1 K in the TPC, as required. The real detector will require the removal of 500 W, hence needing five thermosyphons connected to five cold plates, like the one shown here, submerged in the HFE-7000 [6]

More refined engineering will be required and will have to include a study of the conductive heat losses through the thermosyphons both during operations as well as in the event of power failures, when the passive stability of the large thermal mass should be disturbed as little as possible. The possibility of cooling convectively through the HFE-7000, if confirmed by further engineering, is very valuable, as it may allow the thermosyphons to be built out of a metal other than copper, reducing conductive heat losses.

#### 4.3.3.2 Cool-down by HFE-7000 recirculation

Recirculation of the HFE-7000 will be used at cool-down in order to efficiently couple and transfer the substantial power required. This method also achieves thorough mixing of the fluid during the cool-down, where a large amount of heat is removed from the system. Depending on the result of further engineering, it may be desirable to stop the HFE-7000 recirculation before starting the liquefaction of the xenon in the detector, and trim the final temperature for detector filling using the thermosyphons, as explained above. As shown in Figure 4.41, in this system the heat will be removed from the HFE-7000 by an external heat exchanger (HX). In this way, the construction of the HX becomes conventional in the sense of the radioactivity content of the materials. Radon, possibly introduced in the system by external elements in the recirculation, will decay after the external loop has been valved off at the end of the cool-down. A system to maintain the HFE-7000 pressure, essential to transfer the load across the TPC vessel to the inner vessel of the cryostat, requires negligible flow and has been already tested with success in EXO-200. We also note that the HFE-7000 recirculation system has to be integrated with the filling system, as sketched in Figure 4.41. Since the HFE-7000 contracts by about 30% during the cool-down, more fluid has to be constantly added while maintaining the pressure within a safe envelope. The pump required to circulate the HFE-7000 must be able to work at low temperature and have a sealed fluid volume, as is typically achieved with a magnetic drive. Proper design is required to operate in a region of parameters safe against cavitation, because of the low vapor pressure of the HFE-7000.

[6] We note, however, that only 250 W out of 500 W originate in the TPC, the rest being directly coupled to the inner vessel of the cryostat.

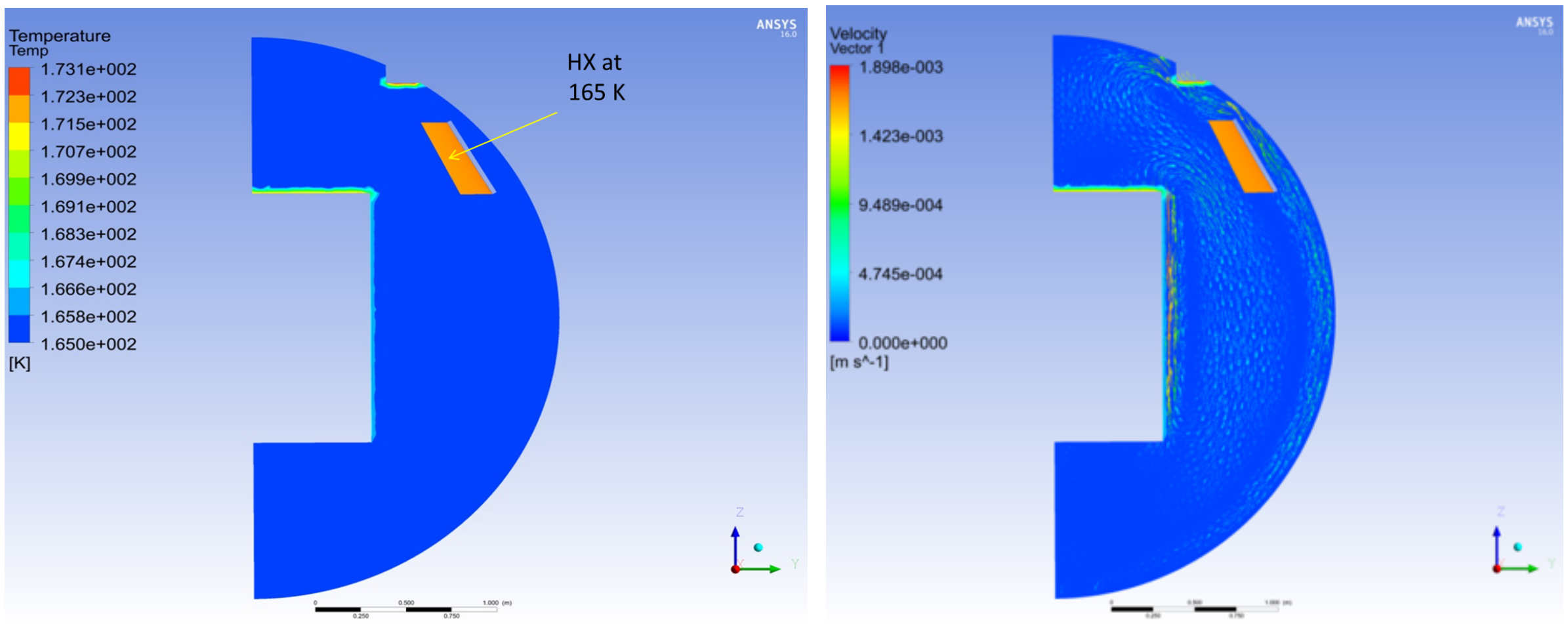

**Figure 4.40:** 3D model of the steady state cooling of the detector using a thermosyphon in thermal contact with the HFE-7000 (plate at top-right in the models). Heat generated inside the TPC is removed by conduction through the thin copper vessel and convection of the HFE-7000. The temperatures obtained by the thermofluid simulation are shown for 100 W produced inside the TPC (left). The fluid velocity field during operation (right). Only a 1/4 segment of the detector is modeled and no LXe convection is included.

The thermal performance of the recirculation system is driven by the HX, the volumetric flow achievable and the refrigeration power available to the heat exchanger. With a simplified model based on the NTU theory of HX [52], the governing equation is:

$$\frac{dT_i}{dt} = -\frac{\dot{m}}{M}(T_i - T_f)(1 - e^{-\mathrm{NTU}}), \tag{4.3}$$

where $M$ is the total HFE-7000 mass, $\dot{m}$ is the mass flow through the HX, and $T_i$ and $T_f$ are the initial and final temperatures, respectively. NTU is the Number of Transfer Units defined as

$$\mathrm{NTU} \equiv \frac{UA_S}{\dot{m}C_P}, \tag{4.4}$$

which gives a direct measure of the efficiency of the HX. Here $A_S$ is the surface area of the HX, $C_P$ the specific heat of the HFE-7000 and $U$ is the series of the thermal conductances through heat exchange process. In our case, $U$ is dominated by the coefficient of heat transfer of the HFE-7000, being the lowest. This results in:

$$T_i(t) = T_f + [T_i(0) - T_f]e^{-\beta t}, \tag{4.5}$$

where the time constant $\beta$ is given by

$$\beta = \frac{\dot{m}}{M}\left(1 - e^{-\mathrm{NTU}}\right). \tag{4.6}$$

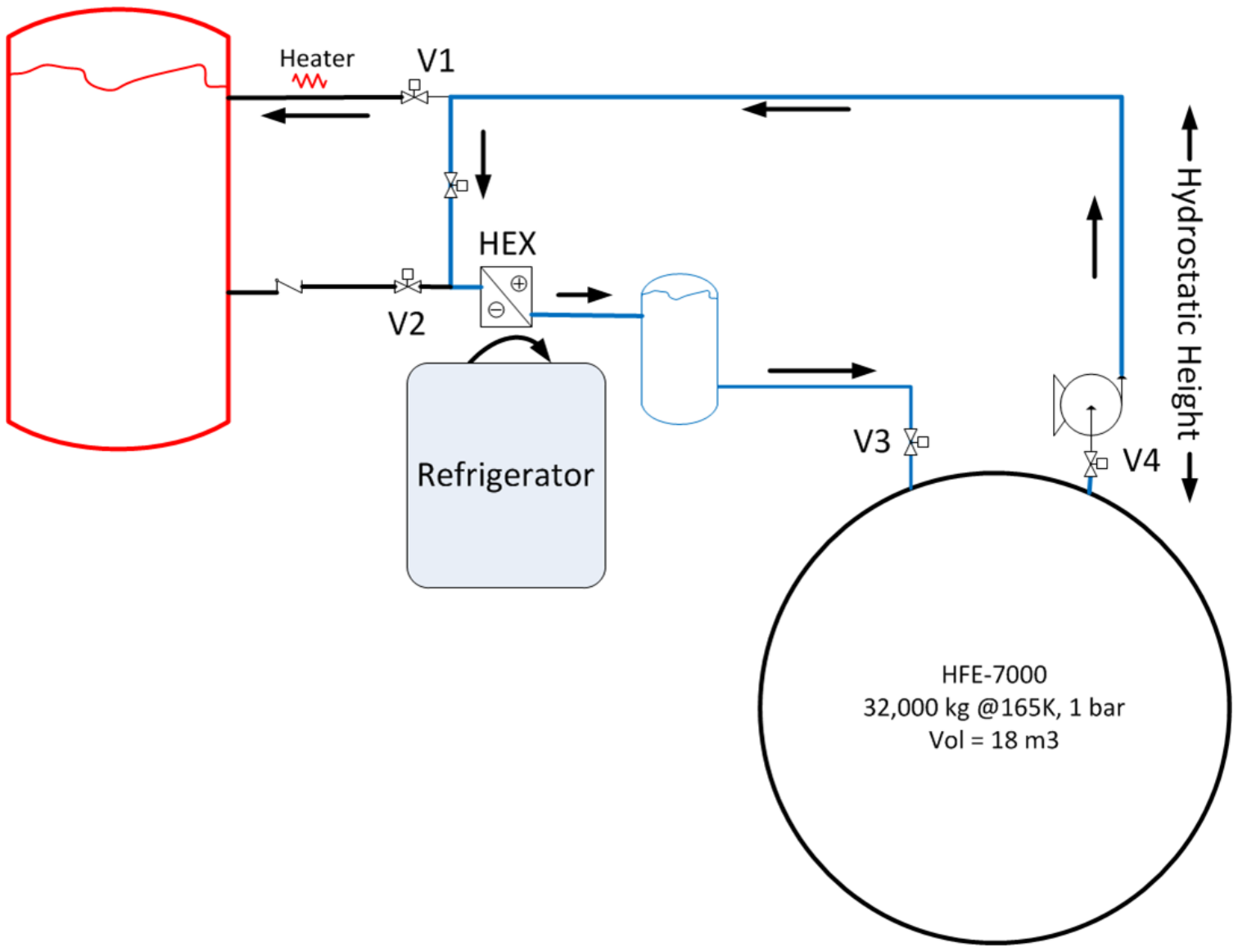

**Figure 4.41:** Schematic diagram of the HFE-7000 recirculation system for nEXO. A HX that is external to the cryostat will be used to cool-down the fluid and the detector. The warm-up heat exchanger is not shown.

A larger mass flow has the effect of shortening the cool-down time, within the limits provided by the maximum available refrigeration power. With nEXO parameters, an HFE-7000 flow of 3 liters/minute is sufficient for a cool-down in the required 30 days. Care will be applied to mitigate the risk of freezing HFE-7000 in the HX.

The HFE-7000 recirculation system can also be used for a relatively rapid (presumably also about 1 month) warm-up operation, where the heat is provided resistively to a dedicated heat exchanger in the recirculation loop (not shown in Figure 4.41).

A large LXe setup is being built by the collaboration for the final testing of HV components. This system, using up to 800 kg of LXe, also employs HFE-7000 that can be recirculated and will be used to validate various design parameters discussed here.

### 4.3.4 Xenon Recirculation and Purification System

Xenon recirculation and continuous purification are required to remove electronegative impurities that inevitably leach into the LXe from various components inside the TPC. While experience from EXO-200, LUX, and XENON exists on the need of this process, it is exceedingly difficult to provide quantitative assessments on the recirculation requirements, since the outgassing rates (and, indeed even the species outgassed) are generally not known. EXO-200 routinely achieves electron drift lifetimes of 3 ms and lifetimes as long at 5 ms have been observed with recirculation rates of $\sim$ 15 slpm. Because of the longer drift length (and to preserve the possibility of running at lower electric field than 400 V/cm) an analysis of the detector calibration (see Section 4.4) has set the goal of $>$ 10 ms on the electron lifetime in the LXe. Two qualitative improvements are expected to make the factor of $\sim$ 3 in electron lifetime possible:

- As described in Section 4.1.1, nEXO will employ very little plastics in the TPC, to limit the outgassing. All LXe detectors built until now for $\beta\beta$ decay and Dark Matter searches used large amounts of Teflon that will not be present in nEXO;
- The surface-to-volume ratio in the nEXO TPC will be more favorable by a factor of $\sim$ 3 than in EXO-200. While the surface of the plumbing in the recirculation system is not known yet, it is also expected that this will scale by a factor which is substantially smaller than the 25-fold increase in the volume of LXe.

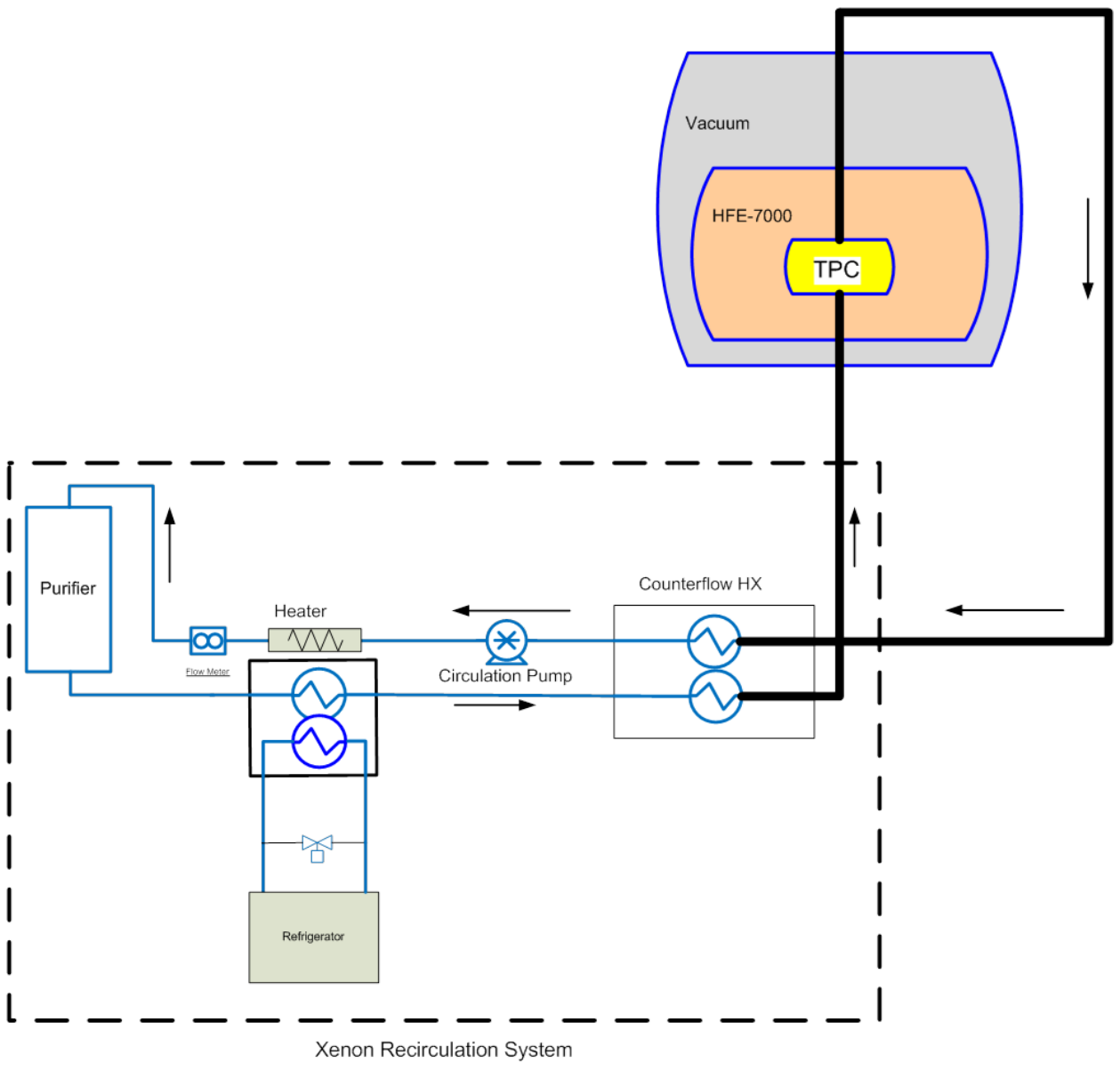

**Figure 4.42:** Schematic process diagram of the xenon recirculation system for nEXO.

Notwithstanding these improvements, at this time it is considered prudent to plan for the same recirculation time for the entire stockpile of xenon as in EXO-200. This corresponds to a recirculation rate of 350 slpm.

Purification is generally done in gas phase[7] through the use of heated zirconium getters, such as those produced by SAES in the Mono-Torr line [53]. These getters have been shown to reduce the concentration of $O_2$, $H_2O$, CO and $CO_2$ to less than 1 ppb in a single pass and generally have a very substantial capacity, so that the limitation usually derives from their flow impedance. Unlike other getter materials, the hot zirconium-alloy medium has been shown not to emanate Radon in problematic quantities. A schematic process diagram is shown in Figure 4.42.

A critical component of the recirculation system is a pump capable of offsetting the impedance of the loop for the required flow of $^{enr}$Xe. The pump has to be exceedingly reliable in terms of leaks to the atmosphere (both not to lose $^{enr}$Xe and not to contaminate the $^{enr}$Xe), non-emanating radon, and non-contaminating the $^{enr}$Xe with electronegative impurities. A novel magnetically coupled piston pump design was conceived and built for EXO-200, with a maximum gas flow rate of 20 slpm. A larger version of this pump, schematically shown at the left in Figure 4.43, has been designed and prototyped jointly with colleagues from the XENON collaboration for the two detectors. The prototypes of this pump have been tested to produce flows of 130 slpm of xenon with a compression of at least 1.1 bar, averaged over one cycle [49], as shown at the right in Figure 4.43. Two or three pumps are expected to be run in parallel to reach the required flow and, equally important, to ensure continuous recirculation during maintenance [8].

[7] Purification in liquid phase is being studied by different groups but, at present, has not been proven at the levels of electronegative and radiological purity required.

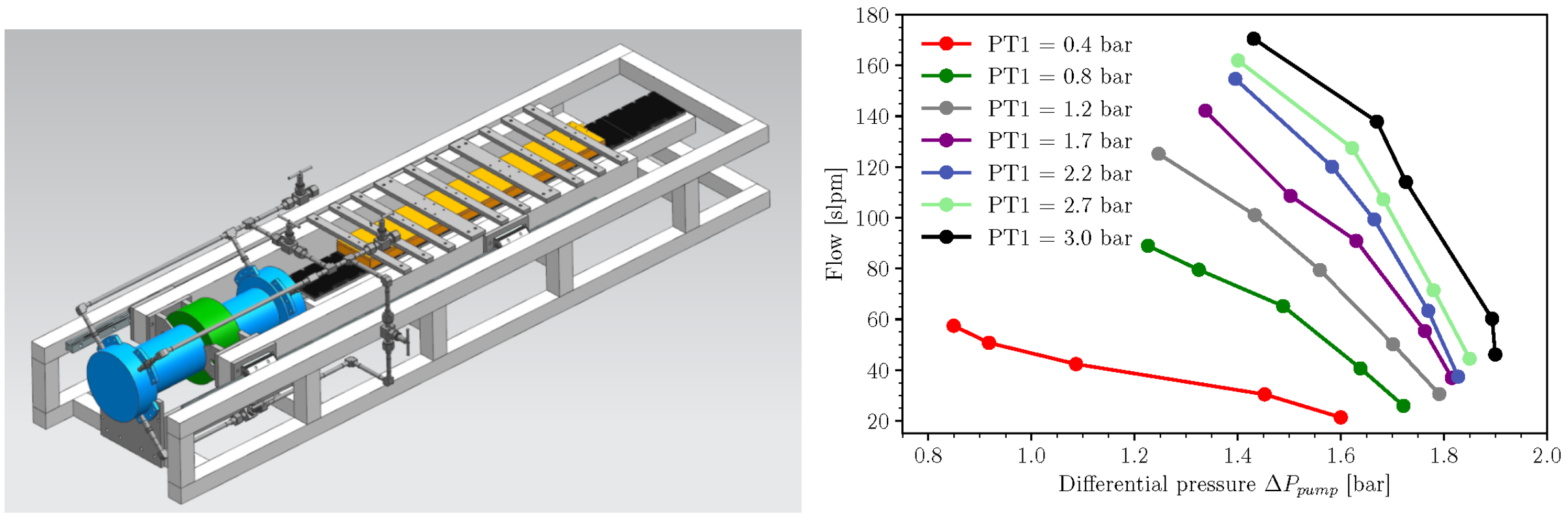

**Figure 4.43:** High capacity magnetically coupled piston pump developed for $^{enr}$Xe recirculation. Left, model of the pump. Right, performance diagram, from [49].

Xenon purification in gas phase requires substantial energy (mainly to produce phase transitions). A counterflow heat exchanger, recovering most of the heat required for vaporization from the condensation process, substantially increases the efficiency of the system and, more importantly here, improves the stability of the process by tying the vaporization and condensation together and only supplying a small amount of cooling and heating from the outside. This is in contrast to the independent boilers and condensers used in EXO-200, where two large (heating and cooling) powers have to be fine tuned in their difference. A counterflow heat exchanger design for LXe is being tested by the collaboration.

While the electron lifetime will be measured in the TPC with great accuracy for the purpose of detector calibration (see Section 4.4), EXO-200 experience shows that, for diagnostic purposes, it is convenient to also provide tools to measure the effective purity of the xenon, in the recirculation loop, possibly at more than one location. A combination of gas-phase and liquid-phase purity monitors is under study and a novel liquid-phase design, sensitive to very long lifetimes in a small mechanical envelope, is shown in Figure 4.44. These external purity monitors are generally not expected to provide an absolutely calibrated electron lifetime, but they have a fast response and can be used to study trends.

A schematic of the purity monitor prototype currently under construction is shown in Figure 4.44. Electrons will be generated from a photocathode using a 5 W Hamamatsu xenon flash lamp and drifted into the switching region. Four Behlke [54] high voltage switches will provide the 20 kHz switching frequency, trapping the electrons for an effective drift length of more than one meter before releasing them for collection.

[8] The EXO-200 pump requires gasket replacement every 12 to 18 months of continuous operation.

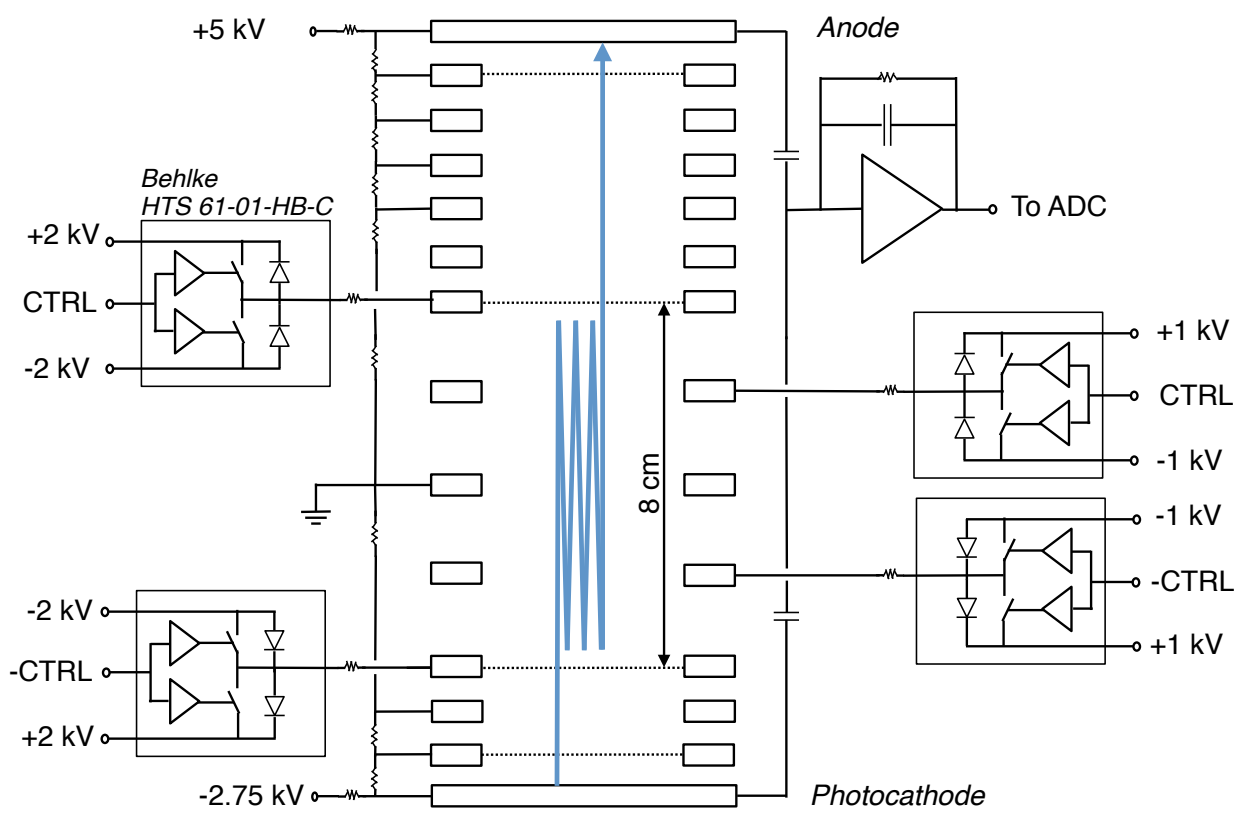

**Figure 4.44:** Diagram of the novel LXe purity monitor being developed. Since a long drift is required to measure lifetimes exceeding $\sim$ 10 ms, this new device uses fast high voltage switches to provide folded trajectories for the electrons and reduce the overall device length. A photocathode illuminated by UV light pulses is used to produce the electrons. Gas phase monitors are already used in EXO-200 and could be upgraded for the use in nEXO.

### 4.3.5 In-Line Radon Trap

The target radon level in the nEXO TPC is set to <600 atoms. For comparison, EXO-200 exhibits, in steady-state, a total of close to 200 Rn atoms [55]. It is unknown if the sources are inside the TPC or in the xenon plumbing. The nEXO TPC is much larger therefore a two-fold strategy was adopted: material screening (see Section 6.5) and removal with an in-line radon trap.

The nEXO facility for Rn emanation measurements, shown in Figure 6.9, includes a xenon recirculation system dedicated to the development of such a radon trap, comprising:

- a recirculation pump of the EXO-200 design;
- a radon counter (ESC, described in Section 6.5);
- a radon source (about 15 Bq of $^{222}$Rn);
- a vacuum-insulated enclosure housing the trap;
- a refrigeration system to control the temperature of the trap.

The trap is based on refrigerated, activated charcoal and is inspired by the results obtained by the XMASS collaboration [56]. A radon reduction factor of 44.7 was demonstrated at -44°C, 280 mbar and a flow of 1 slpm, using a commercial activated charcoal getter [57]. A simulation based on resistive-capacitive chromatography was developed that reproduces qualitatively the system behavior, in particular the oscillations reported by XMASS, from which the average radon advection velocity can be determined.

The focus is currently on addressing the logistical issues associated with lowering the trap temperature. Tests will also be carried out with other types of activated charcoal with the goal of establishing their radon removal ability, radiopurity and mechanical properties. Further investigations will address the need of the experiment to operate the trap at a $\sim$350 slpm flow of xenon near atmospheric pressure.

### 4.3.6 Fluid Mechanics in the TPC

The flow of LXe within the TPC vessel is driven by the continuous recirculation of the xenon as well as convection currents caused by heat sources within the TPC. An understanding of the LXe motion in the detector is important both from the points of view of impurity transport, affecting the electron lifetime, and of the thermal management. Here we list some of the main issues that are currently being studied with dedicated thermo-fluid simulation models, and the impact that they will have on the nEXO design.

**Mixing and turnover time of purified xenon.** It is important that the incoming purified LXe entering the TPC vessel is routed quickly and uniformly throughout the active region of the TPC, displacing the existing liquid which is comparatively loaded with impurities. A short turnover time of the liquid (defined here as the time required to replace 90% of LXe in the active volume) optimizes the equilibrium impurity level. In addition to a short turnover time, uniform mixing of the incoming liquid is also important. Regions that are isolated from the recirculation flow ("dead regions") can build up impurities causing nonuniform response of the charge collection in the detector. Turnover time and mixing are strongly dependent on the recirculation flow rate as well as the number and positioning of the inlet and outlet ports and the various internal components limiting the free flow. The optimization of these parameters involves several design tradeoffs. For example, higher recirculation rate will reduce turnover time but involves larger pipes and hence a higher load of radioactivity. Similarly, a large number of inlet and outlet ports increases the uniformity while also increasing the complexity and, again, the radioactivity load.

**Flow patterns of outgassing impurities.** Outgassing rates vary by material, with plastics typically having significantly higher rates than metals or other inorganic components such as silicon. It is therefore expected that the largest contribution to outgassing will come from the Kapton cables in the TPC. A solid understanding of the LXe flow will inform the location of such cables to optimize the removal of impurities.

**Heat removal.** The removal of the bulk heat produced by the electronics in the LXe is a primary concern. LXe convection cells along the cylindrical walls of the TPC and outside of the field cage will be the primary method for cooling of the photodetectors electronics (100 W). LXe convection in the thin, pancake-shaped region above the charge collection tiles is expected to perform the same function for the charge readout electronics (also 100 W). In both cases, the heat will be transferred to the thin copper vessel containing the LXe and from there conductively coupled to HFE-7000 convection cells. All these heat transfer processes need to be understood, first to verify that they are sufficiently efficient for the task and second to optimize their performance.

**Regions of localized boiling.** Even at the nominal operating pressure of 1.5 bar, xenon is in liquid phase only in a range of 11 K. Without adequate heat dissipation (either through LXe flow or by conductive sinking to the TPC vessel) localized sources of heat from the electronics can cause boiling. The resulting gas bubbles can cause HV breakdowns and are generally to be avoided. Thermal simulations will provide requirements for the largest power density in the TPC, also depending on the orientation with respect to gravity and the surrounding materials.

**Distribution of internal calibration sources.** As discussed in Section 4.4, the use of some radioactive sources to be dissolved in the LXe for the purpose of detector calibration is under study. The radionuclides will follow the flow patterns of the LXe and their spatial distribution during the calibration needs to be well understood. Additionally, some of the calibration sources under consideration have relatively short half-lives and fluid simulations are needed to evaluate the activity loss between the injection location and the bulk LXe in the TPC.

Fluid modeling of the LXe within the TPC vessel is divided into two efforts. The first, which uses the SOLIDWORKS modeling package [48], is focused on studying LXe flow patterns to quantify the mixing and turnover time, while the second uses the COMSOL Multiphysics package [18] to study the propagation of radioactive sources dissolved in the LXe. Both simulations use the same NIST [43] temperature-dependent LXe fluid properties and can track fluid pressure, temperature, and velocity as a function of position in the vessel. HFE-7000 convection will be coupled to this model at a later stage. The use of two different fluid simulation packages allows us to cross-check and compare results, for example comparing the different turbulence models available in each software package. Efforts are also currently underway to benchmark the simulations against experimental data, such as the electron lifetime and ion drift data from EXO-200 [2].

For the SOLIDWORKS-based simulations, a full 3D model of the TPC vessel and internal components is used. Turnover time is estimated by using a second fluid, with identical LXe properties but labeled differently, and calculating the time it takes for the second fluid to displace 90% of the original fluid in the vessel. When steady-state flow fields have been calculated for a given geometry, tracking particles along the flow lines can be used to understand the effect of outgassing at different locations of the detector. Simulations including both the full 3D geometry and the heat loads are extremely computationally intensive and hence a broad set of simplified simulations are currently being evaluated. The effects of the different detector components (such as the SiPM staves and field rings) on the flow pattern are also being investigated. In addition, several different configurations of inlet and outlet fluid recirculation ports have been simulated without heat loads and the results compared with respect to turnover and mixing times (see Figure 4.45). These preliminary studies will be used to select a smaller set of the most promising configurations for the full simulations, including heat loads.

The COMSOL-based simulations currently use a simplified 2D geometry to study the time-dependent distribution of radioactive daughters inserted into the TPC vessel. One such source under consideration is $^{220}$Rn, as discussed in Section 4.4. The longest lived nuclide in the $^{220}$Rn decay chain is $^{212}$Pb, with a half-life of 10.6 hours. Preliminary results showing the velocity and concentration distributions soon after source injection are shown in Figure 4.46. The results of these fluid simulations will be used in conjunction with a secondary simulation of the decays of the $^{220}$Rn daughters to improve the calibration model.

### 4.3.7 Xenon Recovery System

The xenon recovery system for nEXO is still under study and must be developed together with a thorough failure analysis of the entire LXe and HFE-7000 systems. It is worth recalling here that the very large thermal inertia of the system makes its behavior intrinsically stable and, in the event of power loss, a substantial time is available before any action is required. There are, however, catastrophic scenarios, such as the loss of the HFE-7000 (rupture of the inner vessel of the cryostat) or the contact of water with the inner vessel of the cryostat (rupture of the outer vessel), under which the above statement is not valid. It is likely that the analysis will conclude that these scenarios are too drastic to require a backup action.

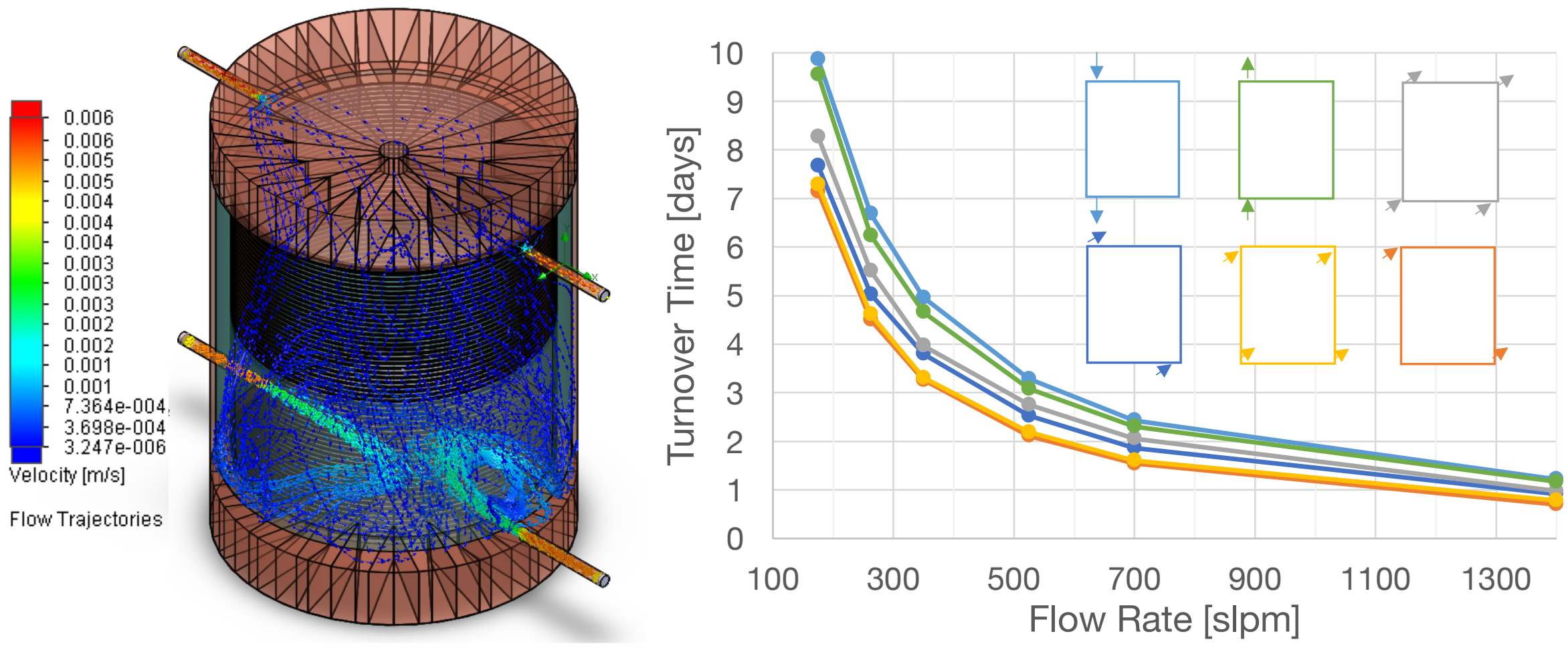

**Figure 4.45:** Preliminary results from the SOLIDWORKS-based fluid simulation. Left: Velocity distribution of liquid xenon within the TPC vessel for a given configuration of two radial inlets at the bottom and two radial outlets at the top. Right: Turnover time as a function of the recirculation flow rate for different configurations of inlets and outlets. Heat loads are not yet included in this simulation.

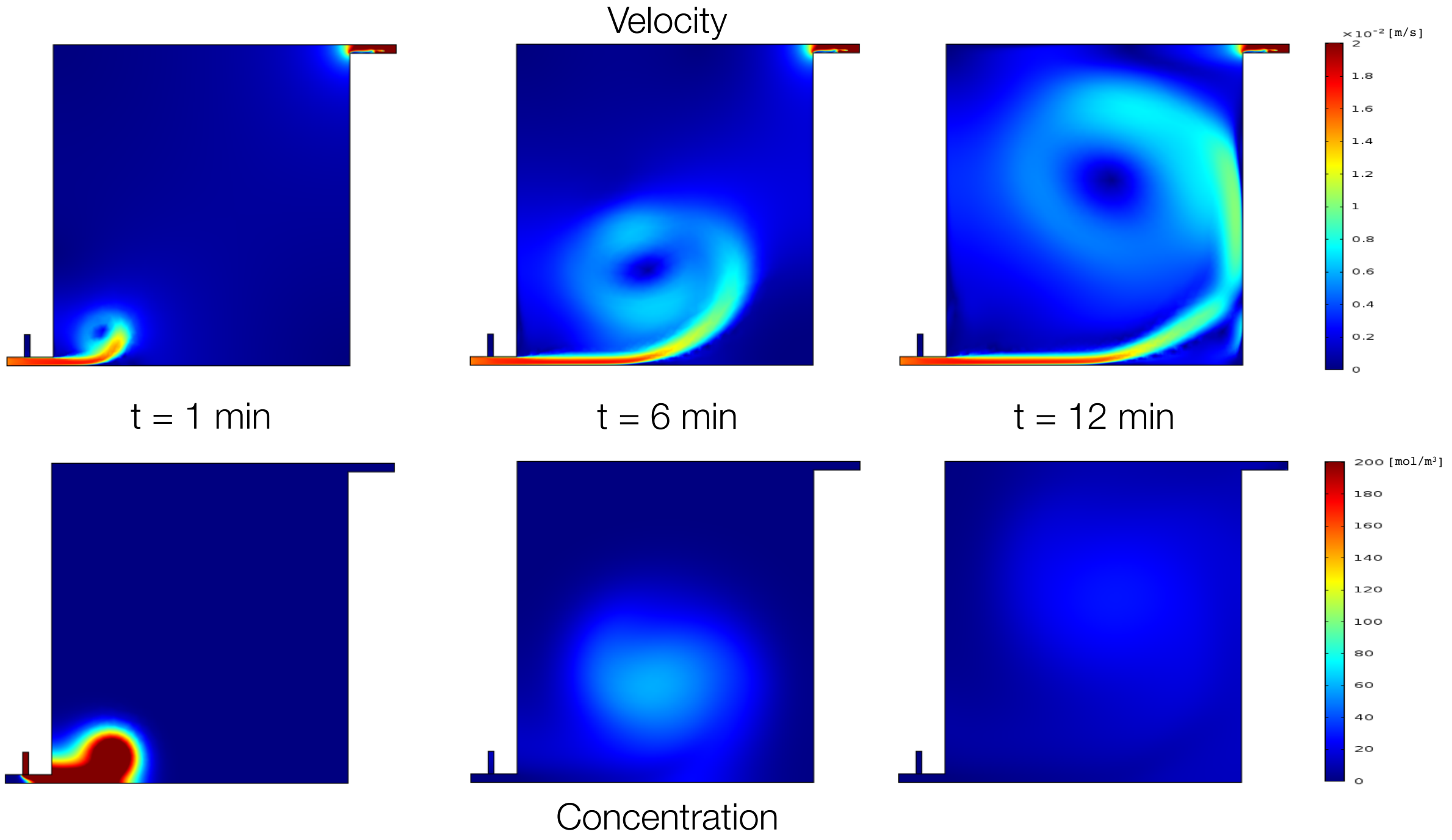

**Figure 4.46:** Preliminary COMSOL-based simulations of LXe velocity (top) and source concentration (bottom), at three different times after source injection in the TPC.

The preliminary assessment is that (both planned and un-planned) xenon recoveries will be intrinsically slow, occurring over days or maybe even weeks. This recovery to high pressure, long-term storage can be achieved either by cryopumping into pressure-capable containers or with commercial compressors, feeding standard cylinders.

A system using compressors has the advantage of using conventional storage cylinders, that can also be used to ship the xenon from the enrichment facility. The compressors, however, require large amounts of uninterruptible power and, being complex electromechanical devices, have failure modes that must be considered. It should be noted that commercial high pressure gas compressors are not generally required to be highly reliable. The cryopumping option is attractive because is passive; particularly as a nEXO $LN_2$ plant and store will likely be available on site. The recovery vessels will most likely need to be custom-made out of stainless steel, as was done, for the same purpose, by both the XENON1T and MEG experiments [45, 58]. Some experience with a cryopumping system will be gained in nEXO by operating the large HV test stand that employs this technique.

### 4.3.8 Control System

The nEXO control system will be responsible for the operation of auxiliary subsystems including cryostat cooling, xenon recirculation, and TPC/cryostat pressure control. The latter will be particularly important because of the limited differential pressure tolerance of the thin-walled LXe vessel. Critical operations carried out through these subsystems will include filling and emptying the detector. The same system will also provide the control, monitoring, and logging interface for users to access data relevant for the detector operation (science data will be collected by a different system, described in Section 4.2.4). Finally, the control system must be highly redundant and reliable and able to recognize critical situations, autonomously acting to protect the detector and the $^{\rm enr}{\rm Xe}$.

A variety of hardware platforms are used for the control system in other experiments based on LXe. EXO-200 and LZ are of particular interest. Experience with the former will inform the nEXO design, while the latter provides an alternative model addressing similar goals and potential issues.

The slow control for both experiments is based on a combination of Programmable Logic Controllers (PLCs) and control computers, communicating with instrumentation, each other, and users via databases and distributed code. The EXO-200 slow control uses National Instruments Compact FieldPoint PLCs, running control logic written in LabVIEW. Nearly all I/O for the $\sim$ 700 channels is carried out by the PLCs through onboard modules at 1 Hz sampling rate. Control of redundant critical system components is divided between the two PLCs. This embedded system communicates with outside users through SQL databases containing the system data and control parameters. A distributed Graphical User Interface (GUI) lets users view system data and, from approved control centers, change control parameters.

The LZ slow control design combines a robust PLC (Siemens SIMATIC S7-410H [59]) with a control server running the Ignition software platform [60]. The PLC controls the critical subsystems while the Ignition server communicates with the PLC, directly interfaces with non-critical hardware, allows user control, and stores system data in an SQL database for user access. The PLC includes two redundant CPUs with independent connections to the I/O modules, which also allows for "bumpless" software and hardware changes during detector operation.

The number of channels in nEXO would be somewhat larger than in EXO-200, driven by both the increase in the scale of the system (e.g. multiple xenon recirculation pumps instead of one) and new subsystems (e.g. HFE-7000 recirculation), but this is expected to be only a modest increase. More important are issues of reliability and the ability of upgrading software or hardware without the need to restart the PLCs. EXO-200 also encountered a few instances where realtime logic stopped executing on one of the PLCs for unknown reasons and system upgrades require a full system restart, involving risks.

## 4.4 Calibration System

### 4.4.1 Overview

A comprehensive calibration strategy must build into the detector design a set of capabilities to achieve two overarching and interrelated goals:

- achieve a complete understanding of the detector response to double-beta decays over the full range of energies throughout the detector volume, including the absolute energy scale and the energy and position resolution functions,
- accumulate the required calibration data to monitor time variations in the calibration parameters at the requisite level without a significant loss of efficiency for low background data collection.

The required tasks can be broken down into two categories:

- a "baseline" calibration to understand the ionization and scintillation responses, the charge-light anti-correlation and the electron drift parameters over the entire fiducial volume,
- a set of validation and cross-check data required to demonstrate a complete understanding of the detector response to both signal and background events.

The focus of nEXO activities to date has been to establish the conditions for the conventional techniques required to achieve the baseline calibration. This has entailed the optimization via Monte Carlo simulations of the activities and placement of several $^{228}$Th sources, and also establishing the framework by which to evaluate the quality of the scintillation lightmap i.e. the detailed 3-D spatial variations in the scintillation light response in the TPC active volume. In the following subsections, we briefly summarize the main baseline calibration to monitor the energy scale, the electron lifetime, the charge-light anti-correlation and the scintillation light map. We conclude by briefly describing some new ideas to improve continuous monitoring with minimum loss in efficiency, as well as additional validation, cross-check and monitoring capabilities which might impact technology and material choices in the detector's conceptual design.

### 4.4.2 Electronics Calibration

The gain of the nEXO front-end readout electronics can be determined with the on-board calibrators. As shown in Table 4.3, the ASIC design specifies an on-board calibrator for each charge channel with an absolute precision of 0.2%. This level of precision is achievable due to the uniformity of the CMOS fabrication process, which has been demonstrated for example in the LAr ASIC designed by the BNL group. A typical charge calibration run will involve the injection of different amounts of charge into the front end across the entire dynamic range. The calibration data can be used to measure the gains and characterize possible non-linear responses of the preamps and ADCs. Since each run only takes 5 - 10 min, taking a run each day will allow tracking the performance of the electronics with minimal loss of the livetime.

Besides the on-board calibrators, the gains of the charge channels can also be measured using source calibration data. For example, EXO-200 uses the double escape peak of the $^{228}$Th source for the electronics gain measurement. The method requires the collection of a large amount of calibration data and, therefore, can only be performed occasionally during running. However, it will provide a useful cross-check of the channel gains measured by the on-board calibrators.

### 4.4.3 Baseline External Source Calibration

The purpose of external $\gamma$ source calibration is to provide periodic monitoring with adequate accuracy of the energy scale, the charge-light anti-correlation and the electron lifetime in the TPC. The large size of the nEXO TPC has been shown to be of great importance to reduce and properly classify the most important backgrounds. However, this same feature makes the calibration with external sources challenging. Simulation work has shown that a useful energy calibration can be achieved using external $\gamma$-ray sources.

The baseline external source calibration plan involves regular two-hour simultaneous deployments of six $^{228}$Th sources in guide tubes around the TPC, located as illustrated in Figure 4.47. Four sources are deployed outside the barrel of the TPC, precisely midway between the anode and the cathode, separated by 90° azimuthally (labeled as PX, PY, NX and NY). Two additional sources are deployed behind the cathode and the anode respectively, along the central axis (labeled as PZ and NZ), primarily for more precise electron lifetime measurements. As described in Section 4.4.3.1 below, a Monte Carlo simulation was used to determine that the optimal source activity is 850 Bq for the four sources outside the barrel and 85 Bq for the remaining two, under the conservative requirement that no more than two events occur within the maximum $\sim 1$ ms drift time, to minimize ambiguities due to event pile up.

With this configuration, 260 SS $^{208}$Tl events are expected in the inner one tonne region (aka "deep events") in a two hour run. This allows for the determination of the peak position at the 0.1% level. In addition, we estimate a systematic shift in the energy scale of $\lesssim 0.15\%$, due to the finite precision of the electron lifetime determination and its tracking over time. We conservatively assume these two effects to combine, obtaining an overall energy uncertainty of $< 0.5\%$. This is substantially inflated in recognition of the early stage of our work and the general challenges related to obtaining absolute calibrations at levels much below 1%. Uncertainty in the energy scale resulting from systematic errors in the $\gamma$ source calibration, or a potential systematic difference in the energy response for $\beta$ and $\gamma$ events can also be constrained by fits to the $2\nu\beta\beta$ spectrum. This procedure was effectively employed in EXO-200, where such systematics have a negligible impact on the sensitivity [4].

#### 4.4.3.1 Optimal Source Activity

While using higher activity sources reduces calibration time, such sources also produce events so close in time that they occur in the same drift time window. While these events cannot be deconvolved using time information only, they could be disentangled with full energy and position reconstruction, using a technique which is yet to be fully developed. Without any details on the exact triggering and cluster matching algorithm, we can conservatively assume that if one deep event is accompanied by no more than one shallow event, the two events can be deconvolved and the deep event can be used for calibration.

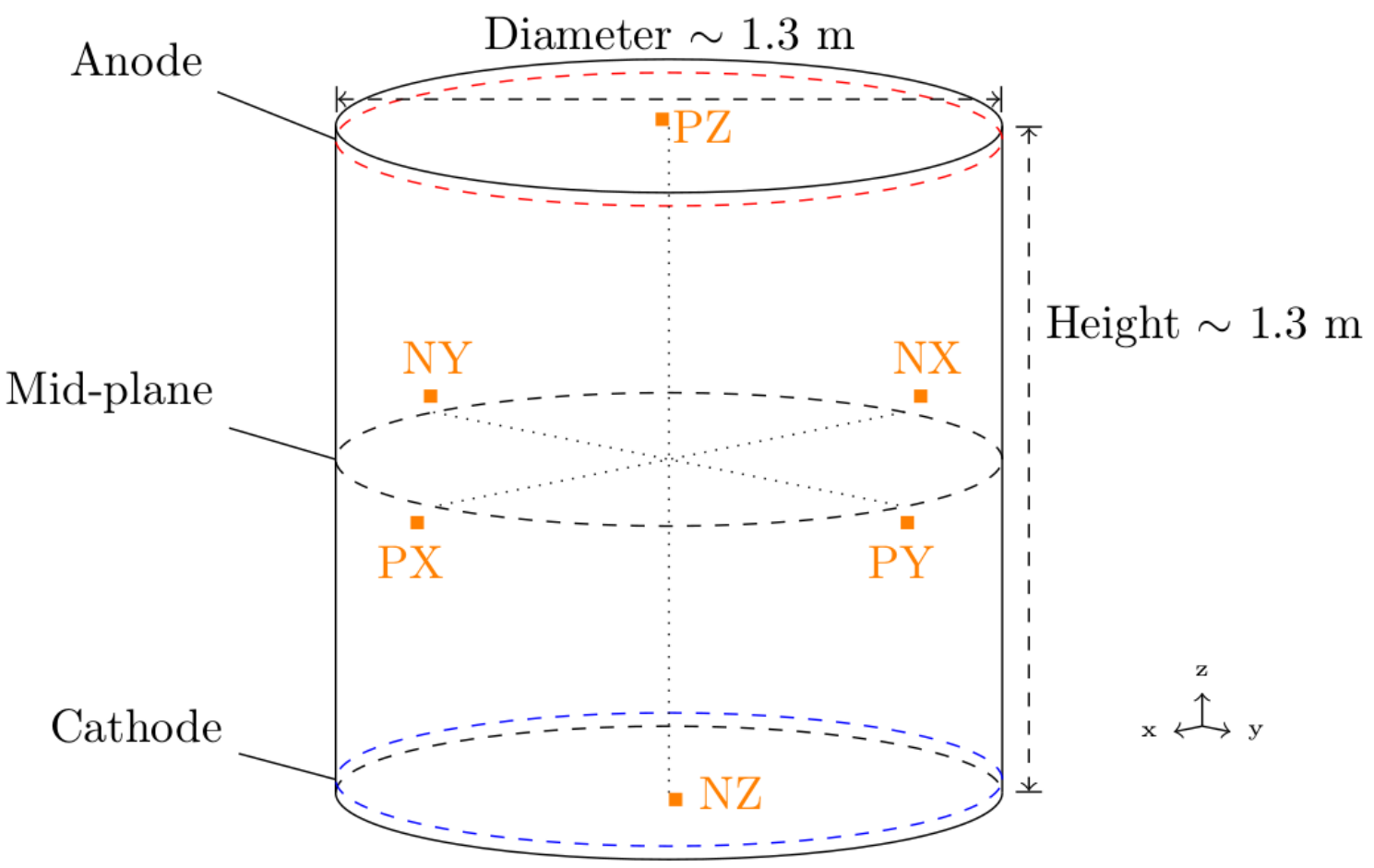

**Figure 4.47:** Source positions (shown as orange squares) around the TPC.

Under this assumption, the calibration time ($T$) required to accumulate $N$ usable deep events as a function of the total activity of the sources ($A$) is,

$$T(A) = \frac{N}{\beta A(1 + \alpha At)e^{-\alpha At}} \tag{4.7}$$

where $\alpha$ is the fraction of $^{228}$Th disintegrations that deposit energy anywhere in the TPC; $\beta$ is the fraction of $^{228}$Th disintegrations that produce a deep event; $t$ is the drift time window assumed to be 1 ms. Using a GEANT4 Monte Carlo simulation of the TPC, $\alpha$ and $\beta$ are determined to be 0.463 and $1.99 \times 10^{-5}$ respectively. $T(A)$ is shown in magenta in Figure 4.48. Also shown in Figure 4.48 are the calibration times required in two other possible scenarios dubbed *pessimistic* (red) and *optimistic* (blue). In the *pessimistic* scenario, where any accompanying shallow event will render the deep event unusable, the required time can be calculated to be $T_{\text{pess}}(A) = \frac{N}{\beta Ae^{-\alpha At}}$. In the *optimistic* scenario where any number of accompanying shallow events can be removed, the required time is $T_{\text{opt}}(A) = \frac{N}{\beta Ae^{-\gamma At}}$ where $\gamma$ is the fraction of $^{228}$Th disintegrations that deposit energy in the inner 1000 kg region. Using the same Geant4 simulation, $\gamma$ is determined to be $2.10 \times 10^{-2}$.

As seen in Figure 4.48, the optimal source activity is 3.5 kBq where $T(A)$ attains a minimum. The sources at PX, PY, NX and NY are allocated 850 Bq each. PZ and NZ are allocated a lower activity of 85 Bq each because they could produce false deep triggers if a deep trigger scheme considers central charge tile hits as a trigger condition. All sources would be simultaneously deployed to achieve the total optimal source activity. Under these conditions, we expect that a calibration run should last less than two hours, which is reasonable as long as the detector does not require one such calibration more frequently than once every few days, as in the case of EXO-200.

#### 4.4.3.2 Electron Lifetime

To study the calibration needs for electron lifetime measurements, a Monte Carlo simulation is used to model electron drifts in the TPC at various electron lifetimes ($\tau$ = 2, 5, 10, 20 and 50 ms).

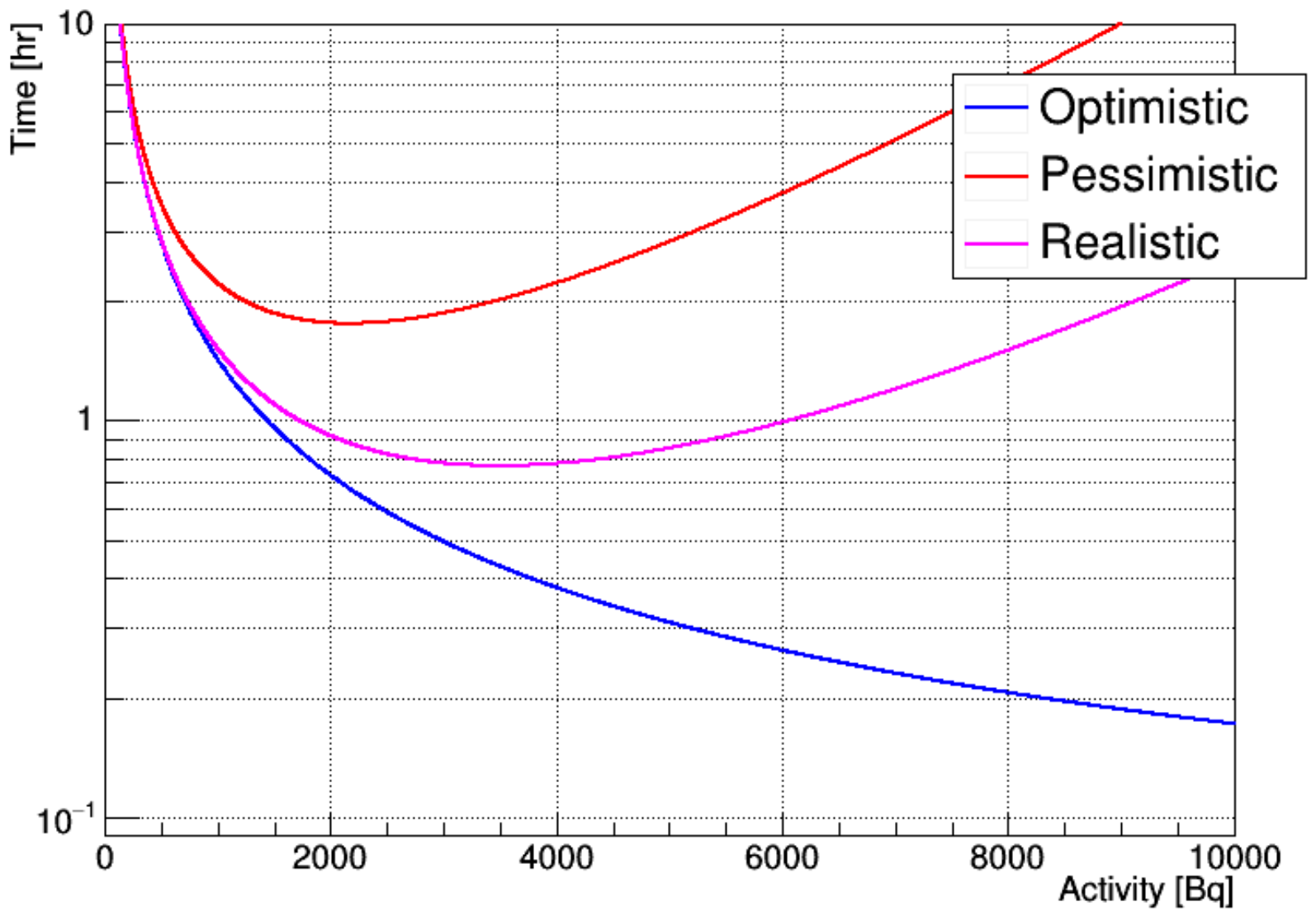

**Figure 4.48:** Time required to accumulate 100 usable deep events in different scenarios. Blue: *Optimistic* scenario ($T_{\text{opt}}(A)$). Red: *Pessimistic* scenario ($T_{\text{pess}}(A)$). Magenta: *Realistic* scenario ($T(A)$). See text for explanation.

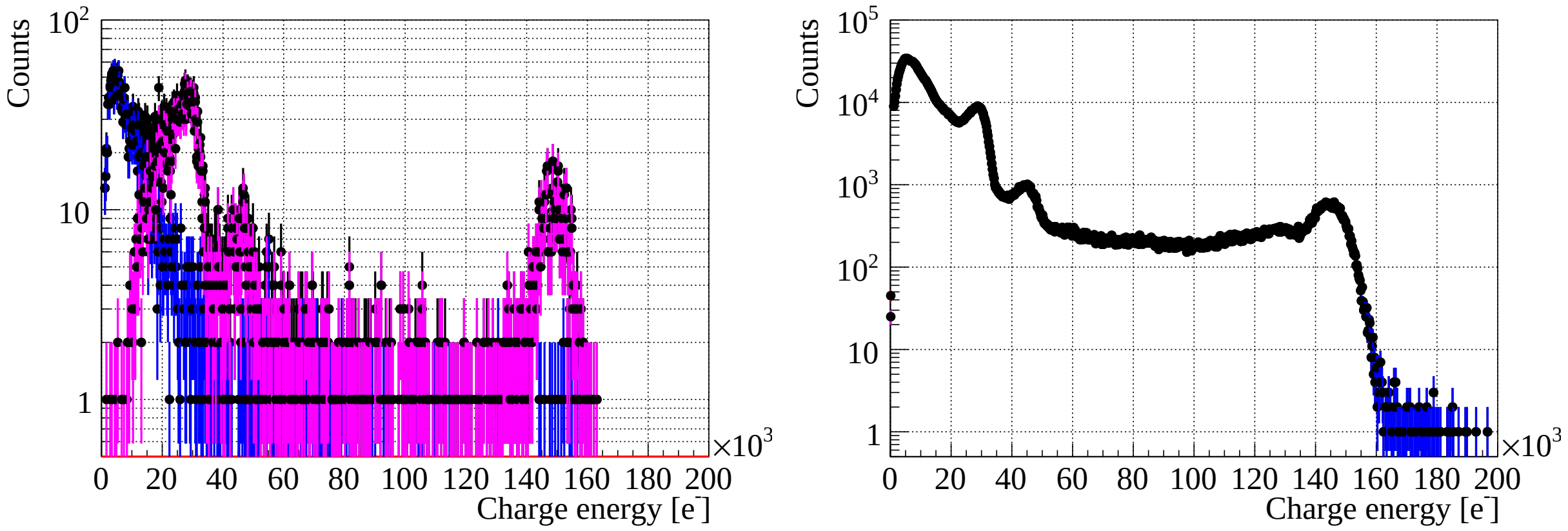

**Figure 4.49:** Examples of 2.6 MeV peak fits in TPC slices at two different locations: near the anode (left) and midway between the electrodes (right). Blue: Hits due to PX, NX, PY, and NY; Magenta: Hits due to NZ; Black: All hits. The electron lifetime is set to 10 ms in these examples.

The production of ionization electrons is simulated with NEST. Single site events are found using a simple clustering and reconstruction algorithm. To estimate the electron lifetime, the events are first sliced into 20 bins along the z-axis (discarding events near the electrodes). Then, in each Z-slice, the peak corresponding to $^{208}$Tl-induced 2.615 MeV $\gamma$s is fitted with a Gaussian, as shown in Figure 4.49 at two different locations in the TPC. By distributing the sources at six locations around the TPC as explained above, we can ensure that each slice has sufficient photoelectric events in the $^{208}$Tl peak. The contribution from sources at different locations are shown with different colors in the figure.

The charge energy of the Th full absorption peaks produced at different Z positions are fitted with an exponential function to extract the electron lifetime ($\tau$), as shown in Figure 4.50 for two different lifetimes (5 ms and 20 ms). Applying this procedure to two hour $^{228}$Th source calibration data, the inverse electron lifetime ($1/\tau$) can be determined with a statistical uncertainty of <0.004 (ms)$^{-1}$, independent of the value of $\tau$. Table 4.7 lists the fitted electron lifetimes compared with the true values in the Monte-Carlo. The result shows that 10 ms lifetime can be measured to better than 4%. One can then study how the uncertainty in the lifetime determination would affect the absolute energy scale for a 10 ms lifetime. Assuming that the statistical uncertainty is proportional to $1/\sqrt{t}$ where $t$ is the calibration time, the systematic shift due to electron lifetime correction as a function of calibration time can be obtained, as shown in Figure 4.51. With two hour calibration data, the systematic shift can be reduced to 0.15%, thus having minimal impact on the charge energy measurement.

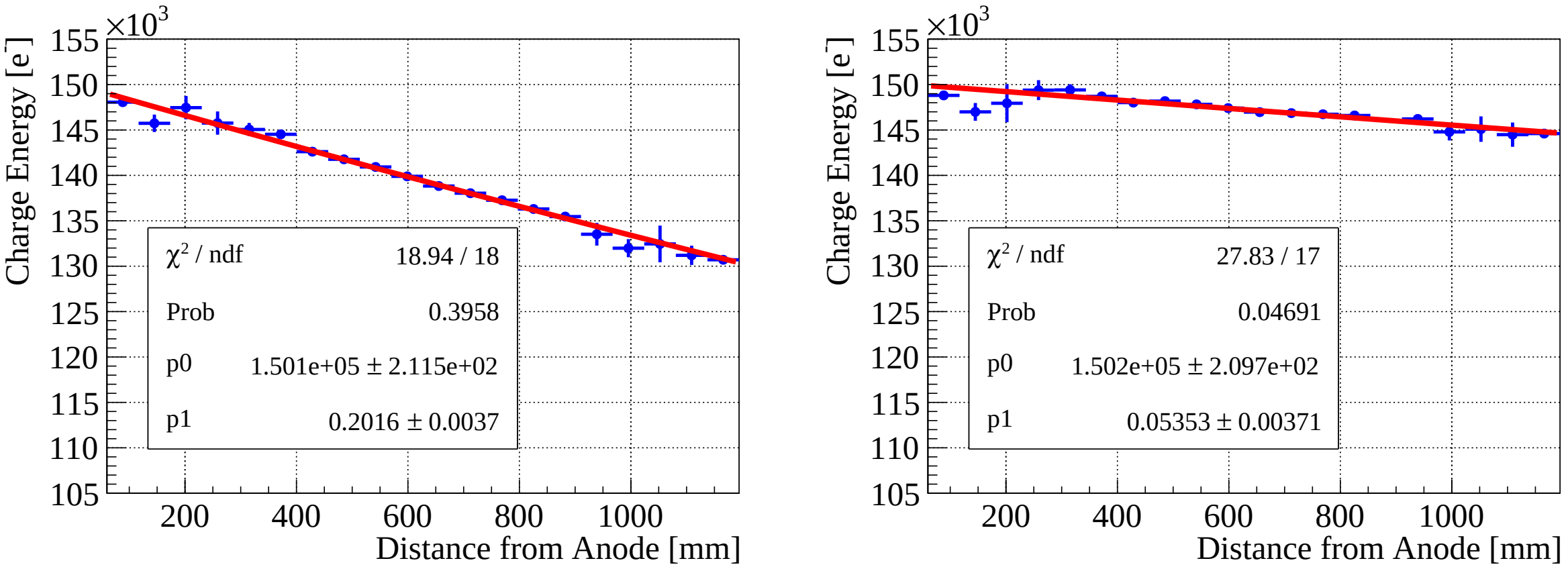

**Figure 4.50:** The electron lifetimes are extracted by fitting the energy of the $^{208}$Tl peak using events at different TPC locations to an exponential decay curve. Each panel represents an example fit to simulated data with 5 ms (left) and 20 ms (right) lifetimes. The horizontal bars indicate the extent of the spatial slices.

| True Lifetime $[ms]$ | Inverse Lifetime $[(ms)^{-1}]$ | Fitted Inverse Lifetime $[(ms)^{-1}]$ |
|---|---|---|
| 2 | 0.5 | 0.4989 ± 0.0035 |
| 5 | 0.2 | 0.2016 ± 0.0037 |
| 10 | 0.1 | 0.1015 ± 0.0038 |
| 20 | 0.05 | 0.0535 ± 0.0037 |
| 50 | 0.02 | 0.0212 ± 0.0031 |

**Table 4.7:** The fitted (inverse) lifetimes from simulated calibration data compared with the values fed into the Monte Carlo.

For longer lifetimes, although the precision of the lifetime measurements decreases, the charge measurement accuracy actually improves because the overall correction is smaller. If the lifetime is much shorter than 10 ms, other effects such as non-uniform impurity distributions are likely to dominate. Given that EXO-200 has already achieved 3 to 5 ms lifetime, we expect that 10 ms lifetime can be reached in nEXO by minimizing the use of plastics in the TPC design (see Section 4.1.1). The above study assumes that the detector has no spatial variation in electron lifetime.

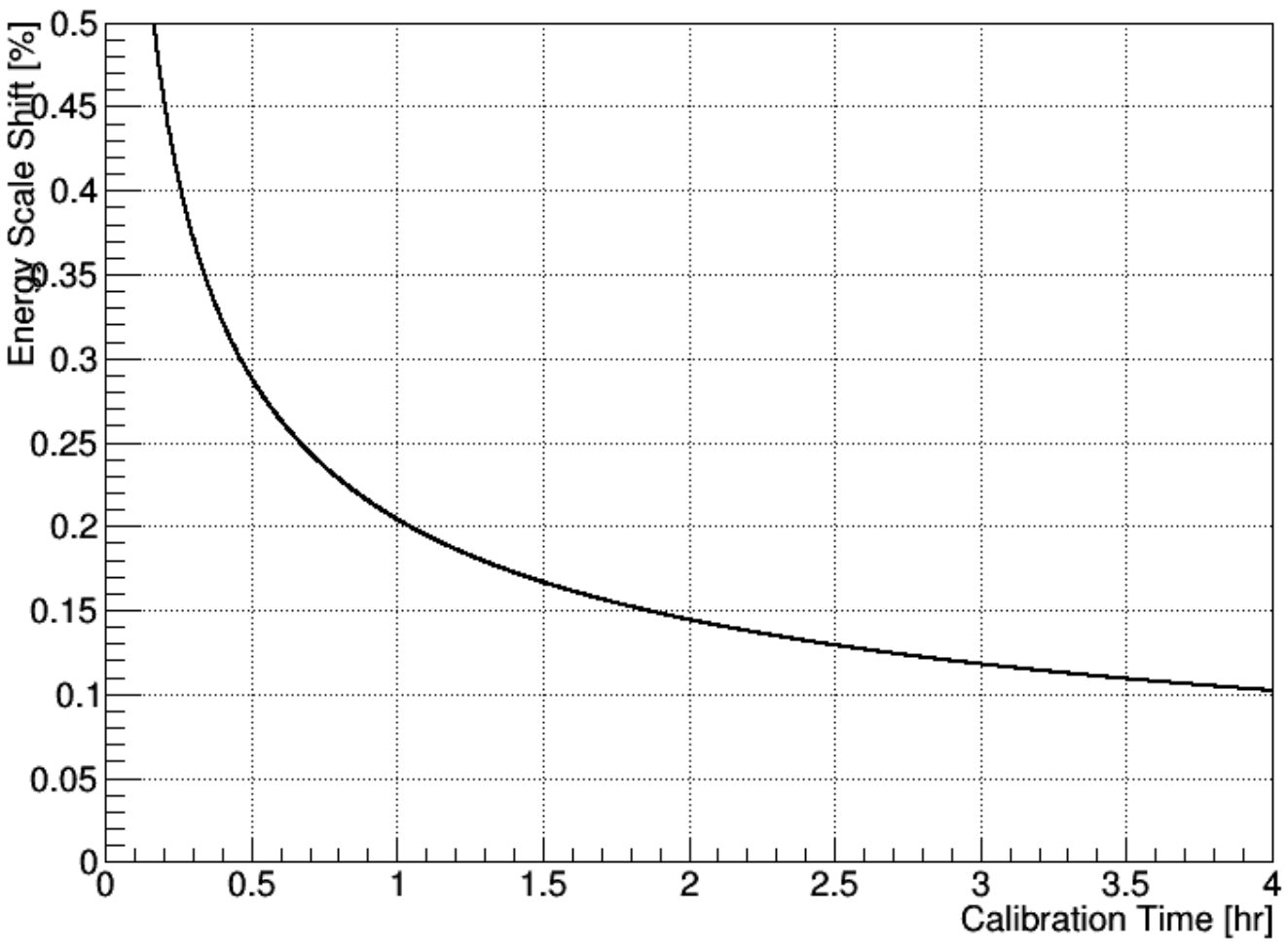

**Figure 4.51:** Energy scale shift as a function of calibration time, assuming optimal source activity in the *realistic* scenario. With a calibration time of 2 hours, the systematic shift in energy scale due to the electron lifetime correction is under 0.15%.

Such an assumption appears to hold true for EXO-200 where no statistically significant spatial variation was observed in the data. For a large detector like nEXO, it is conceivable that spatially dependent impurity levels can develop over time. Additional studies will be carried out to understand the impact of possible electron lifetime non-uniformities on the charge measurements.

#### 4.4.3.3 External Source Calibration Campaign

Besides $^{228}$Th source calibration every few days, source calibration campaigns with various external sources will be used to measure non-linearity in the detector energy response. Non-linear responses have been observed in large LXe detectors. For example, the energy response for EXO-200 is best modeled by a quadratic function. Typical sources will include $^{60}$Co, $^{137}$Cs, $^{226}$Ra and $^{228}$Th. For EXO-200, quarterly 5 day source calibration campaigns are sufficient to establish the energy response curve across the entire energy range. For a larger detector like nEXO, more calibration data will likely be needed, but much stronger sources can also be used. Additional studies will be conducted to establish the complete baseline source calibration campaign plan for nEXO.

### 4.4.4 External Source Hardware

The external source calibration described above is to be achieved by inserting sources in copper guide tubes that wrap around the LXe vessel. The external source hardware consists of a) two guide tubes (GT); b) source-cable assemblies (SCA); and c) deployment-retraction manipulators (DRM). The design of these elements is inspired by the hardware developed for EXO-200, which uses a single guide tube and performs satisfactorily. We first present the EXO-200 design for reference, followed by a discussion of some specifics relevant for nEXO.

The EXO-200 guide tube is a permanent structure and to minimize its contribution to the background, the whole system was designed around the small, 0.7 mm diameter sources commercially available. The sources are sealed in a miniaturized stainless steel capsule crimped to an 8 m long cable. The cable is made from stainless wire rope on which eight hundred 6 mm diameter Teflon beads are held 8 mm from each other by crimped stainless steel tube sections. The beads reduce friction with the guide tube and offer grip for a sprocket kept outside the detector to deploy and retract the source-cable assembly. The sprocket is held in a sealed polycarbonate cassette where the source is parked when not in use. The tail cable is kept in a sealed Teflon tubing permanently attached to the cassette for storage. For source deployment, the cassette is mounted on a load-lock mechanism permanently attached to the guide tube. To prevent damage to the source assembly, the sprocket axis is equipped with a torque-limiting, custom-designed clutch. To prevent the accumulation of $^{210}$Pb or ice by air entering the cold tube, the load-lock is equipped with a purge gas system. The EXO-200 calibration hardware was used for over a thousand deployments over 5 years. Torque-distance data was collected for each source deployment for quality control purposes. No measurable wear of the hardware over time, nor substantial variability between sources or operators were observed. EXO-200 data suggests that the length of the cable should be kept below 15 m and the sum of the bend angles made by the guide tube in its path be less than 1000 degrees. For nEXO, a configuration of two guide tubes, each conforming to one of the TPC end-caps is a possible design keeping in mind these constraints. A double-ended calibration tube may be adopted whereby the action of pushing the source cable in can be aided by aspirating gas from the other end of the tube.

### 4.4.5 Scintillation Light Map

nEXO needs to calibrate the light collection efficiency as a function of event position. This calibration corrects the portion of the event energy measurement that depends on the total light collected. To determine this correction, nEXO will produce calibration events throughout the TPC at a known energy or energies and observe how the measured light of these events varies according to event position. The final result of the calibration is a light response function (LRF) that describes the measured light response as a function of energy and position.

The LRF must be measured to high precision and accuracy, as uncertainty in the LRF contributes to the energy resolution. nEXO's nominal target is 1% LRF accuracy over the entire TPC volume, which is expected to have a subdominant effect on the overall energy resolution. The total effect of LRF precision on nEXO's capabilities is an area of active investigation, and preliminary work show that there is only minor benefit to further improving beyond the 1% LRF target. Assuming nEXO reaches its 1% overall energy resolution goal with 1% LRF error, improving to zero LRF error would improve nEXO's energy resolution to 0.94%, while if nEXO fails to calibrate the LRF to better than 2% error, the overall energy resolution is estimated to rise to 1.17%. However, the LRF precision will likely vary over location in the TPC. Further studies are needed to determine, for example, whether nEXO's sensitivity depends more on the LRF precision in the center of the TPC or towards the edges of the TPC.

As already emphasized in previous sections, discussing the external source calibration, $\gamma$ sources external to the TPC limit the number of calibration $\gamma$'s reaching the center of the detector. A source dissolved in the LXe or with greater penetrating power than $\gamma$s will not face this difficulty. A monoenergetic source is capable of determining the LRF at a point using fewer events than a source producing a broad spectrum. To weigh the effect of these tradeoffs, we have developed the nEXO light map calibration framework. This software takes as input a hypothetical true LRF and a simulated calibration run and outputs the measured LRF the experiment would produce with such a calibration run as well as a comparison between the measured and true LRF. The framework is designed with a flexible machine-learning based interpolation to determine an LRF at all points from calibration events at discrete points.

The framework has been exercised on two example calibration methods using a single example true LRF. The LRF was generated starting from an optical simulation of an ideal TPC and then altering that by depressing the light collection in a particular region. This sort of localized variation in light collection is consistent with the observations of EXO-200. The two example calibration methods evaluated were the baseline calibration (described in Section 4.4.3) and a calibration based on neutron inelastic scattering.

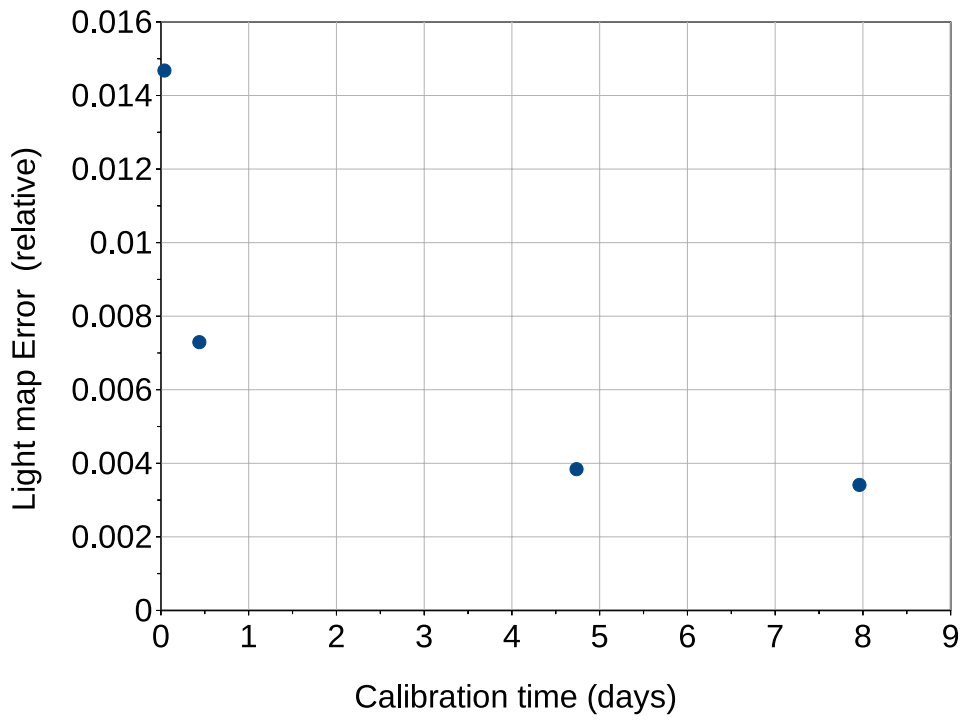

**Figure 4.52:** Results of a trial of the light map framework evaluating the baseline external calibration method using an ideal, uniform detector response.

The baseline calibration was evaluated assuming the scenario described in Section 4.4.3.1. In the case of a smoothly varying LRF, an uncertainty of 1% (0.5%) can be achieved in $< 1$ day ($\simeq 5$ days). These results are shown in Figure 4.52. While it is likely that such a large fiducial mass will have a fairly uniform lightmap response especially in the inner volume, experience from EXO-200 indicates that substantially more statistics might be required to account for non-uniformities in detector geometry as well in the detector response. Further studies incorporating realistic detector variations will be required to determine the duration of the external source calibration campaigns required to achieve the requisite LRF accuracy.

### 4.4.6 Emergent Ideas

The baseline calibration gives a measurement of the drift velocity, a light map, and the overall resolution function in charge, light, and charge-light anti-correlation at regular time intervals, in a manner similar to that done in EXO-200. It may be possible to determine if there is spatial dependence in the charge energy in some parts of the detector, but for the inner 1000 kg, for example, only one average number will be generated. Since nEXO is a substantially larger detector than EXO-200 with longer electron drifts, it is prudent to consider additional ideas that might provide finer spatial and/or temporal measurement of the key calibration quantities.

Here we discuss the possibility of injecting $^{220}$Rn or $^{222}$Rn in the LXe, with the purpose of generating events deep in the TPC. This method may provide a more uniform event distribution in the detector and could be used to occasionally map, over the entire volume, the scintillation collection and the LXe purity. Another possibility being investigated is to have a continuous monitoring system relying on calibrated laser pulses driving judiciously placed photocathodes or tuned to produce localized charge clusters from multi-photon ionization of LXe. We elaborate briefly on these ideas in the following.

#### 4.4.6.1 Radon Injection

Injection of the short-lived isotope $^{220}$Rn and the longer-lived $^{222}$Rn have been successfully demonstrated as an internal calibration source in dual phase time projection chambers [61–64]. In these works, radon is mixed with xenon gas in the circulation loop used for purification. For $^{220}$Rn, as the decay chain ends with stable $^{208}$Pb, no long-lived radioactivity contamination is introduced from this isotope injection, assuming that no other radionuclide is transported by the gas flow, as has been demonstrated [61].

Due to the short 55.6 second half-life of $^{220}$Rn, an optimization of source activity and path length for the injection is required, but the technique has been shown to provide a uniform distribution of $^{220}$Rn decay products throughout a liquid xenon volume up to 3.5 tonnes [62, 63]. Each of the steps of the decay chain are clearly identifiable, allowing spectroscopy using $\alpha$s of different energies, $\beta - \gamma$ sum energies, and delayed coincidence of multiple decay products [61–63]. Additionally, the 10.6 hour half-life of $^{212}$Pb separates the calibration into two phases, a grow-in phase when the source is being injected, then a second phase when the injection stops, and only the $^{212}$Pb decay chain is present. This has been used to demonstrate a time dependent identification of the full decay chain, whereby the movement of the isotopes can be followed through the detector [61, 62]. The grow-in and subsequent decay of the intermediate $^{212}$Pb requires several half-lives between the start of a calibration and the return to low-background conditions, which sets the timescale for this calibration to several days, and is thus only implemented at a frequency on the order of months [63].

For $^{222}$Rn, the injection is simplified due to the longer half-life of 3.8235 days, which would allow a weaker source and more thorough mixing in the xenon [64]. Since all decays in the early part of the chain lead to the long-lived nuclide $^{210}$Pb with a 22.2 year half-life, only these isotopes in the early portion can be used for calibration. But this still provides several distinct $\alpha$s that can be used for a light map. However, estimates of the integrated buildup of $^{210}$Pb due to such a calibration are at the level of the total background budget. Thus, $^{220}$Rn is considered as a first option and in the event it is unfeasible after further study, $^{222}$Rn will be a secondary option, with the mandate that a filtering method be implemented to reduce the $^{210}$Pb, in turn inducing ($\alpha$,n) background, to an acceptable level.

#### 4.4.6.2 Laser Driven Photocathodes

Two other emergent ideas that employ laser pulses, injected through optical fibers, are being considered for charge and purity calibration on a finer time and spatial scale than the primary calibration method. If demonstrated to be viable, these would complement but not replace the other calibration methods. The first laser method involves injection of calibrated electron pulses generated by the photoelectric effect from a pulse of UV light on a gold photocathode surface. The concept is illustrated in Figure 4.53 (left), along with a diagram of the apparatus where tests are ongoing (center). Laser light at 248 nm or 262 nm passing through a fiber and a transparent photocathode generates photoelectrons on a semi-transparent gold coating on the cathode surface. The charge of the initial pulse of electrons is measured in the test apparatus by the induction signal on a grid 1 mm from the cathode. A second pulse is measured at the anode. The ratio of the two pulses gives a measure of the electron attenuation due to impurities. Electron drift velocity, longitudinal diffusion and purity are among the measurements being made in this test setup.

A second emerging laser method is based on multiphoton ionization of liquid xenon by pulsed UV lasers. Two- and three-photon ionization of LXe, e.g., with 266 nm and 355 nm beams, have been investigated within the nEXO Collaboration and cross sections have been measured. A concept for nEXO calibration by multiphoton ionization is shown in Figure 4.53 (right). UV laser beams could be brought in by optical fibers attached to the field cage insulators and then propagated in a parallel or focused beam geometry. While a parallel beam would provide a relatively uniform line of charge in any desired direction, our plan is to investigate how to provide a fo-cused beam to provide a localized source of charge. Due to the nonlinear multiphoton ionization process, significant charge is only created near the focus. Another possibility is to create localized charges with parallel beams by a phase dispersion method on a femtosecond laser that comes to constructive interference at a desired (and variable) distance along the beam. Various beam geometries are conceivable. It is worth noting that the attenuation length for these laser beams is estimated to be on the order of 7 m in LXe.

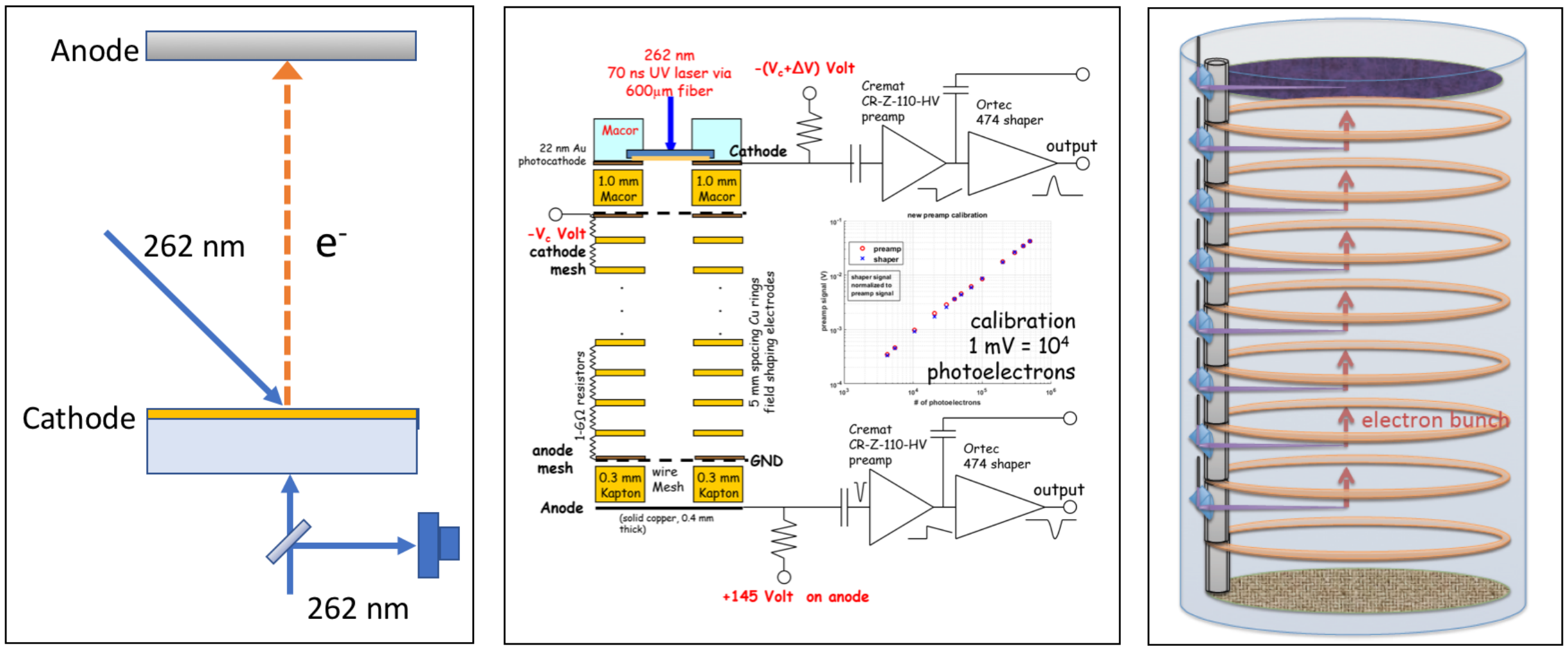

**Figure 4.53:** Concept of laser calibration by photoinjection of charge (left); SBU-BNL test setup (center); possible concept for calibration by multiphoton ionization of LXe (right).

Since these two methods do not involve movement of sources and are rapid, they could be implemented at regular intervals, perhaps hourly, between baseline calibrations. These methods are also particularly attractive for probing the inner 1000 kg as they do not suffer from the $\gamma$-ray attenuation limitations of the primary calibration. The background impact of the required calibration hardware will have to be carefully studied.

## 4.5 Outer Detector[9]

Large water tanks are commonly used to shield against external backgrounds and veto the passage of muons. This approach is attractive because of cost considerations, hermeticity and the possibility of being instrumented as Cherenkov counters. While the low density of water increases the required thickness, its hydrogen content makes it ideal for neutron moderation. The light nuclei in water also minimize neutron production by cosmic-ray spallation.

A tank of ultra-pure deionized water has been identified as shielding of choice for nEXO. The cryostat, containing the TPC, will be submerged and placed at the center of this water volume, serving three purposes: (I) shield against $\gamma$ rays originating from radioactive decays in the walls of the underground cavern; (II) moderate and stop neutrons also produced in the walls and (III) detect cosmic radiation, i.e. muons, passing through the water and potentially producing correlated events in the TPC. Items (I) and (II) require passive shielding only, while the detection of cosmic muons requires instrumentation. In addition to rejecting cosmogenic backgrounds, the direct measurement of incident muons allows the validation of the Monte Carlo simulation of cosmogenic backgrounds. While the exact location of the experiment is still under discussion, here for sake of concreteness, we assume the detector to be installed in the Cryopit at SNOLAB. Table 4.8 shows measured and calculated muon and neutron yields at this location.

[9]We are grateful to D. Sinclair who contributed the initial draft of parts of this section and other useful comments.

| | | |
|---|---|---|
| Total muon flux | $(0.326 \pm 0.035)\mu$ m$^{-2}$day$^{-1}$ | [65] |
| | $0.27\mu$ m$^{-2}$day$^{-1}$ | [66] |
| Measured thermal n flux | $4144.9 \pm 49.8(\text{stat}) \pm 105.3(\text{syst})\text{n m}^{-2}\text{day}^{-1}$ | [66] |
| Estimated fast n flux | 4000 n m$^{-2}$day$^{-1}$ | [66] |

**Table 4.8:** Neutron and muon fluxes at SNOLAB. References are listed for each value.

In order to estimate the minimum thickness of water required, Figure 4.54 shows the calculated rate of background from external $\gamma$ rays, as a function of the radius of the water tank (resulting in a certain water thickness). Here the background rate is defined as described in the figure caption and the external activity is dominated by the shotcrete on the walls, with $^{226}\text{Ra}(^{238}\text{U}) = 1.11 \pm 0.13$ ppm $= 40.6$ Bq/kg and $^{232}\text{Th} = 5.56 \pm 0.52$ ppm $= 22.6$ Bq/kg [67]. Radon is assumed to be emanated by the stainless steel of the water tank and released by recoil, at a rate of 262 atoms/day for the entire water tank [68], corresponding to a 10 ppb concentration of $^{238}$U in the stainless steel. In the simulation, this radon was uniformly distributed in the water. The radon curve scales differently from the one for external backgrounds because of the uniformity of the radon source in the water. Backgrounds from external neutrons are expected to be negligible for all water thicknesses considered here. External backgrounds, attenuated by the water, are to be compared with the sum of all internal backgrounds and become negligible for radii greater than $\sim 4.5$ m.

Photomultiplier tubes (PMTs) will be installed on the outer walls of the water tank, where they are shielded by the water with respect to the central detector, to detect the Cherenkov light produced by muon tracks. The information collected in this way will be added to the data stream so that correlation with events in the TPC can be studied offline (see e.g. [4]).

### 4.5.1 Water Tank

Applying some safety margin, Figure 4.54 shows that any tank radius greater than 4.5 m would suffice to make external backgrounds negligible. However, a preliminary cost analysis shows that the savings from building a tank that is smaller than the full 13 m diameter available in the Cryopit are rather modest. This is particularly true because it is probably convenient to build the tank to the full height of 14 m, so that its top “deck” is level with the upper access drift. In addition, a larger tank could later accommodate an upgraded version of the detector. At the same time, a smaller diameter tank would have the advantage of leaving room on the side for the location of the liquid portion of the Xe recirculation system and possibly of the HFE-7000 system. This could allow for smaller liquid heads on the TPC and the inner vessel of the cryostat, simplifying some of the operations (see Section 4.3.1). More advanced engineering of the LXe and HFE-7000 handling systems will be required before freezing the diameter of the water tank. In addition, details of the tank will depend on the actual location chosen for the experiment.

For the purpose of this document, we assume a 13 m diameter, 14 m tall tank, shown in Figure 4.55, notionally filling most of the volume of the Cryopit. We note that the preliminary cost analysis found that at the SNOLAB location a free-standing structure is not more expensive than a tank resting against the rock and is more convenient in terms of accessibility and location of auxiliary infrastructure.

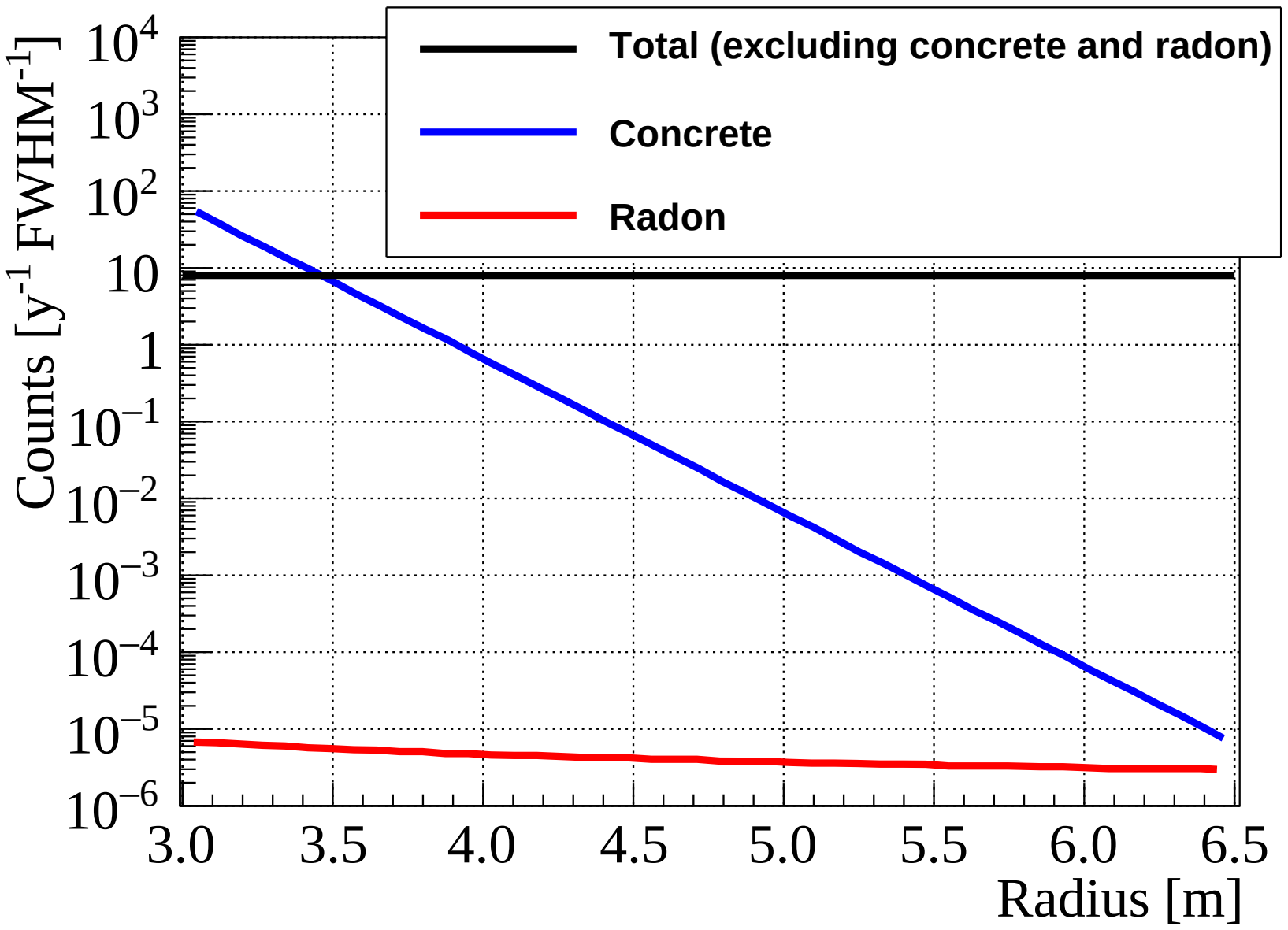

**Figure 4.54:** Background rate from external $\gamma$ rays and radon dissolved in the water as a function of the water tank radius and, hence, shielding thickness. While, as discussed, a universal background figure cannot be quoted for nEXO, where signal and background are simultaneously calculated by a fit to the data, for concreteness here we show rates which are calculated in one FWHM around the $\beta\beta$ decay $Q$-value, conservatively using the entire fiducial volume of LXe. In the calculation the total height, $h$, of the water tank is related to its radius, $R$, indicated by the abscissa, as $h = 2 \times R + 1$ m. The horizontal line represents the upper bound of all combined backgrounds, produced inside the water tank, in the same energy region and in the full fiducial volume.

The tank is anticipated to have an integral deck on top, with a central hatch to access the cryostat, as shown in Figure 4.56, in addition to various other penetrations (not shown). The cryostat, containing the TPC, will be supported with a truss from the deck, on which much of the process equipment will be located. An artist's view of the configuration of water tank, detector, and support equipment is shown in Figure 3.1. Power and signal cables are routed vertically, from the top of the cryostat to the deck, discussed in detail in Section 4.3.2. A similar routing is also envisaged for cooling services, HFE-7000 and Xe pipes, unless the level solution is chosen, whereby all LXe and HFE-7000 handling occurs on the side of a (smaller) tank, roughly at the same level as the cryostat. The system pressures of Xe and HFE-7000 are discussed in Section 4.3.1.

Depending on the fabrication technique selected for cryostat and TPC, the integral deck may have to carry additional load during assembly and testing. These various load scenarios must be considered in the design of water tank structure and integrated deck. It is expected that the water tank will be built out of stainless 304L steel welded in situ, because of its size [69]. The assembly sequence has to be defined once the dimensions of cryostat, TPC and support structure have been finalized. It is possible that the tank can be initially used as clean space in which the cryostat and, potentially, the rest of the detector are built, to then be hoisted up to their final configuration.

Mining activity causes rock bursts in the vicinity of SNOLAB. A study found a 9% probability of a Nuttli Magnitude 4.3 event occurring within a distance of 160 meters at least once during any 10 year period (see [69] for details and references). This corresponds to a peak velocity of 800 mm/sec leading to a loading approaching $20g$ without mitigation. More work is required to understand this risk and design an appropriate seismic damping system, as has been done for others experiments at SNOLAB.

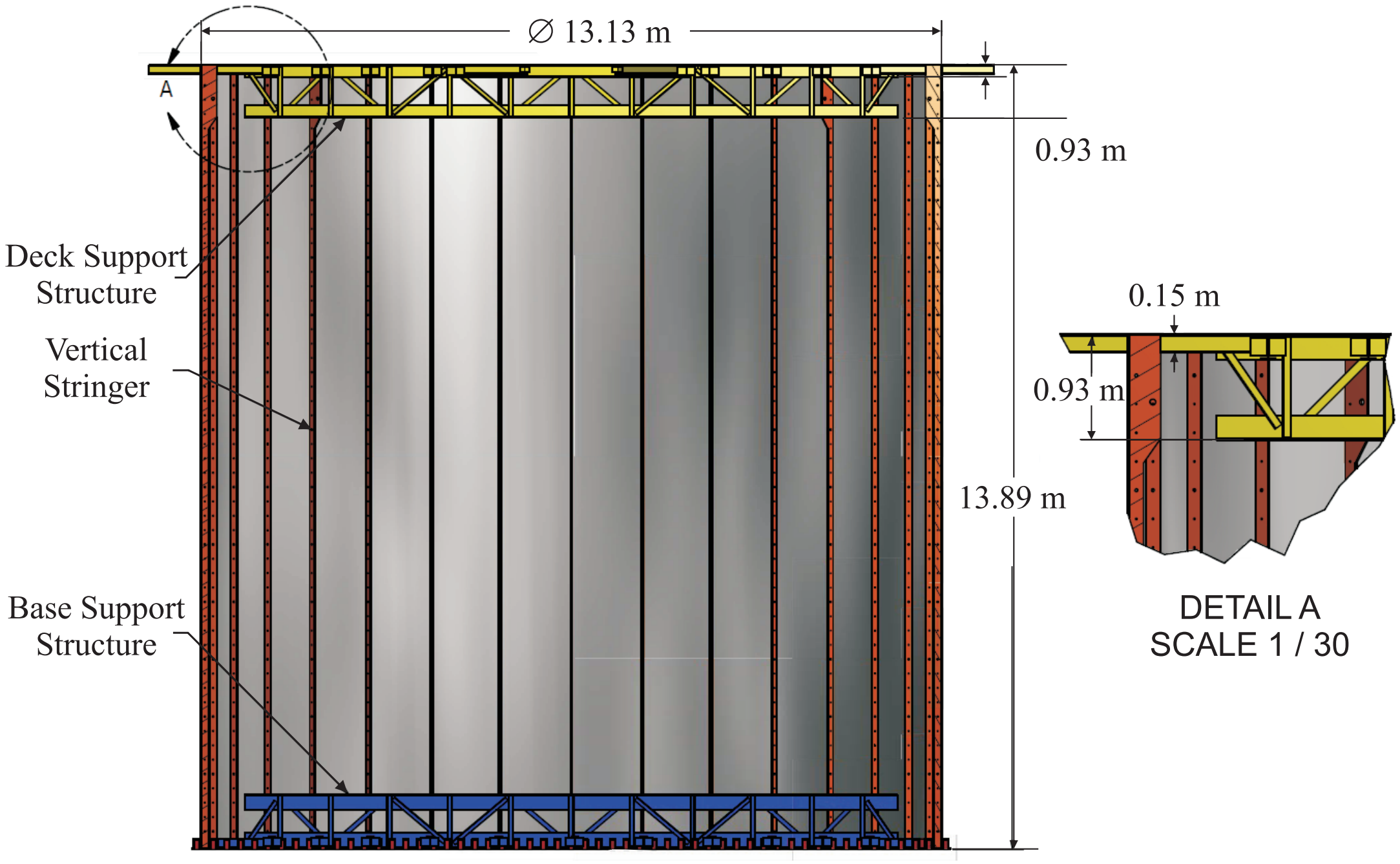

**Figure 4.55:** Cross-section view of a pre-conceptual design of the water tank. Support structure of the integral deck as well as the base structure are shown in yellow and blue, respectively. Figure taken from [69].

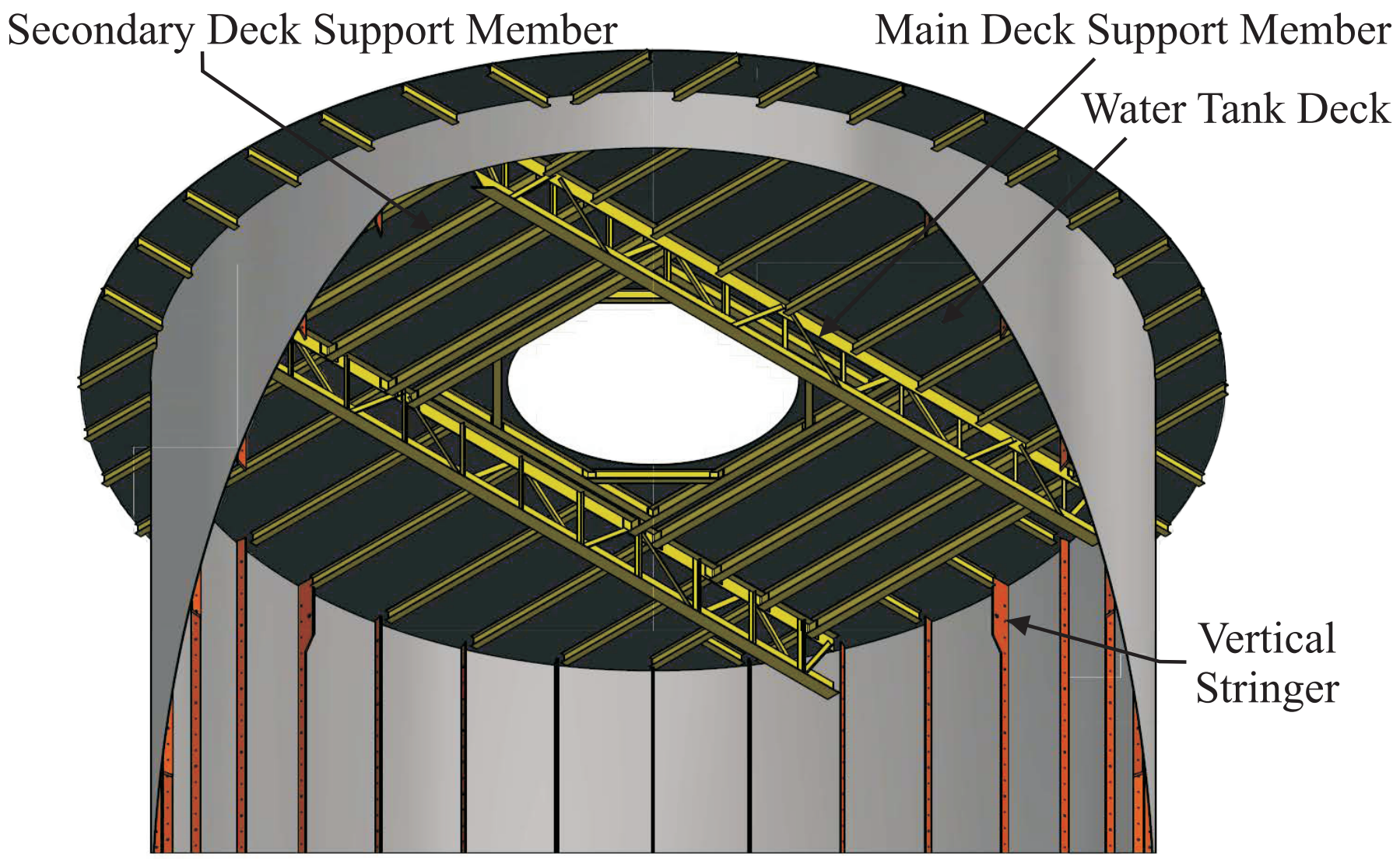

**Figure 4.56:** Cutout view of the water tank, showing the integral deck. The opening will allow the installation of cryostat and TPC as well as installation of read-out infrastructure for the detector. Figure taken from [69].

### 4.5.2 Photomultiplier System

The water tank will be instrumented with PMTs to record Cherenkov light and hence tag traversing muons. The PMT coverage has not been optimized yet but, based on the experience of similar detectors, it is expected that on the order of 100 8-inch PMTs will be required to obtain a simple muon tag. Higher PMT coverage would be required in order to measure the properties of passing muons.

About 500 8-inch PMTs (Hamamatsu 10-stage model R5912) complete with waterproof base assemblies are available from the Daya Bay experiment [70, 71] and reserved for the use in the nEXO muon veto. Each PMT has been pre-assembled with a $\mu$-metal magnetic shield [72] and a bracket, and mounted on a so-called ”Tee” support structure, as shown in Figure 4.57. The latter is then fixed onto a wall module. These 8-inch PMTs were tested to withstand a pressure of up to 7 atm, several times the maximum pressure in the nEXO veto counter.

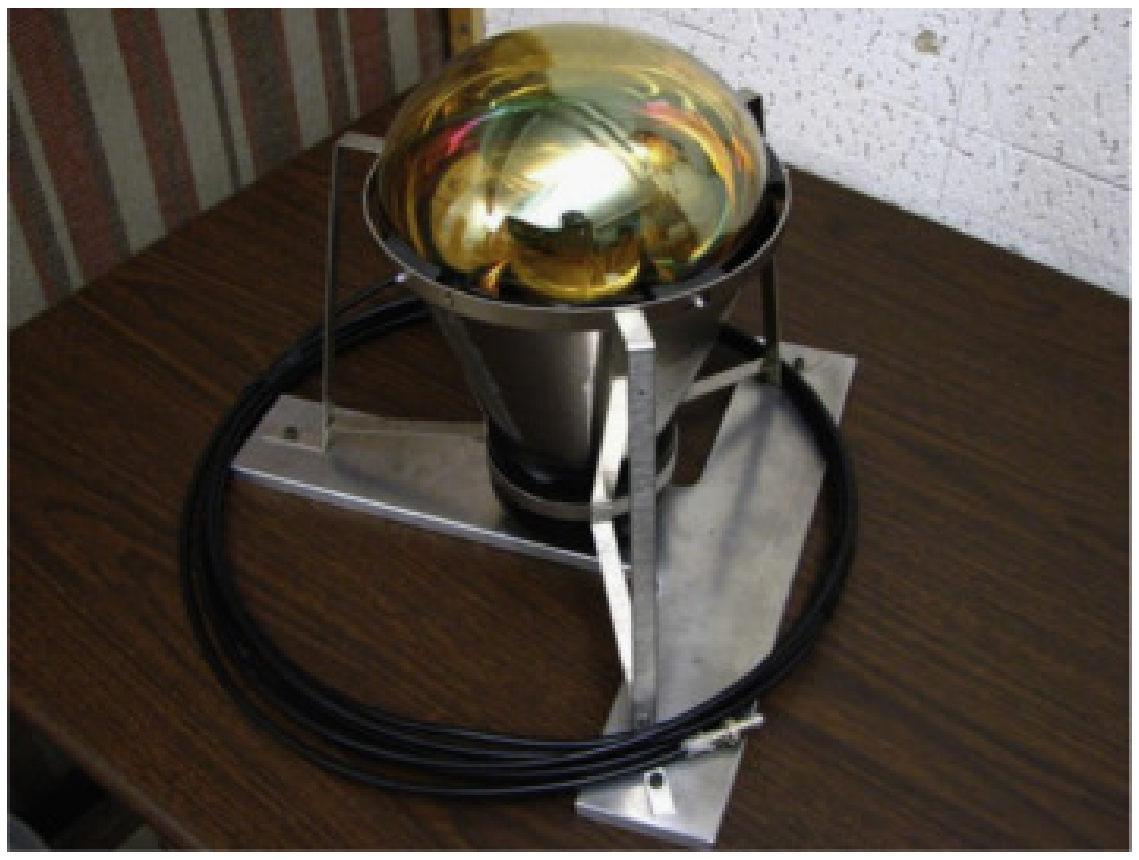

**Figure 4.57:** A waterproof PMT assembly, including PMT, base, cable, magnetic shield, bracket and supporting structure.

The inward-facing PMTs will be mounted either directly on the tank walls or on frames. All surfaces inside the water tank will be covered with highly-reflective material to improve the photon collection efficiency. For instance, Daya Bay uses a highly reflective multi-layer film formed from two layers of 1082D Tyvek bonded onto a layer of polyethylene, for which the reflectivity in air is more than 96% for wavelengths from about 300 to 800 nm. The reflectivity is 99% in water [73], and the reflectance is diffuse with a small specular component. Using all 500 PMTs available and distributing them evenly on the lateral surfaces and the top and bottom disk areas, one could achieve a density of $\approx$ 0.5 m$^{-2}$, corresponding to a photocathode coverage of $\approx$ 2%. With such coverage, the muon tagging efficiency is expected to be well above 99%, based on the experience from similar detectors. Detailed Monte Carlo studies will be performed to optimize the PMT configuration and define performance of the veto detector.

The Daya Bay PMTs are designed for positive high voltage and equipped with a single 50 $\Omega$ coaxial cable that is compatible with ultra-pure water. Failure rates of $\simeq$ 2% in five years were observed. The properties of the PMTs are listed in Table 4.9.

Timing and gain stability of the PMTs will be monitored by a pulsed-LED system. In addition, a periodic trigger can be implemented in the DAQ system to perform continuous PMT gain calibration from dark noise.

| Quantity | Specification |
|---|---|
| Peak Quantum Efficiency | $>$ 25% at 420 nm |
| | 8% at 300 nm |
| | 12% at 320 nm |
| | 1% at 600 nm |
| Gain | $10^7$ |
| Single Photoelectron Resolution | Peak-to-valley ratio $\geq$ 2.5 |
| Magnetic Field Sensitivity | Gain, resolution and timing constant to $\geq$ 20% |
| | in the presence of transverse magnetic field of 450 mG |
| Photocathode Uniformity | better than 15% at 420 nm |
| Pulse Linearity | better than 5% over the dynamic range of 0–1 nC |
| Dark Pulse Rate | $\leq$ 10 kHz above 0.25 photoelectron at 20°C |
| Pre-Pulsing | $<$ 5% of the charge in the 10 ns preceding a pulse |
| After-Pulsing | $<$ 10% of the charge in the 10 ns following a pulse |
| Gain Stability | $<$ 5% drift per week |
| | $<$ 10% drift over a period of one year |
| Temperature dependence | $<$ 1% $°C^{-1}$ |
| Rise Times | $<$ 5 ns for single photoelectron pulse |
| Fall Times | $<$ 10 ns for single photoelectron pulse |
| Transit Time Spread (FWHM) | $<$ 3 ns for a single photoelectron |
| Radioact. Specifications per PMT | $^{40}$K: 2.7 Bq, $^{232}$Th: 0.5 Bq, $^{238}$U: 0.64 Bq |
| Radioact. Specifications per Base | $^{40}$K: 0.14 Bq, $^{232}$Th: 0.28 Bq, $^{238}$U: 0.20 Bq |

**Table 4.9:** Specifications of Daya Bay PMTs, listed for a gain of $10^7$.

### 4.5.3 Veto Readout Electronics

The complete PMT readout system from the Daya Bay experiment can also be reused, including the high voltage modules, front-end electronics (FEE) [74], HV-signal decouplers, and cables. The system was designed to:

- Determine the charge from each PMT signal.
- Provide precision timing information for muon tracking.
- Generate multiplicity (nPMT) triggers, i.e. record the number of PMTs exceeding a preset threshold (*e.g.* 0.25 p.e).
- Generate total energy (ESUM), also to be used for triggering
- Digitize at 1 GS/s sum waveform of groups of 32 PMTs to provide a redundant crosscheck of the primary data.

While, at present, this system appears to be overkill for the purpose of the nEXO veto, it is available at no (or negligible) cost and is well characterized; hence it is considered the primary design. The most important parameters of the PMT front end are listed in Table 4.10.

A simplified circuit diagram of the electronic readout system, showing its main functions, is given in Figure 4.58.

Each FEE VME board accepts 16 PMT signals and performs time and charge measurements. The number of channels over threshold and the total charge observed by the FEE board is fed to the trigger system for a fast trigger decision. After collecting information from all readout boards, a trigger signal may be generated and distributed to each FEE board and used as a common stop

| Quantity | Specification |
|---|---|
| Full Charge Dynamic Range | 1.6 - 1800 pC |
| 0.5cm Fine Range | 1.6 - 160 pC |
| Coarse Range | 160 - 1800 pC |
| Shaping time (to 1% of peak) | $< 325$ ns |
| Error on sampled peak | 4% at 40 MS/s |
| ADC granularity | 10% of a single p.e. |
| ADC resolution | 12 bits (on each range) |
| ADC Sampling Rate | 40 MS/s |
| Disc. Threshold | $\geq 0.25$ p.e. (programmable for each channel) |
| Frame length | 0-1.3 $\mu$s |
| Time Bin | 1.56 ns |
| Timing Precision (RMS) | <1 ns |
| Multi-hit Separation | Yes |
| Multi-hit Resolution | 50 ns |

**Table 4.10:** PMT frond-end readout electronics specifications.

for the TDCs. This also initiates the readout of the ADC and TDC data.

For each PMT channel, a precision discriminator provides the start for a TDC that is obtained as part of a Field Programmable Gate Array (FPGA). The RMS timing resolution is better than 0.5 ns. Also for each PMT the charge is measured with a dual-range system and digitized at 40 MS/s. Data is then processed by the same FPGA (for each 16-channel card), also providing range selection, peak finding, data pipelining, pedestal subtraction, nonlinearity corrections, and data buffering. Self testing and calibration is accomplished using a programmable pulse, generated by a fast on-board DAC chip and sent as a calibrated input signal simultaneously to all channels on the board.

16 front-end boards are housed in a single 9U VME crate, additionally containing one trigger board, and one fan-out board. Therefore the maximum contingent of 500 PMTs can be readout by two crates.

In a detector like nEXO, data in the veto counter and in the TPC are, to a large extent, independent from one another (although, of course, the interest is in the rare correlated events). Hence, in nEXO the veto readout is expected to collect data asynchronously from the TPC, for offline re-synchronization (this is also the way the EXO-200 plastic scintillator-based veto detectors are handled). The substantial functionality of the front-end system described may be used for additional purposes, such as to further study and constrain background models, as will be studied in the future.

### 4.5.4 Water Purification System

The purpose of the water system is to produce and maintain the purity of the water with respect to radioactive contamination and optical transmission. To accomplish this, the water will be continuously recirculated through a system that can remove ions that may be leached from the detector materials. Organic substances that may shed from the detector structure will also be removed by the same system. It is important to control any biological activity that could lead to bio-films,

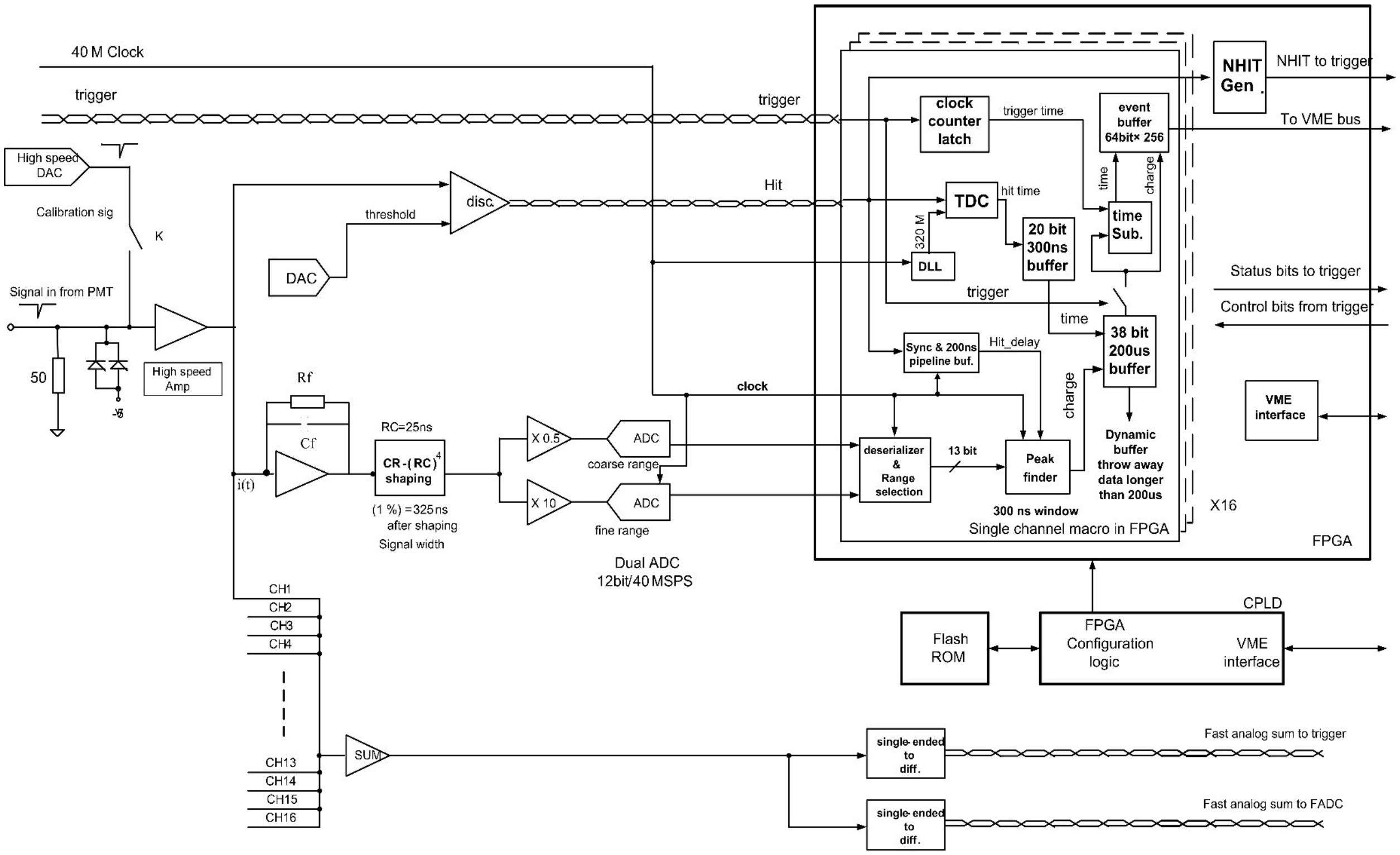

**Figure 4.58:** PMT front-end block diagram.

particularly on the photomultiplier tubes.

Effects of $^{238}$U, $^{232}$Th, and $^{40}$K impurities on the ROI background rate have been investigated [75]. Assuming 1 ppt U, 1 ppt Th, and 1 ppb K in water, the veto detector contributes 0.43% of the total estimated background. The K contamination only affects the low energy part of the spectrum; even at the 1000 ppb level the contribution on the $2\nu\beta\beta$ background is negligible.

Based on radioassay studies of purified water by other experiments [76, 77], concentrations of $< 0.1$ ppt U, $< 0.1$ ppt Th are readily achievable. For nEXO the 2448 keV $\gamma$ from $^{214}$Bi decays in the $^{238}$U chain is an important concern. Rn permeation by means of diffusion or through small cracks will have avoided. To address this specific concern, all systems will be made from stainless steel and all joints will be welded or will use metal seals (e.g. Conflat or VCR). All water pumps will be magnetically coupled to avoid Rn ingress along shaft seals. The surface areas of the water purification system and tubing are small compared to the area of the water tank, thus, radon emanation from these systems is expected to be negligible compared to emanation from the water tank.

The requirement on the water temperature depends on the desired dark rate of the PMTs, which is temperature-dependent. A lower water temperature (e.g. 12°C for SNO or 10°C for KamLAND and SuperKamiokande) can be desirable to reduce the dark rate of the PMTs, but in the case of the nEXO veto, with the primary goal of detecting cosmic-ray muons producing a substantial amount of light in the water, this is a less important consideration. For water temperatures below the dew point (around 10°C at SNOLAB), thermal insulation of the nEXO water tank would be required.

At SNOLAB it is likely that nEXO could receive pure water from the SNOLAB water plant, while at other sites a pure water supply system would be required. In any case a recirculation, purification and monitoring system will be required. The water system will require process control, compressed air for pneumatics, electrical power, space, and cooling power to remove excess heat. In addition, a drainage system has to be considered to discharge $\sim 2000$ m$^3$ of water over a period of a few weeks.

The total volume of the tank is 1800 m$^3$. A flow rate of 100 l/min would recirculate the water in 12 days, which appears adequate, based on the experience from other large water Cherenkov detectors. A monitoring system could sample the water for Rn at various locations in the veto tank and in the recirculation system.

#### 4.5.4.1 Pipe Runs

Assuming the Cryopit location at SNOLAB, the distance from the edge of the veto tank to the process equipment can be taken as 46 m (i.e. the length of the utility drift). For the flow rate of 100 l/min the pressure drop over 46 m is $\simeq 1.8$ bar ($\simeq 0.2$ bar) for 1-inch (1.5 inch) diameter tubing. For electropolished pipes this figure may be slightly smaller, so that it appears that 1.5-inch pipe would be conservative. The monitoring systems, running at a lower flow will employ smaller pipes.

#### 4.5.4.2 Recirculation System

The concept for the recirculation system has the following components:

- *Nanofiltration System*. This system serves to remove thorium and organic material from the water. For a sizing calculation a Synder TFC membrane system is assumed. This is a tight nanofiltration membrane with a molecular weight cutoff of about 150 Daltons, 40% rejection of monovalent salts, and 99% removal of sulphate salts. We assume the use of 5 units of 8-inch modules housed in an existing R/O housing. Assuming 800 l/m$^2$ [78] such a system can process 100 l/min at the nominal 8 bar pressure. A second set of smaller modules would be used to further concentrate the impurities into an ion exchanger (and possibly radium or thorium specific resins) and fine filter.
- *Pumps*. Two single-stage, magnetically coupled centrifugal pumps are anticipated to achieve the required pressure head and a third pump would re-pressurize the water after the nanofiltration system. A fourth pump would recirculate the purified concentrate into the feed of the reverse osmosis purifier.
- *Ultra-Violet Light UV1*. A 185 nm UV irradiation cell aimed at breaking up organic material will be installed in the circulation system. Cells sized for flows similar to the one of interest here were already used in the SNO detector.
- *Ion Exchanger IX2 and IX3*. These are nuclear grade mixed bed ion-exchange tanks. IX2 and IX3 are two resin tanks connected in series. The first is the "worker" and takes out most of the impurities. The second, called "polisher", sees a much lower load and produces the cleanest water. When the "worker" is loaded, it is regenerated, the "polisher" replaces it as the "worker" and the fresh or regenerated system becomes the "polisher". The IXs are assumed to be housed in stainless steel tanks and will be valved to allow either of them to be regenerated.

- *Ultra-Violet Light UV2*. This is a conventional 256 nm UV sterilizer to kill any organisms that might grow on the IX materials.
- *Membrane Degas*. At this early stage it is assumed that the degasser will consist of 3 Liqui-cell[10] modules operated at vacuum followed by three cells operated as a regasser using boil-off nitrogen.
- *Final Filter*. The final filter is a 0.2 $\mu$m filter in a stainless housing.

The IX beds are in the concentrate side of the nanofiltration unit. Since the beds are a possible source of radon, this arrangement minimizes the risk of introducing contamination. A small feed and bleed tank is required to maintain water level in the water tank.

#### 4.5.4.3 Monitoring System

The system to monitor radioactive contaminants could be copied from the SNO light-water monitoring skid. We may replace the spray degasser with a Liquicell degasser. The functions of the monitor system are (I) to measure the $^{224}$Ra and $^{226}$Ra concentrations by trapping these species on MnOx and then measure the evolved Rn using the electrostatic counters on surface and (II) to measure the $^{222}$Rn concentration by degassing the water, separating the Rn from the water vapor by a cold trap, and then transferring the radon to a Lucas cell by nitrogen cooled trapping.

[10] http://www.liquicel.com/

# 5 Facilities

nEXO will require a sufficiently deep underground site to properly operate. While the overburden of SURF [1, 2] at the 4850′ level is expected to be sufficient for the experiment, the deeper SNOLAB site further reduces risk associated with cosmogenic backgrounds and minimizes shielding requirements [3]. Hence, for the purpose of the present discussion, the Cryopit at SNOLAB will be considered the primary location. Details of this section would need to be revised should a different location be chosen; yet it is worthwhile to provide a cursory discussion of facilities to understand the level of complexity required.

The strict requirements on contamination and quality control will require special attention and support from the site material handling system. This includes a sealed transport system, similar to the one already used in the construction of the SNO detector. The Cryopit has access from both the top level and the bottom one, as shown in Figure 5.1. Access for most of the construction will occur from the top, which is within the clean area of the laboratory. Most of the infrastructure, like fluid handling systems, electronics and cryogenics will be located on the deck, although it is possible that the HFE-7000 and the LXe recirculation systems may be located on the outside of the veto tank, roughly at the same height as the TPC, to minimize liquid heads. This option would require a smaller diameter veto tank, as discussed in Section 4.5. Some equipment, such has the $LN_2$ plant may be located in the lower drift, outside of the clean envelope, to simplify maintenance.

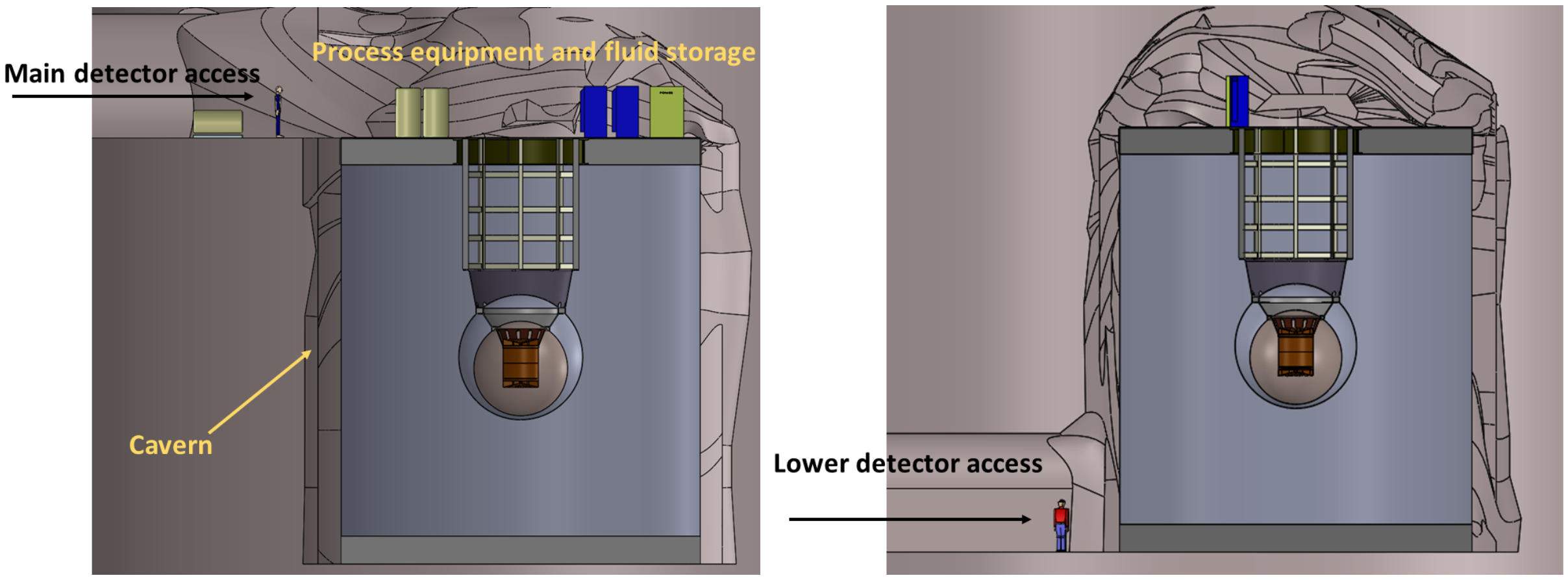

**Figure 5.1:** Access to the nEXO detector in the Cryopit at SNOLAB. Left: main access from the drift at top level. The drift is expected to be roughly level with the deck above the veto tank, from where the cryostat is supported. Most other services are located on the deck. Right: lower access, which is normally sealed because it is connected to the “dirty” portions of the mine. Some equipment may be installed in the lower drift, near the main heat exchangers for SNOLAB. The lower drift is also used for emergency personnel egress.

| Item | Required space | Comments |
|---|---|---|
| Veto Tank | 15 m diameter, 20 m high | includes space around tank and dome headroom |
| Other underground space | 300 $m^2$ | Maybe be reduced as some process equipment may be located on the deck |
| Surface space | 100 $m^2$ | Assumes TPC fabrication and welding underground |

**Table 5.1:** Space requirements for the nEXO detector.

| Equipment | Normal power | Max outage duration | Backup power |
|---|---|---|---|
| $LN_2$ plant | 60 kW, 3Φ, 480V | 1 week | 0 |
| Control system | 5 kW | 0 | 5 kW |
| Readout electronics and DAQ | 5 kW | infinite | 0 |
| Veto system | 3 kW | infinite | 0 |
| Xe recirc. pumps | 5 kW | 0 | 5 kW |
| Xe boiler and condenser | 15 kW | 0 | 15 kW |
| Vacuum pumps 1 | 2 kW | 0 | 2 kW |
| Vacuum pumps 2 | 3 kW | infinite | 0 |
| HFE-7000 pump(s) | 3.6 kW | infinite | 0 |
| HEPA filters | 5 kW | infinite | 0 |
| Getters | 7.5 kW | 0 | 7.5 kW |
| Total | 114.1 kW | – | 34.5 kW |

**Table 5.2:** Operations electrical power estimates.

The interface between nEXO and the laboratory will be key, both during construction and operations. Document systems, drawing interface control, engineering interface support will all need to be established with the site. Agreements for document review, approval, and storage will need to be made. Facility work control requirements will need to be integrated with the nEXO processes. Project management of the nEXO project and the site will need a coordination liaison office to support the transfer of information and work control between the two. This will be essential also to properly coordinate EH&S issues.

While discussions on nEXO requirements have begun with the management of SNOLAB, here we provide a terse list of functions and activities that will be required. During installation, in the underground, support will be needed for:

- Subsystem assembly;
- Machining in the underground location;
- Special assembly (copper e-beam welding and carbon-fiber winding);
- Material and subsystem interim storage;
- Utility connections to the detector subsystems;
- The detector location will need to be finished to a cleanliness level similar to that of other

| Equipment | Power |
|---|---|
| Carbon composite winder | 25 kW, 3Φ - 480V |
| Ebeam welder | 5 kW |
| Machining | 15 kW |
| HEPA filtration | 10 kW |
| Crane(s) | 10 kW |
| Battery powered equip | 5 kW |
| Total | 70 kW |

**Table 5.3:** Construction electrical power estimates.

clean areas at SNOLAB. The precise cleanliness level is still to be determined;
- Transportation of parts and materials from the surface to the detector site.

Above ground facilities will also be needed to:
- Receive parts;
- Stage parts and provide assembly space;
- Provide personnel office space;
- Provide meeting space;
- Support offsite communication;

Exact infrastructure requirements will be formulated as the initial engineering design is accomplished. Here we provide some preliminary estimates, based on the present understanding of the detector and previous experience with other similar projects.

Space requirements are provided in Table 5.1. Preliminary electrical requirements for the detector operations are provided in Table 5.2. Separate power will be required during detector construction. This is estimated in Table 5.3. No backup power is required during the construction phase, except for safety purposes and this is expected to be provided by site operations.

Additional utilities required are as follows:
- HVAC (includes clean room facility sq m TBD);
- Compressed air (pressure / flow TBD);
- Nitrogen - gas (pressure / flow TBD);
- Nitrogen - liquid (43,000 l storage);
- Uninterruptible power (see Table 5.2);
- Gbit network access;
- Material lifting and handling;
- Material transport (receiving facility, surface to collar, downshaft, bottom-landing to SNOLAB, “carwash”);
- Ultra-high purity water (1800 $m^3$ at veto fill time, see Section 4.5);
- Veto counter water disposal;
- Standard DI water supply and disposal;
- General storage.

Document control between the host facility and nEXO will need to be coordinated. As will safety and security needs. Control documentation for this will be established.

# 6 Trace Analysis and Quality Control

## 6.1 Overview

Radioassay of detector materials is a central aspect of demonstrating feasibility, since the sensitivity projection is based on it. This activity is intimately connected with the Monte Carlo and engineering efforts. It enables the preliminary detector engineering to be performed with appropriately selected low activity materials. The nEXO radioassay program gives and receives guidance from the Monte Carlo simulations by defining materials of interest and provides the overall normalization for the background model. In return the simulation-derived hit efficiency, coupled with a rate allowance for each component, defines to what level of radio purity a certain material needs to be tested. This close coordination between radioassay and simulation efforts allows us to focus the material measurements on the most important background contributors, thus, maximizing our impact on optimizing sensitivity. Currently our focus is on bulk activities, as we assume that surface activities can be remediated using appropriate cleaning procedures. This working hypothesis is based on experience gained during the building of the EXO-200 experiment. Measurements of radioactivity removed from surfaces by etching showed a sharp reduction after cleaning with solvents and treatment with dilute acids. The fact that the observed EXO-200 background rate and composition agrees with expectations derived from bulk radioactivity measurements alone indicates that the surfaces are not strong sources of background. The combination of material measurements and simulations provides guidance to the detector design effort. This is done by defining how much of each material is consistent with the sensitivity goal of nEXO. The entire process is, thus, a complex feed-back loop.

As the detector design, the composition and makeup of its components are not yet frozen at this early stage of development; the material and parts acceptance criteria necessarily need to be "fuzzy". Following the approach of the successful EXO-200 program, a SS background event rate of currently $R_b = 3.6 \times 10^{-4}$ counts/(kg·y) in the inner 2000 kg of $^{\rm enr}$Xe and within $\rm Q_{\beta\beta}$±FWHM/2 forms the figure of merit for materials assessment. The collaboration has recently published a detailed assessment of experiment sensitivity, in which we demonstrate that this background goal is consistent with the scientific goals of nEXO [1]. "Small" components contributing less than 1% of this rate are assumed to be "acceptable". In the framework of our design effort this signals that such materials or components are no longer a high analysis priority; we assume this to be proof of feasibility that a material with the required properties can be obtained. A few major components/materials are given a larger contamination allowance. Copper, to be used in the construction of major mechanical detector components, is a prime example. The commercially available material, assumed by the current design, contributes about 30% of the predicted background rate; the sum of all components still obeying the overall requirement. Cases like this still leave room for background reduction, however, at the expense of adding complexity to the project (for ex-ample, by embarking on an electro-forming program, such as that successfully undertaken by the Majorana Demonstrator Collaboration [2]). Figure 6.1 provides a graphical representation of the current fractional detector-background composition. As can be seen, there is no single dominant background component. We use this information in the nEXO radioassay program to assure the assignment of analysis priority is well balanced.

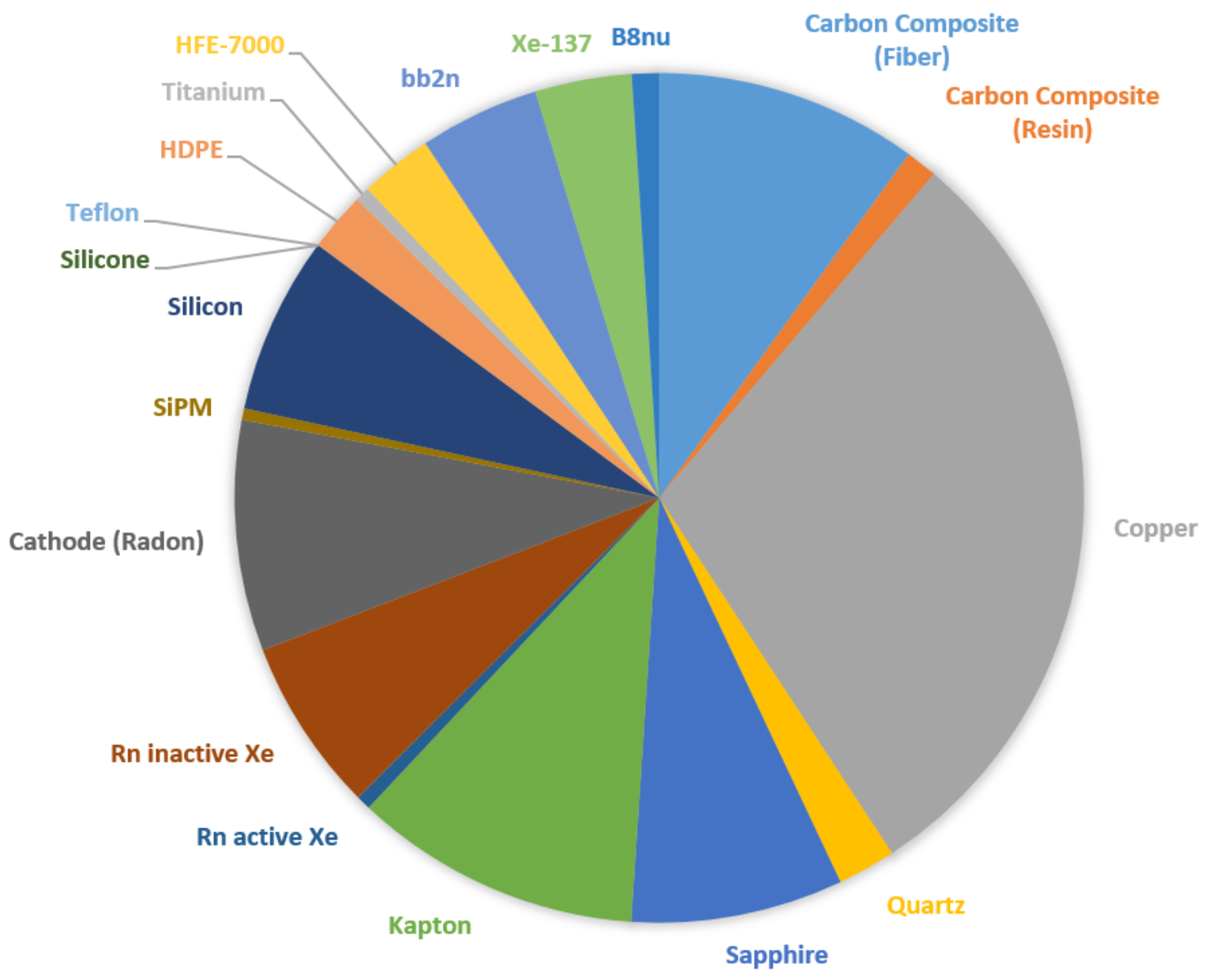

**Figure 6.1:** Fractional-rate contribution of each material currently included in the nEXO background model. The rates are evaluated using the benchmark condition discussed in the text. This assessment of fractional background contributions results from the combination of the simulation and radioassay efforts, it allows a reasoned allocation of assay resources.

The nEXO radioassay group strives to provide radioactivity data for *all* materials and components relevant to the nEXO R&D effort. This aspiration for completeness reduces the risk of project delays due to time consuming material searches, or need for redesign of complex detector components, once the detector design and construction are underway. A proactive radioassay program also helps advance the pursuit of using radio pure materials in the design development by seeking and finding advantageous low background materials. The EXO-200 radioassay program, the template for the nEXO approach, provided hundreds of material measurements, of which some 300 measurements have been published in two dedicated papers [3, 4]. As a benchmark, at the time of writing the nEXO materials database, described in Section 6.7.1 reports about 130 material measurements.

The nEXO radioassay program is being conducted at multiple collaborating nEXO institutions worldwide. In addition, commercial services are utilized whenever required and/or cost effective. The following assay techniques have been utilized (benchmark sensitivities given in square brackets, for Th/U in techniques 1-5):

**Above ground and shallow-overburden low-background $\gamma$-ray spectrometry** The University of Alabama group operates two above-ground dual purpose, shielded HPGe detectors, each equipped with cosmic-ray veto systems. Pacific Northwest National Laboratory offers access to a multi-crystal counting system at shallow overburden. [300/150 ppt]

**Underground low-background $\gamma$-ray spectrometry** Laurentian University and SNOLAB (Sudbury, Canada) have multiple HPGe detectors underground at SNOLAB. For most of them nEXO can request counting time. The nEXO groups at the University of Alabama and the University of South Dakota are bringing on line a new low-background Ge counting station underground at SURF. nEXO can also request time on additional counting stations at SURF, currently dedicated to counting LZ components. [2.3/1.2 ppt achieved with multi-kg samples]

**Inductively coupled plasma mass-spectrometry (ICPMS)** This method is practiced by the nEXO groups at Pacific Northwest National Laboratory, the Center for Underground Physics (Daejeon, Korea) and the Institute for High Energy Physics (Beijing, China). [routine: 1/1 ppt, 0.008/0.01 ppt achieved with pre-concentration][5]

**Glow-discharge plasma mass-spectrometry (GDMS)** nEXO utilizes the semi-commercial analysis service of the National Research Council of Canada. Their service is prompt and priced such that setting up an independent analysis capability appears unnecessary. [10/10 ppt]

**Neutron activation analysis (NAA)** The nEXO group at the University of Alabama utilizes the research reactor at MIT, and UA's two Ge detectors mentioned above to perform analyses on mainly non-metallic samples. [routine: 1/1 ppt, 0.02/0.02 ppt achieved with sample pre-concentration]

**Radon out-gassing** The nEXO group at Laurentian University can measure trace radon activities via electrostatically-boosted solid state detection. The nEXO group at the University of Alabama utilizes liquid scintillation counting. The Laurentian group is leading the nEXO radon analysis and reduction program. The group not only operates sensitive counters but is also responsible for the development of a method to chromatographically remove radon from xenon. [3 atoms/($m^2$ d)]

**$\alpha$-counting** The nEXO group at the University of Alabama is operating several solid state Si $\alpha$-detectors, including one 30 $cm^2$ low-background device. This device has sufficient sensitivity to survey surfaces to the level required to meet the sensitivity goals of nEXO. Pacific Northwest National Laboratory operates multiple $\alpha$-screeners. Assuming radon-daughter plate-out rates reported in [6], derived air exposure limitations are rather relaxed. However, this assertion is being scrutinized by the collaboration. [$^{210}$Po: 30 mBq/$m^2$]

These methods complement one another. ICPMS and NAA offer the best sensitivity, even sufficient for the innermost and therefore most demanding detector components. However, converting the concentrations of nuclides at the head of the natural radioactivity decay chains, as measured by these methods, into background rates in nEXO requires the assumption of chain equilibrium. Under conditions where secular equilibrium may not be assumed, these techniques can only estimate, rather than rigorously predict, the expected background rates. $\gamma$-spectroscopy with HPGe detectors directly determines the Th and U-chain activities relevant for the background: $^{208}$Tl and $^{214}$Bi. $\gamma$-ray spectroscopy is further used to probe for short lived activities (cosmogenic or man-made). Radon counting directly probes backgrounds originating from this nuclide. The success obtained by EXO-200 in exploiting the same amalgamation of techniques validates its use for nEXO. The agreement between radioassay-derived rate-predictions and observed background rates justifies this assumption for EXO-200 after the fact.

The nEXO radioassay program follows the EXO-200 example. Here we summarize the validation of the EXO-200 radioassay derived background predictions by means of the data-derived background composition. This comparison focuses on the $0\nu\beta\beta$-background rate and serves as a justification of the chosen approach. A detailed discussion has been published in [7]. Just as proposed for nEXO, the EXO-200 radioassay program was fashioned around a Monte Carlo and radioassay driven system of background predictions. Utilizing a similar mix of assay techniques as described here, the EXO-200 pre-data taking Monte Carlo simulation predicted 90% CL ranges for SS counting rates in the energy interval $Q_{\beta\beta} \pm 2 \cdot \sigma_{\beta\beta}$ of: 0.9-10.3 counts/yr for $^{232}$Th and 6.3-26.8 counts/yr for $^{238}$U. The analysis of 477.6 days of EXO-200 data yielded 10.3-13.9 counts/yr for $^{232}$Th and 5.3-7.1 for $^{238}$U, in agreement with predictions. Feeding the EXO-200 radioassay data into the data-tuned GEANT4 detector simulation, instead of the schematic pre-data GEANT3 simulation, results in rate predictions of: 0.5-7.7 counts/yr for $^{232}$Th and 2.0-9.5 counts/yr for $^{238}$U. All ranges show overlap, indicating that the radioassay-derived normalizations are consistent with the data and that a simplified Monte Carlo model, built before the detector construction, can reliably predict background rates even for an experiment as complex as EXO-200.

The EXO-200 background prediction utilized the assumption of decay chain equilibrium for materials similar to those to be used in nEXO. The agreement of predicted and measured rates therefore validates this assumption. The EXO-200 background model contains Monte Carlo-derived probability density functions for all relevant experiment components. These are fitted with free floating normalizations, just like described for nEXO in Section 3.3.4, to reproduce the observed spectrum. These data-derived normalizations carry information on the activity content of their respective components. This data is independent of the radioassay. During the EXO-200 data analysis it was observed that components of comparable proximity to the active xenon typically result in highly correlated component activity fits. The EXO-200 data analysis therefore only determines well summed activities of component groups of comparable proximity. As a result of these fit degeneracies, the radioassay data provides more restrictive assay results than EXO-200 data since each component is identified separately. One notable exception is copper, used in substantial quantity close to the EXO-200 active mass. For copper the fit and the radioassay data are at least comparable in their restrictive power. The copper constraints are: $^{232}$Th: <3.8 ppt (EXO-200 assay) and <14.3 ppt (EXO-200 data); $^{238}$U: <3.7 ppt (EXO-200 assay) and <5.9 ppt (EXO-200 data). Analysis of the peak-integral ratio of the 2.615 MeV $\gamma$-peak from $^{208}$Tl (a $^{232}$Th daughter) decay with the summation peak at 3197 keV shows that the $^{208}$Tl activity observed in EXO-200 must be located in a component further away than the TPC copper. In the current nEXO background estimate copper is the largest single background contributor. Note that the current nEXO radioassay constraint for copper from the same manufacturer is with Th: $0.13\pm0.06$ ppt and U: $0.25 \pm 0.01$ ppt substantially more stringent than what can be inferred from EXO-200. Direct counting of a large amount of this copper, using an underground Ge-detector (the VdA detector mentioned later), yielded Th <2.3 ppt and U <1.2 ppt assuming strong constraints on chain equilibrium.

In order to understand in which direction to further develop the nEXO radioassay program it is interesting to consider the fractional background contributions "by source". Figure 6.2 shows that $^{238}$U is responsible for about 67% of the benchmark background rate. Clearly, assay of U must play an important role for nEXO. Table 6.1 lists the detector materials and activities entering into the current background estimate. The nEXO material assay data are recorded and stored in an online database, described in Section 6.7.1. The data catalog systematically tracks measurement values and uncertainty to enable the computation of a total background rate, with an appropriate error. This effort is described below. For convenience Table 6.1 lists 90% CL limits whenever the value is consistent with zero at 90% CL, and assumes the errors are normally distributed. The limit conversion uses the "flip-flop" method [8] to avoid inflating the sensitivity for unphysical negative concentrations. For measurements near the limit of sensitivity, which is often the case for nEXO samples, seemingly unphysical results are encountered and dealt with in a consistent fashion, as described in [8].

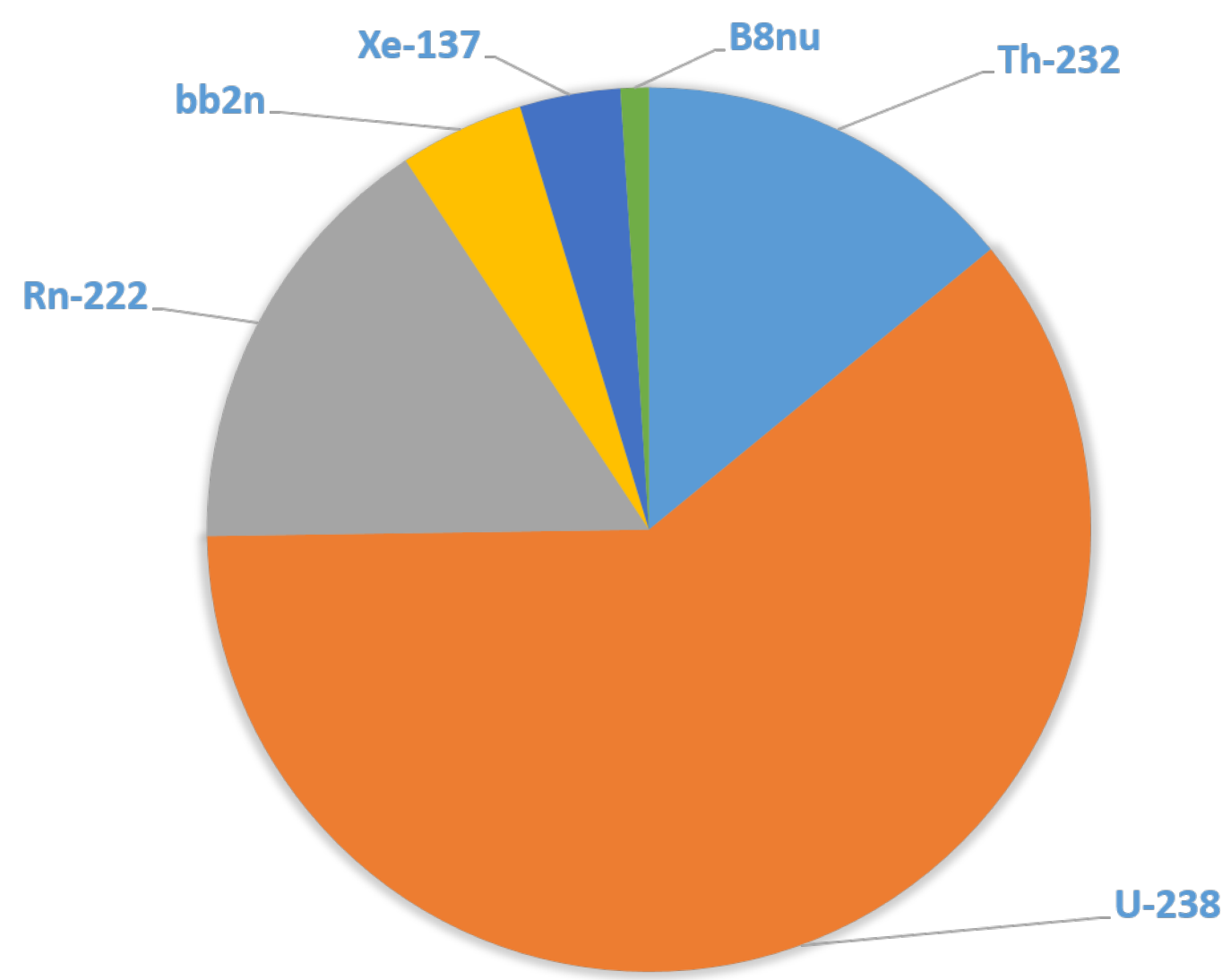

**Figure 6.2:** Fractional rate contribution of all background components currently included in the nEXO background model. The rates are evaluated using the benchmark conditions mentioned earlier in the text.

| **Material** | **Supplier** | **Method** | **K [ppb]** | **Th [ppt]** | **U U [ppt]** | **$^{60}$Co [$\mu$Bq/kg]** |
|---|---|---|---|---|---|---|
| Copper | Aurubis | ICPMS/Ge/GDMS | <0.7 | 0.13±0.06 | 0.26±0.01 | <3.2 |
| Sapphire | GTAT | NAA | 9.5±2.0 | 6.0±1.0 | <8.9 | - |
| Quartz | Heraeus | NAA | 0.55±0.04 | <0.23 | <1.5 | - |
| SiPM | FBK | ICPMS/NAA | <8.7 | 0.45±0.12 | 0.86±0.05 | - |
| Epoxy* | Epoxies Etc. | NAA | <20 | <23 | <44 | - |
| Kapton* | Nippon Steel | ICPMS | - | <2.3 pg/cm$^2$ | 4.7±0.7 pg/cm$^2$ | - |
| HFE* | 3M HFE-7000 | NAA | <0.6 | <0.015 | <0.015 | - |
| Carbon Fiber | Mitsubishi Grafil | Ge | 550±51 | 58±19 | 19±8 | - |
| ASICs | TSMC | ICPMS | - | 25.7±0.7 | 13.2±0.1 | - |
| Titanium | TIMET | Ge | <3.3 | 57±5 | <7.3 | - |
| Water | SNOLAB | Assumed | <1000 | <1 | <1 | - |

**Table 6.1:** Materials, analysis method and radioactivity concentrations entering the nEXO background model. Data for entries marked with a * were taken from the EXO-200 materials certification program. Data for titanium are from Table VI of Ref. [9]. Limits are stated at 90% CL and are computed using the "flip-flop" method [8]. The water purity previously achieved at SNOLAB [10, 11] exceeds the nEXO requirements and therefore constitutes proof of principle.

The following sections introduce the various analysis efforts in detail.

## 6.2 ICPMS

Inductively coupled plasma mass spectrometry (ICPMS) is a major analytical technique employed by many ultralow background rare-event physics experiments for detector material validation. Unlike other radiometric assay techniques, ICPMS directly detects the atoms of $^{232}$Th, $^{238}$U, and $^{nat}$K, through the quantitative mass separation of these isotopes from the material matrix.

Samples are typically introduced to the ICPMS in the form of an aqueous aerosol solution. A flow of gas (usually argon) converts the liquid sample into a fine aerosol. A portion of the sample aerosol is then directed through the center of a high temperature argon plasma torch, where the material is atomized and ionized. All samples must be background subtracted using a blank (the dilute solution containing the preparation reagents without sample) and normalized to a standard since the ionization efficiency depends on the element. The latter is achieved by adding pure isotopes to the sample that are naturally not present. This method is called isotope dilution. Sample preparation prior to analysis depends on the material composition of the sample. Many common detector materials (e.g., copper, titanium, stainless steel, Kapton) can be dissolved in acids in order to bring the sample into solution for introduction to the ICPMS. For other materials impervious to acid dissolution (e.g., many polymeric materials, like PTFE), removing the polymer material matrix via ashing in a high temperature furnace and then reclaiming the refractory analytes (Th, U, and K) in acidified solutions works well [12, 13]. For solid samples impervious to acid dissolution and dry ashing (e.g., sapphire, SiC), samples can be ablated by a laser to create a fine aerosol that is swept to the ICPMS in a carrier gas for analysis. This last method, laser ablation (LA)-ICPMS, requires a material standard with a known amount of Th, U, and K in the same (or similar) matrix for quantification, which unfortunately must often developed as part of the assay method development. However, LA-ICPMS is extremely useful for measuring surfaces of materials and can provide depth profiling information as the laser ablates deeper and deeper into the surface.

The exquisite sensitivity of the ICPMS technique stems from the control of radioimpurities throughout the sample handling and dissolution process. It goes without saying the purest reagents are required not to limit the measurement sensitivity. Furthermore, techniques of clean sample preparation are paramount in ensuring reproducibility. The groups performing ICPMS for nEXO have demonstrated the capacity for achieving these levels of rigor.

In order to avoid instrument instability and mass interference effects, analyte pre-concentration and matrix removal techniques are oftentimes employed when large quantities of dissolved solids are present in solutions. Ashing, evaporation and ion exchange are examples for methods employed for this purpose. With these efforts ppt and ppq (parts per quadrillion, $10^{-15}$) sensitivities can be achieved, often beyond what is possible using radiometric counting techniques.

As described previously, there are three collaborating facilities providing ICPMS capability to nEXO: Pacific Northwest National Laboratory (Richland, WA, USA), the Center for Underground Physics (Daejeon, South Korea), and the Institute for High Energy Physics (Beijing, China).

### 6.2.1 Pacific Northwest National Laboratory

Pacific Northwest National Laboratory (PNNL) has a long history of ICPMS method and instrument development. For over a decade the nEXO group at PNNL has developed ultralow background material assays capabilities. The ICPMS group at PNNL has collaborated with $0\nu\beta\beta$ and dark matter collaborations, including Majorana Demonstrator, (Super)CDMS, COUPP/PICO, SABRE and DarkSide-20k. PNNL currently provides numerous assays for nEXO.

PNNL has a dedicated cleanroom facility for sample assay and preparations (Figure 6.3), including Class 10 laminar flowhoods, a large quartz tube furnace, two microwave digestion systems, and two ICPMS instruments, dedicated to measuring ultralow backgrounds samples. These ICPMS instruments include an Agilent 8800 triple quadrupole ICPMS and Agilent 7700 ICPMS. All measurements of $^{232}$Th and $^{238}$U at PNNL are quantified using known small amounts of added $^{229}$Th and $^{233}$U tracers to track all sample preparation efficiencies (i.e., isotope dilution). Quantitation by isotope dilution is the most accurate and precise quantitation method for ICPMS. Beside ultrasensitive U and Th detection limits using the Agilent 8800 triple quadrupole instrument, PNNL can achieve stringent potassium detection limits from which $^{40}$K can be inferred. Such sensitivity was previously unattainable using more conventional (single quad) ICPMS instruments [14]. Thus far, numerous assay campaigns in support of nEXO have been conducted on a variety of materials. Some of these materials include SiPMs, copper, resistive cathode materials, polymers, Kapton, Cirlex, and sapphire. Detection limits for Th and U are typically better than 1ppt for assays of almost all materials, and the turnaround time on a given sample is on the order of a few days. Methods for the ultrasensitive analysis of copper, lead, polymers, and NaI scintillators have been published (See refs [5, 12–14]).

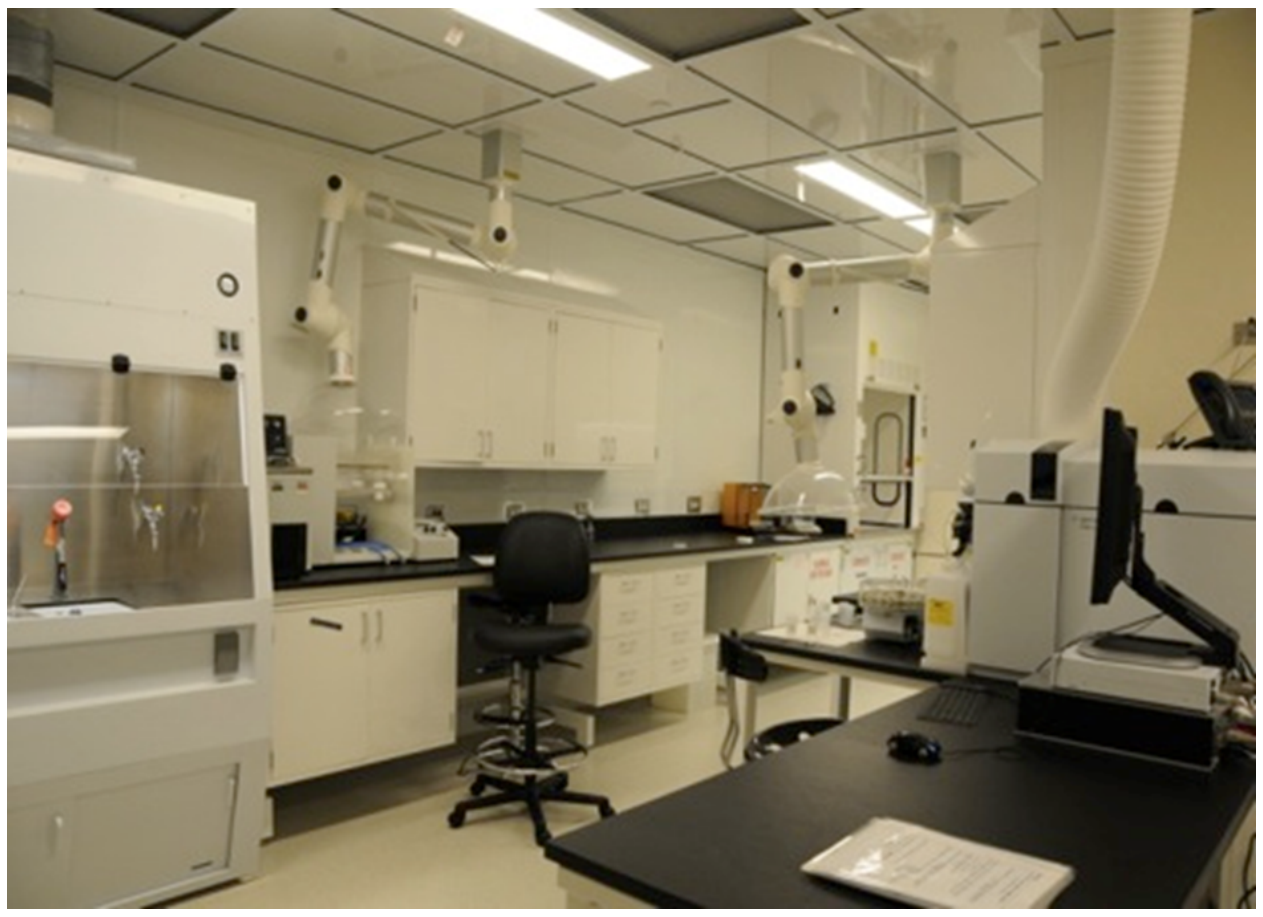

**Figure 6.3:** The cleanroom assay lab at PNNL with one of the ICPMS instruments in the front right.

### 6.2.2 Center for Underground Physics

The Center for Underground Physics (CUP) at IBS in Korea has a wide variety of laboratory facilities related to control of backgrounds for rare-event experiments. In particular CUP contributes part-time use of, and information exchange with nEXO related to, its ICPMS laboratory. The lab space contains a 20 m$^2$ soft-walled unidirectional flow cleanroom operating at about 180 air changes per hour, designed as an ISO 6 (class 1000) cleanroom. It houses an Agilent 7900 ICPMS instrument with cold-plasma, He collision cell, and gas-dilution options installed. It also includes three linear meters of chemical hood space, and an in-house acid distillation system. The lab is supplied from an 18 M$\Omega$ building-wide water supply as input to a Millipore Advantage A10 water-purifier. CUP has recently added a Milestone ETHOS Easy microwave digestion system increasing capability to digest difficult samples consistently and efficiently.

Work at CUP has until recently largely focused on optimizing sensitivity for measurements using simple sample preparations under high throughput demand, with a history of experience connecting back to the EXO-200 assay work. CUP has recently begun exploring the use of the microwave digestion system in combination with chemical/resin separation techniques to enhance U and Th sensitivity to levels significantly below 1 ppt, and has been exploring the use of tracers (isotope dilution) to track purification processes. These techniques are similar to the sample pre-concentration performed at PNNL. CUP is also involved in significant efforts to use tracer elements for purification studies.

CUP further operates a 14 detector HPGe array and an XIA Ultra-Low 1800 alpha counter, providing possibilities to improve understanding and constraints on the lower decay chain. This could be of particular interest for application to copper.

### 6.2.3 Institute for High Energy Physics

Institute for High Energy Physics (IHEP) has over ten years of experience using ICPMS, with most of the main work focusing on biological and environmental samples. The assay of ultralow background materials has just been initiated starting in 2016 and the ICPMS group at IHEP has been collaborating with the nEXO and JUNO experiments. In the last two years, a co-precipitation method has been developed and optimized to separate Th and U from copper. Co-precipitation denotes the controlled removal of the soluble analyte from solution by means of a suitable agent (e.g. $ZrCl_4$). All work is carried out in a Class 10,000 clean-room, including a water purification system and a Thermo Fisher iCAP Q ICPMS instrument. Blank samples are prepared simultaneously to estimate the background introduced by the sample preparation processes (cleaning, dissolution in suitable acids). $^{233}$U is used as tracer to determine the yield. Measurements of $^{238}$U at IHEP are quantified using standard-addition, isotope dilution, or external calibration methods, as required. Detection limits below 1 ppt of $^{238}$U in Cu have recently been achieved. Similar assays will soon be performed for $^{232}$Th after the acquisition of a $^{229}$Th standard has been completed. A Milestone ETHOS UP microwave digestion system is in place and can be used to assay quartz, ASIC chips, SiPMs and other materials that are difficult to dissolve.

## 6.3 Low Background Counting

Direct $\gamma$-ray counting of materials is the only technique capable of assessing the background impact of $^{232}$Th and $^{238}$U natural radioactivity decay chains, *independent* of assumptions about chain equilibrium. High purity germanium (HPGe) detectors can observe the full energy deposition of $\gamma$-rays from about 40 keV up to several MeV with very good energy resolution (typically 1–2 keV FWHM from 122–1332 keV). This energy resolution gives the analyst the ability to identify decaying isotopes through their uniquely identifying $\gamma$-ray emission energies. HPGe detectors, with sufficiently low intrinsic radioactive backgrounds, can identify and quantify essentially all of the $\gamma$-rays emitted by any of the isotopes in the natural radioactivity decay chains of thorium and uranium.

nEXO has several well shielded, high efficiency, high-purity germanium-detector $\gamma$-ray counting facilities that can be used for direct $\gamma$-ray counting. nEXO currently operates two counters located at the University of Alabama (UA), one located underground at each SURF and SNOLAB. Additionally, nEXO can request spare counting cycles on four other counting stations underground at SURF[1] and four other counting stations located underground at SNOLAB.

[1] A fifth dual-HPGe detector counting-facility is currently under construction underground at SURF that will be able to further reduce external backgrounds by requiring coincident energy deposition in both detectors. The MJD/Legend Collaborations are in the process of relocating a sixth counting station to this facility as well.

### 6.3.1 Above Ground Counting at the University of Alabama

UA operates two $\gamma$-ray counting stations, Ge-II and Ge-III, in an above-ground laboratory[2]. These counting stations utilize 60% and 100% relative efficiency p-type coaxial HPGe detectors. The end-cap surrounding Ge-II is constructed out of copper while that of Ge-III is constructed out of ultra-low-background aluminum for greater transparency to low-energy $\gamma$-rays. Both HPGe detectors are mounted horizontally on a vertical cold-finger. This orientation permits the direct line-of-sight separation of the cryostat from the dewar vacuum; the cryostat vacuum is maintained by high purity activated charcoal, specially selected for low radon outgassing. The HPGe detectors are centrally positioned within a roughly 47 l sample chamber that is surrounded by 5 cm of copper shielding that is further surrounded by 20 cm of lead. The inner sample chamber is flushed with boil-off $N_2$ from their respective dewars. At least the first 24 h of counting time is usually reserved for the displacement of radon gas in the sample chamber. If the sample itself contains trapped laboratory air, this artificial background may be remediated by a time series analysis.

In order to successfully operate as low-background counters, both counting stations are equipped with efficient cosmic-ray veto systems. Both are surrounded on all sides by 5 cm thick plastic scintillator panels that serve to veto background events within 150 $\mu$s of the passage of nearby cosmic-rays. Both detectors are shown in the upper row of Figure 6.6; both are fully shielded and with the front 20 cm of lead and cosmic-ray detector removed. For large sample masses, these counting stations can achieve a few hundred ppt sensitivities to the natural thorium and uranium decay chains. However, the first priority for these detectors is for the measurement of neutron activated samples. They are ideally suited for this purpose due to their convenient access, which is essential for these time critical measurements, and their residual instrument specific background are less of a concern as the NAA activation excites isotopes to an activity level above those present in the detectors themselves. Due to the relatively high transmission of low-energy $\gamma$-radiation, Ge-III further is a useful tool for testing $^{210}$Pb surface activities created by radon daughter deposition. The device is currently extensively used for studies of radon daughter removal from surfaces for the LZ project. UA has intense radon sources and can perform such cleaning studies that likely will have to be done for nEXO materials at a later time.

### 6.3.2 Shallow-depth Counting at PNNL

The CASCASDES low-background, 14-crystal HPGe array [15] resides in the PNNL shallow underground laboratory [16]. The CASCADES system targets high detection efficiency sample counting of $\gamma-$ray activity. Background levels of the system permit measurement of $\sim$ 130 and 90 ppt levels of $^{232}$Th and $^{238}$U in samples, respectively. For physics experiments such as nEXO, the CASCADES system is an excellent material pre-screening instrument, available to be used in advance of committing valuable deep underground counting time on the collaboration's most sensitive HPGe detectors. The system is primarily maintained and supported through applied research programs. Experience has shown requesting physics research screening time on the CASCADES system is usually a 2-3 week turn-around time considering a typical week-long queue duration, 3-5 days of counting, and several days to complete analysis reporting.

### 6.3.3 Ge-IV Counting Station at SURF

[2] Ge-I is not sufficiently shielded for low-background counting

Ge-IV (100% relative efficiency) is under construction in the Black Hills State University Underground Campus (BHUC) of SURF (see Figure 6.4). The Ge-IV counting station is modeled after Ge-II and, in particular, Ge-III. Due to the depth of its host facility SURF (4300 m.w.e.), it has been constructed only from materials verified to be of acceptably radiopure materials. These include every component of the HPGe detector except for the germanium crystal itself, oxygen-free high-conductivity copper (OFHC), and Doe Run lead. Doe Run lead is well known for particularly low levels of radioimpurities compared to other types of modern lead. Further, at its depth, a cosmic-ray veto system is not required for the Ge-IV system. SURF, and particularly the area in which Ge-IV will be located, is subjected to relatively high levels of radon, varying between 50 and 300 Bq/m$^3$. As this is expected to be the dominant background for this counting station, the entire shield will be wrapped in aluminized Mylar, Nylon or solid metal and, along with the inner sample chamber, flushed with $N_2$.

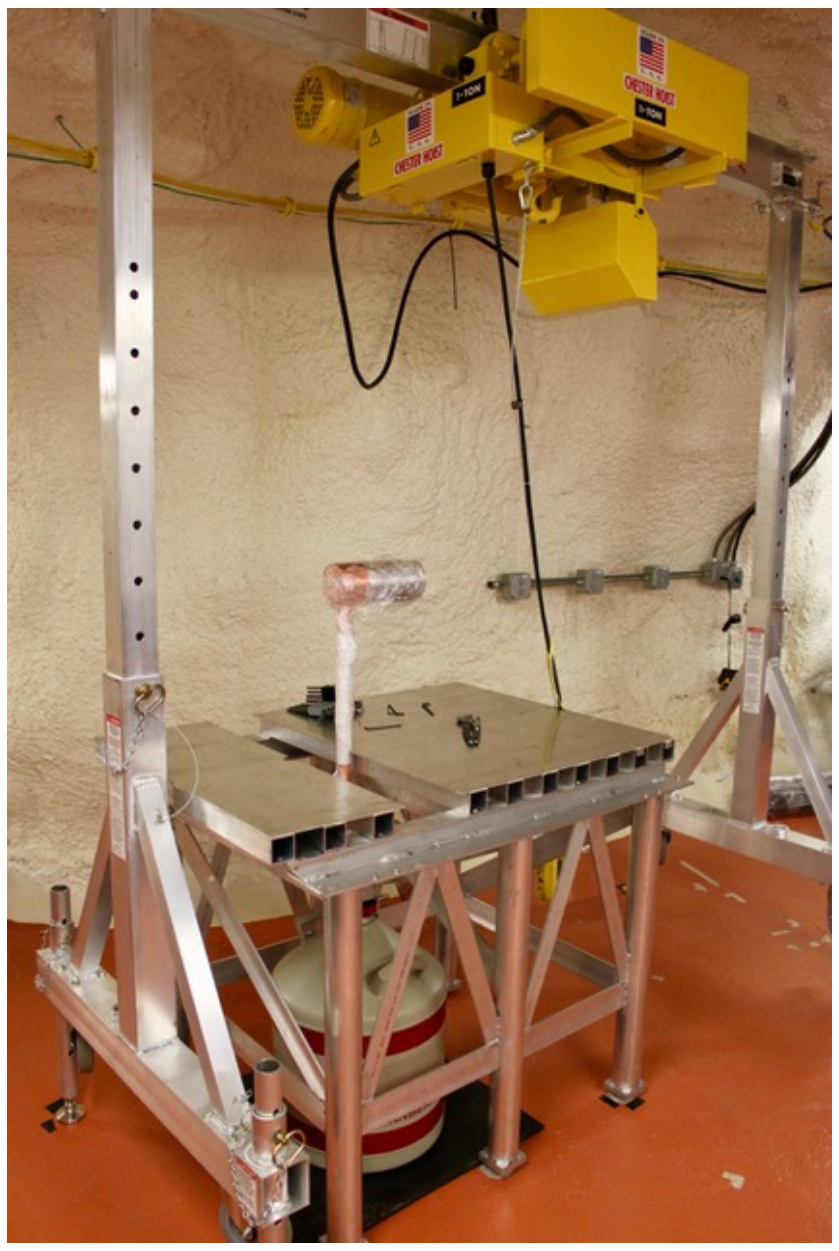

**Figure 6.4:** The Ge-IV HPGe detector, dewar, support table, and gantry crane as installed underground at SURF (prior to shield assembly).

Based on the measurement of the residual natural radioactivity in its components, Ge-IV is anticipated to have sensitivities to the natural thorium and uranium decay chains on the order of a few ppt (for large sample masses). Additionally based on these estimates, counting longer than 2–3 weeks is unlikely to improve the sensitivity of Ge-IV to radioimpurities in a typical sample. The throughput of Ge-IV is therefore expected to be approximately 20 samples per year, with additional counting time devoted to large and high risk materials.

### 6.3.4 Other Counting Stations at SURF

The Black Hills State University Underground Campus (BHUC) currently contains four coaxial well-type HPGe counting stations (see [17] for details). The Morgan and Maeve systems both employ 80% relative efficiency p-type HPGe detectors surrounded by 20 cm of Doe Run lead and 1.3 cm of OFHC copper. Maeve has, additionally, a 1.3 cm layer of ∼300 year old inner lead shielding. SOLO employs a small p-type HPGe detector surrounded by 5 cm of 19$^{th}$ century low-radioactivity "German" lead surrounded by an additional 20 cm of Doe Run lead. Mordred employs a 60% relative efficiency n-type HPGe detector that is shielded by 5 cm of OFHC copper and 20 cm of Doe Run lead. All of the detectors are wrapped in aluminized Mylar and purged with $N_2$ gas to eliminate radon within the outer lead shield. SOLO is owned by the LZ Collaboration while the others are all part of the Berkeley Low Background Facility of Lawrence Berkeley National Laboratory. They are all operated by the BHUC facility that maintains a sample counting queue. These detectors will be heavily subscribed during construction of the LZ experiment. However, nEXO samples will be able to utilize spare counting time in this queue in exchange for spare cycles on Ge-IV and are likely to become more available as LZ construction comes to completion.

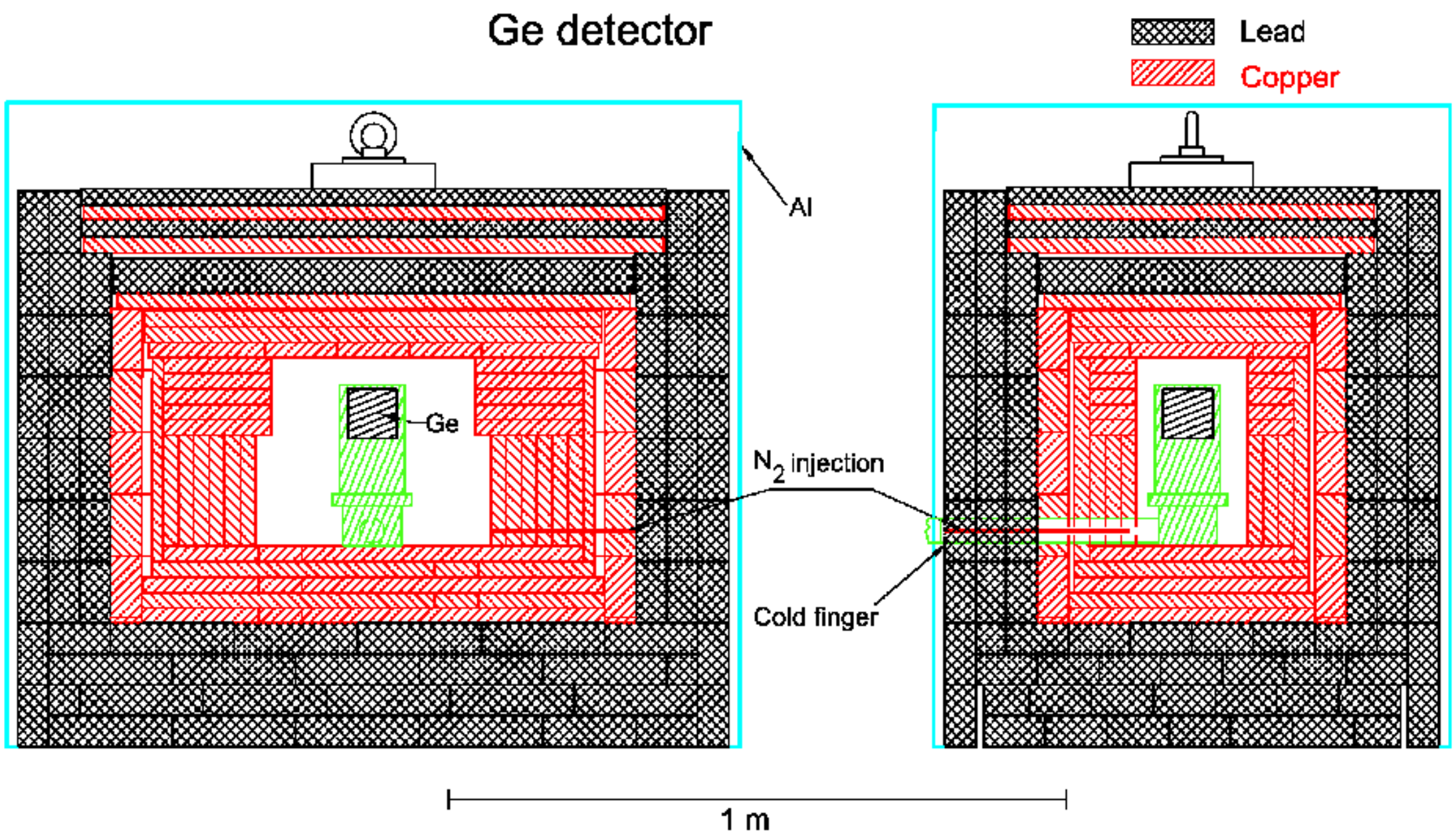

**Figure 6.5:** The VDA germanium detector in its copper (red) and lead (grey) shielding. The aluminum radon containment box is also shown (blue).

### 6.3.5 VDA Germanium Detector at SNOLAB

The VDA detector is a p-type coaxial germanium detector with a Ge volume of 400 ml. It was built and operated by Université de Neuchâtel in 1998. The detector was extensively used in support of the EXO-200 materials screening program. It was operated in the Vue-des-Alpes tunnel (hence the VDA acronym) in the Swiss Jura mountains until 2015, when it was relocated to SNOLAB's 6800 ft level campus, to be integrated in their future Low Background Counting facility and benefit from the larger overburden. The geometry is shown in Figure 6.5. The germanium crystal is housed in a cryostat made from highly purified Pechiney aluminum. All materials entering in the construction of the detector were themselves selected for low activity. The detector is protected against local activities by a shielding composed of 15 cm of OFHC copper and 20 cm of lead. The shielding is contained in an air-tight aluminum box which is slightly overpressurized with boil-off nitrogen from the detector's $LN_2$ dewar, thus preventing radioactive radon gas from entering the detector volume. As shown in Figure 6.5, a small volume is free around the sensitive part of the detector in order to position samples to have the best possible solid angle. Small samples were placed on top of the cryostat whereas larger ones were arranged on top and around the cylindrical part of the cryostat. This arrangement has the extra advantage that it reduces self absorption of gammas in the sample.

Data taking is started one day after closing the shielding to ensure that all radon gas is flushed out. Normally data are accumulated for a period of one week for each sample. Detector backgrounds limit the effectiveness of longer accumulation times. In critical cases, (measurements of lead and copper) data were accumulated for periods as long as a month. The contribution of a sample is obtained by subtracting the background spectrum taken without any sample. For each sample, the geometry is entered in a simplified way into a GEANT3 based Monte Carlo simula-tion which contains also the detector configuration. The acceptance as a function of energy of the full energy gamma peaks is computed and used to translate the observed intensity of a transition, or the upper limit on it, into a specific activity. The computed acceptance is cross checked by exposing the detector to calibrated gamma sources. The sensitivity depends on the sample mass and configuration, which affect the solid angle and the self absorption. For transitions with several gamma emissions contributing in parallel, or by cascade, the peak intensities are combined. The best sensitivity was achieved in the Vue-des-Alpes tunnel with copper samples with masses of several kilograms: $< 35$ pgTh$_{\rm equiv}$/g and $< 5$ pgU$_{\rm equiv}$/g, respectively [3]. The systematic uncertainty is dominated by the acceptance, and is estimated to be of order 10 %.

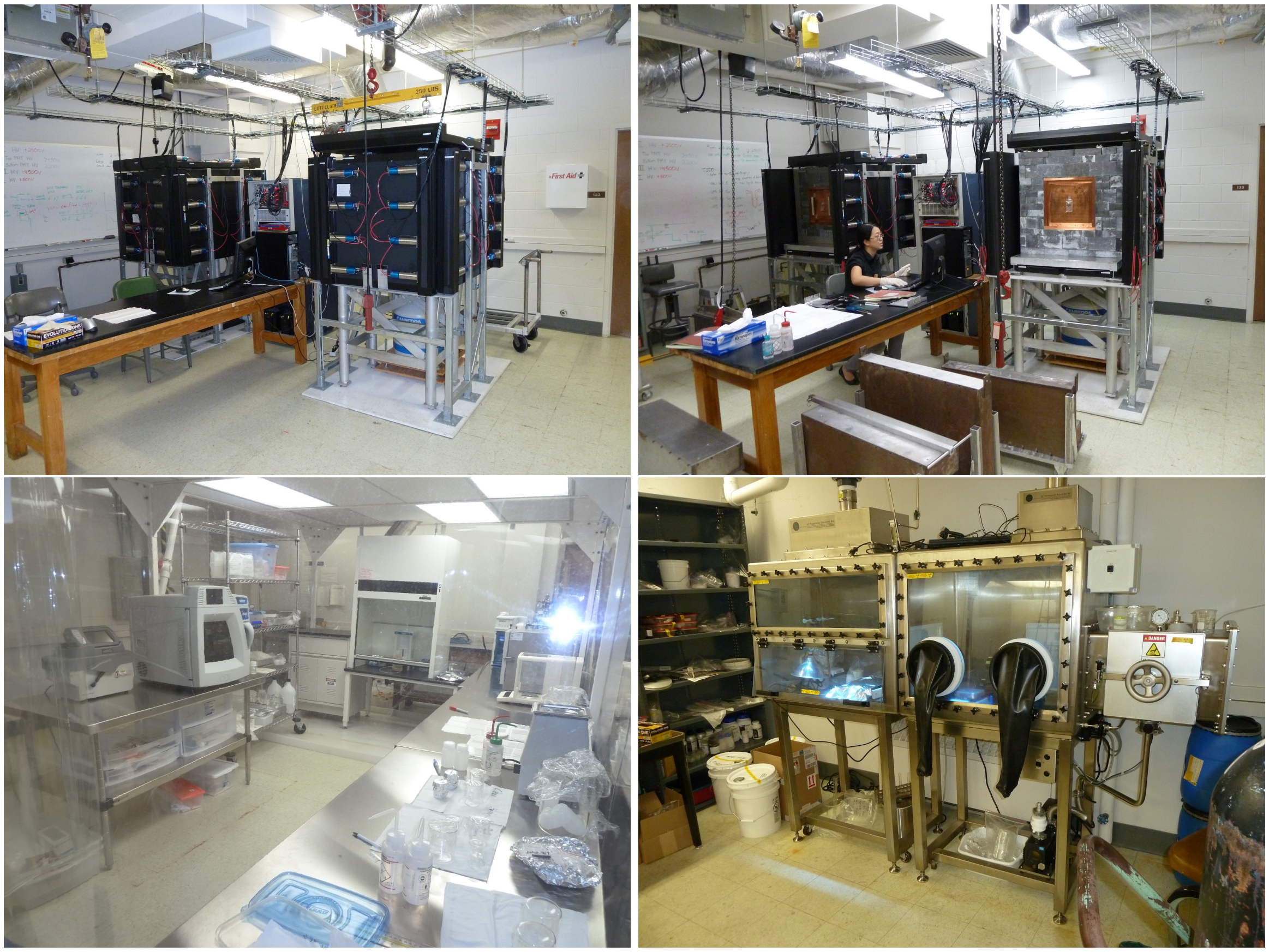

**Figure 6.6:** The upper row shows (left) the shielded low background Ge detectors operated at the University of Alabama and their cosmic ray veto systems. The right panel gives a view of the inner copper shields. The lower row depicts the clean room (left) and the handling facility for open radioactivity (right).

The background was measured at regular intervals in the VDA tunnel although it was found to be very stable. Initial measurements indicated a slightly lower background rate at SNOLAB compared to the VdA lab.

## 6.4 Neutron Activation Analysis

The decay rate $R$ of a radioactive nuclide depends on the number of atoms $N$ of its species present in a sample and its mean life time $\tau$: $R = N/\tau$. For long-lived $^{40}$K, $^{232}$Th and $^{238}$U detection by counting decays, therefore, requires large samples to boost $N$. Making samples very large has limited advantage because of self-absorption of the emitted radiation in the sample and the diminishing solid angle. The low decay rate is the main reason why low background counting is rarely able to determine effective $^{232}$Th and $^{238}$U concentrations below 1 ppt; the method runs out of detectable decays. Neutron Activation Analysis (NAA) addresses this problem by shortening $\tau$. It utilizes the capture of thermal and epi-thermal neutrons by stable or long-lived nuclides. If a sufficient fraction of the target nuclei is transmuted (depending on the capture cross section and therefore the nuclear species) the decay rate is boosted. For $^{41}$K, $^{232}$Th, and $^{238}$U this requires the use of high neutron-flux nuclear research reactors. The activation products ($^{42}$K, $^{233}$Pa and $^{239}$Np) have short half-lives boosting the decay rate, but are sufficiently long-lived for the transport from the reactor to the counting site. In the procedure used by nEXO, samples are prepared at the University of Alabama (UA), irradiated at the MIT research reactor (MITR) and returned to UA for counting typically within 24-36 h after end of activation. This is sufficiently fast to detect even the short-lived activation product $^{42}$K ($T_{1/2}$=12.36 h). Utilization of the higher-flux reactor HFIR at Oak Ridge National Laboratory could be explored. The measured decay rates are then combined with tabulated neutron capture cross sections and measured neutron fluxes to determine the concentration of the parent species in the sample. Ref. [4] gives a detailed description of how this analysis is performed quantitatively.

NAA is applicable to materials where the matrix does not form long-lived radioactivity after neutron capture which creates background when counting the activated sample. This condition is typically fulfilled for non-metals although there are certain metallic species (e.g. Al, sapphire) that can be assayed via NAA. From this perspective NAA complements ICPMS, which requires dissolving the sample in acid. NAA is relatively robust against sample contamination as it requires only little pre-activation sample preparation. Just as ICPMS, NAA determines the long lived heads of the decay series. Similar to ICPMS, sample surfaces need to be carefully conditioned prior to analysis.

The group at the UA has been performing NAA for a number of years, for the KamLAND and EXO-200 experiments and now for nEXO. The group operates a sample preparation clean area, has handling facilities for activated samples, has all required health and safety related handling licenses, and maintains three high resolution Ge $\gamma$-spectrometers. Some of these assets are shown in Figure 6.6. MITR is a 6 $\mathrm{MW_{th}}$ tank-type reactor, delivering a reactor power-dependent thermal neutron flux of up to $5.5 \times 10^{13}$ cm$^{-2}$s$^{-1}$ when the insertion tube near the fuel element is used. A remote activation tube with a neutron flux of about $7 \times 10^{12}$ cm$^{-2}$s$^{-1}$ can also be utilized, offering a more than factor 300 reduction in the flux of fast-fission neutrons. This reduces metal-induced side activities that can create background which limits the counting sensitivity. The lower thermal neutron flux can be compensated by longer activation times. This technique also permits analysis of certain weakly metal-doped semiconductors.

Thus far 10 activation campaigns have been performed with various nEXO related samples. These NAA campaigns were devoted to various SiPM species (different types, different manufactures and different stages within the production process), SiPM base silicon, quartz, sapphire, and silicon wafers. Sub-ppt sensitivities for Th have been achieved for some samples. The sensitivity to U is often limited to few ppt due to the short half life of $^{239}$Np ($T_{1/2}$=2.36 d), resulting in substantial sample-related backgrounds due to the co-activation of non-radioactive impurities or dopants. The group is exploring possible ways to reduce such background, for example by means of $\gamma - \gamma$ coincidence counting. However, the utility of such an approach has not yet been demonstrated. The best sensitivity achieved for an EXO-200 sample (by pre-concentration) is 0.015 ppt for both Th and U in the heat transfer fluid HFE-7000.

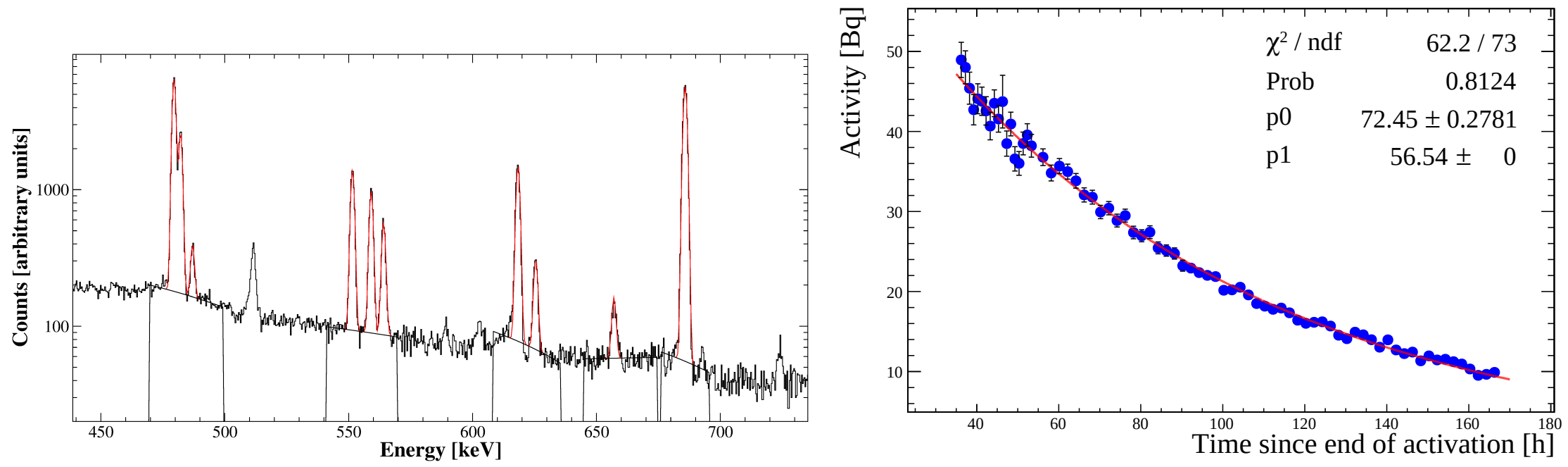

**Figure 6.7:** Energy and time information obtained during the counting of activated samples. The left panel shows a typical energy spectrum, recorded by one of the Ge-detectors. The right panel shows the time dependence of the $^{239}$Np lines, as determined for a sample. The combination of both observables greatly enhances the certainty of our nuclide determination. Both figures were taken from ref. [4].

Energy and time differential measurements, enhancing the sensitivity and allowing unambiguous nuclide identification, are the standard analysis tools. The activated samples are typically counted for about a month to cover a sufficient number of decays of $^{233}$Pa ($T_{1/2}$=26.98 d). This means that a fast sample turnaround is difficult to achieve. Figure 6.7 shows an example of the analysis of the energy and time information. Custom analysis software is being used for this purpose. The UA group is collaborating with CUP on the maintenance and development of this software.

## 6.5 Radon Emanation Measurements

The $^{enr}$Xe target is a key component of the detector in which the radon concentration must be controlled very carefully. While it was determined that a trap to remove the radon from the xenon was not necessary for EXO-200, such a trap is deemed necessary for nEXO and described in Section 4.3.5. In addition, it is essential to have the means for assessing the radon emanation from materials in contact with the xenon, so that the presence of sources of radon can be reduced via a careful material selection process. The large body of heat transfer fluid, assumed to be stagnant, is less of a concern but the radon level in it must be kept as low as possible. Radon emanation measurements are routinely carried out at UA and Laurentian University, where higher sensitivity counters are also being developed.

### 6.5.1 Facility Operated by Laurentian

Laurentian operates a facility that utilizes low background electro-static counters (ESCs) for material screening (six counters) and for the development of a radon trap (one counter). Figure 6.8 shows a schematic of these counters. These counters were initially developed for the assay of $^{224}$Ra and $^{226}$Ra in the SNO light and heavy water systems [18, 19]. Their sensitivity was improved to serve the needs of the EXO–200 materials screening program.

The counters are all built in the same fashion and consist of 10 l chambers in which an electrostatic bias of typically -1000 V is applied to attractively precipitate charged radon daughters onto a 18×18 mm$^2$ PIN photodiode utilized for alpha spectroscopy. The counting chamber is part of a recirculation loop that forces carrier gas through the container holding the sample, followed by a 0.22 $\mu$m filter to protect the chamber from fine particles possibly shed by the sample. The ESCs are typically operated at a pressure of 25 mbar and a flow of 0.3 slpm. The pressure and flow are monitored throughout the run as they influence the counting efficiency. The counters are calibrated for use with dry nitrogen, argon and xenon. Sample containers are chosen from a selection of columns made from low background polypropylene resin to optimize the transport and volume efficiency. The largest column has a volume of 1 l. Four columns can be mounted in parallel for larger samples using compact manifolds.

A typical radon emanation measurement requires two counting runs, one with the sample and the other with exactly the same configuration but no sample ("blank"). Results from the blank run include the counter background and are subtracted from the sample run results to produce the net emanation rate of the sample.

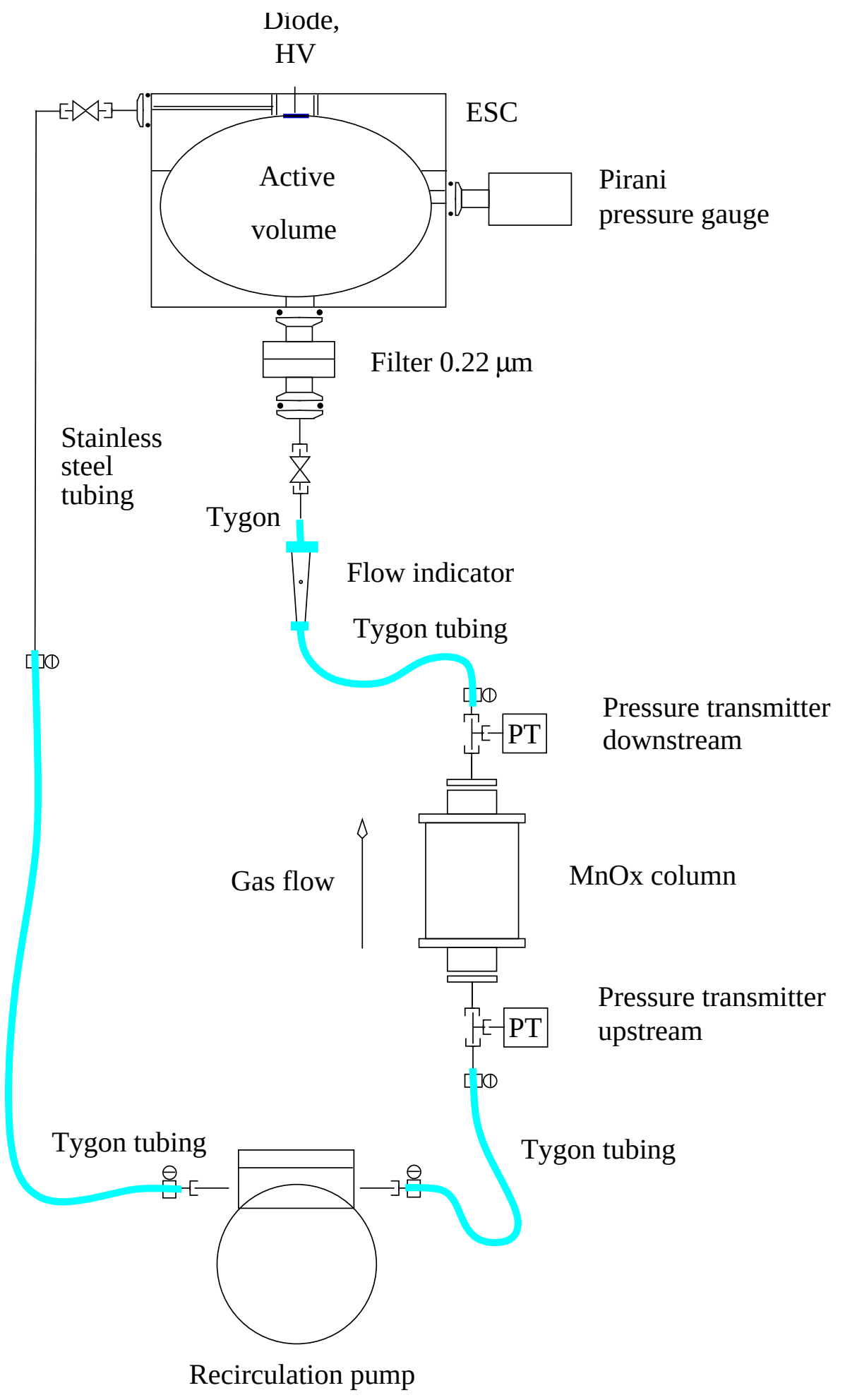

**Figure 6.8:** Schematic of the electrostatic counters used at Laurentian.

Typical background rates are about 11±2 $^{220}$Rn/day and 80±5 $^{222}$Rn/day, but vary counter by counter, depending on their history. A feature of this assay technique is that the areal or specific sensitivities obviously depend on the sample geometry and composition, which both affect the fraction of radon produced by the sample that escapes. From the uncertainties on the background rates, sample production rates of 5 $^{220}$Rn/day and 10 $^{222}$Rn/day can be observed at the 2$\sigma$ level above the background. The observed rates are related to the area or mass of the samples as appropriate.

To reach this sensitivity, a good separation of the *supported* (the desired radon production rate) and *unsupported* (radon atoms present at the time the run is started) components is desired, which is achieved in about two weeks for $^{222}$Rn, see [19] for more details. Therefore a single sample requires four weeks of counting time for low background work. This represents 13 samples per year per counter, assuming no equipment failure. The rupture of the recirculation pump's diaphragm represents the principal mode of failure, which happens every two years per pump on average. Other failure modes are power outages lasting longer than 20 min (the current UPS capacity), that either disrupt the activity curves and/or cause data loss. Note that the Rn-analyses reported in Figure 9.1 constitute only about 41% of the sample measurements performed by the Laurentian group in the reporting period.

Small, metallic samples that do not distort the electric field can be placed directly inside the counting chamber. In this configuration the background, and hence the sensitivity, improves by a factor of two in both isotopes.

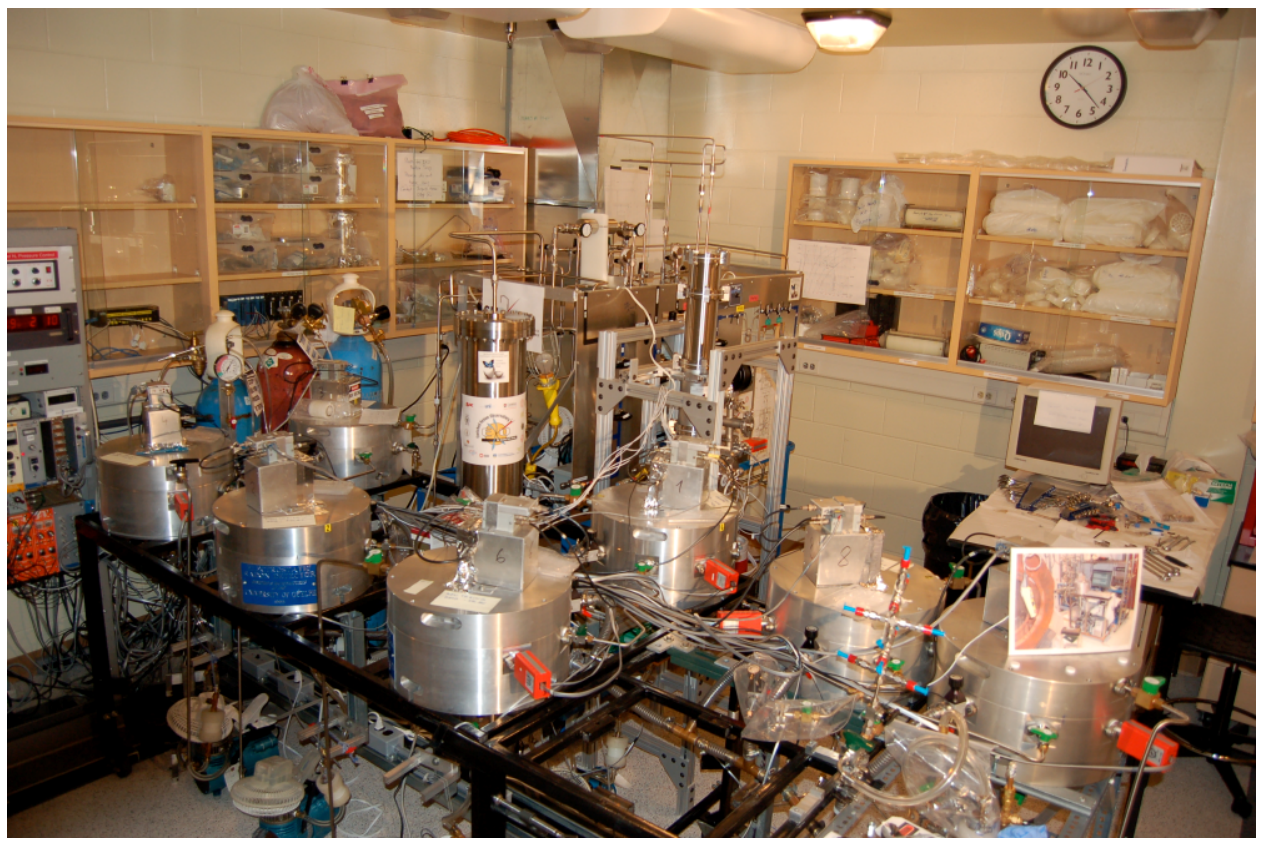

**Figure 6.9:** Radon emanation counting and trap development facility operated by Laurentian at SNOLAB

The group is designing higher sensitivity detectors with a combined reduction in background by at least a factor of ten, and an increase in sample size by also at least a factor of 10. To keep the transport and volume efficiency high, larger sample sizes imply a larger counting chamber. Reducing the background while increasing the chamber size requires cleaner materials. To inform material selection, we are investigating how much each component of the current counters contributes to the background. Another very important element of this effort has been the development of a physical model of ElectroStatic Counters of arbitrary shape to inform the design of larger counters. This work shows that a large counter with 800 l volume is able to achieve a high counting efficiency even for $^{220}$Rn, whose daughter $^{216}$Po half-life is only about 150 msec. For the support of designing larger, lower background counters, the simulation software is now ready for application. To validate the simulation we must compare its predictions with data acquired with a calibrated $^{222}$Rn source that is on hand; a $^{220}$Rn calibration source must still be procured. The plan is to use the higher sensitivity design both *ex-situ* for material screening, and *in-situ* to verify that the plumbing meets expectations.

The Laurentian radon emanation facility (as shown in Figure 6.9) is hosted by SNOLAB in their surface building's clean room laboratory. The group also measures samples for SNOLAB users and the low background community in general, on an as-needed basis.

### 6.5.2 Facility Operated by Alabama

A facility at Alabama was built to measure emanation rates for the LZ screening program. The facility will become available to nEXO when LZ measurements wind down. Two vacuum chambers allow for the radon to accumulate before it is transferred to a liquid scintillator for counting the $^{214}$Bi-$^{214}$Po coincidence signals. The minimum detectable Rn activity for a 0.1 m$^2$ sample area is: 80 atoms/(day· m$^2$) @90% CL for the UA chamber #1, and 144 atoms/(day· m$^2$) @90% CL for the UA chamber #2. The UA radon emanation system was cross calibrated with a rubber sample previously counted at Laurentian. UA can count 2 samples per month with two emanation chambers, considering the emanation duration, counting time, calibration time and blank measurement time.

## 6.6 Alpha Counting

$^{210}$Po radioactivity, attached to component surfaces due to radon daughter plate-out, is a concern for nEXO as it can result in the emission of energetic neutrons from ($\alpha$,n)-reactions, in case the $\alpha$-particles are emitted into a low-Z material. FLUKA and GEANT4 based estimates indicate that no more than 56 mBq/m$^2$ $^{210}$Po can be present on the TPC copper in order for neutrons produced in the HFE-7000 to not contribute more than 0.01 SS events/y in the inner 3 tons of xenon.

The UA group operates several semiconductor $\alpha$-detectors, one of them a large (30 cm$^2$) Ortec low background device, shown in the left panel of Figure 6.10. Measurements of the detector background, coupled with a Monte Carlo generated counting efficiency for a reasonable sample geometry, indicate a 90% CL detection limit of about 10 mBq/m$^2$, after one month of counting. The right panel of Figure 6.10 shows the $^{210}$Po $\alpha$-spectrum detected with a Teflon test patched exposed to lab air during 73.7 days of counting. The activity determined for this sample is 27 $\pm$ 4 mBq/m$^2$. Commercial devices of this type seem adequate to screen nEXO samples to the required sensitivity. The number of detectors will have to be boosted to cope with the routine screening of nEXO samples.

PNNL maintains a robust, above ground $\alpha$-counting capability for the resident applied radiochemistry program. Of interest to the nEXO experiment is a bank of 24 commercially available CANBERRA Alpha Analyst counters. Typically analysis sensitivities of $\sim$150 mBq/m$^2$ can be reached within 7 days of counting. Precision alpha spectroscopy often requires substantial sample preparation such as acid dissolution and electro-deposition to create planar counting planchets. The chemistry expertise, including characterization of sample preparation to counting efficiency, is available to the nEXO collaboration if need arises to screen for indications of $\alpha$-active isotopes in materials of interest to the experiment.

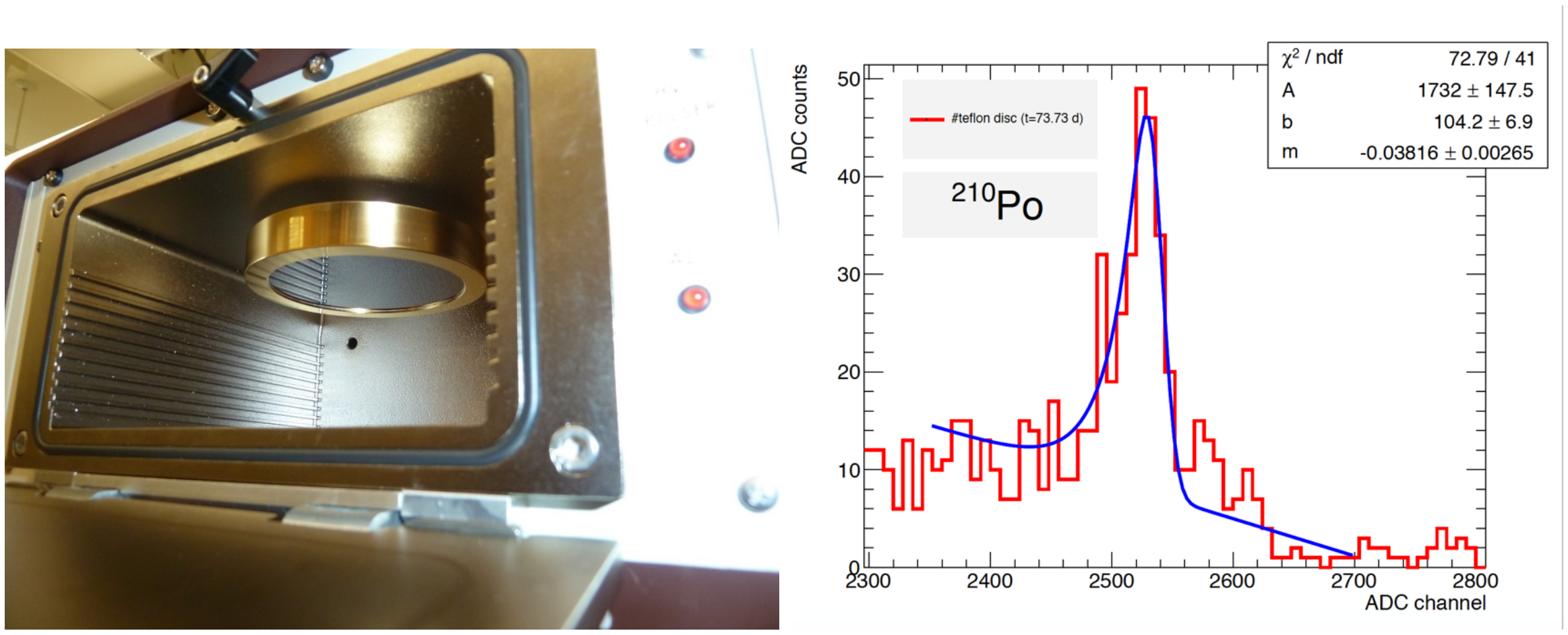

**Figure 6.10:** Left: low background $\alpha$-detector operated at UA. Right: $^{210}$Po $\alpha$-spectrum determined for a Teflon test patch exposed to lab air.

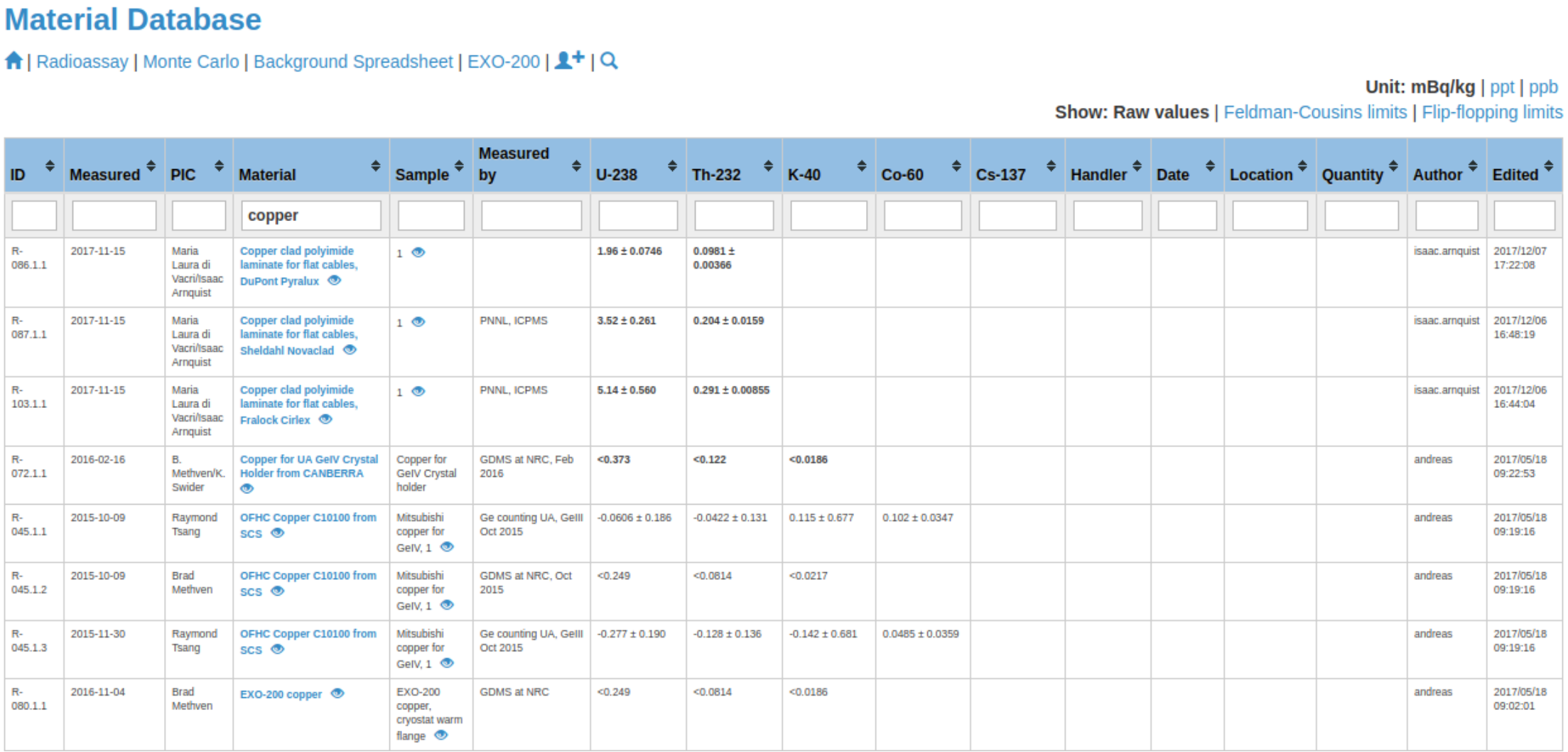

Material Database

| Radioassay | Monte Carlo | Background Spreadsheet | EXO-200 | |

Unit: mBq/kg | ppt | ppb

Show: Raw values | Feldman-Cousins limits | Flip-flopping limits

| ID | Measured | PIC | Material | Sample | Measured by | U-238 | Th-232 | K-40 | Co-60 | Cs-137 | Handler | Date | Location | Quantity | Author | Edited |
|---|---|---|---|---|---|---|---|---|---|---|---|---|---|---|---|---|
| | | | copper | | | | | | | | | | | | | |
| R-086.1.1 | 2017-11-15 | Maria Laura di Vacri/Isaac Arnquist | Copper clad polyimide laminate for flat cables, DuPont Pyralux | 1 | | 1.96 ± 0.0746 | 0.0981 ± 0.00366 | | | | | | | | isaac.arnquist | 2017/12/07 17:22:08 |
| R-087.1.1 | 2017-11-15 | Maria Laura di Vacri/Isaac Arnquist | Copper clad polyimide laminate for flat cables, Sheldahl Novaclad | 1 | PNNL, ICPMS | 3.52 ± 0.261 | 0.204 ± 0.0159 | | | | | | | | isaac.arnquist | 2017/12/06 16:48:19 |
| R-103.1.1 | 2017-11-15 | Maria Laura di Vacri/Isaac Arnquist | Copper clad polyimide laminate for flat cables, Fralock Cirlex | 1 | PNNL, ICPMS | 5.14 ± 0.560 | 0.291 ± 0.00855 | | | | | | | | isaac.arnquist | 2017/12/06 16:44:04 |
| R-072.1.1 | 2016-02-16 | B. Methven/K. Swider | Copper for UA GeIV Crystal Holder from CANBERRA | Copper for GeIV Crystal holder | GDMS at NRC, Feb 2016 | <0.373 | <0.122 | <0.0186 | | | | | | | andreas | 2017/05/18 09:22:53 |
| R-045.1.1 | 2015-10-09 | Raymond Tsang | OFHC Copper C10100 from SCS | Mitsubishi copper for GeIV, 1 | Ge counting UA, GeIII Oct 2015 | -0.0606 ± 0.186 | -0.0422 ± 0.131 | 0.115 ± 0.677 | 0.102 ± 0.0347 | | | | | | andreas | 2017/05/18 09:19:16 |
| R-045.1.2 | 2015-10-09 | Brad Methven | OFHC Copper C10100 from SCS | Mitsubishi copper for GeIV, 1 | GDMS at NRC, Oct 2015 | <0.249 | <0.0814 | <0.0217 | | | | | | | andreas | 2017/05/18 09:19:16 |
| R-045.1.3 | 2015-11-30 | Raymond Tsang | OFHC Copper C10100 from SCS | Mitsubishi copper for GeIV, 1 | Ge counting UA, GeIII Oct 2015 | -0.277 ± 0.190 | -0.128 ± 0.136 | -0.142 ± 0.681 | 0.0485 ± 0.0359 | | | | | | andreas | 2017/05/18 09:19:16 |
| R-080.1.1 | 2016-11-04 | Brad Methven | EXO-200 copper | EXO-200 copper, cryostat warm flange | GDMS at NRC | <0.249 | <0.0814 | <0.0186 | | | | | | | andreas | 2017/05/18 09:02:01 |

**Figure 6.11:** A screenshot of the materials database showing a selection of radioassay measurements of various copper samples. The figure deliberately represents a screenshot from the system to show the functionality and is not necessarily fully readable in this form.

## 6.7 Materials Database

### 6.7.1 Purpose and Functionalities

The CouchDB-based materials database is created to act as a central repository for radioassay data, and to a provide a tool for background calculations. The database consists of three main parts:

- **Repository for radioassay measurements**
  The radioassay measurement results of candidate materials for detector construction are the primary information stored in the database. Material provenance and supplementary information are also stored in the same documents or as attachments. New attributes for supplementary information can be easily added when such needs arise in the future.
  Radioassay records can be tagged for easy retrieval, in addition to keyword search. All previous versions of each radioassay record remain in the database for accountability tracking.
  The repository allows radioimpurity concentrations to be input in different units (e.g. ppt and mBq/kg) and it can automatically convert between them. In case of inconclusive radioassay measurements, Feldman-Cousins upper limits can be calculated.
  Engineers designing the detector can use the database to select materials suitable for the parts they are designing. Background impacts can be assessed using the automatically generated background spreadsheet (described later in this section).
  Figure 6.11 shows a screenshot of the database. Figure 6.12 shows the number of radioassay measurements performed, broken down by radioassay method, since the inception of the database.
- **Repository for Monte Carlo simulation results**
  Hit efficiencies calculated by the Geant4 detector simulation are stored as probability density functions (PDFs) in ROOT format since they are required to calculate the experiment background. Uploads of these PDFs can be automated so as to allow a higher level of automation in the simulation processing pipeline.
The database displays these PDFs in the browser with ROOT TCanvas-like interactivity (as shown in Figure 6.13).

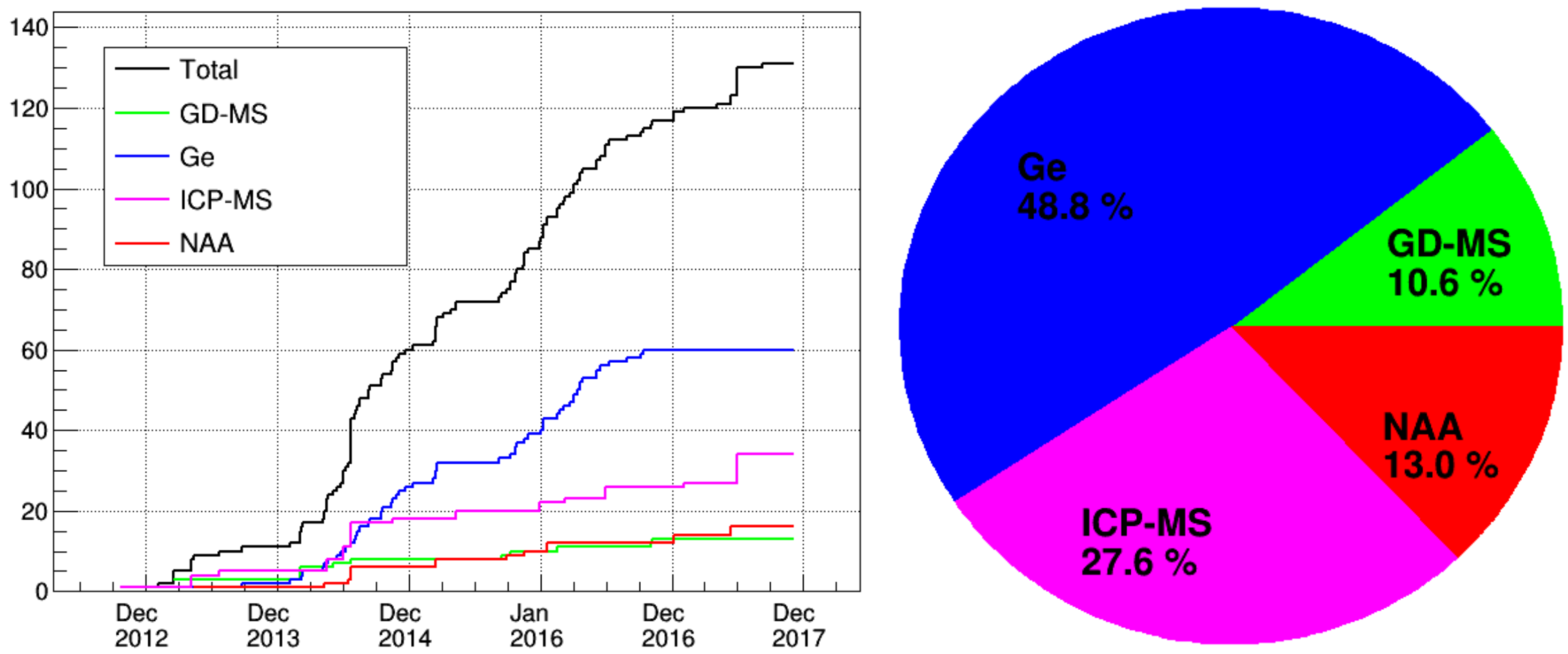

**Figure 6.12:** Speed and throughput of radioassay measurements using different methods. (left) Cumulative number of radioassay measurements performed as a function of time. (right) Breakdown by method as of December 2017.

- **Background estimate spreadsheet**
  The database can generate Excel spreadsheets that calculate the background contributions from all detector parts.
  On the background spreadsheet generation page, the user can define a detector model by specifying the mass, the radioassay record and the hit efficiency associated with each part.
  The software will then collect all necessary data from the repositories and will generate an interactive Excel spreadsheet as the output.
  This provides a quick estimate of the total background and its sources.

The database and all its related software are hosted on a server at the University of Alabama. The repositories are backed up every week at multiple locations.

### 6.7.2 Future Development

Currently, at the planning and design phase, the materials database has been solely and extensively used for documenting information about probable nEXO detector construction materials. As the nEXO experiment progresses to the construction phase, we foresee newer information crucial for a low background experiment that will need proper documentation.

The construction phase will involve transport of nEXO detector materials to the experimental site, partial assembly at the surface and complete assembly in the underground location where the experiment will be housed. An example of newer documentation than presently used are records of screws, adhesives, solder, and other binding material that are needed to assemble a complete detector. These additional materials could be packaged into individual plastic bags and tracked using a barcode. Radon outgassing properties of the detector materials will also be recorded. Knowledge of exposure to $^{210}$Po surface activities and cosmic rays will enable determination of expected background activity. Exposure conditions can be tracked using new attributes for all materials. Details of cleaning methods and chemical solutions used to remove surface impurities from detector materials will be recorded.

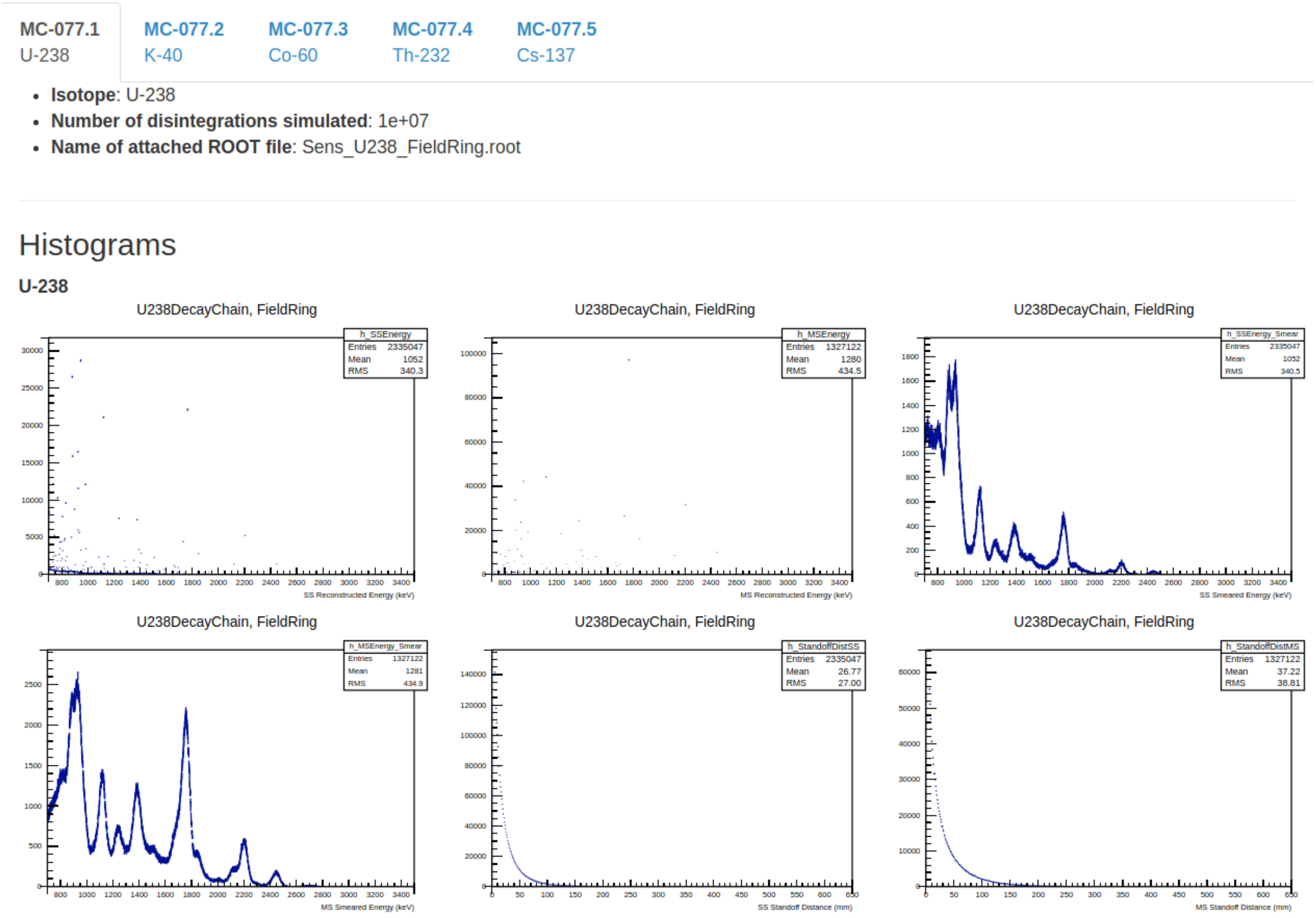

**Figure 6.13:** An example of Geant4-simulated PDFs shown in the materials database. The figure deliberately represents a screenshot from the system to show the functionality and is not necessarily fully readable in this form.

We have an initial idea of the breadth of new information that will need documentation during the construction phase. While it is possible to expand the existing materials database to store this information, it will likely become complicated and unfriendly to the average collaborator. Alternatively, we intend to explore commercial products that can efficiently store and track information during the construction phase. We will eventually converge on a practical choice to store detailed records of nEXO detector materials in the near future.

# 7 Xenon Production

nEXO will require about five tonnes of Xe, enriched in the isotope 136 to a level of 90% ($^{\mathrm{enr}}\mathrm{Xe}$) in the primary design. While the total amount is not particularly large with respect to other scientific projects (e.g. LZ uses 7,000 kg of Xe), the isotopic enrichment requirement implies a more complex logistics already practiced, on a smaller scale, for the EXO-200 and the KamLAND-Zen experiments.

Natural Xe is industrially extracted from the atmosphere, where it is present at a $87\pm1$ ppb concentration by volume [1], with a yearly production of about 50 tonnes world wide. Because of the energy cost of distillation, Xe production is usually achieved as a by-product of oxygen production and is economically viable only in conjunction with large air liquefaction plants. Industrial uses of Xe include high performance incandescent light bulbs, plasma displays and spacecraft propulsion (mostly for attitude control and station-keeping). While the first two uses are currently in decline, the use for propulsion is increasing and medical uses such as anesthesia may emerge as the main application in the long term. Scientific use, mainly for dark matter and $0\nu\beta\beta$ decay searches, is substantial but usually regarded as a transient. As already done for large dark matter detectors, the procurement for nEXO will have to be carefully planned, over a period of time and using more than one supplier, to avoid disrupting the market.

There are nine isotopes of Xe which are stable for $\beta$ decay. Of these, two ($^{134}$Xe and $^{136}$Xe) are double-beta decay emitters, although the Standard Model $2\nu\beta\beta$ decay has only been observed in $^{136}$Xe ($T_{1/2} = 2.165 \pm 0.016(\mathrm{stat}) \pm 0.059(\mathrm{sys}) \times 10^{21}$ yr [2]). The concentration of the isotope 136 in natural xenon is 8.9% [3] and gas ultracentrifugation is usually considered the separation method of choice. The process is simplified by the fact that Xe is already gaseous at STP. Russia has capacity sufficient to separate material resulting in at least 1000 kg/yr of $^{\mathrm{enr}}\mathrm{Xe}$. The enrichment capacity of China is under investigation.

It is expected that natural Xe will be sourced from several companies and shipped out for enrichment. The depleted fraction will then be returned to the suppliers, possibly paying a restocking fee, while the $^{\mathrm{enr}}\mathrm{Xe}$ will be retained for the experiment.

# 8 Data Management Plan

The nEXO collaboration is governed by a “Collaboration Agreement” including provisions regulating the access to data and the dissemination of results. The overarching rule and intent is that all data collected is accessible to all collaborators and that all results are published in peer-reviewed literature in a timely fashion.

Principal investigators, research scientists, post-doctoral associates, and graduate and undergraduate students working on this research program are strongly committed to compliance with policies on the preservation, dissemination, and sharing of research data put forward by DOE, NSF and other agencies outside of the US. This data management plan addresses the specific requirements of the nEXO research effort. In addressing the data management challenges we will make use of the extensive knowledge base and support for data management available at the participating universities and laboratories and at the institutions where research is performed.

The following goals, compliant with best practices and the requirements from various agencies (see e.g. [1]), are set for nEXO:

1. Research data should be accessible to the public for all published results.
2. The origin of all published results should be recorded so that results can be reproduced.
3. The raw, simulated and processed data, code repository, database and associated documentation should be managed, maintained and archived for a reasonable period of time. Long-term preservation of some form of the data is desirable and will be implemented, consistent with the level of funding and effort dedicated to it.

## 8.1 General Data Management Policies

The principal investigators and senior investigators will take several steps to ensure compliance with the various data management policies of the agencies and institutions involved. This applies, in particular, to all policies and other laws regulating intellectual property, including the acquisition and maintenance of licenses for all commercial software used by the experiment. Similarly, collaborators will be made aware that results from nEXO are intended for publication in the open literature.

## 8.2 nEXO-Specific Data Management Policies and Activities

It is expected that the nEXO research work will generate a variety of data in different formats, primarily originating from experimental data taking, data analysis, and simulation. The collaboration management will provide resources to assist researchers in the development of good practices for handling research data and will issue guidelines for the storage, preservation, and dissemina-tion of data. Formats for the data and metadata will be documented and stored as appropriate to enable data preservation, sharing and archiving.

### 8.2.1 Published Results and Supporting Data

Results from the experiment will be published in peer reviewed journals and made available through the arXiv [2] and nEXO's webpage [3]. Data for publication figures and tables will be supplied as ROOT files and/or flat text files as appropriate. We will use services provided by INSPIRE-HEP [4] and publishers such as Physical Review that support ancillary data preservation.

In order to ensure reproducibility of the results, the team will take advantage of modern tools for maintaining versioning of the simulated and processed data as well as software, databases, and documentation. Besides the nEXO GitHub software repository, the nEXO materials database, based on the open source database software CouchDB, will be used to support simulation work. External software packages like ROOT, GEANT4, python and cmake will likely be used in the context of the proposed work. These are widely used, versioned and supported by physics and software communities.

### 8.2.2 Documentation

Documentation of the work performed under this proposal will be maintained in a document repository based on the open source SeedDMS tool [5] (technical notes, presentations, etc.). A wiki service and email lists will also be used. The origin of all processed files (original file, software version, database version, etc.) will be recorded to enable reproduction of the results. The document server, the wiki and email lists are hosted, maintained and backed up by LLNL IT personnel and regularly backed up through the institutional backup system.

### 8.2.3 Experimental Data

Experimental data will be saved in widely adopted and well-supported formats, such as HDF5 or ROOT for processed and summary data files.

## 8.3 Access, Sharing, and Archiving

Research data and results produced as part of the proposed activities will follow the general open access model, while protecting the intellectual property of the institutions and researchers involved. Business confidential, proprietary information, patents, and inventions will follow the rules and regulations of collaborating institutions and funding agencies. Data will not contain any personal identifiable or other confidential or proprietary information, and therefore will not require special security measures to protect confidentiality.

All team members and collaborating institutions will have open access to all data.

Requests for raw data from the scientific community and the public will be discussed and managed on a case-by-case basis by the collaboration management. Supporting data for published results will be made available along with the publication as described above.

Data will be archived for a reasonable amount of time following the conclusion of the research activity and publication of results. At least one copy of the raw data will be archived.

# 9 Towards a Project

The construction of nEXO will be organized as a conventional project, following DOE guidelines. It is expected that there will be substantial non-US contributions to the project and it is agreed that the non-US components will be fully integrated as parts of the US project, following the same practices and regulations. This is in order to ensure proper and uniform execution. Particular attention will be given to the interface with the management (technical and otherwise) of the underground laboratory.

A fully documented project structure will be put in place at the start of the project. Here we briefly review the technical work required to complete a Conceptual Design Report (CDR). The classification of tasks used for the present document may not directly map on top level work-breakdown structure (WBS) that will be established later.

## 9.1 High Voltage

Further work on HV is intended to retire risk and provide input to the CDR. Since HV is connected to most other systems in the TPC, this process has many complex interdependencies.

A substantial component of the R&D will be devoted to understanding the physics consequences of running at a field which is lower than 400 V/cm. The results of this study, based on input from other developments (e.g. in the area of photodetectors), simulation, and specific laboratory tests, will be required to define the configuration of the TPC and, in particular, the diameter and length of the field cage. This input is, obviously, required to engineer the TPC. In parallel to this, a better understanding of the conditions of breakdown in LXe, along with limiting factors and mitigating techniques will be required. This component of the R&D involves the use of the various HV test rigs in the collaboration, including the full scale LXe system, currently under construction. Part of this task includes experimentation with SiPMs, including the test of protective grids, although it is hoped that some combination of proper engineering and resistive electrodes will render those unnecessary.

These activities will also connect with further FEA analysis of the high field regions and, more specifically, the initial design of the feedthrough, which is of some urgency because of the constraints on dielectrics both from the radioactivity as well as electron-lifetime point of view. This work will contribute to the decision between cold or warm seals in the feedthrough, as discussed in Section 4.1.2. Potential feedthrough materials, in particular dielectrics, will be assessed for radioactive contaminations, as appropriate. Research on resistive materials and, more specifically, intrinsic silicon for the electrodes of the field cage, is in full swing and its results will have a substantial impact on the detector design (not just in the area of HV). This involves on the one hand the work on electrical and breakdown properties of the materials and, on the other, preliminary mechanical design of a polygonal field cage, in order to establish feasibility and provide a basis for the engineering. It should be pointed out that a copper field cage is still the primary option and R&D results will be required to make resistive components, with their attractive advantages, a realistic option for engineering.

A separate effort, tightly coupled with the R&D on electronics, will be required to understand the level of HV filtering required and the current return paths in the TPC. Careful work on this subject is essential for low noise operations. While this kind of study was carried out for EXO-200, much of it will have to be repeated because of the internal electronics, involving potentially different current returns and noise properties. The results of this work may influence not only the electronics but also the mechanical design of the field cage and the charge collection tiles.

The validation of the HV design will include full-scale component testing in LXe. This will involve the construction of a cathode and a few field shaping rings, with appropriate dielectric standoffs and field grading resistors and the feedthroughs. All materials should be identical to the production ones, except that, while material types will be certified for low radioactive contamination, the quality control in this area may be relaxed in the interest of time and budget.

## 9.2 Photodetectors

The preliminary R&D on photodetectors performed by nEXO established feasibility, demonstrating that at least one type of SiPM, the LF version of the device from FBK, matches the optical and electronic requirements of the experiment with sufficiently small radioactivity. The R&D also started investigating a number of VUV optics issues, including reflective coatings on electrodes and anti-reflective coatings on the SiPMs. In order to reach the stage at which full system engineering can be done, some more work is needed in all these areas.

On the device front we expect to commission nEXO-specific runs of devices from FBK and, pending negotiations, Hamamatsu. The new FBK devices will further push the photodetection efficiency, while possibly, at the same time, decreasing the rate of correlated pulses. This will provide some safety margin that will be useful in the engineering phase, when trade-offs between different parameters and cost will be explored. One of these prototyping runs is also expected to produce devices in which all connections are brought to the back of the SiPM using through silicon vias. While the technology exists and we do not expect show stoppers, this demonstration is necessary before establishing a baseline design, as signal access from the back side has a number of advantages. Antireflective coatings will also be further explored for commercial SiPMs. The result of this work will directly impact the overall scintillation light readout, as it may improve the performance of the system or slightly ease some of the photodetector requirements. Radioactive contamination testing of FBK and Hamamatsu devices will continue, as appropriate.

On the optics front, the collaboration expects to gain experience with reflective coatings for metallic electrodes and on the VUV optical properties of silicon. In the area of reflective coatings the challenge is to gain confidence that reliable and tested techniques exist, hence retiring risk and informing the direction of the engineering. Silicon electrodes will be further explored under the HV R&D, but, in parallel, samples will be studied from the point of view of VUV reflectivity. It is essential that these two activities proceed in parallel, since silicon electrodes are attractive because of their high resistivity and reflective coatings are only practical if they do not alter substantially the resistivity at low frequency.

Finally, the optical simulation work will be further strengthened, by better integrating it with the rest of the simulation and, possibly, by adopting GPU-based ray tracing, which runs substan-tially faster than equivalent GEANT4 code. This will speed up the turnaround time for simulations, something that may be important when testing a relatively large number of possible combinations of electrode structures, and reflective and anti-reflective coatings on various components.

In parallel with this, we expect that the R&D on "digital" SiPMs will continue and, in fact, ramp up at the Université de Sherbrooke with funding from Canadian sources. While this is at the same demonstration stage the "analog" VUV SiPMs were some time ago, this work is still worth pursuing because of its potentially high value for the detector. An important feature of digital SiPMs is the lower power consumption and dissipation. While the charge collection electronics is at the top of the detector (where the anode tiles are located) and thermally connected to the top copper "lid" of the vessel by convection cells in a thin and isolated layer of LXe, SiPM electronics will be more distributed along the vertical staves supporting the photodetectors around the barrel of the TPC and relaxing the thermal management requirements is likely to simplify the engineering or reduce the materials budget, cost, or a combination of both.

## 9.3 Electronics

Section 4.2 illustrates in detail the progress made to date in retiring major risks related to the cryogenic electronics for nEXO. The challenges are, in no particular order, the ability to design a charge readout system within the power, radiopurity and resolution constraints, the understanding of the relationship between power, SiPM sensitive area and electronics noise in the light readout system, the ability to find suitable external, passive components to be used in the electronics system and the feasibility of integrated technologies for the design of the electronics front-ends. In all cases feasibility has been demonstrated, either by acquiring and testing commercial components or by analytical techniques, coupled with prototype testing.

Before the conceptual design can be completed, more R&D is required in the following areas. The charge readout system is at an advanced enough stage where a design is being implemented on silicon. Tests will reveal the performance of this implementation and it is likely that further iterations will be necessary to refine the design. An important step in this process is coupled to the optimization of the charge detector pitch and number of strips, with impact on the capacitance and scale of integration at the level of the ASIC. The light readout channel is at an earlier stage of development, having reached confidence that a system can be designed within the requirements from the physics and the properties of SiPMs in hand. A front-end architecture has also been developed and, to some extent, optimized. We expect to couple this information with the data derived from the testing of the charge ASIC as input for the design of a first scintillation ASIC prototype. In parallel, we plan to continue the development of alternative front-end schemes that may allow further reductions in power consumption and system noise. The latter may allow for more efficient grouping of sensors, reducing the cable multiplicity and hence background.

It is also expected that work will begin to study and define triggering schemes and data acquisition, to optimize the performance over an extended range of data rates, from negligible, in the case of physics runs, to over 1 kHz (for certain calibrations). This work will be complemented by a careful study of possible data reduction techniques, to be applied at the front-end level. These schemes will have to balance the risk that is always associated with sparsification to the advantage of fewer cables and hence lower background. Work on system design, also connected with the ground returns, relevant in connection to HV, will also be carried out.

## 9.4 Charge Collection Tiles

A number of tasks need to be completed in order to reach the point where the engineering of the charge collection tile system is possible. Some of these tasks are intimately connected with other elements of the R&D. More tests in LXe with single tiles and a substantial array of SiPMs are needed to verify that the required energy resolution can be achieved. In terms of tile fabrication there are advantages in transferring the signals to the back, for readout. This is being prototyped by using through-quartz vias as well as metalized traces wrapping around two edges. In addition various ideas about producing ground planes on the back side of (part of) a tile are being explored for the purpose of shielding the charge collection strips from electromagnetic noise produced by the readout chip. Two "free parameters" are available to the general optimization of the charge readout system, with impact on many interfaces. The first parameter is the quartz thickness, which increased from the current 300 $\mu$m would provide better mechanical robustness and lower capacitance to the structures on the back side, in exchange for a larger contribution to the radioactive background. The other parameter is the strip pitch, which increased from the current 3 mm decreases the number of channels and their capacitance, with gains in the area of radioactivity and heating (fewer channels), at the expense of a (possibly slight) degradation of the position resolution and, hence, background discrimination. The simulation will play an important role in this optimization.

In a related area, the integration of a tile with its readout chip and the discrete components required, will have to be completed. This work will then merge with the engineering of the mechanical support for the tiles and of the interconnection system, allowing the signal and power transfer between tiles and external devices. Finally, more work will be required in the area of material qualification and radioactivity testing. For the time being, feasibility has been demonstrated by showing that quartz of the required purity can be procured and, independently, that electrically working tiles can be fabricated on quartz wafers. It is important to prototype working tiles using only radioclean materials.

## 9.5 TPC Mechanics

Several important items need to be explored in the area of the mechanics of the TPC. The design of the cathode is still at a very preliminary stage and more work is required to fully balance trade-offs between different techniques prior to detailed engineering. This will require the investigation of materials (aluminized mylar films, solid metallic films, meshes), as well as mechanisms for mounting the tensioning of the films that, in all cases, are too thin to be self supporting. Long tensioning members (ideally 130 cm long sapphire rods) supporting the TPC field cage will be purchased and, if this turns out to be too difficult, a design in which the field cage is built out of two or more sub-assemblies then spliced together needs to be developed. Ideally a full scale mockup, maybe full scale only in the $Z$ direction, would be built and tested at low temperature, to verify its mechanical behavior. The resistors to be used to grade the voltage in the field cage are well understood because, at least in principle, they can be identical to the ones used in EXO-200. Yet, the idea of making those resistors an integral part of the sapphire or quartz spacers between rings needs to be tested. Of course this design would have to be altered, should the field cage be built out of intrinsically resistive components.

So far, little work has gone into the development of staves holding the SiPMs. The idea of an interposer providing the mechanical support for clusters of SiPMs with their readout and signal routing will have to be prototyped. Work in this area will not only serve as basis for the engineering of the light collection system, but will also inform the overall field cage design, because, along with the HV R&D, this will better define the radial constraints and clearances. As always, this work will be accompanied by materials screening.

Finally, some preliminary work of integration is required, given the large number of components and complexity of the TPC assembly. This is distinct from the final integration and has the only goal of informing the engineering of possible undiscovered challenges. This process will also connect to the calibration development, so that potential needs in that area can be taken into account early in the design process.

## 9.6 Electrical Connections and Signal Transmission

Preliminary R&D on signal transmission and interconnections has focused on demonstrating proof-of-principle cable technology suitable for high-speed ($\sim$Gbit/s) digital data transmission. This work has demonstrated cables with sufficiently low radioactivity and outgassing to meet requirements for transfer of signals from the cold electronics out of the LXe vessel. The primary candidate for digital data transmission cables are differential microstrips patterned on thin copper clad Kapton laminates. While measurements of the electrical properties of prototype cables and characterization of the radiopurity of the construction materials indicate that they can meet the requirements for nEXO, further work is required to study potential alternatives prior to detailed engineering design. Alternative designs for the cold electronics that include analog signal transmission would require further development of cabling and interconnection techniques, for example, the chemical purity and associated out-gassing of analog micro-coaxial cables are being evaluated in a LXe purity monitor.

In addition to flat Kapton cables, other techniques that could offer improved radiopurity will be investigated for use along the anode and SiPM staves (e.g., signal transmission on rigid quartz interposers), where flexible materials are not required. Further work to develop polyimide films with lower U/Th content may also ease radiopurity requirements on cables and interconnections.

R&D to optimize the SiPM and charge readout modules will also proceed as the cold ASICs are further developed. The optimum module size and number of required interconnections within each module will be determined once the channel counts and power requirements are finalized. The required number of connections and mechanical design will dictate corresponding requirements on bonding techniques and their failure rate. Design of electrical feedthroughs for transferring cables through the LXe vessel will also be investigated as total channel counts and wiring requirements are further determined.

Cryogenic testing of candidate bonding techniques will be pursued to ensure candidate techniques have sufficient reliability. While characterization of the radiopurity of possible bonding materials (including wire bonds and alloys for bump bonding, including lead-free solders) are promising, further detailed screening will be required. Surface contamination of cabling and interconnection materials that may occur during fabrication will also require further characterization.

## 9.7 Refrigeration and Cryogenics

Relatively little R&D has been required in the preliminary phase in the area of cryogenics, owing to the success of the techniques and process engineering implemented in EXO-200 and almost directly applicable to nEXO. In the next phase, we foresee work in this area, to better define the type and amount of equipment required.

The development of the large xenon recirculation pump will continue, with the goal of establishing the reliability and gaskets lifetime, as the scaling of these parameters from the smaller EXO-200 pump are not known. This work will result in a design for the pump and assessment of the number of pumps and spares required. The switched-HV purity monitor will be tested and characterized, resulting in a design also using materials compatible with the radon emanation requirements of the experiment. Similar work will proceed to revise and improve the design of the gas-phase purity monitors already in use on EXO-200.

Substantial work will be required in the study of the thermodynamics and fluid mechanics of the HFE-7000 and the xenon in both liquid and gas phase. This work will inform the process engineering and the design of the TPC and cryostat, providing essential information on the heat removal and matter (hence impurities) transport. Ideally a full FEA model of the two fluid loops would be coded, using an appropriate process simulation tool. This would allow a better understanding of the stability of various feedbacks. However in the past this has proven a daunting task and it is unclear what fidelity is to be expected.

Finally, the interface between fluid handling and structural considerations will have to be sufficiently understood to reach a decision on whether the LXe and HFE-7000 handling systems will be located on the deck or on the side of the cosmic-ray veto tank. This will substantially affect the design of the tank, especially its diameter, and the general infrastructure at the experimental site.

## 9.8 Calibration System

After some early R&D efforts, mainly based on simulation, to demonstrate the feasibility of calibrations sufficient for the full exploitation of the detector's capabilities, the next R&D phase will be directed towards better understanding the constraints imposed by the calibration systems onto other areas of the detector's design.

As already discussed, the calibration using external sources will simultaneously deploy six sources with judiciously chosen strengths and positions in order to ensure that a sufficient number of useful events are accumulated from the inner one tonne of the detector to monitor daily variation of the energy scale and electron lifetime with no more than a 10% loss to the live-time of the experiment. The next step, which is now ongoing, is to devise a full simulation of the sources with time-ordered events and then carry out event by event reconstruction. The goal is to ensure that the large number of "shallow" events at the periphery of the active volume does not degrade the quality of the event reconstruction for the "deep" events that are critical to monitor liquid xenon properties for accurate reconstruction of the position and energy. The outcome of this study will allow us to understand the impacts on the trigger and DAQ systems, providing input for their engineering.

Tests will be carried out to understand the potential problems that might be encountered in deploying sources via copper guide tubes wrapped around the TPC vessel. EXO-200 experience suggests that the length of the cable/tube system should be kept below 15 m and therefore a configuration of two guide tubes might be used to cover the entire TPC. The possibility of using a double-ended tube with a cable that can be used to pull the sources through (as opposed to simply pushing, like in the EXO-200 case) needs to be investigated, as it would provide a more robust system. However, issues related to the pulling cable and its potential radioactive contamination need to be understood. This requires the construction of a simple mockup (unless cryogenic operation is required).

Dedicated simulation will be used to evaluate the efficacy of various calibration methods to obtain an accurate scintillation light map. It is possible that the best method will be to inject a controlled amount of radon isotope into the liquid xenon once every few months, something that would have an impact on various aspects of the xenon recirculation system. In order to get a better handle on the source activity required to obtain sufficient statistics, a test plan is being devised to inject $^{220}$Rn and $^{222}$Rn into the EXO-200 apparatus as part of the end-of-run set of detailed studies.

One of the emergent ideas for monitoring time variations of the energy scale and the electron lifetime is to inject laser pulses via optical fibers onto special regions of the TPC cathode which would be fitted with small gold photocathodes. This concept is currently being tested in a dedicated setup and early results have demonstrated feasibility. The next step is to devise a new setup that will have sufficient diagnostics to determine whether the calibrated amounts of charge injected by this method is stable at the required sub-1% level.

## 9.9 Cryostat and TPC Vessel

The preliminary work done on the mechanics of the TPC vessel and the low background cryostat was aimed at assessing feasibility and providing a material inventory, to be used in the estimate of the detector's background and sensitivity. Future R&D is required to reach the point where the engineering can be completed.

In the area of the cryostat, having established the feasibility of using carbon composites, the screening of more samples of fibers, resins and completed parts is desirable to assess variability between batches (particularly of fibers) and of the contamination potentially introduced by the fabrication process. The results of this process will result in engineering specifications for the weaving process, including the clean room. This, in turn, has an impact on the design of the water tank that, as mentioned, is likely to be initially used for this purpose. In addition all the penetrations will have to be understood so that their initial layout in the carbon composite can be studied. This is important, because in a reliable and efficient (i.e. light weight) composite structure the fiber orientation has to be carefully arranged around "singularities" in the vessel(s). A more detailed study of cryogenic effects on the inner vessel needs to be carried out. This is particularly important near flanges and penetrations. In addition, input from the fluid mechanics simulation and the scenarios studied by process control, will be used to refine the pressure specs of the cryostat vessels and produce a revised design, with more fidelity, particularly on the amount of materials used.

The TPC vessel requires more studies, both in the area of determining the optimal material and in structural analysis. Electron-beam welding must be confirmed as being able to implement the geometries being considered and the mechanical properties of test welds measured. Port geometries and placement need to be refined and engineering safety margins better understood, so that a second round of finite element analysis can be performed, resulting in better fidelity on the quantity of material required. This is particularly important, because the TPC vessel has a large impact on the background in the detector. The possibility of using electroformed copper, as well as the more conventional high purity material from the supplier used for EXO-200, will also be explored.

Given the tight tolerances and exotic materials and manufacturing technologies, specific studies will be required to verify the practicality of the possible assembly techniques and sequences.

## 9.10 Water Shield and Veto

The water tank used for the cosmic-ray veto detector is rather conventional. However, a second iteration of the design will be needed, once the location of the liquid recirculation loops is decided. More work is also required to better understand and mitigate the effects from the local high frequency seismicity at SNOLAB, due to relaxation of stresses in the mine. Of course, this is a site-specific issue and hence will be carried out in parallel with more work in the area of site selection.

## 9.11 Trace Analysis and Quality Control

The material assay effort needs to be further developed such that it will be able to support the project development and construction phases. This development will proceed along two main lines:

- After the initial effort to investigate the existence of certain construction materials, we expect a substantial increase in the required throughput before and during the construction of nEXO, as illustrated by the EXO-200 radioassay campaign. Figure 9.1 shows the number of sample measurements performed during the development and construction of the EXO-200 experiment. nEXO will be larger than EXO-200 and, although the number of its principal components will be comparable, the larger number of components will require more sampling with respect to EXO-200. At present, we envisage a factor 2 to 3 larger sample demand than EXO-200 and the resulting increase in equipment and personnel will have to be supported by the project.
- Some tools and methods will need to be enhanced for some specific needs deriving from the types of materials to be used in the detector. R&D will study the following issues:
  1. The systematic determination of $^{210}$Po surface activity (created by radon daughter plate-out) by means of $\alpha$-spectroscopy.
  2. $^{26}$Al will need to be systematically tracked in the event substantial quantities of sapphire will be required by the nEXO TPC.
  3. Radon outgassing data will need to be systematically collected. This effort is highly component-specific. Design of the nEXO Xe piping must precede this effort. The materials database is being augmented to systematically store the resulting data.
  4. The nEXO materials database will have to evolve to allow the tracking of the cosmic-ray and air exposure of all nEXO components. This is important as not every component can be extensively counted.
  5. Following the experience gained during the EXO-200 construction, the group will help investigate how to use its ICPMS and NAA capabilities to perform sensitive testing for surface contamination.

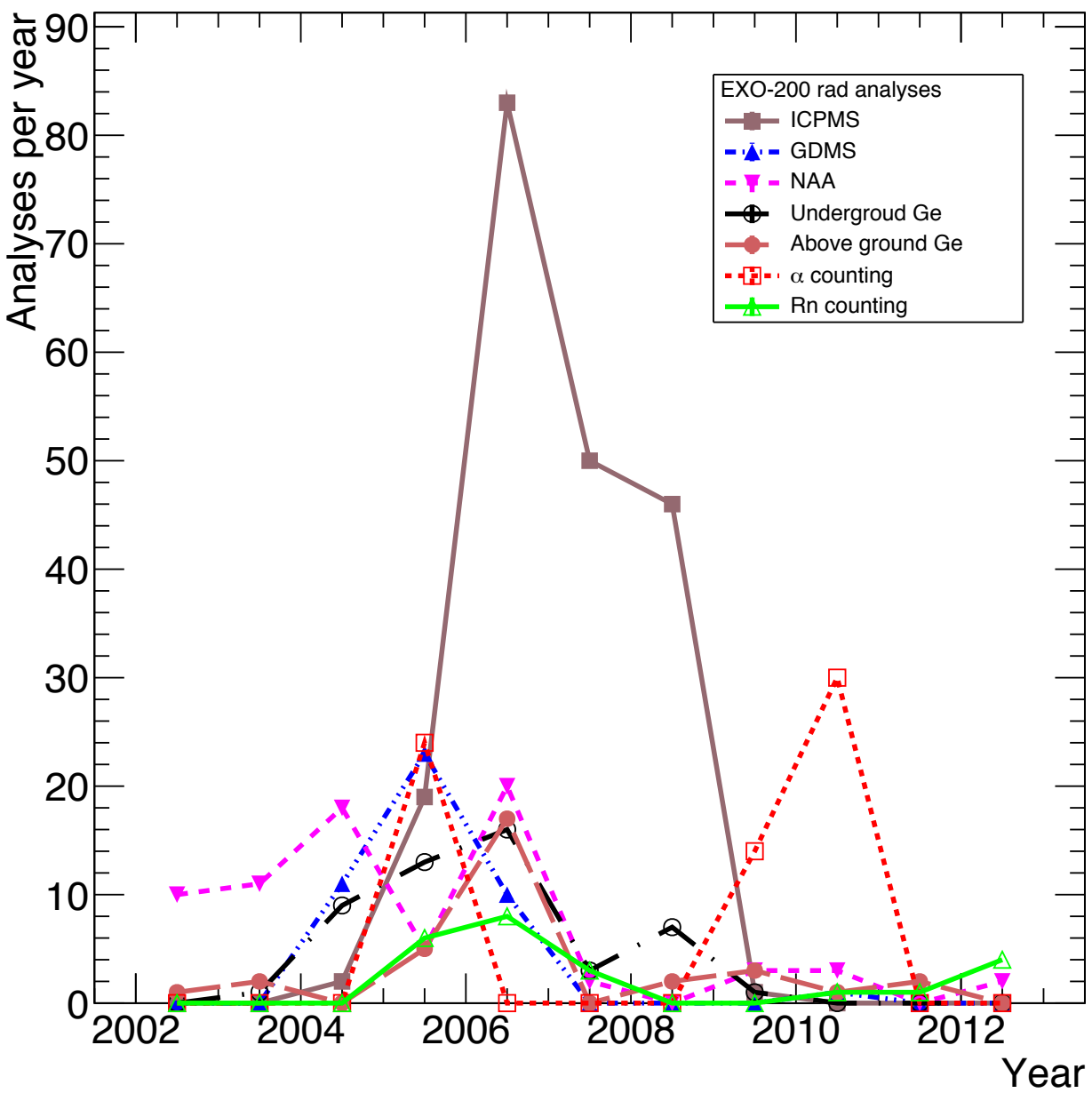

**Figure 9.1:** The magnitude of the materials analysis effort, as a function of time, during the preparation of EXO-200. Different lines represents various methods used.

## 9.12 Simulation and Data Analysis

Data simulation and analysis have important roles to play in the next phase of the R&D. Indeed simulations support and inform the experiment design by establishing its impact on the ultimate physics goal of $0\nu\beta\beta$ detection. Development of data analysis techniques is necessary to ensure that experiment information is extracted to its fullest, thus maximizing the experiment's sensitivity.

The model used for the estimate of the sensitivity will be kept up to date so that major "releases" of the detector design and geometry can be properly analyzed and compared. While some of this work can be carried-out with existing code (and the evolving detector geometry), we envisage several directions along which new software developments will occur. The fidelity of the modeling of the detector is improving because of the better understanding of various phenomena (e.g. electron diffusion in LXe) and of components (e.g. better models of the electronics). This is generally validated using both EXO-200 data and data from various nEXO-specific test beds (e.g. the prototype LXe TPC where signals from the charge collection tile prototypes are collected). nEXO is also in close collaboration with NEST [1] to improve the simulation of LXe ionization and scintillation yields at the nEXO relevant energies.

The sophistication of the data reconstruction and analysis is also increasing. For example, the choice and combinations of parameters used in the sensitivity analysis described in Section 3.3 can be tuned to optimize the discrimination of signal from the background and thus the sensitivity reach. Recent work in EXO-200 [2] has shown that the use of a boosted decision tree (BDT) to identify minute differences between electron and $\gamma$-ray events, resulted in a 15% increase of sensitivity.

Finally, the software infrastructure will be upgraded to use the SNiPER framework [3] that has been developed by our IHEP colleagues for use in the JUNO experiment [4, 5], adapted from the GAUDI framework [6, 7] (used by Fermi [8], Daya Bay [9, 10] and LZ [11]) . The framework provides the same file formats and data structures for simulation and real data, while using a structure that makes it easy for new algorithms to be implemented in data analysis by independent researchers without disturbing the baseline analysis chain for all users. Straightforward access to the data for both offline and online data-monitoring tools is also provided.

The ultimate goal of the R&D in the area of simulation and analysis is to prepare a full-scale mock data challenge incorporating a full simulation of the detector, data acquisition and analysis. The mock data challenge requires the necessary tools and environment to test the entire software machinery necessary to perform a sophisticated experimental data analysis. The simulation will produce realistic ionization and scintillation waveforms for signal and background events occurring in the detector, which are then sent to a simulated trigger and data acquisition followed by the reconstruction of events in software, eventually resulting in a half-life measurement. In running this mock data challenge, we will ensure that all pieces to the simulation and data analysis framework are working properly before the actual data arrives.